\documentclass[twocolumn]{aastex63}
\usepackage{multirow}
\usepackage{comment, longtable}
\usepackage{savesym}
\savesymbol{tablenum}
\usepackage{siunitx}
\restoresymbol{SIX}{tablenum}
\usepackage{amsmath}
\usepackage{graphicx}
\graphicspath{{images/}}
\usepackage{natbib}
\usepackage{graphicx}	
\usepackage{url}
\usepackage{siunitx}
\usepackage{amssymb}
\usepackage{color}
\usepackage{hyperref}
\usepackage{braket}
\usepackage{float}
\usepackage{physics}
\usepackage{relsize}
\usepackage{soul}
\usepackage{gensymb}
\renewcommand{\arraystretch}{1.5}
\usepackage{fancyvrb}
\usepackage{verbatim}
\usepackage{tablefootnote}
\usepackage{lipsum}
\usepackage [english]{babel}
\usepackage [autostyle, english = american]{csquotes}
\usepackage{longtable}
\usepackage{soul}

\usepackage{subfigure}

\newcommand{\kms}{km\,s$^{-1}$}

\renewcommand{\degree}{\ensuremath{^\circ}}

\newcommand{\hi}{\mbox{H\,{\sc i}}}

\newcommand{\orcid}[1]{\href{https://orcid.org/#1}{\includegraphics[width=8pt]{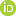}}}

\submitjournal{ApJ}
\shorttitle{MALS data release I} 
\shortauthors{Deka et al.}
\begin{document}

\title{The MeerKAT Absorption Line Survey (MALS) data release I: Stokes I image catalogs at 1 -- 1.4\,GHz} 

\correspondingauthor{P. P. Deka}
\email{parthad@iucaa.in}

\author{P. P. Deka \orcid{0000-0001-9174-1186}}
\affil{Inter-University Centre for Astronomy and Astrophysics, Post Bag 4, Ganeshkhind, Pune 411 007, India}

\author{N. Gupta \orcid{0000-0001-7547-4241}}  
\affil{Inter-University Centre for Astronomy and Astrophysics, Post Bag 4, Ganeshkhind, Pune 411 007, India}

\author{P. Jagannathan \orcid{0000-0002-5825-9635}}  
\affil{National Radio Astronomy Observatory, Socorro, NM 87801, USA}

\author{S. Sekhar}  
\affil{South African Radio Astronomy Observatory, 2 Fir Street, Black River Park, Observatory 7925, South Africa}
\affil{National Radio Astronomy Observatory, Socorro, NM 87801, USA}

\author{E. Momjian}  
\affil{National Radio Astronomy Observatory, Socorro, NM 87801, USA}

\author{S. Bhatnagar}  
\affil{National Radio Astronomy Observatory, Socorro, NM 87801, USA}

\author{J. Wagenveld}  
\affil{Max-Planck-Institut f\"ur Radioastronomie, Auf dem H\"ugel 69, D-53121 Bonn, Germany}

\author{H.-R. Kl\"ockner}  
\affil{Max-Planck-Institut f\"ur Radioastronomie, Auf dem H\"ugel 69, D-53121 Bonn, Germany}

\author{J. Jose}  
\affil{ThoughtWorks Technologies India Private Limited, Yerawada, Pune 411 006, India}

\author{S. A. Balashev}
\affil{Ioffe Institute, 26 Politeknicheskaya st., St. Petersburg, 194021, Russia}
\affil{HSE University, Saint Petersburg, Russia}

\author{F. Combes}  
\affil{ Observatoire de Paris, Coll\`ege de France, PSL University, Sorbonne University, CNRS, LERMA, Paris, France}

\author{M. Hilton}  
\affil{Wits Centre for Astrophysics, School of Physics, University of the Witwatersrand, Private Bag 3, 2050, Johannesburg, South Africa}
\affil{School of Mathematics, Statistics \& Computer Science, University of KwaZulu-Natal, Westville Campus, Durban, 4041, South Africa}

\author{D. Borgaonkar}  
\affil{Inter-University Centre for Astronomy and Astrophysics, Post Bag 4, Ganeshkhind, Pune 411 007, India}

\author{A. Chatterjee}  
\affil{ThoughtWorks Technologies India Private Limited, Yerawada, Pune 411 006, India}

\author[0000-0001-6527-6954]{K.~L.~Emig}  
\altaffiliation{Jansky Fellow of the National Radio Astronomy Observatory}
\affiliation{National Radio Astronomy Observatory, 520 Edgemont Road, Charlottesville, VA 22903, USA}

\author{A. N. Gaunekar}  
\affil{ThoughtWorks Technologies India Private Limited, Yerawada, Pune 411 006, India}

\author{G. I. G. J\'ozsa}  
\affil{Max-Planck-Institut f\"ur Radioastronomie, Auf dem H\"ugel 69, D-53121 Bonn, Germany}
\affil{Department of Physics and Electronics, Rhodes University, P.O. Box 94 Makhanda 6140, South Africa}

\author{D. Y. Klutse}  
\affil{School of Mathematics, Statistics \& Computer Science, University of KwaZulu-Natal, Westville Campus, Durban, 4041, South Africa}
\affil{Astrophysics Research Centre, University of KwaZulu-Natal, Durban 4041, South Africa}

\author{K. Knowles}  
\affil{Department of Physics and Electronics, Rhodes University, P.O. Box 94 Makhanda 6140, South Africa}

\author{J-.K. Krogager}  
\affil{Universit\'e Lyon1, ENS de Lyon, CNRS, Centre de Recherche Astrophysique de Lyon UMR5574, F-69230 Saint-Genis-Laval, France}

\author{A. Mohapatra} 
\affil{Inter-University Centre for Astronomy and Astrophysics, Post Bag 4, Ganeshkhind, Pune 411 007, India}

\author{K. Moodley}  
\affil{School of Mathematics, Statistics \& Computer Science, University of KwaZulu-Natal, Westville Campus, Durban, 4041, South Africa}
\affil{Astrophysics Research Centre, University of KwaZulu-Natal, Durban 4041, South Africa}

\author{S\'ebastien Muller}
\affiliation{Department of Space, Earth and Environment, Chalmers University of Technology, Onsala Space Observatory, SE-43992 Onsala, Sweden}

\author{P. Noterdaeme}  
\affil{Institut d'astrophysique de Paris, UMR 7095, CNRS-SU, 98bis bd Arago, 75014  Paris, France}

\author{P. Petitjean}  
\affil{Institut d'astrophysique de Paris, UMR 7095, CNRS-SU, 98bis bd Arago, 75014  Paris, France}

\author{P. Salas}  
\affil{Green Bank Observatory, 155 Observatory Road, Green Bank, WV 24915, USA}

\author{S. Sikhosana}  
\affil{School of Mathematics, Statistics \& Computer Science, University of KwaZulu-Natal, Westville Campus, Durban, 4041, South Africa}
\affil{Astrophysics Research Centre, University of KwaZulu-Natal, Durban 4041, South Africa}

\begin{abstract}

The MeerKAT Absorption Line Survey (MALS) has observed 391 telescope pointings at L-band (900 -- 1670\,MHz) at $\delta\lesssim$ $+20\degree$.  We present radio continuum images and a catalog of 495,325 (240,321) radio sources detected at a signal-to-noise ratio (SNR)\,$>$5 over an area of 2289\,deg$^2$ (1132\,deg$^2$) at 1006\,MHz (1381\,MHz). Every MALS pointing contains a central bright radio source ($S_{1\,\mathrm{GHz}} \gtrsim 0.2$~Jy). The median spatial resolution is $12^{\prime\prime}$ ($8^{\prime\prime}$). The median rms noise away from the pointing center is 25\,$\mu$Jy\,beam$^{-1}$ (22\,$\mu$Jy\,beam$^{-1}$) and is within $\sim15\%$ of the achievable theoretical sensitivity.  The flux density scale ratio and astrometric accuracy deduced from multiply observed sources in MALS are less than  1\% (8\% scatter) and $1^{\prime\prime}$, respectively. Through comparisons with NVSS and FIRST at 1.4\,GHz, we establish the catalog's accuracy in the flux density scale and astrometry to be better than 6\% (15\% scatter) and 0\farcs8, respectively.  The median flux density offset is higher (9\%) for an alternate beam model based on holographic measurements. The MALS radio source counts at 1.4\,GHz are in agreement with literature.  We estimate spectral indices ($\alpha$) of a subset of 125,621 sources (SNR$>$8), confirm the flattening of spectral indices with decreasing flux density and identify 140 ultra steep-spectrum  ($\alpha<-1.3$) sources as prospective high-$z$ radio galaxies ($z>2$).  We have identified 1308 variable and 122 transient radio sources comprising primarily of AGN that demonstrate long-term (26 years) variability in their observed flux densities. The MALS catalogs and images are publicly available at \href{https://mals.iucaa.in}{https://mals.iucaa.in}.
\end{abstract}


\section{Introduction} 
\label{sec:intro}  

Over the years, there have been several radio continuum surveys at centimeter wavelengths to study both the evolution of AGN and star formation (SF) activity of the Universe, independent of biases due to dust obscuration. 
The extragalactic non-thermal emission at $\sim$1\,GHz arises from {\it (i)} magnetized plasma i.e., radio core, jets and lobes associated with active galactic nuclei \citep[AGN;][]{Padovani17}, and {\it (ii)} relativistic electrons associated with supernova remnants in star-forming galaxies \citep[SFGs;][]{Condon92}. 

The radio emission associated with SF activity is generally fainter and dominates the radio source population only below continuum flux densities of 100\,$\mu$Jy \citep[e.g.,][]{Simpson06, Seymour08, Smolcic17b, Algera_2020, Fangxia21}.  Consequently, radio source population studies have adopted a tiered approach in which  deep small-area  surveys focus on SFGs or radio-quiet quasars, and large area, shallow surveys encompass detections of powerful radio-loud AGN and nearby SFGs. 
The former category includes deepest radio surveys targeting a few deg$^2$ of the sky with exquisite panchromatic coverage and reaching $\mu$Jy level sensitivities \citep[e.g.,][]{Garn08, Smolcic17, Owen18, Mauch20, Heywood22}.

The latter category historically comprised of practically monochromatic surveys covering a large fraction of the entire visible sky, for example, the NRAO Very Large Array (VLA) Sky Survey \citep[NVSS;][]{Condon98} and the Faint Images of the Radio Sky at Twenty Centimeters \citep[FIRST;][]{Becker95}, at 1.4\,GHz. The NVSS observed the sky at declinations north of $\delta>-40\degree$ with a spatial resolution of $45^{\prime\prime}$ and sensitivity of $\sim$0.45\,mJy\,beam$^{-1}$. FIRST survey covered over 10,000\,deg$^2$ of the North and South Galactic Caps with a resolution and sensitivity of $5^{\prime\prime}$ and $\sim$0.15\,mJy\,beam$^{-1}$, respectively, albeit with a lower surface brightness sensitivity than NVSS.  These surveys are complemented by the Sydney University Molonglo Sky Survey \citep[SUMSS;][]{Mauch03} at 843\,MHz (resolution $\sim 45^{\prime\prime}/\sin{\delta}$; sensitivity $\sim$1\,mJy\,beam$^{-1}$) surveying the southern sky at $\delta < -30\degree$ and avoiding the Galactic plane $|b|<10\degree$, the Westerbork Northern Sky Survey \citep[WENSS;][]{Rengelink97} at 325\,MHz surveying the entire sky north of $\delta > 30\degree$ at a $5\sigma$ rms sensitivity of 18\,mJy and resolution of $54^{\prime\prime} \rm cosec\,\delta$, and the first Alternative Data Release (ADR1) of the TIFR GMRT Sky Survey  \citep[TGSS;][]{Intema17} surveying the northern sky at $\delta > -53\degree$ with a resolution of $\sim 25^{\prime\prime}$ and median rms noise of $\sim$3.5\,mJy\,beam$^{-1}$ at 150\,MHz. 

\begin{figure*}[ht]
    \centering
    \includegraphics[width=0.95\linewidth]{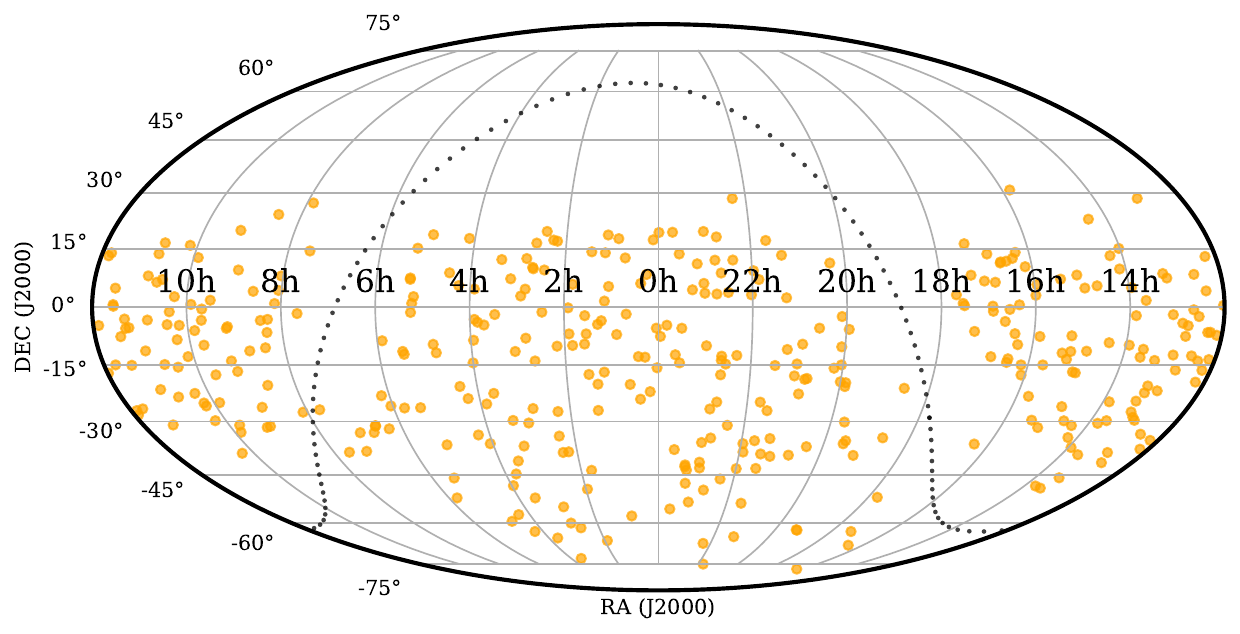}
    \caption{Sky distribution of the 391 MALS pointings observed in L-band shown in Mollweide projection in equatorial coordinates (J2000). The dotted line marks the Galactic plane.  
    }
    \label{fig:skydist}
\end{figure*}

The spectral energy distribution (SEDs) of radio sources derived from combining a large number of surveys at multiple frequencies is a fundamental tool to understand physical processes responsible for the radio emission \citep[e.g.,][]{Rybicki79, Prandoni06, Gasperin18}. For SFGs, SEDs involving measurements at high frequencies ($\nu>$2\,GHz) can be used to disentangle contributions to the radio emission due to free-free emission from H~{\sc ii} regions and synchrotron emission from cosmic ray electrons \citep[e.g.,][]{Condon92, Niklas97, Murphy11, Tabatabaei17, Linden20, Algera21, Stein23}. For radio-loud AGN, the SED and its possible spatial variation can be used to understand the properties of ionized gas, estimate the age of radio plasma and identify young radio sources (age $<10^5$\,yrs) that are still embedded within the host galaxy's interstellar medium \citep[ISM; e.g.,][]{Baum90, Bicknell97, deVries97, Murgia99, Kameno00, Snellen00, Saikia03, DeVries09, Keim19, Ricci19, Odea21}.  
In addition, slow transients and variability of radio emission detected at timescales ranging from seconds, hours, days and years, may be used to study a wide range of phenomena associated with stellar systems, supernovae, gamma-ray bursts (GRBs) and AGN \citep[][]{Cordes04}.

Modern radio telescopes are capable of observing with large instantaneous bandwidths, hence efficiently delivering large radio source catalogs covering a wide frequency range and variability timescales required to address the above-mentioned science cases.  Examples of such ongoing surveys are: the Very Large Array Sky Survey \citep[VLASS;][]{Lacy20, Gordon21} at 2-4\,GHz, the Rapid ASKAP Continuum Survey \citep[RACS;][]{Mcconnell20, Hale21} at 887.5\,MHz, the Low Frequency Array (LOFAR) Two-metre Sky Survey \citep[LoTSS;][]{Shimwell17, Shimwell19, Shimwell22} at 120-168\,MHz, the Galactic and Extragalactic All-sky Murchison Widefield Array survey \citep[GLEAM;][]{Hurley-Walker2017} at 72 - 231\,MHz and its extension GLEAM-X with improved sensitivity and resolution \citep{Hurley-Walker22}, and the Evolutionary Map of the Universe  survey \citep[EMU;][]{Norris21} at 944\,MHz. The spatial resolution of these surveys are $2\farcs5$ (VLASS), $15^{\prime\prime} - 25^{\prime\prime}$ (RACS), $6^{\prime\prime}$ (LoTSS), $100^{\prime\prime}$ (GLEAM; $45^{\prime\prime}$ for GLEAM-X) and $11^{\prime\prime} - 18^{\prime\prime}$ (EMU).  The Stokes-$I$ sensitivities are 69\,$\mu$Jy\,beam$^{-1}$, 0.25-0.3\,mJy\,beam$^{-1}$, 100\,$\mu$Jy\,beam$^{-1}$, 6-10\,mJy\,beam$^{-1}$ (1.27 mJy\,beam$^{-1}$ for GLEAM-X) and 25-30\,$\mu$Jy\,beam$^{-1}$, respectively.  These large area surveys are being complemented by deep small-area surveys such as MIGHTEE \citep[MeerKAT International Gigahertz Tiered Extragalactic Exploration;][]{Jarvis17} which is imaging 20\,deg$^2$ of the sky down-to thermal noise levels of $<$2\,$\mu$Jy\,beam$^{-1}$, although rms in the central regions is limited by confusion at $\sim8^{\prime\prime}$ resolution \citep[][]{Heywood22}. The combination of these surveys will enable large population studies of SFGs and AGN, and the detection of extreme and rare objects.

The MeerKAT Absorption Line Survey (MALS) is observing $\sim$500 pointings, each centered at a radio source brighter than $\sim$200\,mJy at 1\,GHz, with the L-band (900–1670\,MHz) and UHF-band (580–1015\,MHz) of the MeerKAT telescope \citep[][]{Gupta17mals}.  It will deliver a radio continuum catalog of about a {\it million radio sources} from the sky coverage of $\sim1000$\,deg$^2$ at a sensitivity of $\sim$20$\mu$Jy\,beam$^{-1}$. 
The MeerKAT telescope consists of 64 dishes of 13.5\,m diameter located at the Square Kilometre Array (SKA) site in Karoo, South Africa \citep[][]{Jonas16}.
For reference, MeerKAT's field of view i.e., full width at half maximum (FWHM) of the primary beam and spatial resolution\footnote{Based on {\tt robust = -1.3} weighting of visibilities as implemented in the Astronomical Image Processing System \citep[AIPS;][]{Greisen03} used in \citep[][]{Mauch20}.} at $\sim$1\,GHz are $88\arcmin$ and $\sim10^{\prime\prime}$, respectively \citep[][]{Mauch20}. 

While the radio continuum component of MALS will enable a wide a range of radio continuum science associated with SFGs, AGN and clusters of galaxies \citep[see][for details]{Gupta17mals}, its uniqueness lies in that for each pointing the survey will also produce spectral line cubes at a spectral resolution of $\sim$6.1\,\kms.  Consequently, for each radio source brighter than 1\,mJy it will also be possible to search for cold atomic and molecular gas associated with AGN via \hi\ 21-cm and OH 18-cm absorption lines at $0<z<1.4$ and $0<z<1.9$, respectively \citep[e.g.,][]{Gupta21, Combes21, Srianand22}. 
MALS is also enabling the most sensitive and comprehensive search for radio recombination lines nominally arising from hydrogen in ionized gas at $z \lesssim 5$ \citep{Emig23}. These observations enable the direct exploration of the relationship between cold gas, ionized gas, and AGN/SF activity over the redshift range ($0<z<2$) in which maximum evolution in these quantities takes place \citep[e.g.,][]{Hopkins06, Silverman09, Fanidakis12, Heckman14}.

In this paper, we describe the first release of MALS Stokes-$I$ continuum data products for the 391 pointings observed at L-band during the first phase of the survey (see Fig.~\ref{fig:skydist} for the sky coverage). The subsequent MALS observing phases will largely observe in the UHF-band. 
We focus on two spectral ranges in L-band: 976.4 - 1036.5\,MHz and 1350.9 - 1411.0\,MHz, hereafter referred to as SPW2 and SPW9, respectively. 
We utilize the properties of SPW9 images and their comparison with the NVSS catalog, also at $\sim$1.4\,GHz, to demonstrate the quality of the catalog.
The processes presented here lay out the foundation for subsequent L- and UHF-band data releases corresponding to narrowband i.e., SPW specific and wideband continuum products.  For value addition to the community, the SPW2 data products are included in this first data release.
The catalog and initial results from wideband imaging of 10 MALS pointings are presented in \cite{Wagenveld23}. 

This paper is structured as follows. 
In Section~\ref{sec:obs}, we present details of observations and data analysis for the 391 pointings that are part of this data release.  In Section~\ref{sec:cat}, we describe the noise properties of the images, and analyse artifacts. The cataloging procedure, which will also be used for future continuum data releases, is also presented here.
In Section~\ref{sec:accur}, we investigate the accuracy of the astrometry and the flux density scale.  In this context, we make a detailed comparison with NVSS, and elaborate on the primary beam correction.
In Section~\ref{sec:disc}, we use the MALS catalog to determine radio source counts and discuss the completeness of the catalog. Further, we demonstrate the usage of the catalog to potential users by investigating the long-term variability at 1.4\,GHz and spectral indices of the radio source population over 0.3 - 1000\,mJy. 
The results and future prospects are summarized in Section~\ref{sec:summ}. 

Throughout this paper, we use the $\Lambda$CDM cosmology with $\Omega_m$=0.315, $\Omega_\Lambda$=0.685 and H$_{\rm 0}$=67.4\,\kms\,Mpc$^{-1}$ \citep[][]{Planck20}. All the positions are provided in J2000 equatorial coordinates.  The spectral index, $\alpha$, is defined as $S_\nu \propto \nu^\alpha$, where $S_\nu$ is flux density at frequency $\nu$.

\section{Observations, calibration, and imaging}  
\label{sec:obs}   

Each MALS pointing is centered at a radio source brighter than 200\,mJy at $\sim$1\,GHz in NVSS or SUMSS. We have carried out a large spectroscopic campaign using the Nordic Optical Telescope (NOT) and the Southern African Large Telescope (SALT) to measure the redshifts and confirm the nature of 303 AGN candidates identified on the basis of mid-infrared (MIR) colors.  The NOT component of the survey is presented in \citet[][]{Krogager18}.  \citet[][]{Gupta22salt} present the details of the SALT campaign and the selection process of the pool of 650 radio sources based on which approximately 500 pointings are anticipated to be observed at both L- and UHF-bands using $\sim$1655\,hr of MeerKAT time.

The sky coverage of 391 pointings observed at L-band during the first phase of the survey from 2020, April, 01 to 2021, January, 18, is shown in Fig.~\ref{fig:skydist} (see Appendix~\ref{sec:listpoint} for the list).  
For these observations, the 856\,MHz bandwidth of L-band centered at 1283.9869\,MHz was split into 32,768 frequency channels.  This mode of the SKA Reconfigurable Application Board (SKARAB) correlator corresponds to a channel spacing of 26.123\,kHz, which is 6.1\,\kms\ at the center of the band.  
The correlator dump time was 8\,s. For dual, linearly polarized L-band feeds with orthogonal
polarizations labeled X and Y, the data were acquired for all four polarization products: XX, XY, YX, and YY.  
On average 59 antennas of MeerKAT-64 array participated in these observations.

Typically, a single L-band observing run included three targets.  The total on-source time of 56\,minutes on each target was split into three scans of 1120\,s duration at different hour angles to improve the uv-coverage.  Each scan on a target source was bracketed by a 60\,s long scan on a complex gain calibrator.  We also observed 3C\,286, 3C\,138, PKS\,1939–638 and / or PKS\,0408–658 for 5-10\,minutes at the start, middle and end of an observing run for flux density scale, delay, and bandpass calibrations.  Thus, the total duration of an L-band observing run was about 3.5\,hrs, which resulted in a measurement set of $\sim$5\,TB.  There are five exceptions to this observing scheme. Four MALS pointings were observed twice i.e., have a total on-source time of 112\,minutes (see Appendix~\ref{sec:listpoint} for details), and  the time on J183339.98$-$210339.9 (PKS\,1830-211) is 90\,mins.
%

The MALS data were processed using the Automated Radio Telescope Imaging Pipeline (ARTIP) based on NRAO's Common Astronomy Software Applications (CASA) package \citep[][]{Casa22}.  The details are provided in \citet[][]{Gupta21}.  In short, since here we are interested in Stokes-$I$ imaging, for processing we generated a measurement set consisting of only XX and YY polarization products. We also dropped channels at the extreme edge of the bandpass resulting in a measurement set with 30,720 frequency channels.  An initial RFI mask described in \citet[][]{Gupta21} was applied to exclude the frequency channels affected by persistent strong RFI.  After this, wideband model visibilities for the flux density calibrators were predicted, and an initial calibration on a subset of frequency channels (19,000 - 20,000) was performed to identify any malfunctioning antennas and baselines.  For 3C\,286 and 3C\,138, we used models based on {\tt Perley-Butler 2017} \citep[][]{Perley17}.  For PKS\,1939–638, the model based on {\tt Stevens-Reynolds 2016}  was used \citep{Partridge16} whereas for PKS\,0408–658 a model with $S_{\rm 1284\,MHz}= 17.066$\,Jy  and $\alpha=-1.179$ was used. Next, the pipeline proceeded to calibrate the entire band, and performed RFI flagging using {\tt tfcrop} and {\tt rflag} in CASA.  The delay, bandpass, and temporal complex gain calibration solutions were applied to the target source visibilities.  

\begin{table}
\begin{center}
\caption{Details of L-band SPWs.}
\begin{tabular}{lcc}
\hline

SPW Id. & Freq. range    & Image freq.  \\
        &    (MHz)       &     (MHz)    \\ 
 (1)    &     (2)        &     (3)      \\           

\hline
SPW0        &     869.4 -  929.5   &       904.1      \\  
SPW1        &     922.9 -  983.0   &       952.9      \\     
{\bf SPW2}        &     {\bf 976.4 - 1036.5}   &      {\bf 1006.0 }     \\     
SPW3        &    1029.9 - 1090.0   &      1060.3      \\     
SPW4        &   1083.4 -  1143.5   &      1109.7      \\     
SPW5        &   1136.9 -  1197.0   &      1191.7      \\     
SPW6        &   1190.4 -  1250.5   &      1220.8      \\     
SPW7        &   1243.9 -  1304.0   &      1273.9      \\     
SPW8        &   1297.4 -  1357.5   &      1331.6      \\     
{\bf SPW9}        &   {\bf 1350.9 -  1411.0}   &      {\bf 1380.9 }     \\     
SPW10       &   1404.4 -  1464.5   &      1434.4      \\     
SPW11       &   1457.9 -  1518.0   &      1487.9      \\     
SPW12       &   1511.4 -  1571.5   &      1541.4      \\     
SPW13       &   1564.9 -  1625.0   &      1614.5      \\     
SPW14       &   1618.4 -  1671.9   &      1643.0      \\     
\hline 
\end{tabular}
\\
\end{center}
\tablecomments{
Column 1: Spectral window Id.  The SPWs of interest in this paper are highlighted. Column 2: frequency range covered by the corresponding visibility measurement set.  Column 3: reference frequency of the continuum image.
}
\label{tab:spwdef}
\end{table}

After calibration, the spectral line and wideband continuum imaging processes diverge.  For spectral line or cube imaging, we split the continuous band of 30,720 frequency channels into 15 spectral windows (SPWs) labeled as SPW0 to SPW14 (see Table~\ref{tab:spwdef}).  To ensure that no spectral features at the edge of any SPW are lost, the adjacent SPWs have an overlap of $\sim$7\,MHz (256 channels).  
The measurement sets for these SPWs are then processed for continuum imaging with self-calibration and cube imaging.  For each SPW, a continuum data set is generated by flagging RFI-affected frequency ranges and averaging data in frequency by 32 channels to reduce the data volume.  This is then imaged using {\tt robust=0} weighting and {\tt w-projection} algorithm with 128 planes as the gridding algorithm in combination with {\tt Multi-scale Multi-term Multi-frequency synthesis} ({\tt MTMFS}) for deconvolution, with nterms=1 and four pixel scales (0, 2, 3 and 5) to model the extended emission \citep[][]{Rau11}. 
Imaging masks were appropriately adjusted using the {\tt Python Blob Detection and Source Finder} \citep[{\tt PyBDSF\footnote{PyBDSF version 1.10.1};}][]{mohan2015} between major cycles during imaging and self-calibration runs.  
This ensured that at any stage the artifacts in the vicinity of bright sources are excluded from the CLEANing process and the source model.   The relevant details of how this is achieved through {\tt PyBDSF} are presented in Section~\ref{sec:cat}.
Here, we started with high source detection thresholds and gradually reduced these as the imaging progresses through major cycles and self-calibration runs.
Overall, the pipeline performed three rounds of phase-only and one round of amplitude and phase self-calibration.  
The final 6k$\times$6k continuum images with a pixel size of $2^{\prime\prime}$ have a span of $3\fdg3$ for all SPWs, and have been CLEANed down to three times the local rms noise based on a {\tt PyBDSF} mask.

For cube imaging of an SPW, the self-calibration solutions obtained from the continuum imaging are applied to the line dataset and continuum subtraction is performed using the model, i.e., CLEAN components obtained from the last round of self-calibration. The continuum subtracted visibilities are then inverted to obtain spectral line cubes, which may then be deconvolved for line emission \citep[for example, see][]{Boettcher21, Maina22}. 
The wideband continuum imaging utilizing full L-band bandwidth would require {\tt w-projection} algorithm in combination with {\tt MTMFS} for deconvolution, but with nterms=2 \citep[see][]{Wagenveld23}.  

In this paper, we focus on continuum images at 1006.0 and 1380.9\,MHz from the spectral line processing of SPW2 and SPW9.  
For 60.2\,MHz bandwidth, 59 antennas and 56\,mins of integration, the theoretical rms noise for {\tt robust = 0} weighting of visibilities are 22\,$\mu$Jy\,beam$^{-1}$ (SPW2)  and 19\,$\mu$Jy\,beam$^{-1}$ (SPW9).
We use SPW9 images, which are close to the observing frequency of NVSS, to verify the astrometry and flux density scales of MALS. 
For the latter, we make the reasonable assumption that the flux variability due to intrinsic source properties or interstellar scintillations is not a significant factor at 1.4\,GHz (see also Section~\ref{sec:variability}).
The SPW2 images at the low-frequency (1006.0 MHz) end of L-band are used to measure spectral indices of the sources.  Note that we prefer SPW2 over SPW0 and SPW1 for relatively lower RFI and avoiding additional complications due to L-band roll-off.  As previously mentioned, the processes presented in this paper lay the foundation for subsequent L- and UHF-band data releases corresponding to narrowband, i.e., SPW specific data products.  

\section{Image analysis and catalogs}  
\label{sec:cat}   

\subsection{Noise variations in raw images}  
\label{sec:noise}   

\begin{table}
\caption{{\tt PyBDSF} parameters for catalog generation.}
\begin{tabular}{@{}ll}
\hline
{\tt adaptive\_rms\_box} & {\tt True} \\
{\tt adaptive\_thresh} & 100.0 \\
{\tt rms\_box} & (150, 30)\\
{\tt rms\_box\_bright} & {\tt None} \\
{\tt thresh\_isl} & 3.0 \\
{\tt thresh\_pix} & 5.0 \\
{\tt thresh} & {\tt None} \\
{\tt atrous\_do} & {\tt True}\\
{\tt atrous\_orig\_isl} & {\tt True}\\
{\tt atrous\_jmax} & 3\\
{\tt group\_by\_isl} & {\tt True}\\
\hline
\end{tabular}
\label{tab:bdsf_params}
\end{table}

We used {\tt PyBDSF} to generate radio source catalogs from SPW2 and SPW9 images.  
In general, the brightest sources in radio images are often associated with artifacts and raise the rms noise in the vicinity above the theoretically expected value. For reliable detection of sources, {\tt PyBDSF} tackles such noise variations by generating rms maps using a  sliding box of adjustable dimensions, i.e., smaller near brighter sources and vice versa.  The intermediate rms values in the map are then obtained by interpolating between the measurements. 
We performed source finding on `raw' i.e., primary beam uncorrected images obtained from {\tt ARTIP} and use noise properties derived from the rms maps to quantify the impact of bright sources in the field of view. Note that the noise properties of primary beam corrected images and the Stokes-$I$ catalogs are presented in Section~\ref{sec:icat}.
Additionally, we also identify 50 representative pointings -- discussed at the end of this section -- from the sample.  We subject SPW9 images of this representative subset to visual inspection to closely track the possible sources of errors and optimize the {\tt PyBDSF} input parameters.

\begin{figure*}[ht!]
     \begin{center}
        \subfigure{%
            \label{fig:first}
            \includegraphics[width=0.45\textwidth]{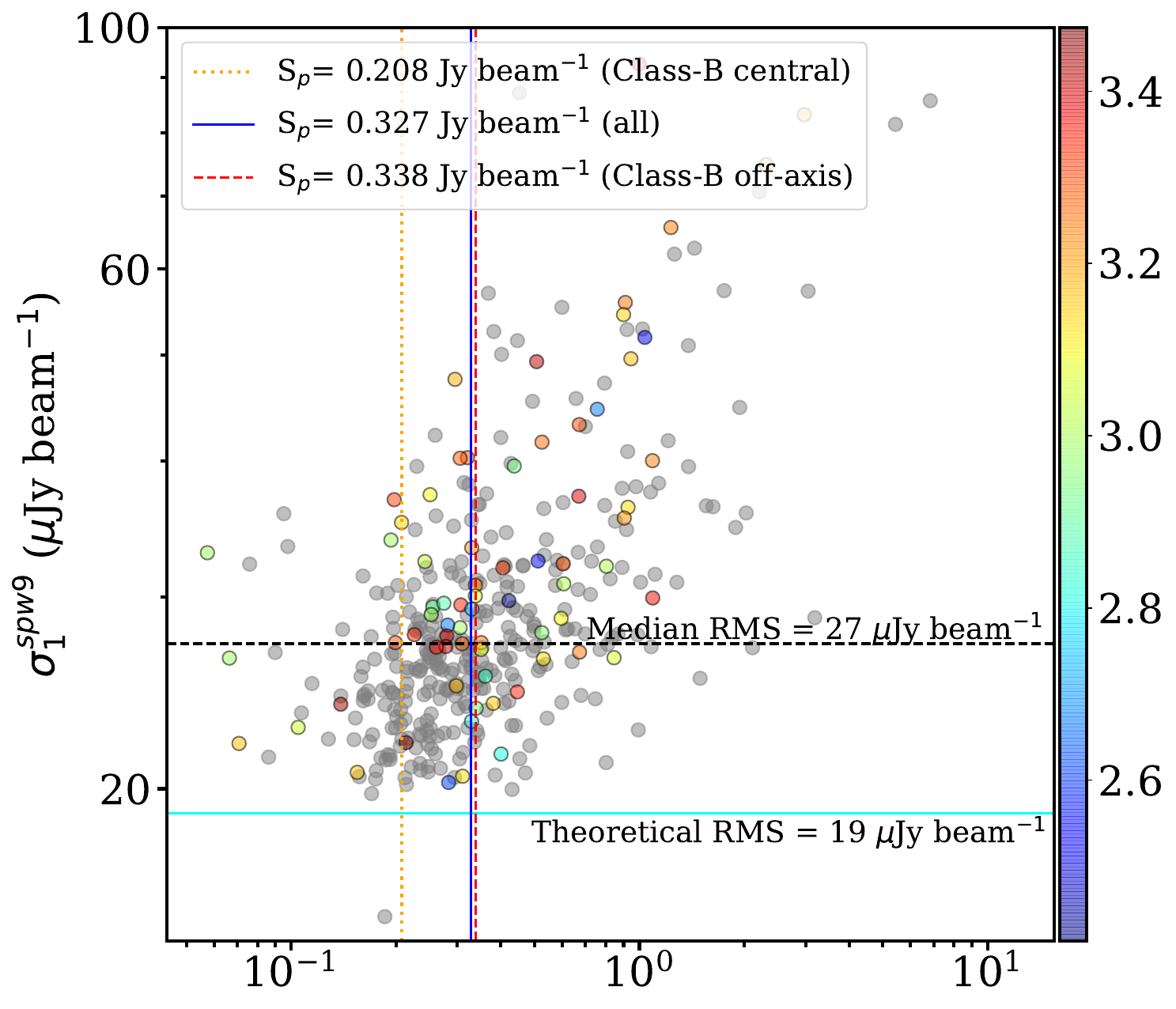}
        }%
        \subfigure{%
           \label{fig:second}
           \includegraphics[width=0.465\textwidth]{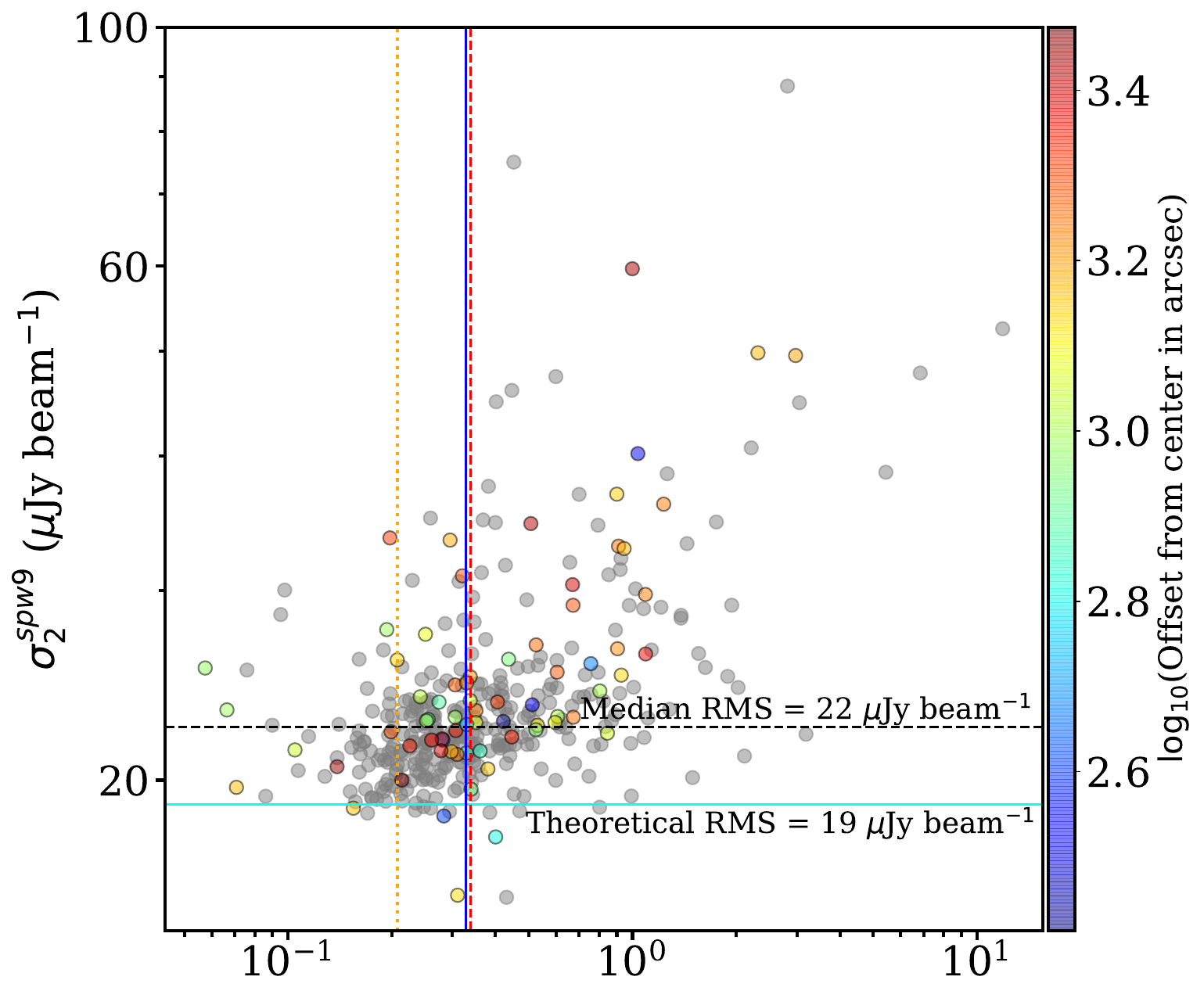}
        }\\ 
        \subfigure{%
            \label{fig:third}
            \includegraphics[width=0.45\textwidth]{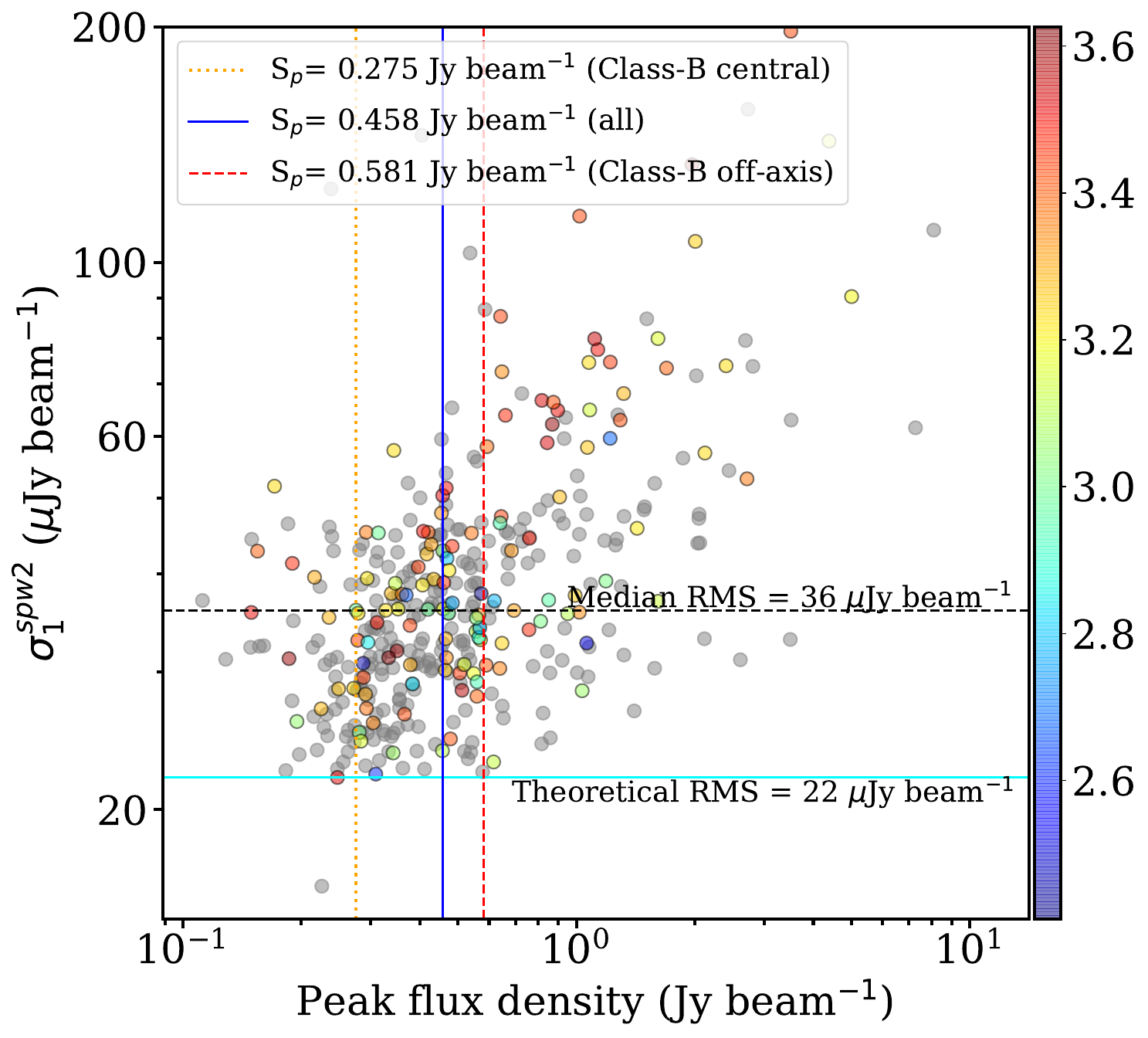}
        }
        \subfigure{%
            \label{fig:fourth}
            \includegraphics[width=0.465\textwidth]{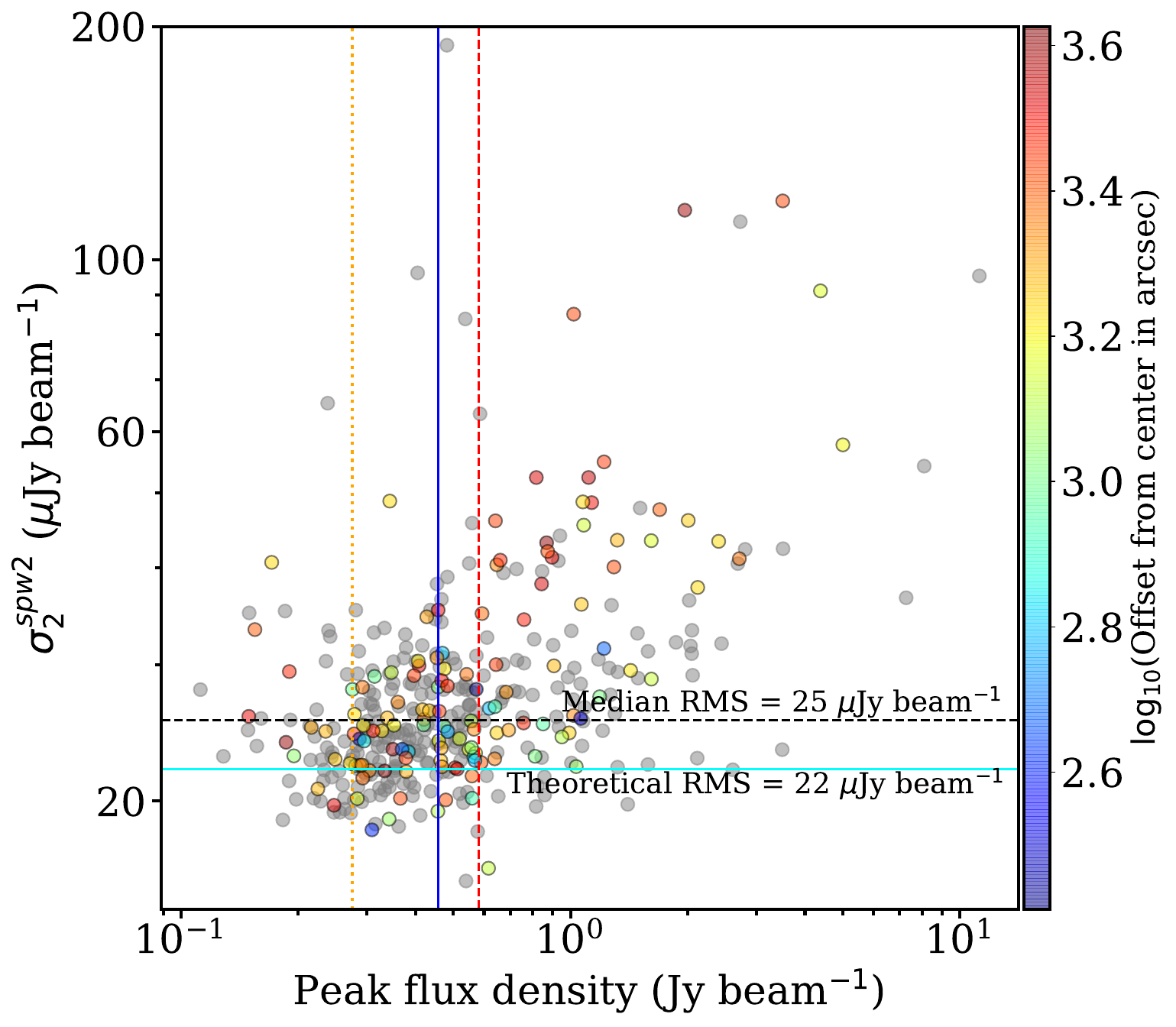}
        }%
    \end{center}
    \caption{%
        The rms measured at one and two times the primary beam FWHM i.e., $\sigma_1$ (left panels) and $\sigma_2$ (right panels), respectively,  as a function of peak flux density ($S_p$) of the brightest source in primary beam uncorrected SPW9 ({\it top panels}) and SPW2 ({\it bottom panels}) images.  
        In the cases an off-axis source is brighter than the central radio source (i.e., Class-B pointings), the points have been color coded with respect to the distance of the source from the pointing center. 
        In each panel, the three vertical lines from left to right mark median flux densities for {\it (i)} central source in Class-B, {\it (ii)} central source in all (391) and {\it (iii)} off-axis source in Class-B pointings.
        Horizontal dashed lines mark theoretical and observed rms noise values. 
        For clarity, in {\it top-} and {\it bottom-left} panels, 3 and 4 points with $\sigma_1$ greater than 100 and 200\,$\mu$Jy\,beam$^{-1}$ have been omitted, respectively.  
     }%
   \label{fig:sigma1sigma2}
\end{figure*}

The key {\tt PyBDSF} input parameters are summarized in Table~\ref{tab:bdsf_params}. The remaining input parameters were set to their default values and the details can be found in the {\tt PyBDSF} documentation \citep[][]{mohan2015}.
We set {\tt adaptive\_rms\_box} = {\tt True} and {\tt adaptive\_thresh} = {\tt 100.0} to allow {\tt PyBDSF} to estimate rms and mean using a smaller box close to bright sources detected at SNR$>100$.
This is based on the visual inspection of SPW9 images from 50 representative pointings which revealed significant artifacts around sources brighter than $>100\sigma$, where  $\sigma$ represents the local rms noise. 
We adopted default values of {\it (i)} {\tt thresh\_isl} = 3$\sigma$ as the threshold to identify the boundary of the island for fitting the radio emission, and {\it (ii)} {\tt thresh\_pix} = 5$\sigma$ as the threshold to detect sources.  
Although, we set the source detection threshold i.e., {\tt thresh\_pix} to 5$\sigma$, the choice of {\tt thresh} = {\tt None}, implied that in general a variable threshold for {\tt thresh\_pix} based on the False Detection Rate algorithm is used \citep[][]{Hopkins02}.  The value of 5$\sigma$ for {\tt thresh\_pix} is used  only when the number of false pixels is less than 10\% of the estimated number of true pixels.

We note that {\tt PyBDSF} runs with default {\tt rms\_box} parameter resulted in box sizes as large as $\sim$900 pixels in a few cases and $\sim$600 pixels in the remaining. Such large boxes over-smoothed the internally calculated rms maps and resulted in detections of imaging artefacts as real sources. Therefore, we experimented with a range of tuples corresponding to {\tt rms\_box} and {\tt rms\_box\_bright} to define the box and step sizes to be used in general and close to bright sources. The tuple  {\tt rms\_box} = {\tt (150, 30)} was found to minimize the number of such artefacts getting fitted.  This is also the value adopted by RACS to obtain optimal results \citep[][]{Hale21}. We found that setting {\tt rms\_box} to significantly smaller values than this results in omission of fainter sources in the vicinity bright sources.
Also, we set {\tt rms\_box\_bright} = {\tt None}.  This implies that we rely on the internal machinery of {\tt PyBDSF} to determine the suitable box and step sizes in the vicinity of bright sources.  We verified that the resultant box and step sizes were significantly smaller than {\tt rms\_box}, and the approach performed better compared to when {\tt rms\_box\_bright} was fixed to any specific value smaller than {\tt rms\_box}.

MeerKAT has excellent surface brightness sensitivity to detect large scale extended radio emission.  Therefore, for better modelling of extended emission we set {\tt atrous\_do} = {\tt True} to turn on the wavelet decomposition module with a maximum of three wavelet scales ({\tt atrous\_jmax} = 3).  
Note that we set {\tt atrous\_orig\_isl} = {\tt True}, to ensure that wavelet Gaussians lie within the islands determined using the original image i.e., prior to any wavelet decomposition. 
Finally, we also set {\tt group\_by\_isl} = {\tt True} to allow {\tt PyBDSF} to group all Gaussians within an island into a single source.  
%

\begin{figure}
    \centering
    \includegraphics[width=0.9\linewidth]{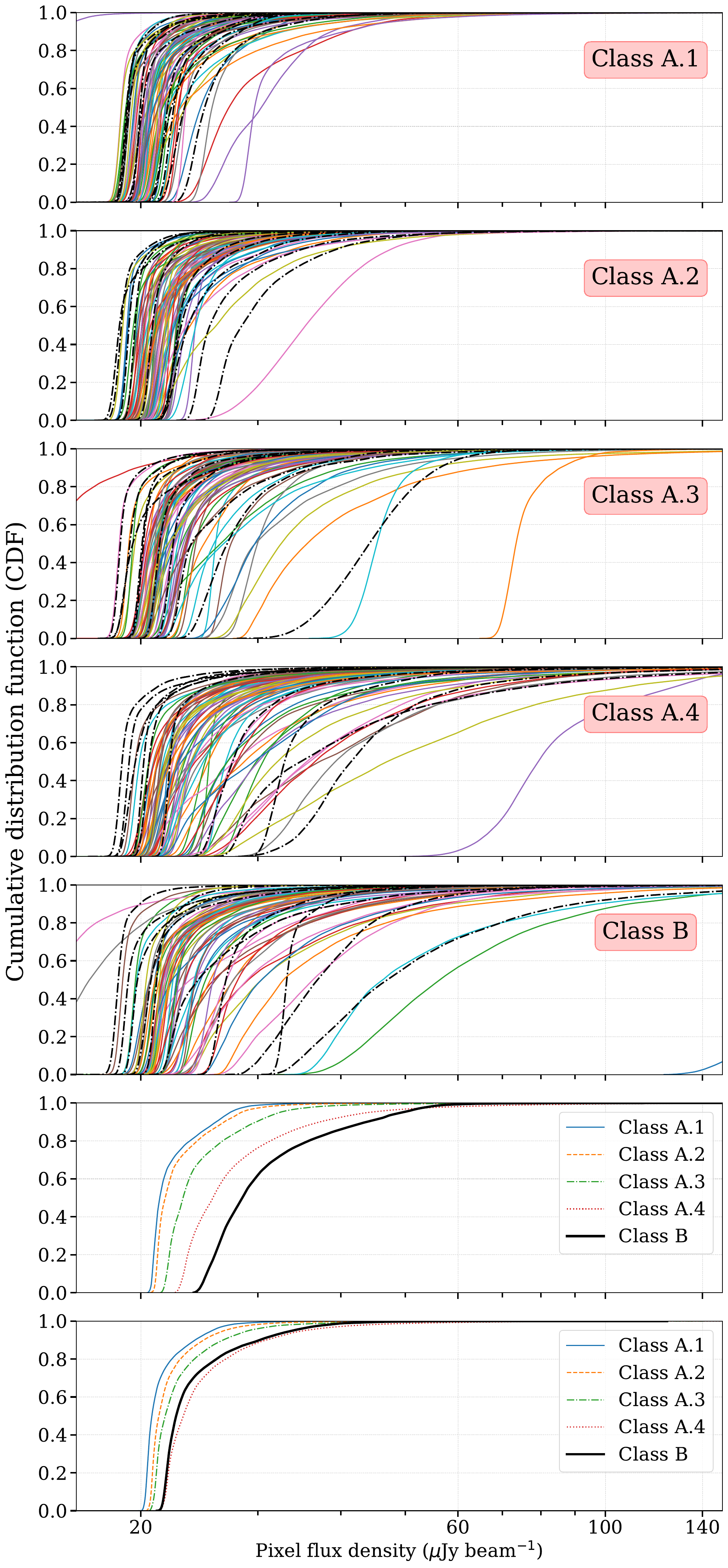}
    \caption{
    Cumulative distribution function of pixels in the rms maps of 391 pointings, arranged as per Class-A.1 to Class-A.4, and Class-B ({\it top} five panels).   The profiles from mean- and median-stacked rms images are shown in the {\it bottom} two panels. The dashed-dotted lines in {\it top} five panels correspond to `representative' pointings selected (see at the end of Section~\ref{sec:noise}).   The curves with lowest rms for Class-A.1, A.3 and B correspond to four pointings with double the integration time (see text for details).
    }
    \label{fig:rms_dist}
\end{figure}

We define two rms measurements using the rms maps from `raw' i.e., primary beam uncorrected continuum images.  We measure $\sigma_1$ and $\sigma_2$ as median rms values using annular rings of 32 pixels wide, at diameters of one and two times the primary beam FWHM. 
These values i.e., $\sigma_1^{spw9}$ and $\sigma_2^{spw9}$ for SPW9 images are provided in Appendix~\ref{sec:listpoint}, and plotted in the {\it top} panels of Fig.~\ref{fig:sigma1sigma2} as a function of peak flux density of the brightest source in the field.  
Note that for 318/391 pointings, hereafter referred to as belonging to Class-A (see column\,6 of Table~\ref{tab:pointings}), the radio source at the pointing center is indeed the brightest source in the SPW9 image.  However, for 73/391 ($\sim 19\%$) pointings, serendipitously, an off-axis source happens to be brighter than the central source. Hereafter, we refer to these pointings as Class-B. In Fig.~\ref{fig:sigma1sigma2}, the points for Class-B pointings are color coded with respect to the distance of the brightest source from the pointing center.

\begin{figure*}[ht!]
     \begin{center}
        \subfigure{%
            \label{fig:first}
            \includegraphics[width=0.4\textwidth]{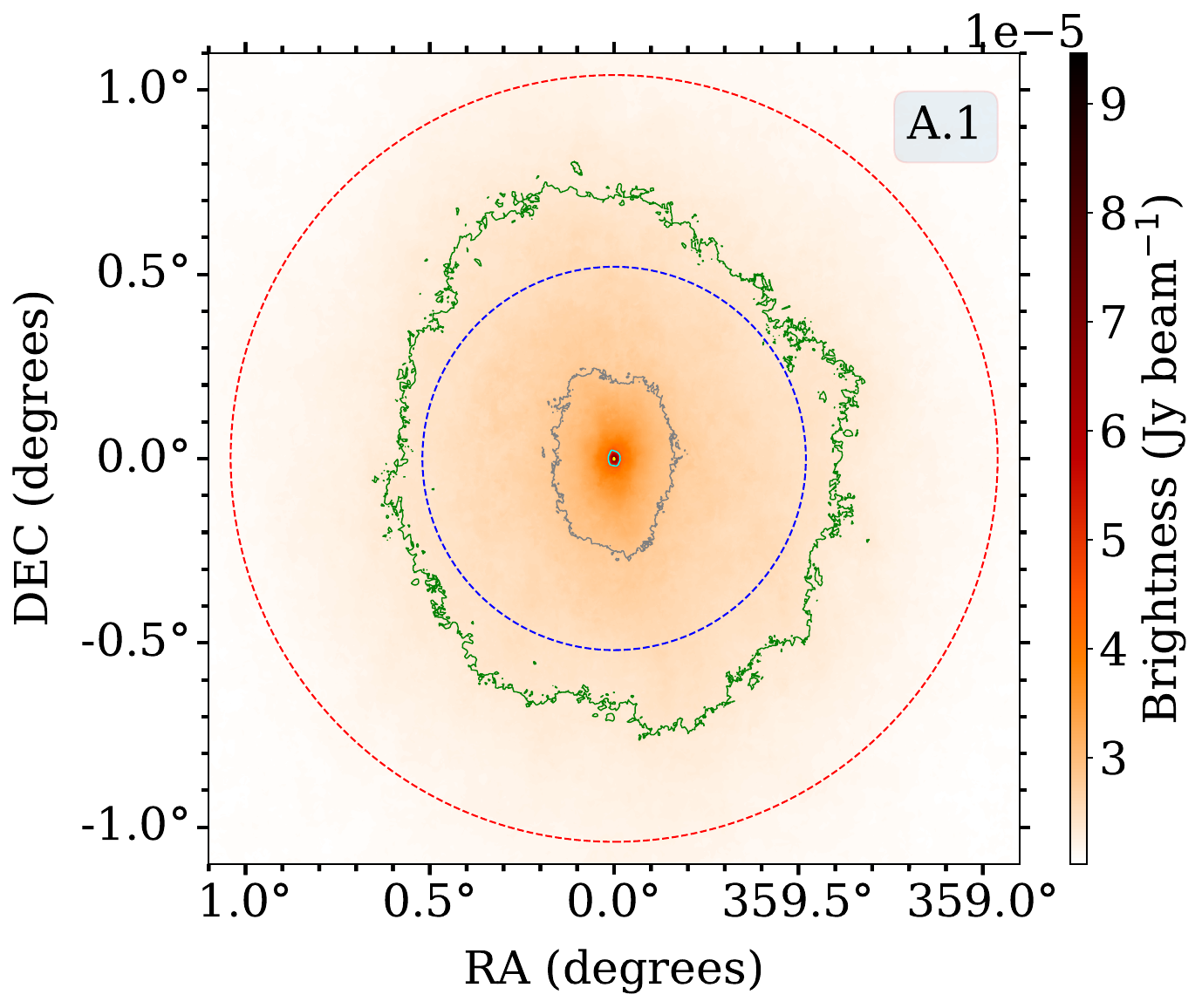}
        }%
        \subfigure{%
           \label{fig:second}
           \includegraphics[width=0.4\textwidth]{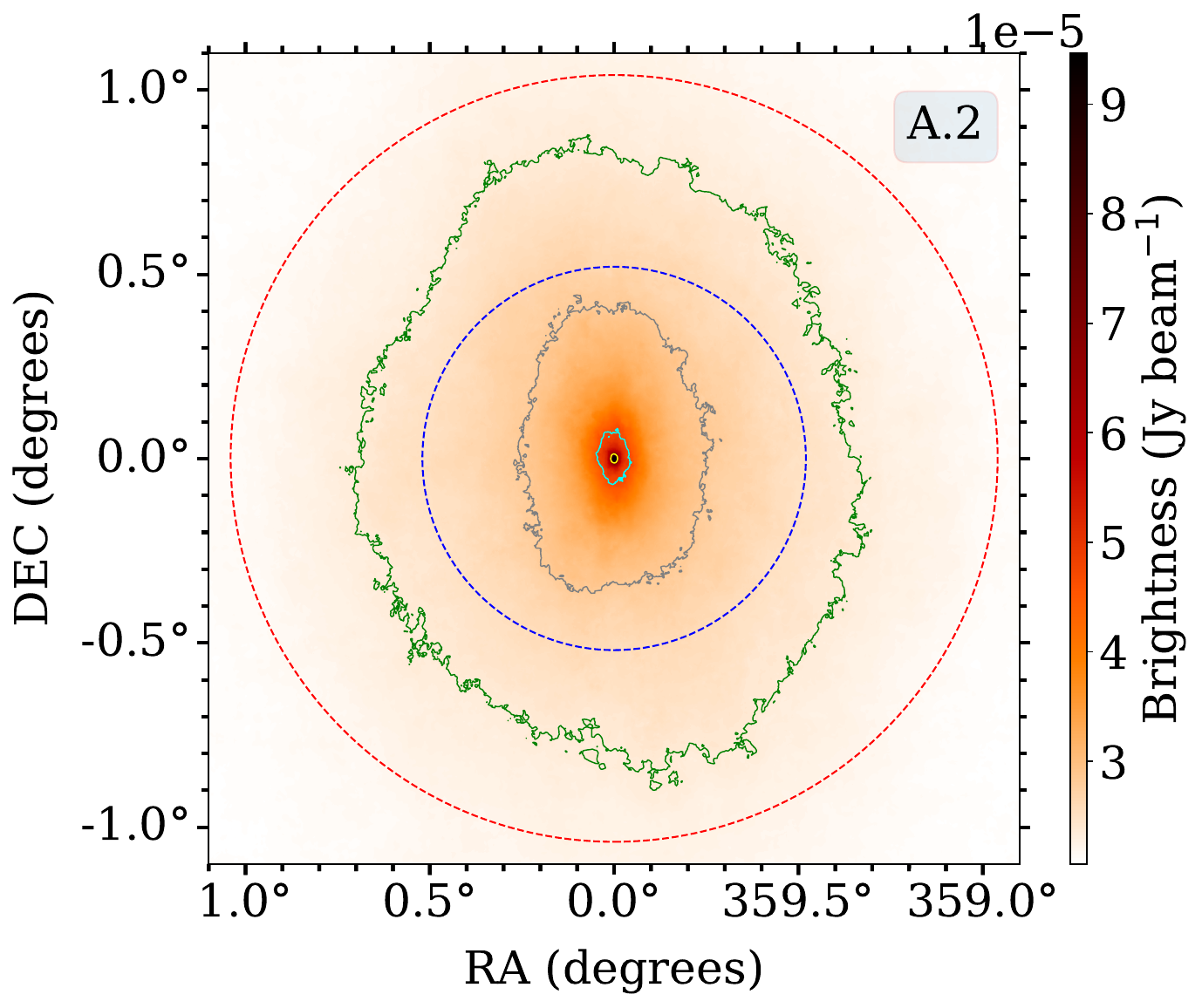}
        }\\ 
        \subfigure{%
            \label{fig:third}
            \includegraphics[width=0.4\textwidth]{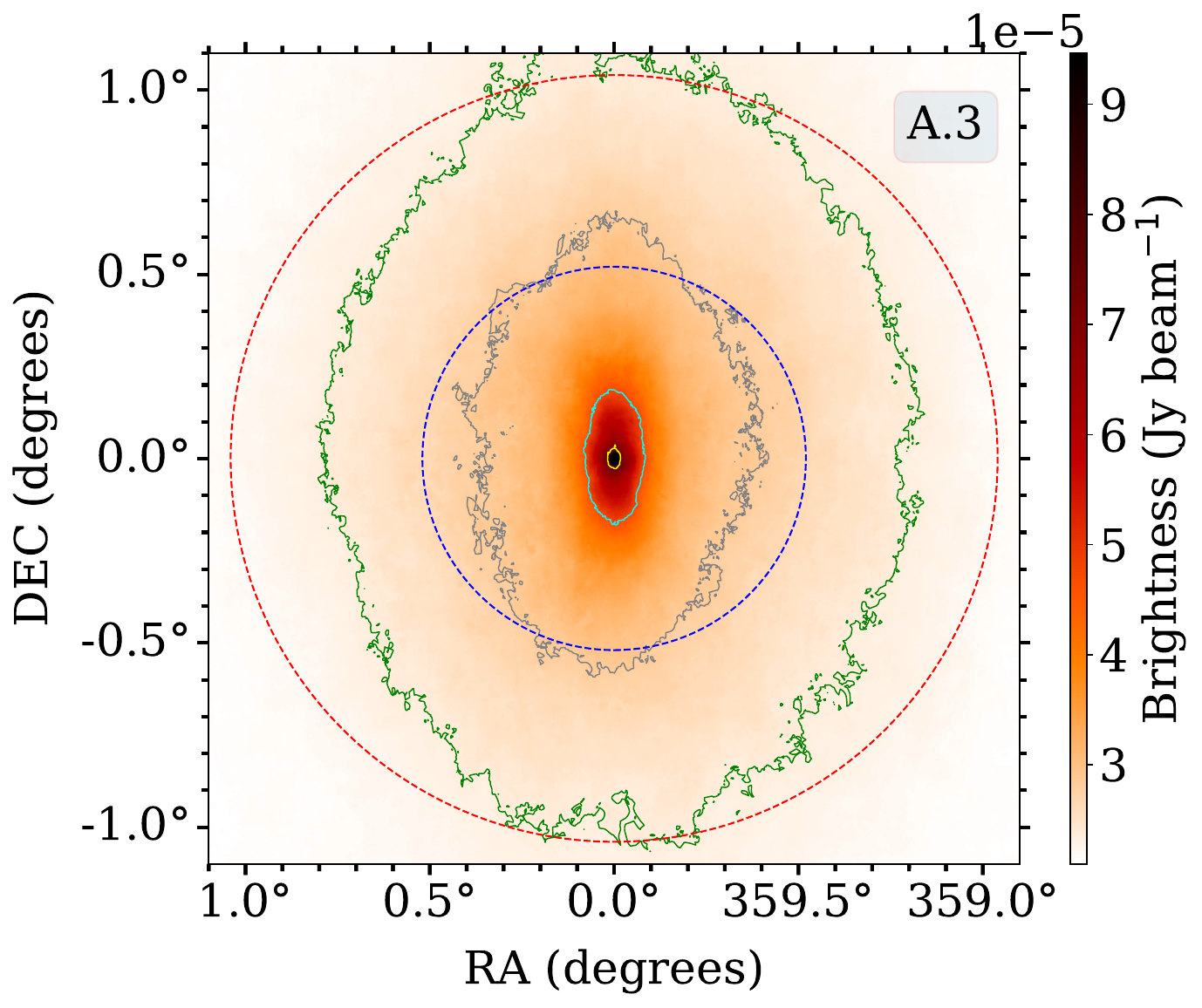}
        }
        \subfigure{%
            \label{fig:fourth}
            \includegraphics[width=0.4\textwidth]{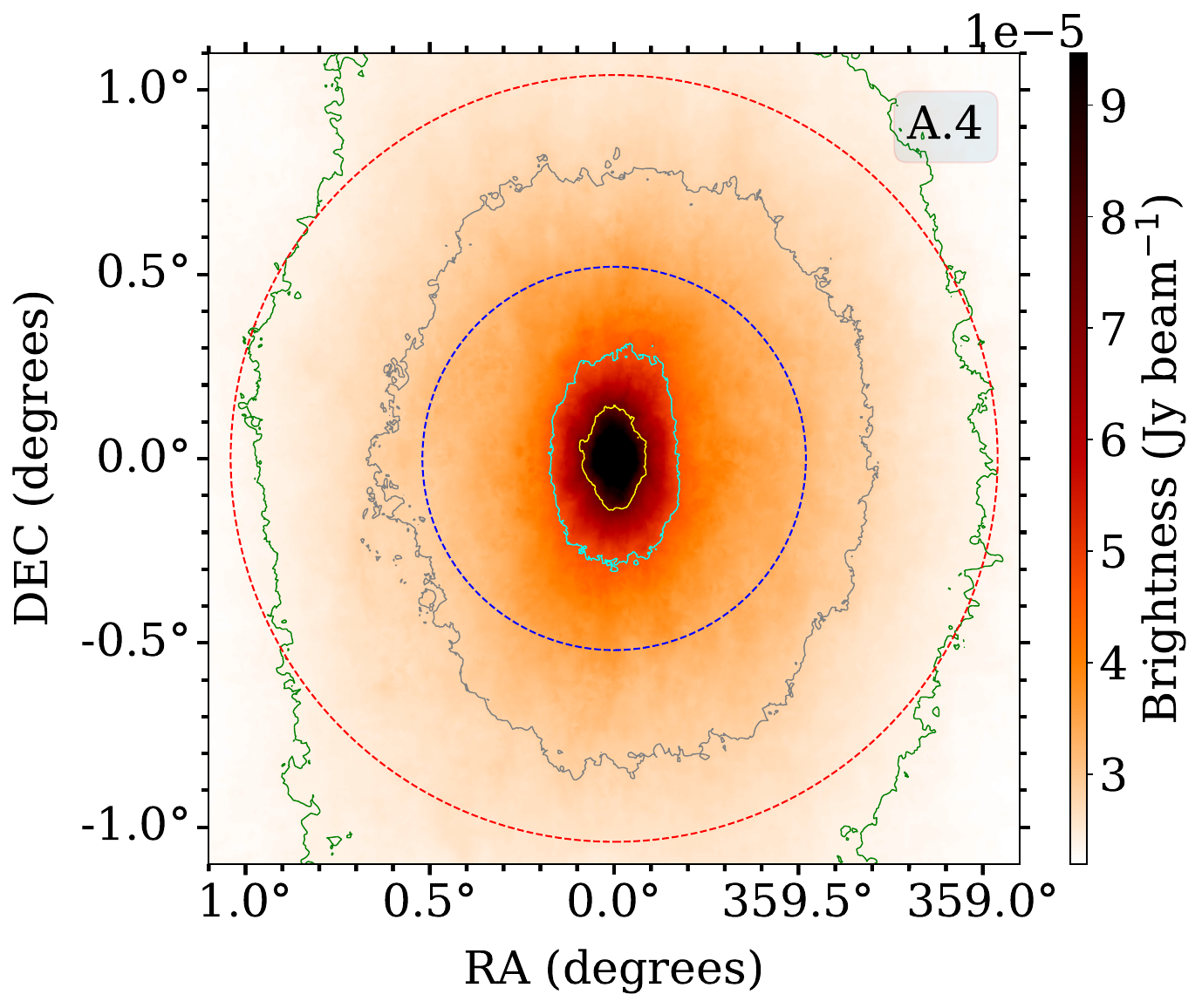}
        }%
        \subfigure{%
            \label{fig:fourth}
            \includegraphics[width=0.4\textwidth]{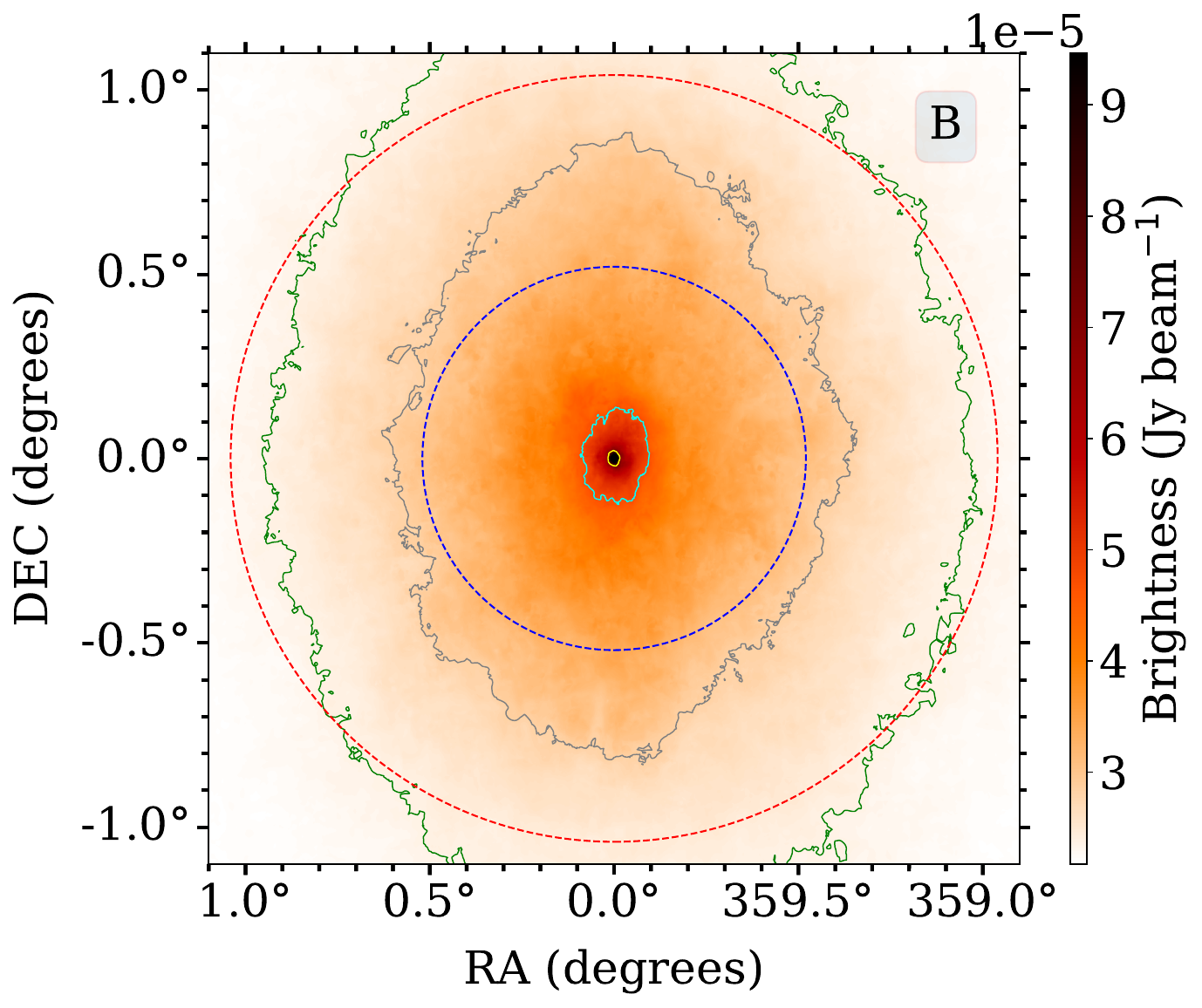}
        }%
    \end{center}
    \caption{%
        Median stacked rms maps for various pointing classes (indicated in the top right labels) based on primary beam uncorrected SPW9 images. The blue and red dotted circles represent diameters of one and two times the primary beam FWHM where $\sigma_1$ and $\sigma_2$ are measured. The colorbar range is saturated at the peak intensity (95\,$\mu$Jy\,beam$^{-1}$) of stacked Class-A.1 map. The contours correspond to 25\% (green), 30\% (gray), 50\% (cyan) and 80\% (yellow) of the same peak intensity.
     }%
   \label{fig:medianStacked}
\end{figure*}

For clarity, in Fig.~\ref{fig:sigma1sigma2} we have omitted 3 and 4 pointings with rms greater than 100\,$\mu$Jy\,beam$^{-1}$ and 200\,$\mu$Jy\,beam$^{-1}$ for $\sigma_1^{spw9}$ and $\sigma_1^{spw2}$, respectively. As discussed below, since $\sigma_1$ is typically larger than  $\sigma_2$, for  $\sigma_2^{spw9}$ and $\sigma_2^{spw2}$ these omissions in the {\it left} panels of Fig.~\ref{fig:sigma1sigma2} translate to the exclusion of 1 and 3 of these pointings, respectively, in the {\it right} panels. 
In terms of above-mentioned classes for $\sigma^{spw9}_{1}$, two of the 3 outliers with $\sigma_1^{spw9}$ = 111\,$\mu$Jy\,beam$^{-1}$ and 138\,$\mu$Jy\,beam$^{-1}$ correspond to Class-A pointings with very strong central radio sources with peak flux densities of  11.9\,Jy\,beam$^{-1}$ and 2.81\,Jy\,beam$^{-1}$, respectively.  Interestingly, the third outlier with $\sigma_1^{spw9}$ = 380\,$\mu$Jy\,beam$^{-1}$ is a Class-B pointing with an off-axis (distance$\sim 0.6\degree$) source of 1.1\,Jy\,beam$^{-1}$.

Overall, Fig.~\ref{fig:sigma1sigma2} clearly demonstrates that both $\sigma_1^{spw9}$ and $\sigma_2^{spw9}$ are correlated with the brightness of the central source. The increase is steeper, as implied by the trend for $\sigma_1^{spw9}$, in the inner regions of the primary beam. 
Overall, the median $\sigma_2^{spw9}$ is only $\sim$15\% higher than the expected theoretical value but the same difference for $\sigma_1^{spw9}$ is $\sim$40\%.  
The dominant role of the central source in this context is further demonstrated in the {\it bottom} panels of Fig.~\ref{fig:sigma1sigma2} showing rms noise for SPW2.  As expected, on average the central radio source is brighter in SPW2 at lower frequency (1006.0\,MHz). Consequently, $\sigma_1^{spw2}$ increases even more rapidly and is about 60\% higher compared to the theoretical rms noise. In comparison, the value of $\sigma_2^{spw2}$ is barely affected. 
%

The brightness of a source and its location within the primary beam can elevate the rms noise in the field through a variety of effects.  Especially, in the case of bright off-axis sources (Class-B pointings) the direction-dependent effects through pointing errors can be the dominant factor.  In order to closely track the possible sources of errors, we subdivide Class-A into four sub-classes (A.1 to A.4) based on the quartiles partitioning the peak flux density range (0.08-11.87\,Jy\,beam$^{-1}$) of the central source in SPW9 images. The peak flux density ranges for these are: A.1 = 0.08-0.23\,Jy\,beam$^{-1}$, A.2 = 0.23-0.32\,Jy\,beam$^{-1}$, A.3 = 0.32-0.49\,Jy\,beam$^{-1}$ and A.4 = 0.49-11.87\,Jy\,beam$^{-1}$.
The effect of the central source in raising the rms floor is apparent from Figs.~\ref{fig:rms_dist} and \ref{fig:medianStacked}.  The cumulative distribution functions (CDFs) of rms noise pixels (Fig.~\ref{fig:rms_dist}), especially the bottom two panels, exhibits an increase in overall rms of the image as the peak flux density of the brightest source in the image increases.  From Class-A.1 to A.4 and B, the distribution starts shifting towards the right and becomes progressively flatter, indicating an increase in the fraction of noisy pixels contributed by brighter sources.   
This is also corroborated by the median stacked rms maps presented in Fig.~\ref{fig:medianStacked} showing that for classes A.4 and B the effect extends well beyond the beam FWHM. Note that for class B, we have not oriented images at position angles of bright off-axis source, hence, the impact of the off-axis sources is smeared out.

Finally, for rigorous artifact analysis involving visual inspection to closely track various systematic errors and the purity of the catalog in Section~\ref{sec:purity}, we identify 50 representative pointings, 10 from each class spanning the typical range of CDF profiles.  These pointings were picked without any visual examination and can be recognized through dashed-dotted lines in Fig.~\ref{fig:rms_dist} and `\_R' in column\,6 of Table~\ref{tab:pointings}.

\subsection{Primary beam correction and Stokes-$I$ catalog}  
\label{sec:icat}   

The `raw' images from {\tt ARTIP} are not corrected for the effects of the primary beam pattern of MeerKAT. \citet[][]{Mauch20} demonstrated that the Stokes-$I$ primary beam response of MeerKAT from holographic measurements is well approximated by a cosine-tapered field illumination function.   We use the publicly available {\tt katbeam}\footnote{\url{https://github.com/ska-sa/katbeam}} module (version 0.1) to generate the primary beam responses at the reference frequencies of SPW-based images (see Table~\ref{tab:spwdef}) and apply these to `raw' images using the {\tt CASA} task {\tt impbcor} to recover intrinsic source properties.  

\begin{figure*} 
\vspace{1.0cm}
\centerline{\vbox{
\centerline{\hbox{
\includegraphics[width=1.0\linewidth]{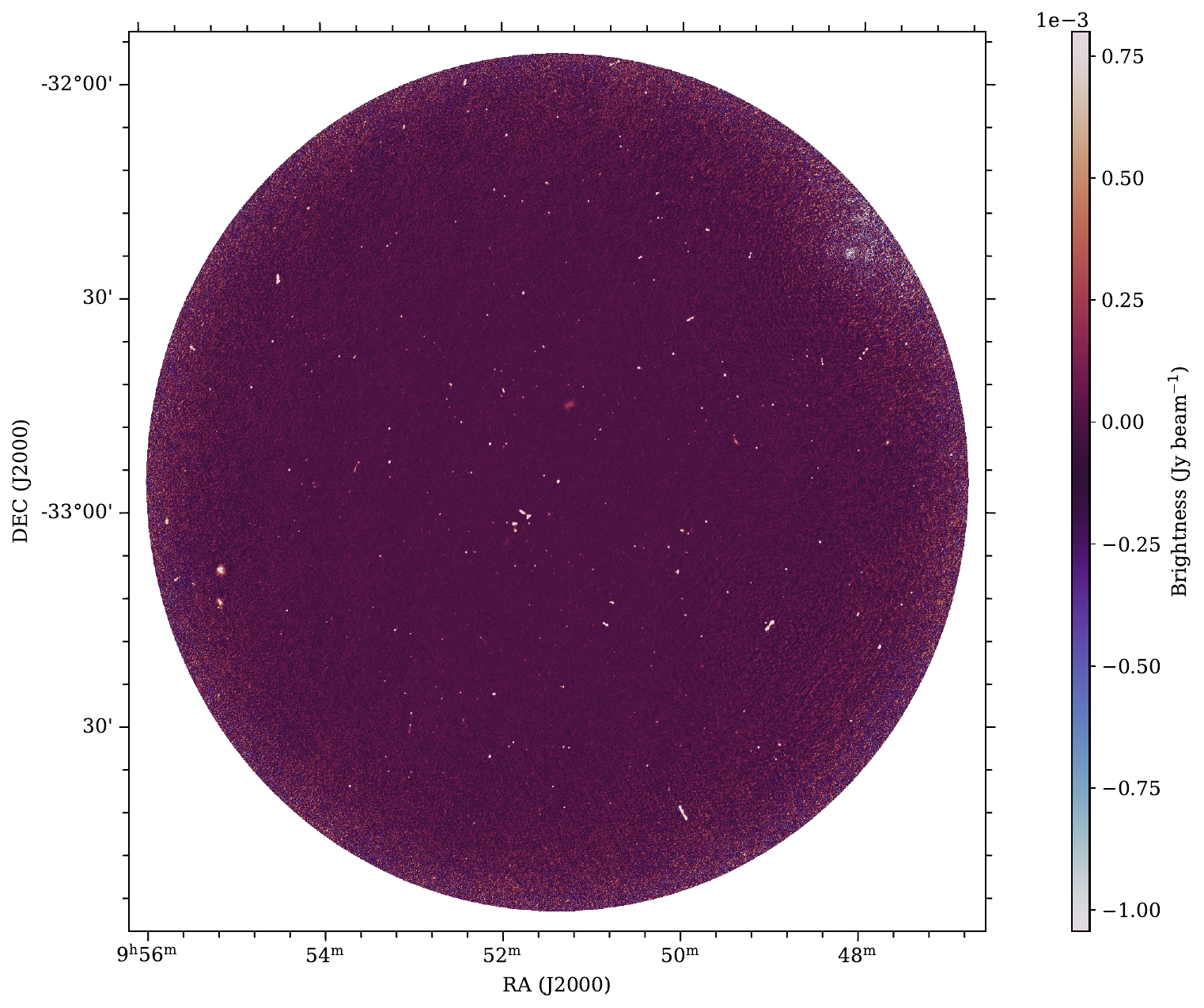}
}} 
}}  
\vskip+0.0cm  
\caption{
	 MeerKAT primary beam corrected L-band SPW9 image centered on the radio source J095123.18-325554.8 (Class-A.1; S$_p\sim$153.0\,mJy\,beam$^{-1}$), with {\tt robust=0} weighting. The rms in the vicinity of the central source is 40\,$\mu$Jy\,beam$^{-1}$ and the restoring beam is $7\farcs7\times6\farcs2$ with a position angle of $-$16$^\circ$. The dynamic range is $\sim$3800.  
} 
\label{fig:fullimgspw9-1}   
\end{figure*}

\begin{figure*} 
\vspace{1.0cm}
\centerline{\vbox{
\centerline{\hbox{
\includegraphics[width=1.0\linewidth]{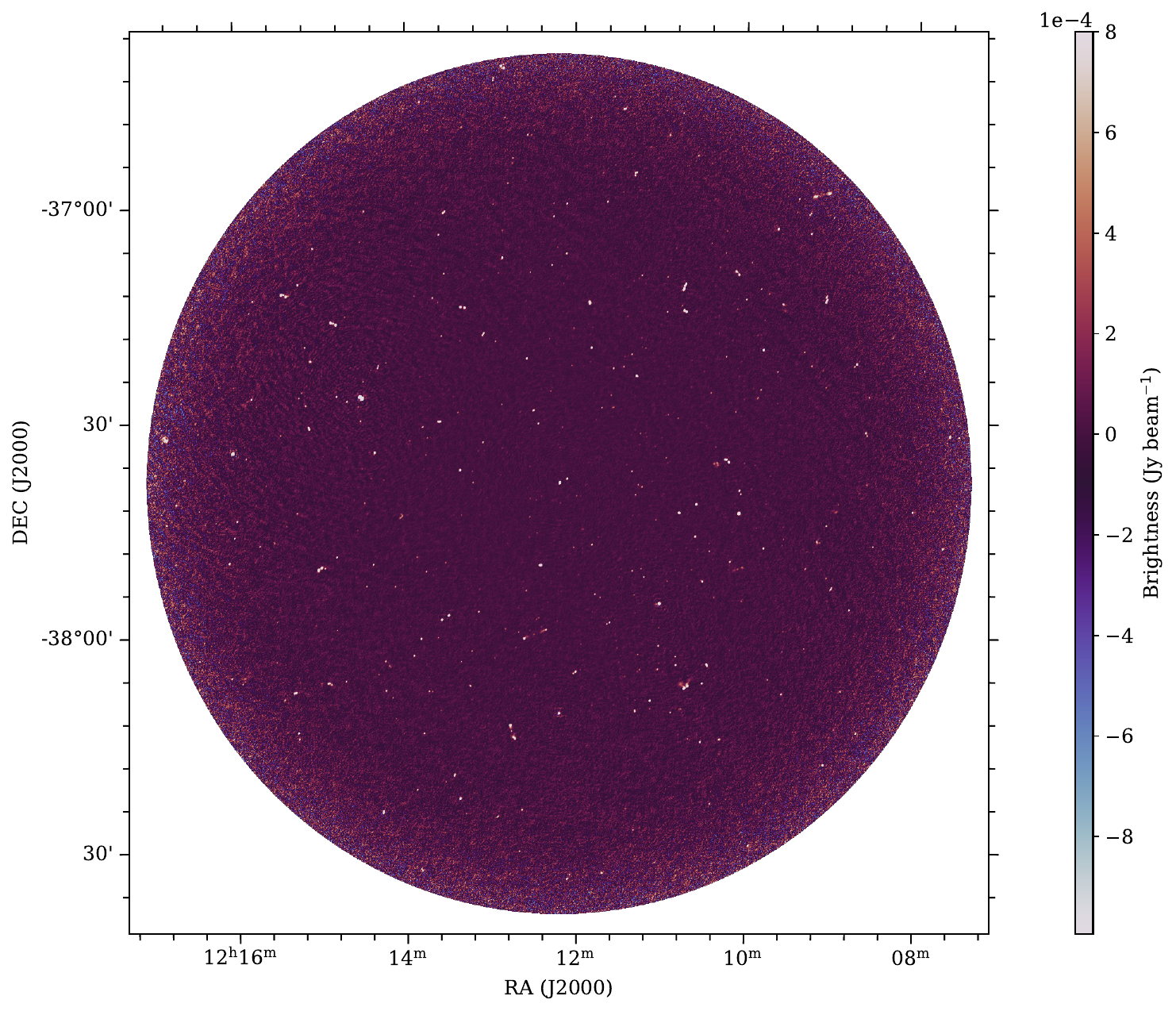}
}} 
}}  
\vskip+0.0cm  
\caption{
	 MeerKAT primary beam corrected L-band image centered on the radio source J121211.89-373826.9 (Class-A.3; S$_p\sim$351.9\,mJy\,beam$^{-1}$), with {\tt robust=0} weighting.  The rms in the vicinity of the central source is 50\,$\mu$Jy\,beam$^{-1}$ and the restoring beam is $9\farcs0\times6\farcs5$ with a position angle of $-24\degree$. The dynamic range is $\sim$7000.  
} 
\label{fig:fullimgspw9-2}   
\end{figure*}

The primary beam gain is often poorly determined in the outermost regions.  Therefore, the usual practice is to cut off the primary beam normalization at 0.2.
However, we adopted a {\tt cut\_off} value of 0.05 which resulted in primary beam corrected SPW9 and SPW2 images of extent $\sim1.92\degree$ and $\sim2.73\degree$ in diameter, respectively. 
Figs.~\ref{fig:fullimgspw9-1} and \ref{fig:fullimgspw9-2} show two SPW9 images as examples -- one of these is from Class-A.1 and the another one is from Class-A.3.  The latter is chosen such that the peak flux density of the central source is close to the median value (327\,mJy\,beam$^{-1}$) for the sample.  The increase in rms noise away from the pointing center due to the primary beam correction is apparent in both the images. Also, the radio sources are detected right up to the edge of the images.

\begin{figure*}
    \centering
    \includegraphics[width=1.0\linewidth]{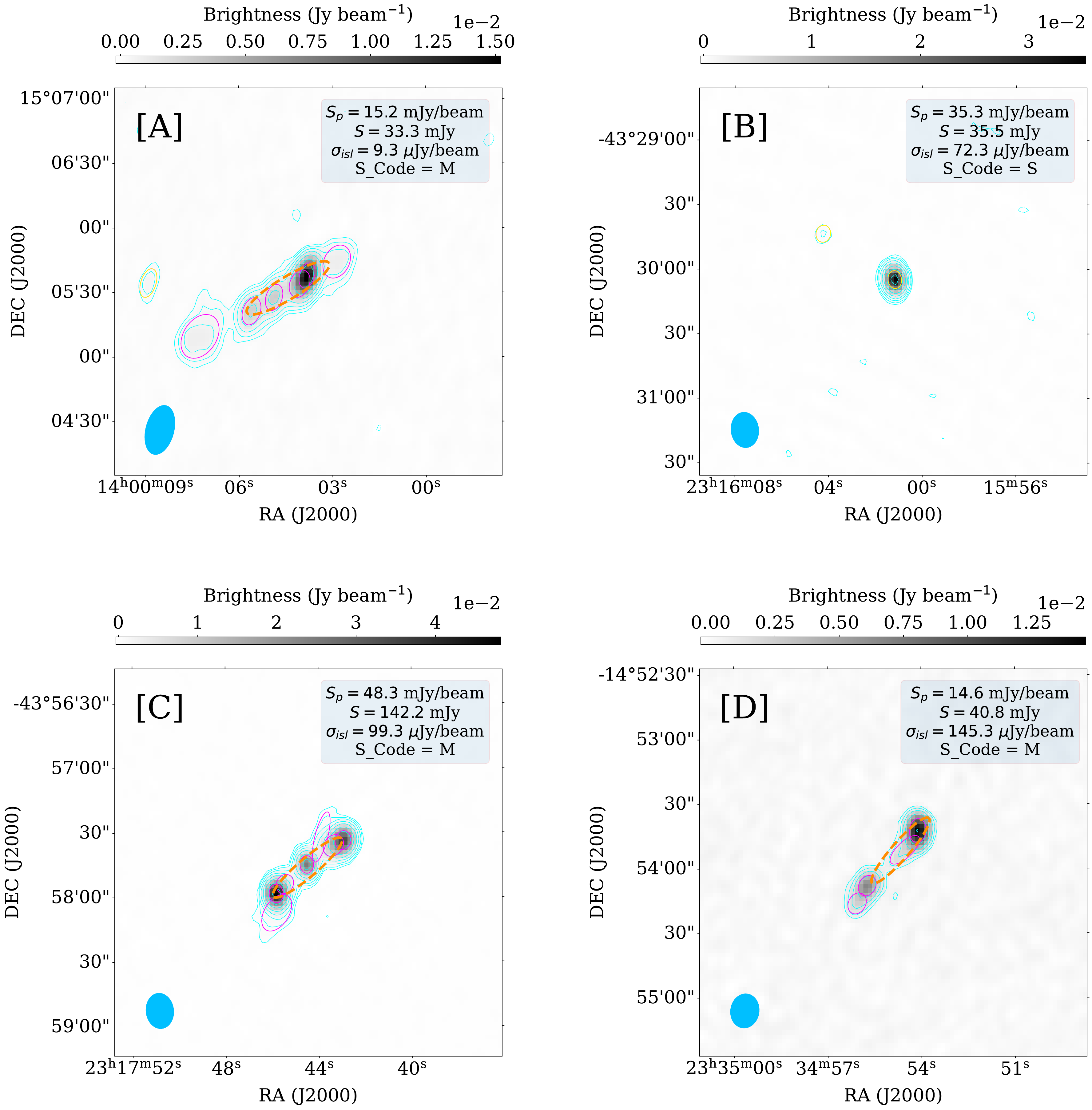}
    \caption{
    Image cutouts ($3\arcmin\times3\arcmin$) exhibiting typical morphology of radio sources detected in primary beam corrected SPW9 images. The contour levels are shown at $3\times${\tt isl\_rms}$\times$(-1, 1, 2, 4, 8, 16, ...)\,mJy\,beam$^{-1}$. 
    The FWHM of major and minor axis of the fit to the source are shown using solid yellow, for {\tt S\_Code = S}, and dashed orange ellipse, for {\tt S\_Code = M} (see Table~\ref{tab:cat_cols} for details).  
    The individual Gaussian components fitted -- 6 in A, 1 in B, 7 in C and 4 in D -- to model the emission are shown as solid magenta ellipses. Note that in panel B, for an {\tt S\_Code = S} type source the yellow ellipse coincides with the magenta ellipse representing the fitted single Gaussian component.
    In panels A and B, another unrelated compact source, in a different island, is also detected. The {\tt isl\_rms} used for plotting contours in these panels is the average of the two islands.  
    The restoring beams are shown as filled ellipses at the bottom left corner of the images.
    }
    \label{fig:src_morph}
\end{figure*}

We note that a primary beam {\tt cut\_off} value of 0.2 will yield images of extent (diameter) $\sim 2.12\degree$ (SPW2) and $\sim 1.49\degree$ (SPW9).  The choice of lower {\tt cut\_off} allows us to detect $\sim20\%$ additional sources.  
The comparison of the properties of these radio sources included in the current and future data releases expands the scope for an independent investigation of the frequency dependent behavior of the primary beam across L- and UHF-bands  (for example, see Section~\ref{sec:pbaccur}).
These sources may also be of interest for various science cases e.g., absorption line search and radio continuum variability, that do not necessarily require the measurement of absolute flux densities.  
The reliability of sources detected in the outermost regions of the primary beam is discussed in Section~\ref{sec:purity}.

We used {\tt PyBDSF} parameters summarized in Table~\ref{tab:bdsf_params} to generate radio source catalogs from primary beam corrected SPW2 and SPW9 images.  In Section~\ref{sec:noise}, we discussed the correspondence between the flux density of the brightest source in the field and noise variations across pointings using `raw' images.  We reexamined the appropriateness of the choice of the same {\tt PyBDSF} parameters for primary beam corrected images.  Of particular interest here, is the modelling of extended emission associated with radio sources.  In Fig.~\ref{fig:src_morph}, we show examples of four radio sources with different morphologies.  The individual Gaussian components fitted to model the radio emission are also shown. In panels A, C and D, the radio source is modelled using multiple components (magenta ellipses), all of which are then grouped to form a single source (thick orange ellipse).  
Such sources are labeled by {\tt PyBDSF} as `M' type  implying a single source fitted with multiple Gaussian components.  A single source fitted with a single component is labeled as `S' (panel B of Fig.~\ref{fig:src_morph}). Note that due to the choice of {\tt PyBDSF} parameter, {\tt group\_by\_isl} = {\tt True}, there are no `C' type i.e., multiple sources within an island in the MALS catalogs.

We used `Source list' and `Gaussian list' catalogs from {\tt PyBDSF} to generate final MALS radio continuum catalogs for both SPW2 and SPW9.  Table~\ref{tab:cat_cols} lists all the columns and also provides a short description of each column.
Columns 1-16, provide overall details of the pointing in which the source is detected.  This includes {\tt Pointing\_id} based on the position of the central source in NVSS or SUMSS, the observing band and date of observation, the version of the primary beam model, the details of flux density calibration, the restoring beam and various rms noise estimates i.e., {\tt Sigma\_1}, {\tt Sigma\_2} and {\tt Sigma\_20}. All these details are common to all the sources detected in a pointing.  
Unlike {\tt Sigma\_1} and {\tt Sigma\_2}, {\tt Sigma\_20} is based on the primary beam corrected images.  Further, while {\tt Sigma\_1} and {\tt Sigma\_2} provide rms at one and two times the beam FWHM, {\tt Sigma\_20} is representative of rms noise coverage in the central region of the images  \citep[see][for details]{Wagenveld23}. Nonetheless, the three noise estimates are correlated.  Typically (median), {\tt Sigma\_20} is 1.8 and 2.2 times {\tt Sigma\_1} and {\tt Sigma\_2}, respectively.
Note that this paper focuses only on sources detected in the SPW2 and SPW9 continuum images but the catalog columns have been defined to support all the subsequent releases based on L- and UHF-band continuum images for individual SPWs or the entire wideband (see column 9 i.e., {\tt SPW\_id}).


The properties of individual sources are provided in columns 17 -- 67.   Since rms noise and systematic errors depend on distance from the pointing center and the proximity to a bright radio source, we provide  {\tt Distance\_pointing}, the distance from the pointing center (column 17), and {\tt Distance\_NN}, the  distance from the nearest neighbour (column 18).  The {\tt PyBDSF} label {\tt S\_code} = `S' or `M' discussed above is provided in column 19.  The number of Gaussian components ({\tt N\_Gauss}) fitted to the source and the maximum separation between the components ({\tt Maxsep\_Gauss}) are provided in columns 20 and 21, respectively. The angular sizes, positions and flux densities of sources are provided in columns 22 -- 56, and the details of individual components are provided in columns 68 -- {86}.  Column 86 provides unique Gaussian component identifiers.
Columns 57 -- 64 provide spectral index measurements.  Columns 65 and 67 are concerned with the reliability of source detection and its morphology, respectively. 
We note that columns 19, 22-42, 44, 48-55, 68-85 are direct outputs from the {\tt PyBDSF} runs. The remaining columns are based on additional analysis discussed in the subsequent sections of this paper.

In total, we detect 240,321 sources consisting of 285,209 Gaussian components from 391 primary beam corrected SPW9 images at 1380.9\,MHz, of which  215,328 and 24,993 are of type ({\tt S\_code})  `S' and `M', respectively.  
On average (mean), we detect 629 and 551 sources in 318 and 73 pointings of type Class-A and B, respectively.  The median dynamic ranges defined as the ratio of peak flux density of the brightest source and $\sigma_1^{spw9}$ achieved at SPW9 are 11,800 (Class-A) and 12,300 (Class-B).
In comparison, the total number of sources in SPW2 images at 1006.0\,MHz is 495,325, with 586,290 Gaussian components.  Of these 441,988 and 53,337 are of type `S' and `M', respectively.  The larger number of sources in SPW2 images can be attributed to larger sky coverage (total~2289\,deg$^2$) compared to that of SPW9 images (total~1132\,deg$^2$). Using a matching radius of 6$^{\prime\prime}$, 205,435 sources were found to be common between SPW2 and SPW9.
%

The catalogs and images for SPW2 and SPW9 can be accessed at \href{https://mals.iucaa.in}{https://mals.iucaa.in}.  Each MALS data release will identify a `reference' SPW and the columns 57 -- 67 based on information from multiple SPWs will be filled only in the `reference' SPW catalog. Since larger number of sources are detected in SPW2, the reference SPW adopted for MALS DR1 is SPW2. 

\startlongtable
\begin{deluxetable*}{p{0.0\linewidth} p{0.15\linewidth} p{0.485\linewidth}}
\newcolumntype{E}{>{\raggedright\arraybackslash}p{25mm}}
\newcolumntype{F}{>{\raggedright\arraybackslash}p{95mm}}
\newcolumntype{C}{>{\centering\arraybackslash}p{20mm}}
\newcolumntype{D}{>{\centering\arraybackslash}p{5mm}}
%
    \tablewidth{0pc}
    \tablecaption{Catalogue column descriptions.\label{tab:cat_cols} }
    \tabletypesize{\footnotesize}
\flushleft{
    \tablehead{
            \colhead{Number}&\colhead{Name}  & \colhead{Description} \\
            \colhead{(1)} &\colhead{(2)} &\colhead{(3)}   
              }              
          }
    \startdata
    \noalign{\smallskip}
    1 & {\tt Source\_name}    & MALS name of the source ({\tt JHHMMSS.ss+DDMMSS.s}) based on its right ascension and declination (J2000). \\
    2 & {\tt Pointing\_id}   & The MALS pointing ID ({\tt JHHMMSS.ss$\pm$DDMMSS.s}) based on the position (J2000) of the central source in NVSS or SUMSS. \\
    3 & {\tt Obs\_date\_U} &   The date and time (UTC) of the start of UHF-band observing block(s) in the format YYYY-MM-DDThh:mm. \\
    4 & {\tt Obs\_date\_L} &   The date and time (UTC) of the start of L-band observing block(s) in the format YYYY-MM-DDThh:mm. \\
    5 & {\tt Obs\_band} &  The observing band: L = L-band and U = UHF-band.\\
    6 & {\tt PBeamVersion}    & The primary beam model ({\tt katbeam} or {\tt plumber}) version used for the primary beam correction (see Section~\ref{sec:pbaccur} for details). All the columns except {\tt Total\_flux\_measured} and {\tt Total\_flux\_measured\_E} in MALS DR1 are based on the {\tt katbeam} model (see also {\tt Flux\_correction}). \\
    7 & {\tt Fluxcal} &   The list of calibrator(s) used for flux density and bandpass calibration of the dataset. \\
    8 & {\tt Fluxscale} &   The flux density scales used for the flux density calibrators. \\
    9 & {\tt SPW\_id}  &   This defines whether the continuum image is made using an SPW or the entire wideband (WB).  The  possible values are LWB-WP, LWB-AWP, UWB-WP, UWB-AWP, LSPW\_\textit{i} and USPW\_\textit{i}; here \textit{i} goes from 0 to 14. For example, LWB and UWB imply L- and UHF-band wideband image, respectively. LSPW\_2 and LSPW\_9 correspond to SPW2 and SPW9 of L-band included in DR1 presented here.  WP and AWP identify the imaging algorithm used for wideband imaging. WP $\implies$ W-Projection algorithm is used to correct for the widefield effect of non-coplanar baselines \citep[][]{Cornwell92} and the primary beam correction is applied after the imaging.  AWP $\implies$ A-term is also included and the wide-band effects of the primary beam are corrected prior to integration in time and frequency for the continuum imaging \citep[][]{Bhatnagar13}. \\ 
    10 & {\tt Ref\_freq}  &   The reference frequency (MHz) of the continuum image. \\
    11 & {\tt Maj\_restoring\_beam}  & The major axis (arcsec) of the restoring beam. \\
    12 & {\tt Min\_restoring\_beam}  & The minor axis (arcsec) of the restoring beam. \\
    13 & {\tt PA\_restoring\_beam}   & The position angle (degrees) of the restoring beam. \\
    14 & {\tt Sigma\_1}  & The rms noise ($\mu$Jy\,beam$^{-1}$) measured from primary beam uncorrected rms image in an annular ring at primary beam FWHM.\\
    15 & {\tt Sigma\_2}  & The rms noise ($\mu$Jy\,beam$^{-1}$) measured from primary beam uncorrected rms image in an annular ring at two times the primary beam FWHM.\\
    16 & {\tt Sigma\_20}  & The rms noise ($\mu$Jy\,beam$^{-1}$) at a cumulative fraction of 0.2 of the rms noise distribution of the primary beam corrected rms image ($\sigma_{20}$; see \citet[][]{Wagenveld23} for details).\\
    17 & {\tt Distance\_pointing}   & The distance of the source (arcmin) from the pointing center. \\
    18 & {\tt Distance\_NN}    & The distance of the source (arcmin) from the nearest neighbour in the field. \\
    19 & {\tt S\_Code$^\ddag$} &  The {\tt PyBDSF} code defining the source structure. {\tt S} = a single source in the island, fitted with a single Gaussian component. {\tt C} = a source with other neighbors within the island, fitted with a single Gaussian component. {\tt M} = a source fitted with multiple Gaussian components.  \\
    20 & {\tt N\_Gauss} & The number of Gaussian components fitted to the source. \\
    21 & {\tt Maxsep\_Gauss} & The maximum separation (arcsec) between the Gaussian components.  This is set to -1 for `S' type sources. \\
    22 & {\tt Maj$^\ddag$} & The FWHM (arcsec) of the major axis of the source. \\
    23 & {\tt Maj\_E$^\ddag$} & The 1$\sigma$ error on {\tt Maj}. \\
    24 & {\tt Min$^\ddag$} & The FWHM (arcsec) of the minor axis of the source. \\
    25 & {\tt Min\_E$^\ddag$} & The 1$\sigma$ error on {\tt Min}. \\   
    26 & {\tt PA$^\ddag$} & The position angle (degrees) of the major axis of the source measured east of north. \\
    27 & {\tt PA\_E$^\ddag$} & The 1$\sigma$ error on {\tt PA}. \\
    28 & {\tt DC\_Maj$^\ddag$} & The FWHM (arcsec) of the deconvolved major axis of the source. \\
    29 & {\tt DC\_Maj\_E$^\ddag$} & The 1$\sigma$ error on {\tt DC\_Maj}. \\
    30 & {\tt DC\_Min$^\ddag$} & The FWHM (arcsec) of the deconvolved minor axis of the source. \\
    31 & {\tt DC\_Min\_E$^\ddag$} & The 1$\sigma$ error on {\tt DC\_Min}. \\   
    32 & {\tt DC\_PA$^\ddag$} & The position angle (degrees) of the deconvolved major axis of the source measured east of north. \\
    33 & {\tt DC\_PA\_E$^\ddag$} & The 1$\sigma$ error on {\tt DC\_PA}. \\
    34 & {\tt RA\_mean$^\ddag$}  &  The right ascension (J2000) of the mean intensity weighted position of all pixels above the island threshold, measured if source is fitted with multiple Gaussians. \\
    35 & {\tt RA\_mean\_E$^\ddag$}   & The 1$\sigma$ error on {\tt RA\_mean} estimated using Equation~\ref{eq:astroer}. \\
    36 & {\tt DEC\_mean$^\ddag$}  &  The declination (J2000) of the mean intensity weighted position of all pixels above the island threshold, measured if source is fitted with multiple Gaussians. \\
    37 & {\tt DEC\_mean\_E$^\ddag$}   & The 1$\sigma$ error on {\tt DEC\_mean} estimated using Equation~\ref{eq:astroer}. \\
    38 & {\tt RA\_max$^\ddag$}  &   The right ascension (J2000) of the pixel corresponding to maximum flux density. \\
    39 & {\tt RA\_max\_E$^\ddag$ }   & The 1$\sigma$ error on {\tt RA\_max} estimated using Equation~\ref{eq:astroer}. \\
    40 & {\tt DEC\_max$^\ddag$ } &  The declination (J2000) of the pixel corresponding to maximum flux density. \\
    41 & {\tt DEC\_max\_E$^\ddag$ }   & The 1$\sigma$ error on {\tt DEC\_max} estimated using Equation~\ref{eq:astroer}. \\
    42 & {\tt Total\_flux$^\ddag$}    & The total integrated flux density (mJy) of the source based on Gaussian component fits; i.e., corrected for primary beam and wideband effects. \\
    43 & {\tt Total\_flux\_E }   & The 1$\sigma$ error on {\tt Total\_flux} estimated using Equation~\ref{eq:fluxer}. \\
    44 & {\tt Total\_flux\_E\_fit$^\ddag$}   & The fitting error on total flux density to be taken into account to obtain  {\tt Total\_flux\_E} (see Section~\ref{sec:fluxscale} and Equation~\ref{eq:fluxer}). \\
    45 & {\tt Total\_flux\_E\_sys}    & The systematic error to be taken into account to obtain {\tt Total\_flux\_E} (see Section~\ref{sec:fluxscale} and Equation~\ref{eq:fluxer}). \\
    46 & {\tt Total\_flux\_measured$^*$}    & The total integrated flux density (mJy) of the source based on alternate primary beam model i.e., {\tt plumber} (Section~\ref{sec:pbaccur}) obtained by multiplying  {\tt Total\_flux} (column\,42) and {\tt Flux\_correction} (column\,56). \\
    47 & {\tt Total\_flux\_measured\_E$^*$}    & The 1$\sigma$ error on {\tt Total\_flux\_measured}. \\
    48 & {\tt Peak\_flux$^\ddag$} & The peak flux density (mJy\,beam$^{-1}$) of the source. \\
    49 & {\tt Peak\_flux\_E$^\ddag$} & The 1$\sigma$ error on {\tt Peak\_flux}. \\
    50 & {\tt Isl\_Total\_flux$^\ddag$}  &   The total integrated flux density (mJy) of the island in which the source is located. \\
    51 & {\tt Isl\_Total\_flux\_E$^\ddag$}    & The 1$\sigma$ error on {\tt Isl\_Total\_flux}. \\
    52 & {\tt Isl\_rms$^\ddag$}  &  The average background RMS noise (mJy\,beam$^{-1}$) of the island in which the source is located. \\
    53 & {\tt Isl\_mean$^\ddag$}  &  The average background mean value (mJy\,beam$^{-1}$) of the island in which the source is located. \\
    54 & {\tt Resid\_Isl\_rms$^\ddag$} &  The average residual background RMS noise (mJy\,beam$^{-1}$) of the island in which the source is located. \\
    55 & {\tt Resid\_Isl\_mean$^\ddag$}    & Average residual background mean value (mJy\,beam$^{-1}$) of the island in which the source is located. \\
    56 & {\tt Flux\_correction}    & The factor to be multiplied to flux density measurements and errors to obtain the values corresponding to the {\tt plumber} beam model (see Section~\ref{sec:pbaccur} and Appendix~\ref{sec:PBratio}).  \\   
    57 & {\tt Spectral\_index} & The spectral index and curvature of the source determined from the wideband {\tt MTMFS} image. For an extended source, a mean value for the pixels above some threshold in the island is reported (see Section~\ref{sec:uss}). \\
    58 & {\tt Spectral\_index\_E}  & The 1$\sigma$ error on {\tt Spectral\_index} (see Section~\ref{sec:uss}). \\
    59 & {\tt Spectral\_index\_spwused}  & The spectral windows used for determining spectral index and curvature using the narrowband multi-frequency synthesis (MFS) images. For example, {\tt [`L:1$\sim$4;7', `U:3$\sim$8;12']} implies Total\_flux from spectral windows 1 to 4 and 7 for L-band, and 3 to 8 and 12 for UHF-band are used. \\
    60 & {\tt Spectral\_index\_spwfit}   & The spectral index and curvature of the source based on {\tt Total\_flux}  from narrowband i.e., SPW-based images. \\
    61 & {\tt Spectral\_index\_spwfit\_E}   & The 1$\sigma$ error on {\tt Spectral\_index\_spwfit}. \\
    62 & {\tt Spectral\_index\_MALS\_Lit}  & The spectral index and curvature based on {\tt Total\_flux} from narrowband images from MALS and measurements from literature. \\
    63 & {\tt Spectral\_index\_MALS\_Lit\_E}   & The 1$\sigma$ error on {\tt Spectral\_index\_MALS\_Lit}. \\
    64 & {\tt Spectral\_index\_Lit}   & The list of external surveys (e.g., VLASS, TGSS) used. For example, [`TGSS-ADR1', `L:2'], flux densities from TGSS ADR1 and SPW2 of MALS L-band are used.\\
    65 & {\tt Real\_source}   & This is a boolean (True or False) indicating whether a source is a real astrophysical source or an artifact. \\
    66 & {\tt Resolved}   & This is a boolean (True or False) indicating whether a source is resolved based on the reliability envelope method (see Fig.~\ref{fig:comp_envlp}). \\
    67 & {\tt Source\_linked} & This is a list of {\tt Source\_name} i.e., other MALS sources to which the source may be linked to. This accounts for the linkages missed by grouping mechanism of {\tt PyBDSF}.\\
    68 & {\tt G\_RA$^\ddag$} &   The right ascension (J2000) of maximum intensity of the Gaussian component. \\
    69 & {\tt G\_RA\_E$^\ddag$}   & The 1$\sigma$ error on {\tt G\_RA}. \\
    70 & {\tt G\_DEC$^\ddag$} &  The declination (J2000) of maximum intensity of the Gaussian component. \\
    71 & {\tt G\_DEC\_E$^\ddag$} &   The 1$\sigma$ error on {\tt G\_DEC}. \\
    72 & {\tt G\_Peak\_flux$^\ddag$}    & The measured peak flux density (mJy\,beam$^{-1}$) of the Gaussian component (using {\tt PyBDSF}). \\
    73 & {\tt G\_Peak\_flux\_E$^\ddag$}   & The 1$\sigma$ error on {\tt G\_Peak\_flux}. \\
    74 & {\tt G\_Maj$^\ddag$}  &   The FWHM (arcsec) of the major axis of the Gaussian component. \\
    75 & {\tt G\_Maj\_E$^\ddag$}    & The 1$\sigma$ error on {\tt G\_Maj}. \\
    76 & {\tt G\_Min$^\ddag$} &   The FWHM (arcsec) of the minor axis of the Gaussian component. \\
    77 & {\tt G\_Min\_E$^\ddag$}  & The 1$\sigma$ error on {\tt G\_Min}. \\
    78 & {\tt G\_PA$^\ddag$} &   The position angle (degrees) of the major axis of the Gaussian component. \\
    79 & {\tt G\_PA\_E$^\ddag$}  &   The 1$\sigma$ error on {\tt G\_PA}. \\
    80 & {\tt G\_DC\_Maj$^\ddag$}  &   The FWHM (arcsec) of the deconvolved major axis of the Gaussian component. \\
    81 & {\tt G\_DC\_Maj\_E$^\ddag$}    & The 1$\sigma$ error on {\tt G\_DC\_Maj}. \\
    82 & {\tt G\_DC\_Min$^\ddag$}  &   The FWHM (arcsec) of the deconvolved minor axis of the Gaussian component. \\
    83 & {\tt G\_DC\_Min\_E$^\ddag$}    & The 1$\sigma$ error on {\tt G\_DC\_Min}. \\
    84 & {\tt G\_DC\_PA$^\ddag$}  &   The position angle (degrees) of the deconvolved major axis of the Gaussian component. \\
    85 & {\tt G\_DC\_PA\_E$^\ddag$} &  The 1$\sigma$ error on {\tt G\_DC\_PA}. \\
    {\bf 86} & {\tt G\_id} & A unique Gaussian component identifier.
    \enddata
    \tablecomments{
    $\dag$: This is unique only for a combination of POINTING\_ID and SPW\_ID. $\ddag$:  This is direct output from {\tt PyBDSF}. $*$: Columns 46 and 47 are based on the {\tt plumber} beam model.  All the other flux density measurements provided in the catalog are based on the {\tt katbeam} model. The measurements corresponding to the {\tt plumber} model can be obtained using {\tt Flux\_correction} provided in column\,56.}
    
\end{deluxetable*}
%
From \href{https://mals.iucaa.in}{https://mals.iucaa.in}, users can download the source catalog (i.e., columns 1 -- 67) of 205,435 sources common between SPW2 and SPW9, as well as 240,321 (495,325) sources corresponding to SPW9 (SPW2).  The Gaussian component catalogs are also available.  These consist of columns 1 ({\tt Source\_name}), 20 ({\tt N\_Gauss}) and 68 -- 85 (Gaussian parameters) from Table~\ref{tab:cat_cols}.  In Tables~\ref{tab:srlcat} and \ref{tab:gaulcat} (Appendix~\ref{sec:cat_cols}), we present first few rows of the source and Gaussian component catalogs, respectively.  
Note that in the current release, columns 3, 57, 58, 65 and 67 in the source catalog are empty.  Column 3 is relevant only for UHF-band whereas 57 and 58 are for wideband images, hence not relevant for DR1 and included only for the completeness.  Columns 65 and 67 require analysis involving images from all the other SPWs, and hence will be provided in a future release. 


\subsection{Purity of the catalog}
\label{sec:purity}  

Catalogs as output from {\tt PyBDSF} are contaminated by spurious sources which could either be due to statistical noise fluctuations or due to bright sidelobes around strong sources. 
To get a handle on these false detections, we followed the simple procedure of inverting (multiply by $-1$) an image and then running source finding on it with the same set of threshold parameters, rms and (inverted) mean maps as were used for the actual catalog generation.  
This method is based on the idea that statistical noise fluctuations are symmetric around the mean and therefore sources detected in the `negative' images due to noise peaks will provide an estimate of the false sources detected in our actual catalogs \citep[e.g.,][]{Intema17, Hurley-Walker2017, Hale21}. 
In the vicinity of bright sources the systematic errors will dominate and the sources detected in the `negative' image may represent an upper limit on the level of spurious sources.

\begin{figure}
    \centering
    \includegraphics[width=\linewidth]{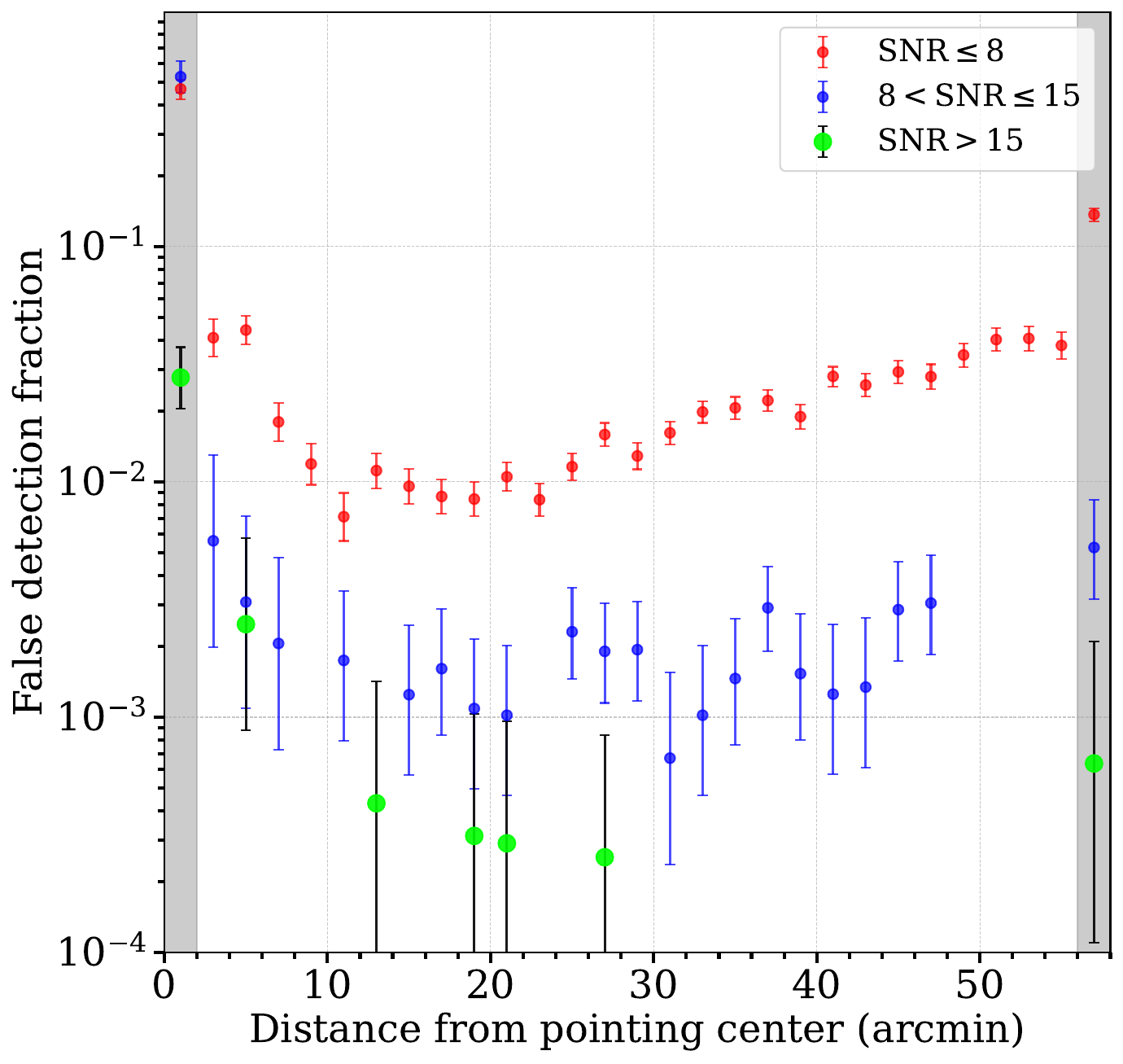}
    \caption{The fraction of `negative' sources with respect to the sources from actual images as a function of distance in three different SNR bins. The binsize is 2\arcmin . Error bars denote 1$\sigma$ Poissonian uncertainties. The shaded portion marks regions with high false detection rate. Note the absence of artefacts with SNR$>15$ in most of the bins, except near the center and at the edges.
    }
    \label{fig:negative_sources}
\end{figure}

From 391 SPW9 pointings, we detect 2,548 `negative' sources, which is merely $\sim$1\% of the sources in the DR1 catalog. The cumulative distribution of artefacts shows a steep dependence on SNR, which saturates near SNR$\approx$8.  About 95\% of the  artefacts lie at SNR$<$8.  Therefore, we consider SNR = 8 as a reasonable cut off to define samples for various analysis. 
Fig.~\ref{fig:negative_sources} shows the fraction of these artefacts as a function of distance from the pointing center in three SNR bins.  As expected the fraction is larger near the pointing center and the edges of the beams (see shaded regions in Fig.~\ref{fig:negative_sources}). We advise caution in using low SNR sources belonging to the shaded region by applying filters corresponding to {\tt Distance\_pointing} parameter. Outside the shaded regions the distribution, even at distances larger than $45^{\prime}$ from the pointing center\footnote{The {\tt cut\_off} = 0.2 usually used for primary beam normalization corresponds to a distance of $45^{\prime}$ from the pointing center (see Section~\ref{sec:icat}).},  is largely uniform and negligible, especially for SNR$>$8.
%

\begin{figure*}[ht!]
     \begin{center}
        \subfigure[Class: A.1]{%
            \label{fig:first}
            \includegraphics[width=0.3\textwidth]{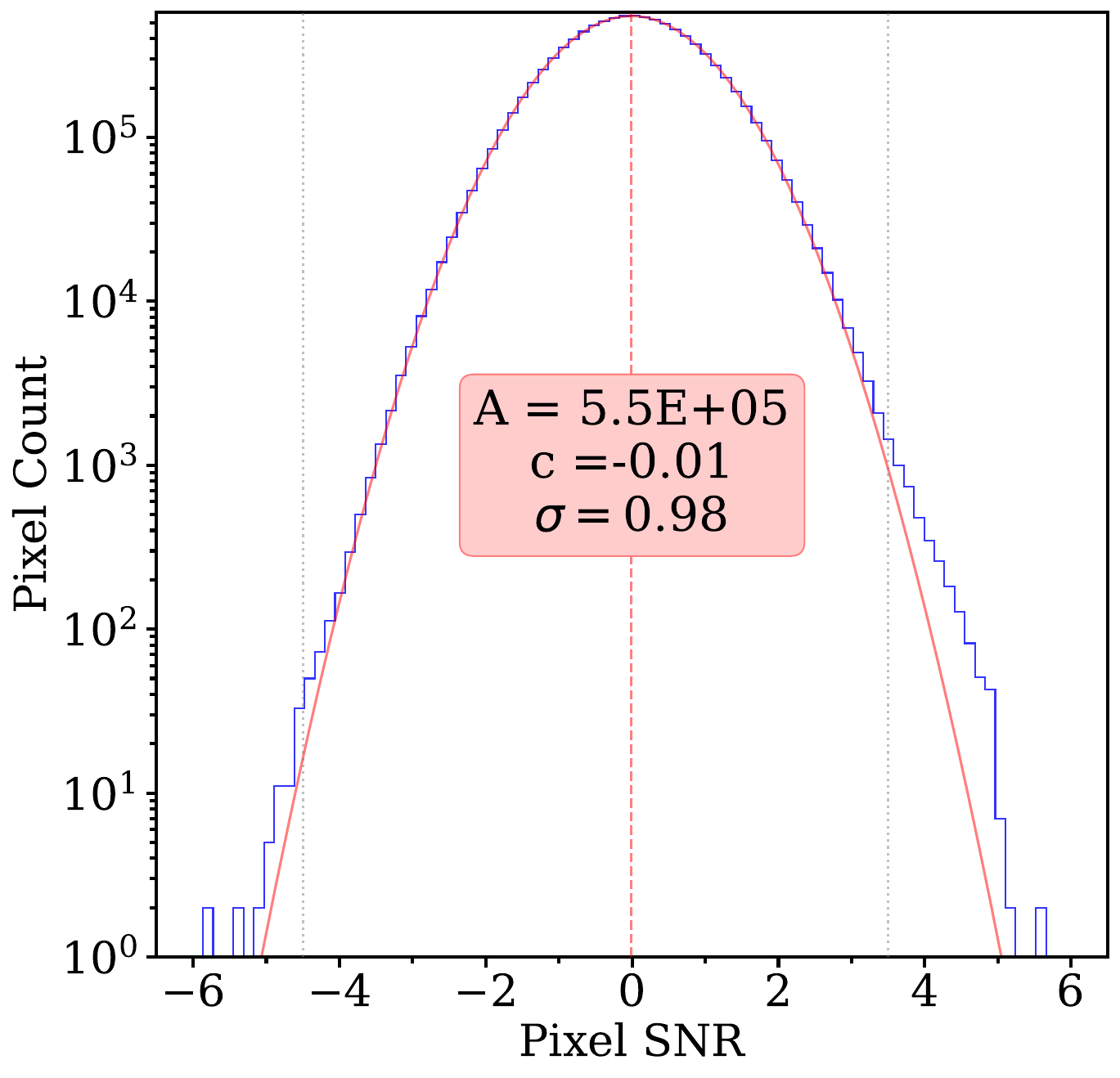}
        }%
        \subfigure[Class: A.2]{%
           \label{fig:second}
           \includegraphics[width=0.3\textwidth]{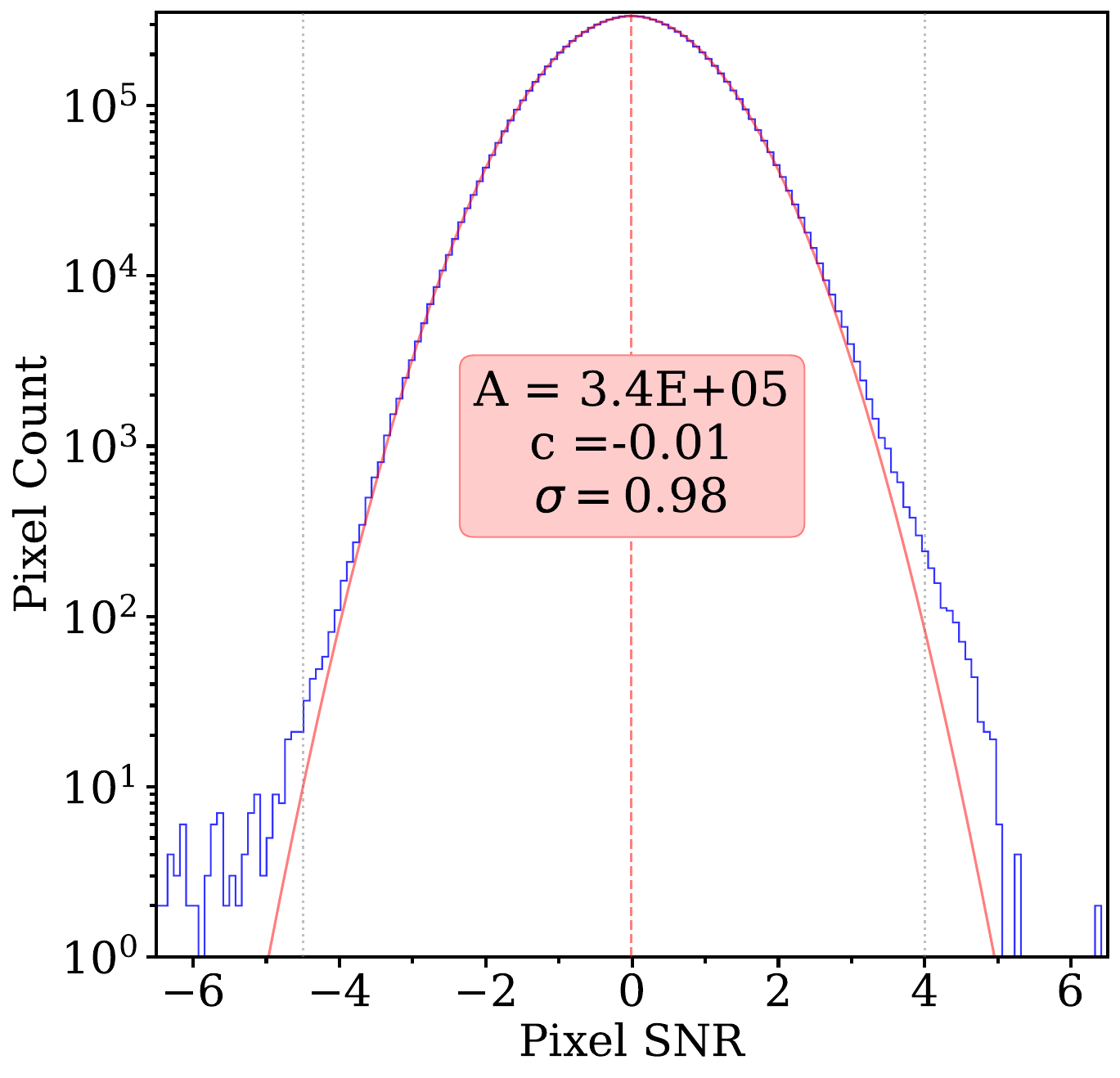}
        } 
        \subfigure[Class: A.3]{%
            \label{fig:third}
            \includegraphics[width=0.3\textwidth]{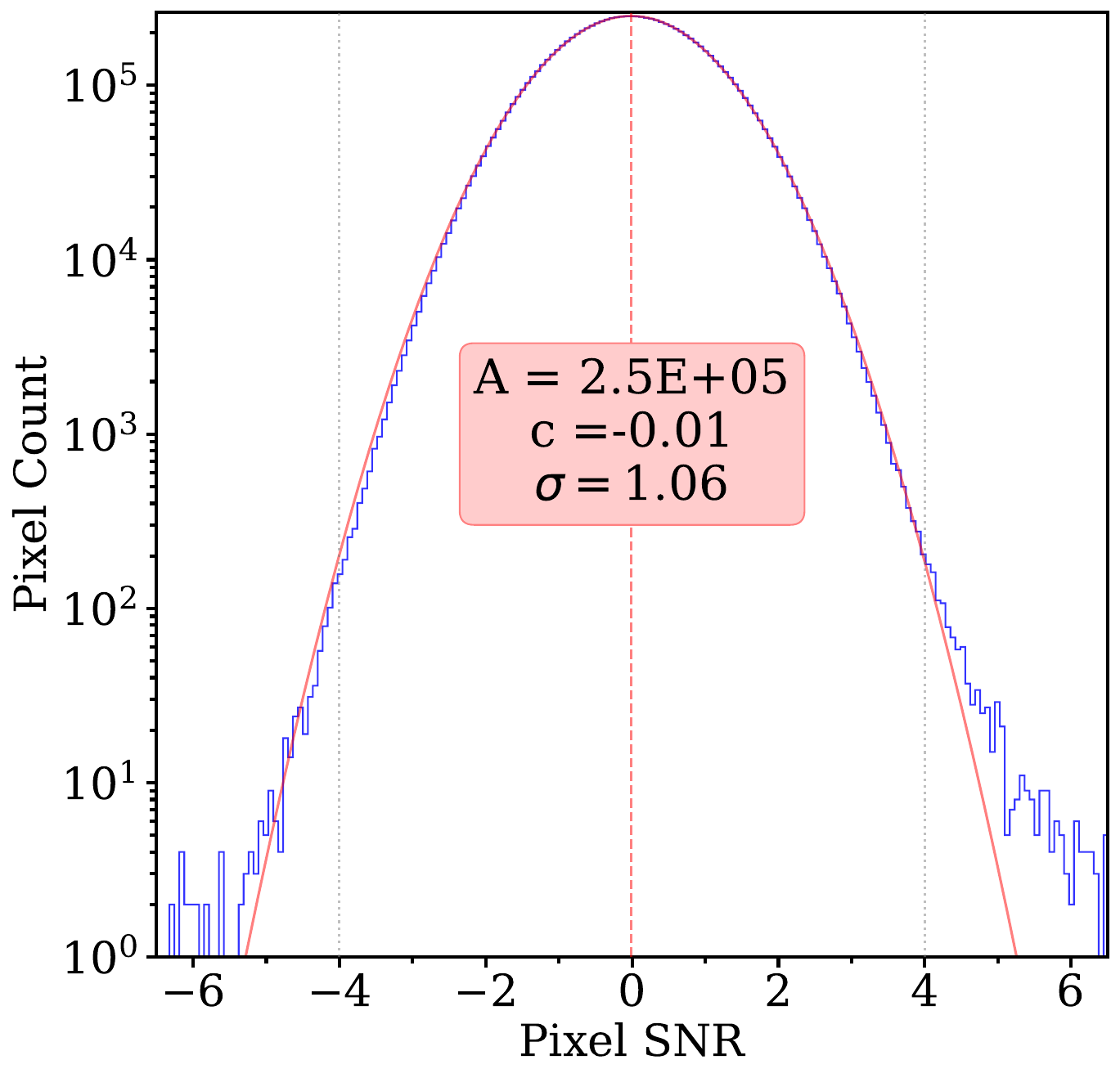}
        }\\%
        \subfigure[Class: A.4]{%
            \label{fig:fourth}
            \includegraphics[width=0.3\textwidth]{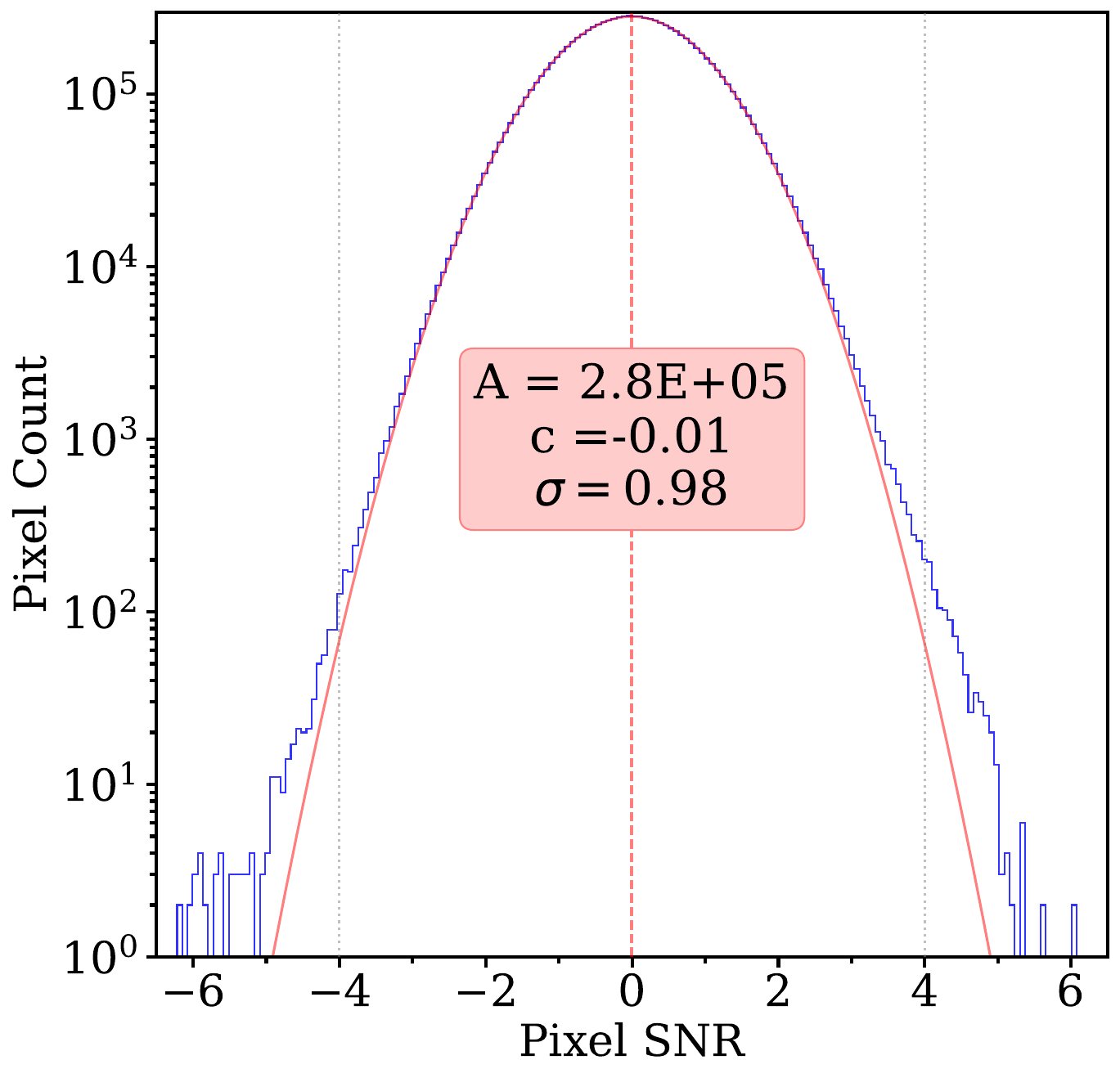}
        }%
        \subfigure[Class: B]{%
            \label{fig:fourth}
            \includegraphics[width=0.3\textwidth]{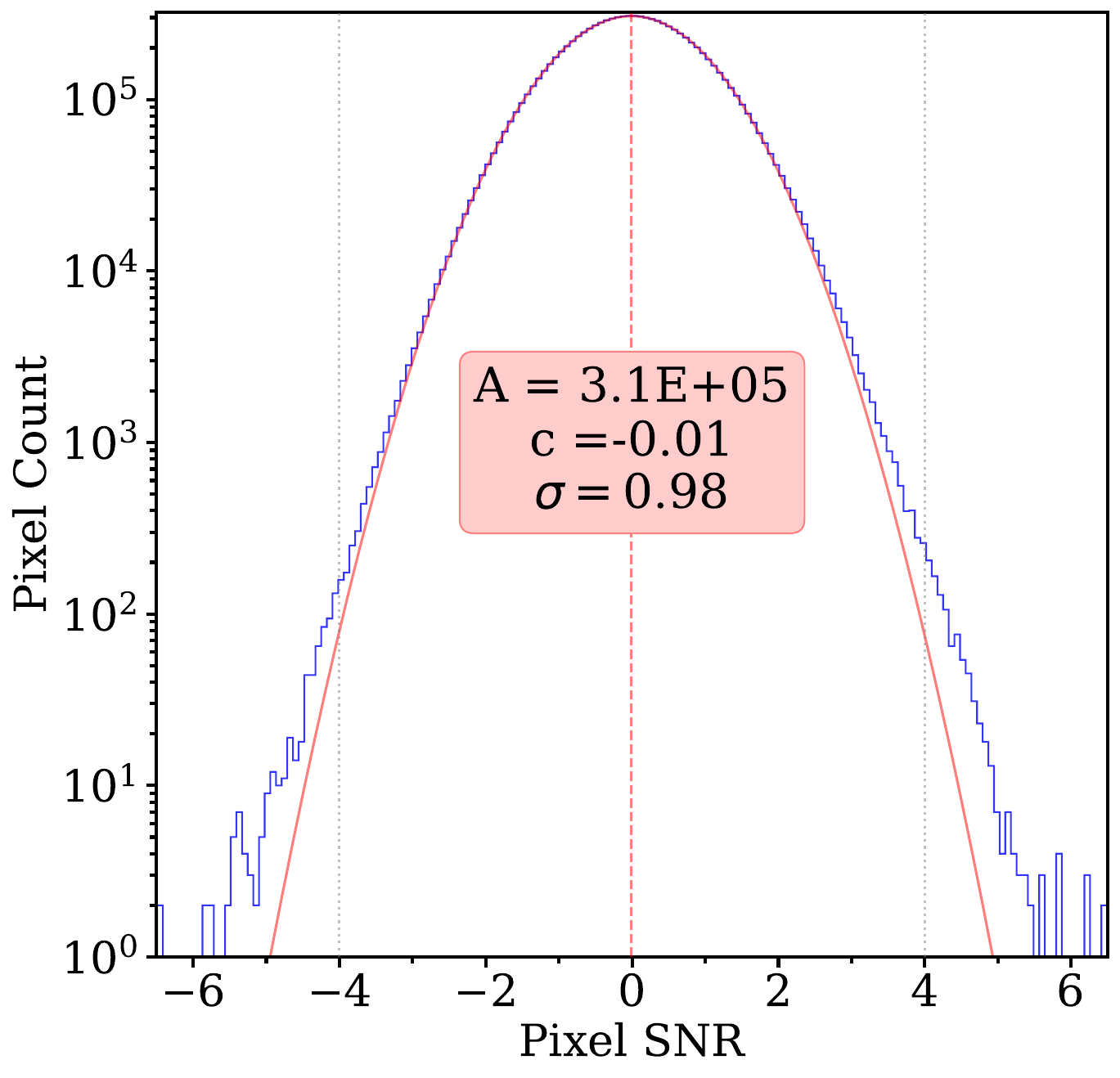}
        }%
        \subfigure[Class: B]{%
            \label{fig:fourth}
            \includegraphics[width=0.3\textwidth]{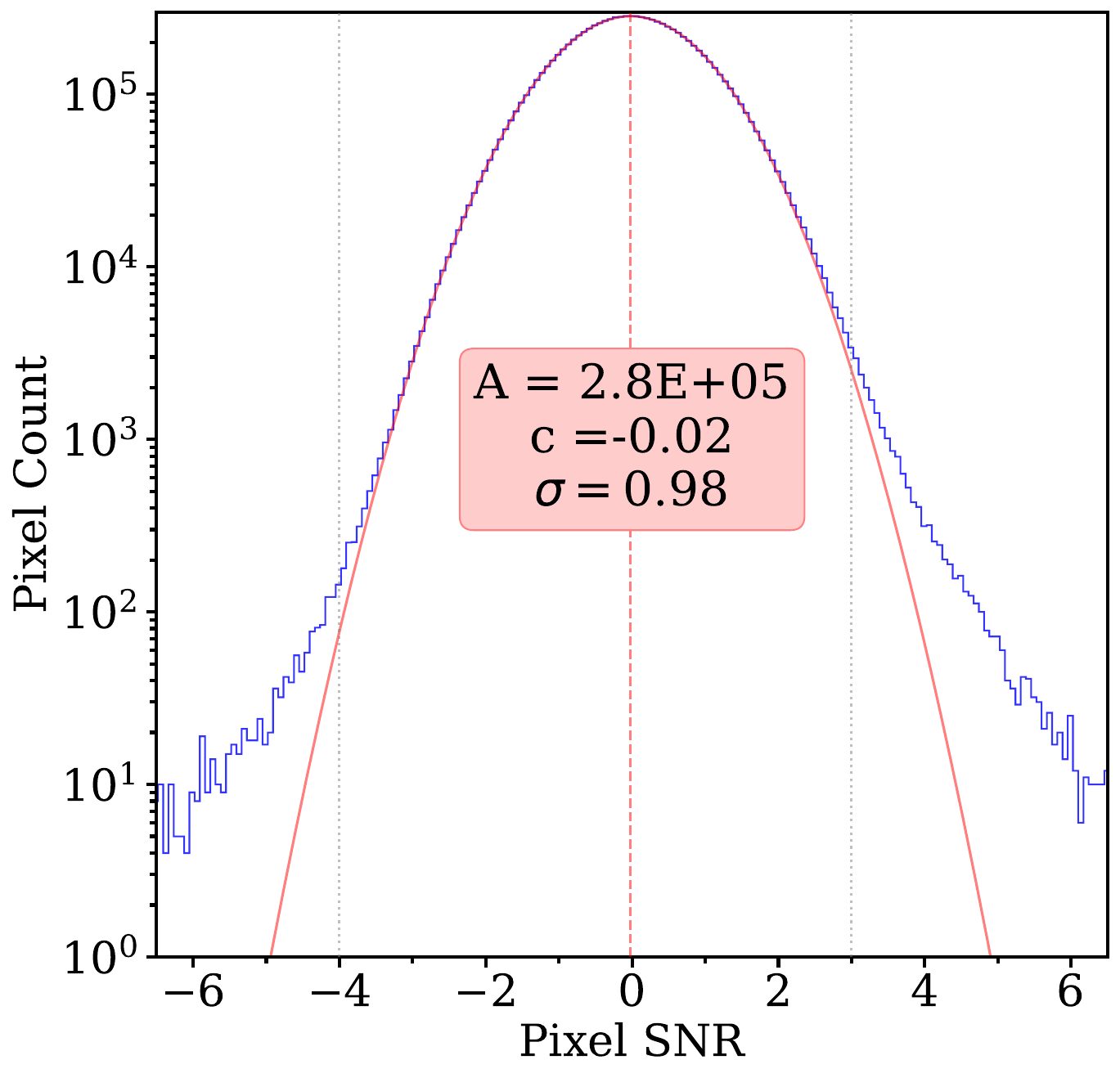}
        }%
    \end{center}
    \caption{%
        Examples of SNR distribution of pixels in the residual images generated by {\tt PyBDSF} for each class of pointings discussed in Section~\ref{sec:noise}.  The sigma-clipped Gaussian ($Ae^{-(x-c)^2/2\sigma^2}$) fitted to the distribution using only the SNR range marked using the dotted vertical lines is also shown.  
     }%
   \label{fig:noise_dist}
\end{figure*}

In Fig.~\ref{fig:noise_dist}, we show the SNR distribution of residual pixels from primary beam corrected images for each class of pointings discussed in Section~\ref{sec:noise}. The residual image is generated by {\tt PyBDSF} after subtracting all the fitted source components from an image. Therefore it should represent only the random background noise, whose SNR distribution is expected to be Gaussian. In each panel, the Y-axis is plotted in {\tt log}-scale to show any deviation from the fitted Gaussian. The dotted grey vertical lines indicate the data range used to fit the plotted Gaussian to the distribution. In majority of the cases only marginal deviation from the fit is seen. In cases where there is a significant excess emission towards the positive side (e.g., bottom right panel in Fig.~\ref{fig:noise_dist}), we inspected them visually and found that the dominant fraction of outlier pixels belong to `empty islands', i.e., islands where there is no Gaussian component fitted to the emission because {\tt thresh\_pix} is less than $5\sigma$. As an additional check, we also ran {\tt PyBDSF} on these images with an additional parameter {\tt incl\_empty=True}, which includes the empty islands in the source ({\tt `srl'}) catalog. Besides this, we also found excess positive pixels due to residual emission left during modelling of complex `M'-type sources as well as due to random positive noise peaks, although the contribution from these two factors are not always appreciable. 

The negative pixels with SNR$\leq$-5 can have three different origins: improper modelling of source emission resulting in negative pixels in the residual image after component subtraction, random negative noise peaks and strong negative peaks near bright sources due to statistical errors related to calibration. The first case affects the measurement of flux densities in poorly modelled (mostly `M'-type) sources. Considering only $\sim 10\%$ of sources in our catalog are of `M'-type, this should not affect our analysis significantly. Still, we recommend the user to compare the {\tt `Isl\_Total\_flux'} and {\tt `Total\_flux'} parameter to judge the quality of Gaussian component fits. The latter two causes of `bright' negative pixels are particularly responsible for contamination of the catalogs through the generation of false sources discussed above and can be eliminated from the analysis by considering sources detected at $>8\sigma$.

\section{Stokes-$I$ properties and accuracy}    
\label{sec:accur}  

We examine the astrometric and flux density accuracy of MALS catalogs by comparing them with NVSS and FIRST at 1.4\,GHz. FIRST is more sensitive and has $\sim$10 times better spatial resolution than NVSS (see Section~\ref{sec:intro}).  But only 119 MALS pointing centers are covered in FIRST.  In comparison, a total of 348 out of 391 MALS pointings overlap with the NVSS sky coverage ($\delta \gtrsim -40\degree$).    
Therefore, for optimal utilization of the available data, we split the analysis into {\it three} parts as following.  In the \textit{first} part involving MALS and NVSS, we consider only the targets at the pointing center.  These are detected at very high SNR ($>$5000), largely compact, and unaffected by errors from the primary beam correction.  In this part, we also include 64 out of 88 gain calibrators observed as part of MALS observations that are common with NVSS.  Like central targets, all these are also bright ($>1$\,Jy at 1.4\,GHz) and at the pointing center.  

As previously mentioned 4 out of 391 MALS pointings i.e., J1133$+$0040, J1142$-$2633, J1144$-$1455 and J2339$-$5523, were observed twice. In the {\it second} part involving only MALS, we compare the properties of all the sources from two observing runs to obtain an estimate of systematic errors in astrometry and flux densities.   The assumption here is that the majority of these sources are intrinsically non-variable within the timescale of observations: $\sim$ 8 days for J2339$-$5523 and $\sim$14 days for the remaining three.  

\begin{figure}
    \centering
    \includegraphics[width=\linewidth]{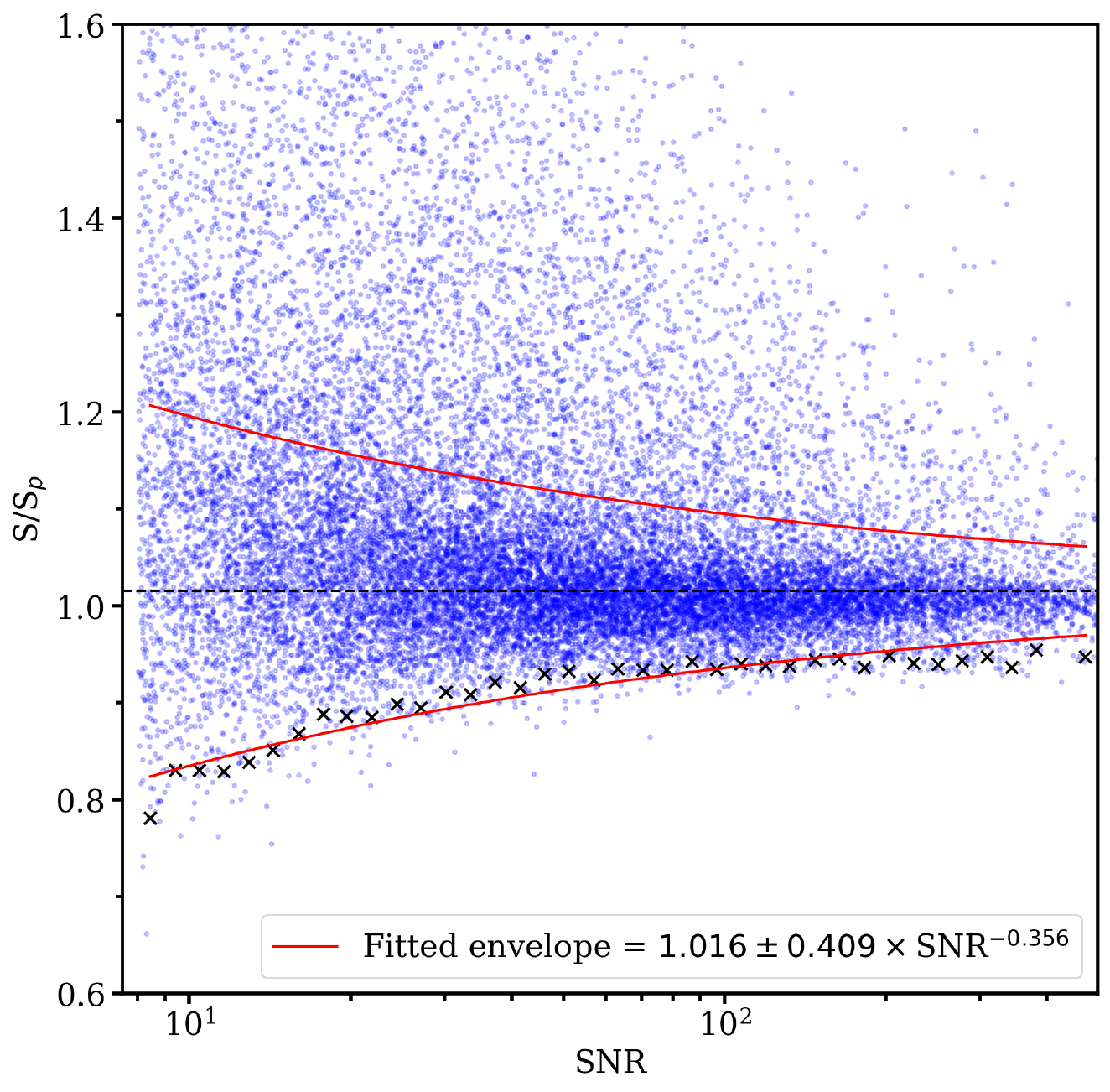}
    \caption{The reliability envelope used to select compact sources from a sample of isolated, single component sources (SNR$>$8) detected in MALS SPW9 images.  The black `$\times$' mark lower envelope encompassing 95\% of these sources with total-to-peak flux density ratio less than one.  The solid red line represents fit to `$\times$' and the reflected envelope. Out of 22,425 sources, 15,834 ($\sim$ 70\%) lie inside the envelope.
    }
    \label{fig:comp_envlp}
\end{figure}

In the {\it third} part, we extend the analysis to off-axis sources detected at SNR$>8$ in MALS SPW9 images. In order to minimize additional uncertainties due to resolution differences between these surveys, we limit the comparison to isolated and compact sources in MALS. 
For isolation, we consider SPW9 sources with no neighbour within $60^{\prime\prime}$ radius i.e., {\tt Distance\_NN} $>60^{\prime\prime}$. The adopted isolation radius is about three times the NVSS resolution ($\sigma$).  It is sufficiently large to exclude sources that are simple in NVSS but split into multiple sources or components in MALS. Such sources will have systematically larger positional and flux density offsets.
The issues arising from differing surface brightness sensitivities of the surveys can be controlled by selecting only compact sources in MALS.  For this we retain only sources with {\tt S\_Code=`S'} and apply the widely used procedure of deriving a SNR dependent `reliability' envelope encompassing 95\% of these sources with total-to-peak flux density ratio less than one \citep[Fig.~\ref{fig:comp_envlp};][]{Bondi08, Smolcic17, Shimwell17, Hale21}.  The derived envelope i.e., fit to `$\times$' in Fig.~\ref{fig:comp_envlp} is then reflected on the other side and all the sources outside the envelope are rejected.  
Note that the increased scatter in Fig.~\ref{fig:comp_envlp} at low SNRs may be  due to the elevated gain errors and noise \citep[for example, see Fig.~7 of][]{Shimwell17}. 
Overall, for MALS-NVSS comparison we obtain a sample of 15,834 compact sources from 22,425 isolated SPW9 sources (SNR$>8$; {\tt S\_Code=`S'}).
Next, we use the envelope method to further reject sources that may be compact in MALS but resolved in FIRST to obtain a sample of 7795 sources suitable for MALS-FIRST comparison.

In passing we note that 43/391 ($\sim 11\%$) pointings with $-72\degree$ $< \delta <$ $-40\degree$ do not overlap with NVSS and FIRST.  Therefore, these are not included in the astrometric and fluxscale comparisons. However, the observing conditions i.e., day-time versus night-time including the telescope elevation ranges as well as the image quality inferred from $\sigma_1$ and $\sigma_2$ (Section~\ref{sec:noise}) of these pointings are similar with respect to the rest.  Therefore,  we do not expect errors associated with these to behave any differently. 

\subsection{Astrometric precision}    
\label{sec:astrom}  
%
%
\begin{figure*}[t]
     \begin{center}
        \subfigure{%
            \label{fig:first}
            \includegraphics[width=0.49\textwidth]{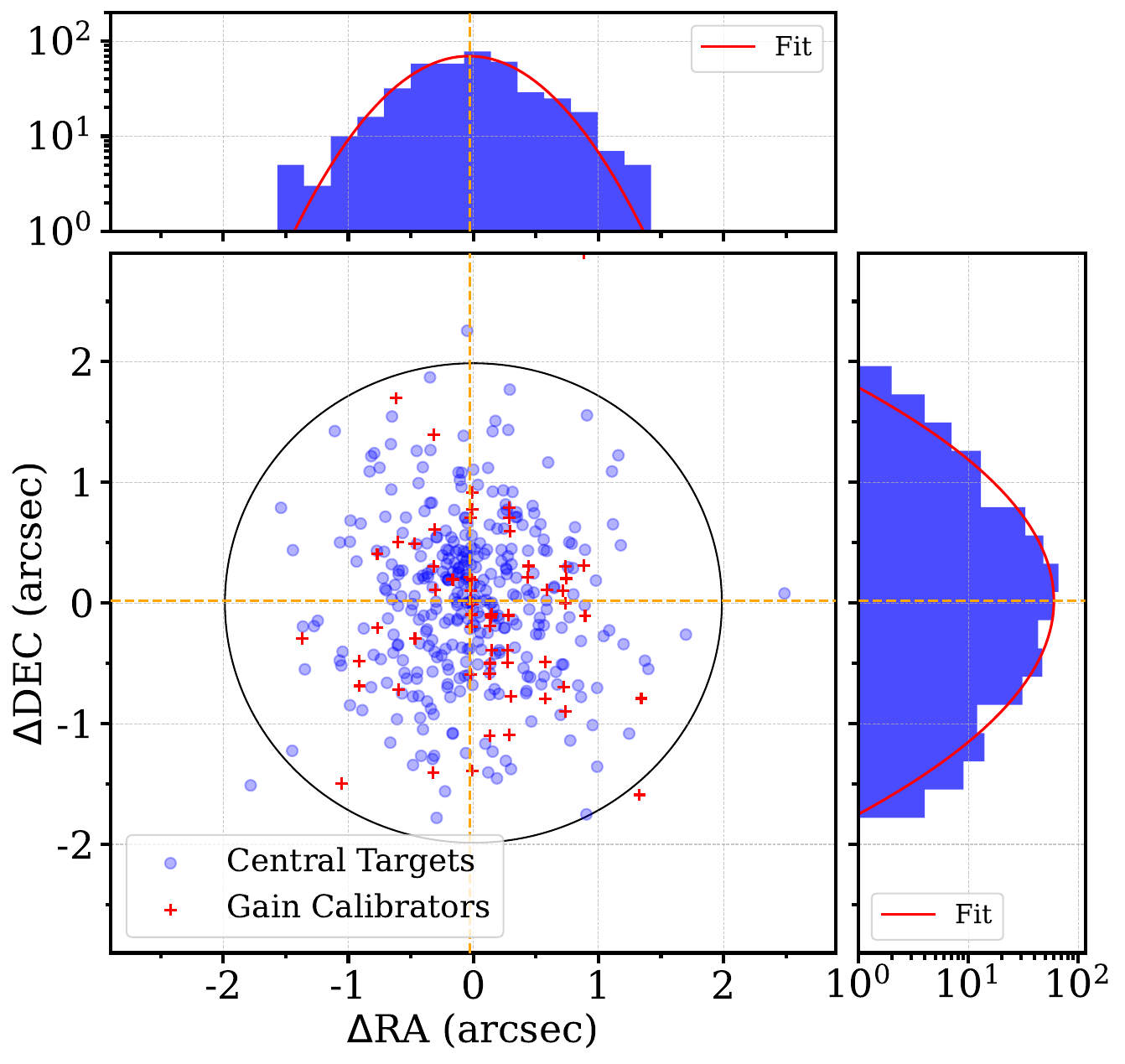}
        }%
        \subfigure{%
           \label{fig:second}
           \includegraphics[width=0.48\textwidth]{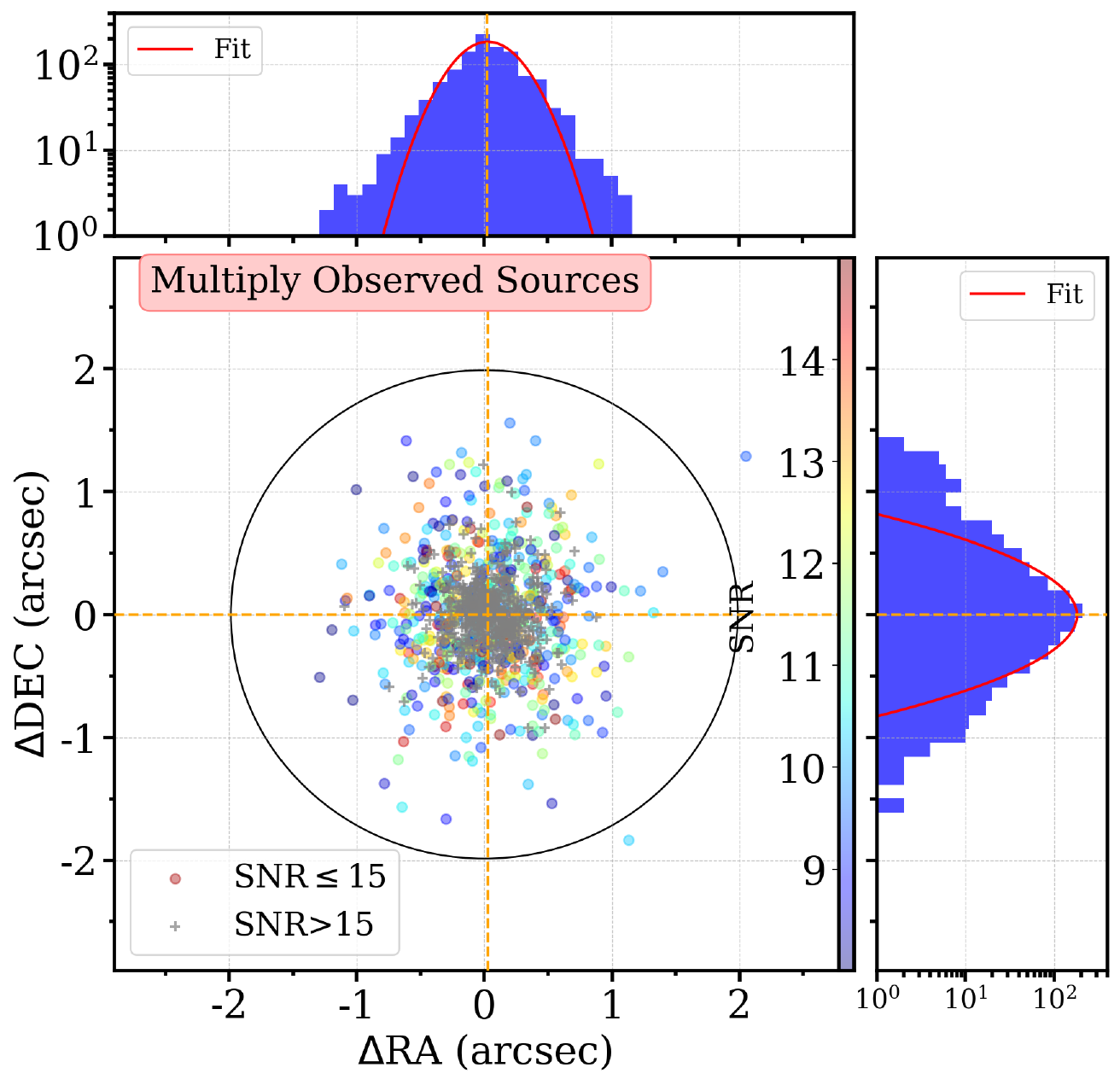}
       }
    \end{center}
    \caption{%
        Astrometric comparison of 345 targets and 64 gain calibrators from MALS with NVSS \textit{(left)} and  1,150 compact sources detected in four twice-observed MALS fields \textit{(right)}. The dashed lines mark median offsets. The circle represents half of the average SPW9 restoring beam FWHM ($8^{\prime\prime}$).
        For clarity, three points have been omitted from the \textit{left} panel (see text for details).
        In the \textit{right} panel, sources with $8\leq$SNR$\leq$15 are color-coded. The histogram distributions of $\Delta$RA and $\Delta$DEC and Gaussian fits to these are also shown.
     }%
   \label{fig:CentralAndMOS_ast}
\end{figure*}
%

The left panel of Fig.~\ref{fig:CentralAndMOS_ast} shows astrometric comparison for central targets and gain calibrators.  The median offsets in right ascension and declination i.e., $\Delta$RA and $\Delta$DEC are $-0\farcs03$ and $0\farcs02$, respectively. The median absolute deviations (MAD) in $\Delta$RA and $\Delta$DEC are $0\farcs32$ and $0\farcs42$, respectively.

More accurate positions for 35 MALS central targets and 71 gain calibrators are available from the VLA Calibrator Manual\footnote{\url{https://science.nrao.edu/facilities/vla/observing/callist}}.  
The median $\Delta$RA = $0\farcs00$ (MAD = $0\farcs04$) and $\Delta$DEC = $0\farcs$01 (MAD = $0\farcs$04) from comparison of these are even smaller.
The histogram distributions of $\Delta$RA and $\Delta$DEC are also shown in Fig.~\ref{fig:CentralAndMOS_ast}.  These are well modelled by Gaussian functions with $\sigma$ = $0\farcs48$ and $0\farcs62$, respectively, and are consistent with the scatter ($\sigma$ = 1.483$\times$MAD) estimated from MADs. 
We note that three central targets i.e., J0110-1648,  J1234+0829, and J2218+2828 have complex morphology in MALS and inappropriate for astrometric comparison. Hence, these have been omitted from Fig.~\ref{fig:CentralAndMOS_ast}.

\begin{figure}
    \centering
    \includegraphics[width=1\linewidth]{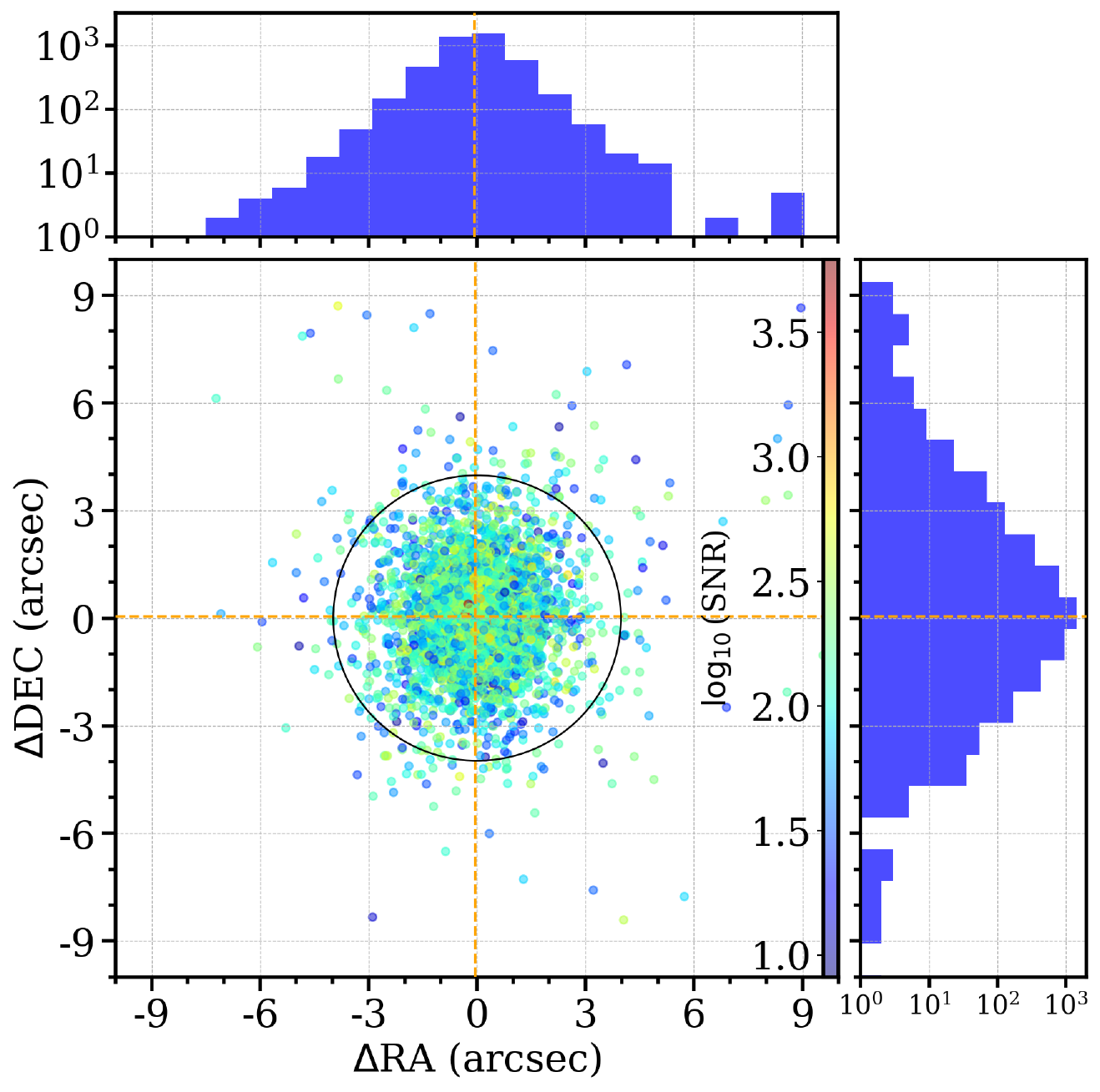}
    \caption{Astrometric comparison of 4,506 compact and isolated field sources from MALS with NVSS.  The dominant fraction ($\sim 95\%$) are within the restoring beam ($\sim 8^{\prime\prime}$; circle). The dashed lines mark median offsets.}
    \label{fig:astrometry_tgts_field}
\end{figure}

In the right panel of Fig.~\ref{fig:CentralAndMOS_ast}, we provide astrometric comparison of 1,150 sources detected (SNR$>8$; {\tt S\_code = `S'}) in four multiply observed MALS pointings.  
The median right ascension and declination offsets between sources detected in the two epochs are $0\farcs03$ (MAD = $0\farcs$17) and $0\farcs00$ (MAD = $0\farcs20$), respectively.
The scatter is slightly larger at lower SNR (see color-coded points for SNR$<15$) and a single Gaussian is not a good fit to the distributions of $\Delta$RA ($\sigma$ = $0\farcs$25) and $\Delta$DEC ($\sigma$ = $0\farcs$25) but consistent within 20\% with the scatter inferred from robust MAD statistics.
Overall, the scatter is quite small compared to the size of the average SPW9 synthesized beam, which, for clarity, is shown as a circle of diameter $4^{\prime\prime}$, i.e., half of its actual size. Also, prior to mixing the sources from individual fields to generate the combined sample of multiply observed sources, we have verified that each of the fields (the 4 fields span a declination range of $\sim 50\degree$) show similar distributions of astrometric offsets and therefore the results reported here are not biased by a particular set(s) of observations. 

\begin{figure}
    \centering
    \includegraphics[width=1\linewidth]{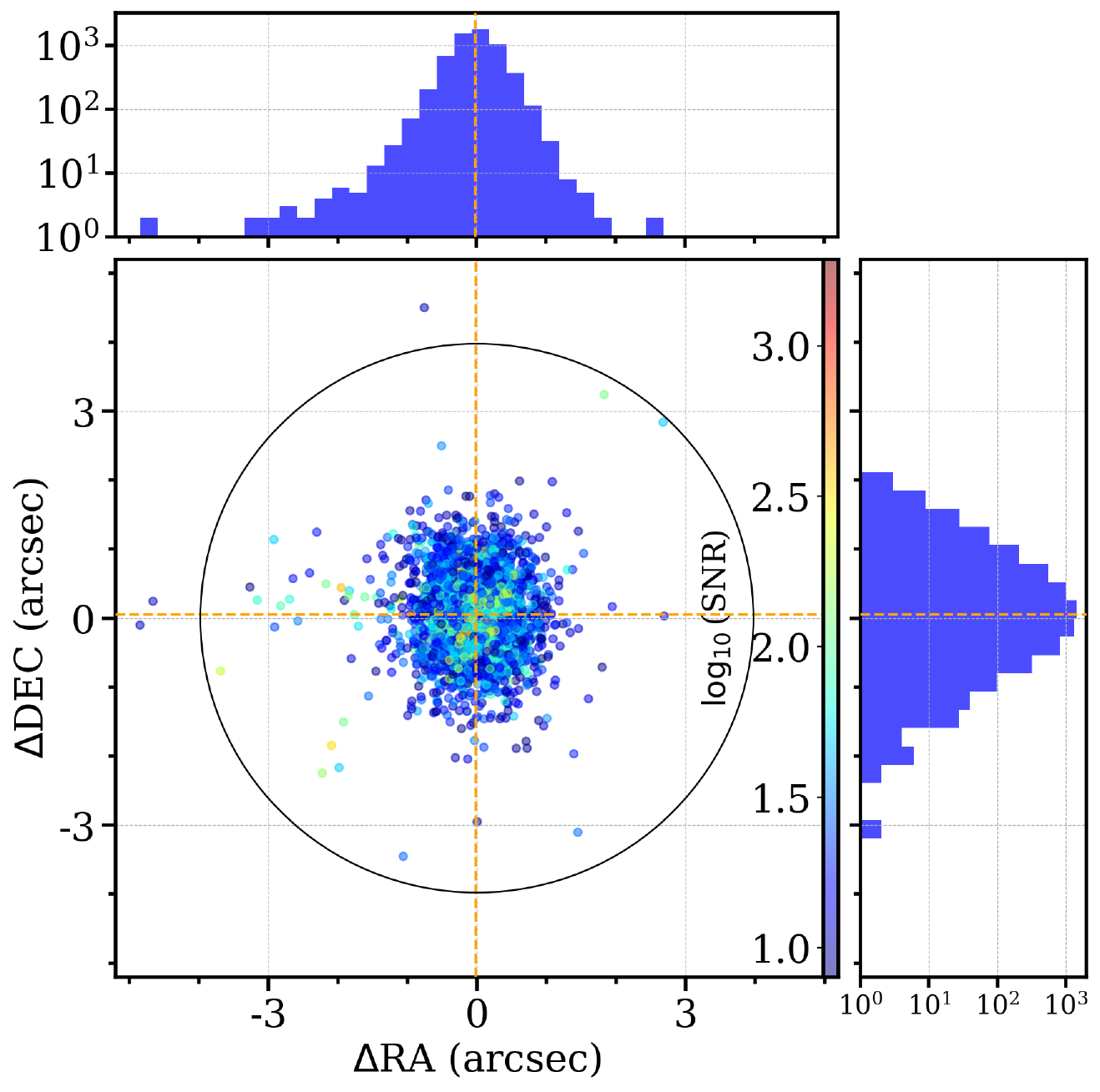}
    \caption{Astrometric comparison of 5,990 sources from MALS with FIRST. The remaining details are same as in Fig.~\ref{fig:astrometry_tgts_field}.}
    \label{fig:astrometry_tgts_fields_FIRST}
\end{figure}

For the {\it third} part of the comparison with NVSS involving all the compact and isolated sources detected in SPW9 images, we consider sources brighter than 10\,mJy in NVSS.  This reduces the scatter introduced by position uncertainties in NVSS which increase from $\sim 1^{\prime\prime}$ at an integrated flux density of 10 mJy to $\sim 6^{\prime\prime}$ at the 5$\sigma$ detection threshold of 2.5 mJy (see Fig. 30 in \cite{Condon98}).  The astrometric comparison of these is shown in Fig.~\ref{fig:astrometry_tgts_field} with individual points color-coded according to their SNR values in {\tt log} scale. The median right ascension and declination offsets are $-0\farcs05$ (MAD = $0\farcs63$) and $0\farcs02$ (MAD = 0$\farcs$78), respectively. We found that $\sim 95\%$ of the sources show a positional mismatch with NVSS that is smaller than MALS SPW9 average restoring beam (see circle in Fig.~\ref{fig:astrometry_tgts_field}). 
From the comparison with sources (SNR$>$10) compact in FIRST (Fig.~\ref{fig:astrometry_tgts_fields_FIRST}), the  median $\Delta$RA and $\Delta$DEC are $0\farcs 02$ (MAD$=0\farcs21$) and 0\farcs 05 (MAD$=0\farcs 29$), respectively.


Overall, small i.e., sub-pixel level values of median $\Delta$RA and $\Delta$DEC indicate that the systematic errors associated with source positions are under control. 
Therefore, we do not apply any offsets to the images or positions reported in the catalog.
The offsets obtained from the comparison with FIRST are comparable to those estimated using multiply observed sources, affirming that these provide a reasonable estimate of the systematic errors associated with the astrometry.
To capture the SNR dependence of $\Delta$RA and $\Delta$DEC (see distributions in Fig.~\ref{fig:astrometry_tgts_fields_FIRST}), we grouped them in bins consisting of 400 or more sources.  We modeled offsets in each SNR bin using a Gaussian and also estimated the scatter ($\sigma$) for each bin using MAD statistics. The two estimates are consistent within 10\%. Since the MAD based estimate is robust to outliers and systematically larger, we adopt this as a slightly conservative contribution to the systematic error ($\sigma_{astrom, sys}$) budget. 
We model the SNR dependent behavior of scatter in $\Delta$RA and $\Delta$DEC  as $\sigma_{astrom, sys}$ = ($1.8\times{\rm SNR}^{-0.8}$ + 0.2) and ($3.1\times{\rm SNR}^{-1.0}$ + 0.3), respectively.  
These offsets level off at SNR$\geq$80 with values of  $\Delta$RA = $0\farcs2$ and $\Delta$DEC = $0\farcs3$.

The errors on ({\tt RA\_mean}, {\tt DEC\_mean}) and  ({\tt RA\_max}, {\tt DEC\_max}) reported in Table~\ref{tab:cat_cols} have been estimated following: 
\begin{equation}
    \sigma = \sqrt{\sigma_{astrom, fit}^2 + \sigma_{astrom, sys}^2} , 
\label{eq:astroer}
\end{equation}
where $\sigma_{astrom, fit}$ is the error in right ascension or declination from {\tt PyBDSF} fitting, and $\sigma_{astrom, sys}$ is the systematic error based on the analysis of offsets in off-axis (for SNR$<400$) and central sources (SNR$>400$). 
In MALS DR1, the same recipe has been used to derive astrometric errors for SPW2.
Note that for SNR$=$8 (15), $\sigma_{astrom, sys}$ =  $0\farcs5$ ($0\farcs4$) and $0\farcs7$ ($0\farcs5$), respectively. 
For the catalog, the astrometric error corresponding to the median SNR of 9 is 0\farcs8. 

\begin{figure}
    \centering
    \includegraphics[width=1\linewidth]{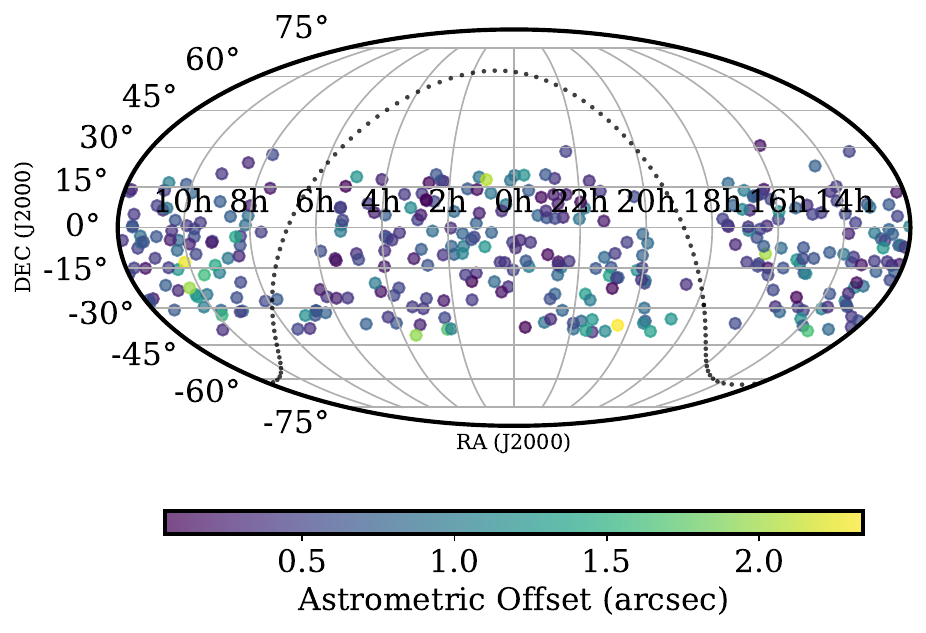}
    \caption{ Sky distribution of MALS pointings -- the points have been color-coded on the basis of median astrometric offsets between MALS and NVSS. The MALS pointings at $\delta <-40\degree$ do not overlap with NVSS and, hence, are absent in this plot.}
    \label{fig:astrometry_mollweide}
\end{figure}

Finally, to investigate variations in astrometric accuracy across the survey footprint, we estimated astrometric offsets for each pointing as the median of $\sqrt{(\Delta \text{RA})^2 + (\Delta \text{DEC})^2}$ for all the compact and isolated sources within the pointing. Due to better overlap with the MALS footprint, we use NVSS for this purpose. The results are shown in Fig~\ref{fig:astrometry_mollweide}. Clearly, there are no significant deviations across the survey footprint -- neither in RA nor in Dec for individual pointings or when grouped into bins of $10\degree$. We also do not find any relationship between the offsets and the flux density of the central source.  The two pointings with most extreme offsets of $\sim2\farcs5$ are  J1007$-$1247 and J2023$-$3655.

\subsection{Flux density scale}    
\label{sec:fluxscale}  
%
\begin{figure*}[ht!]
     \begin{center}
        \subfigure{%
            \label{fig:first}
             \includegraphics[width=0.46\linewidth]{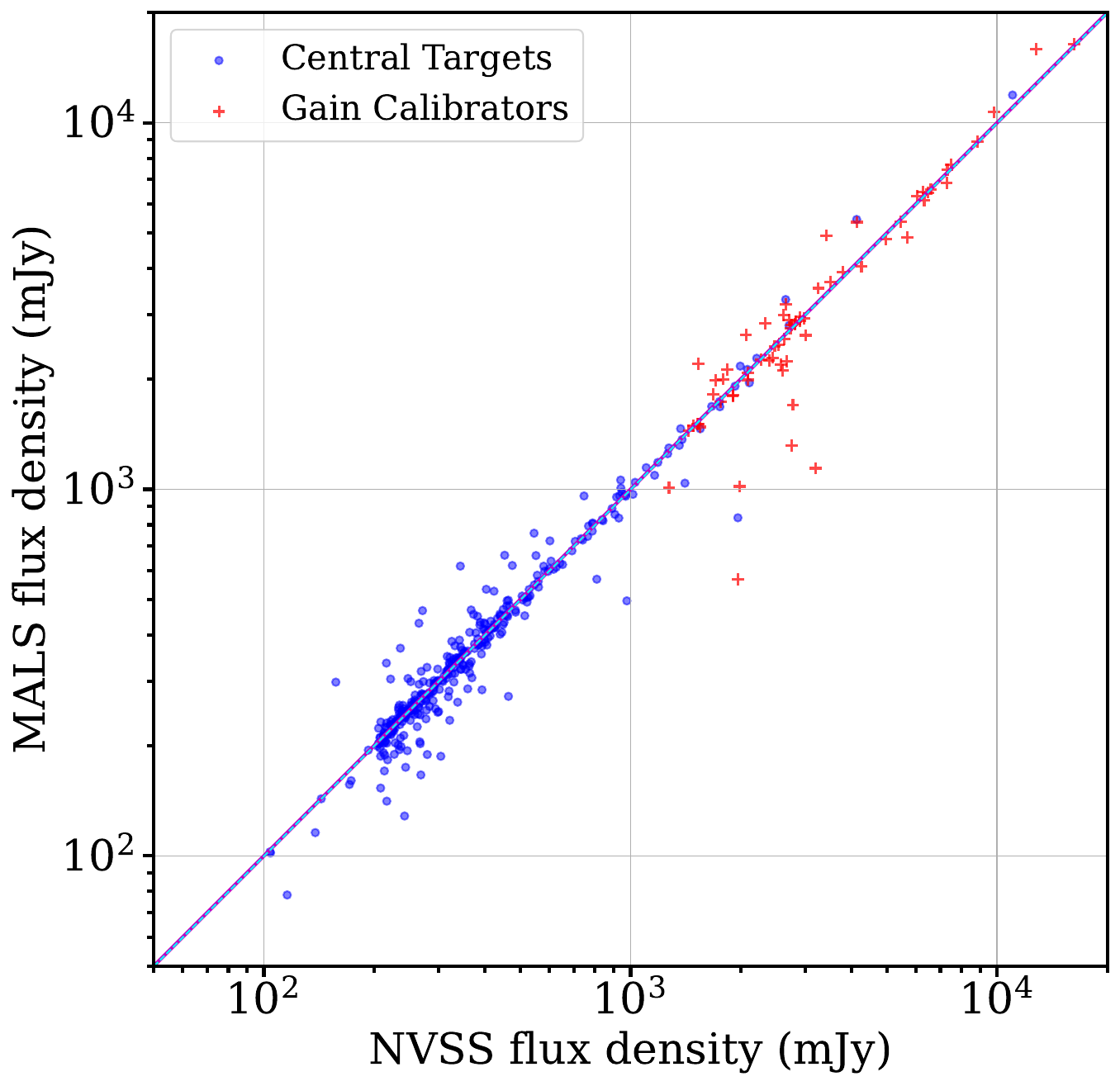}
         }%
        \subfigure{%
           \label{fig:second}
           \includegraphics[width=0.512\textwidth]{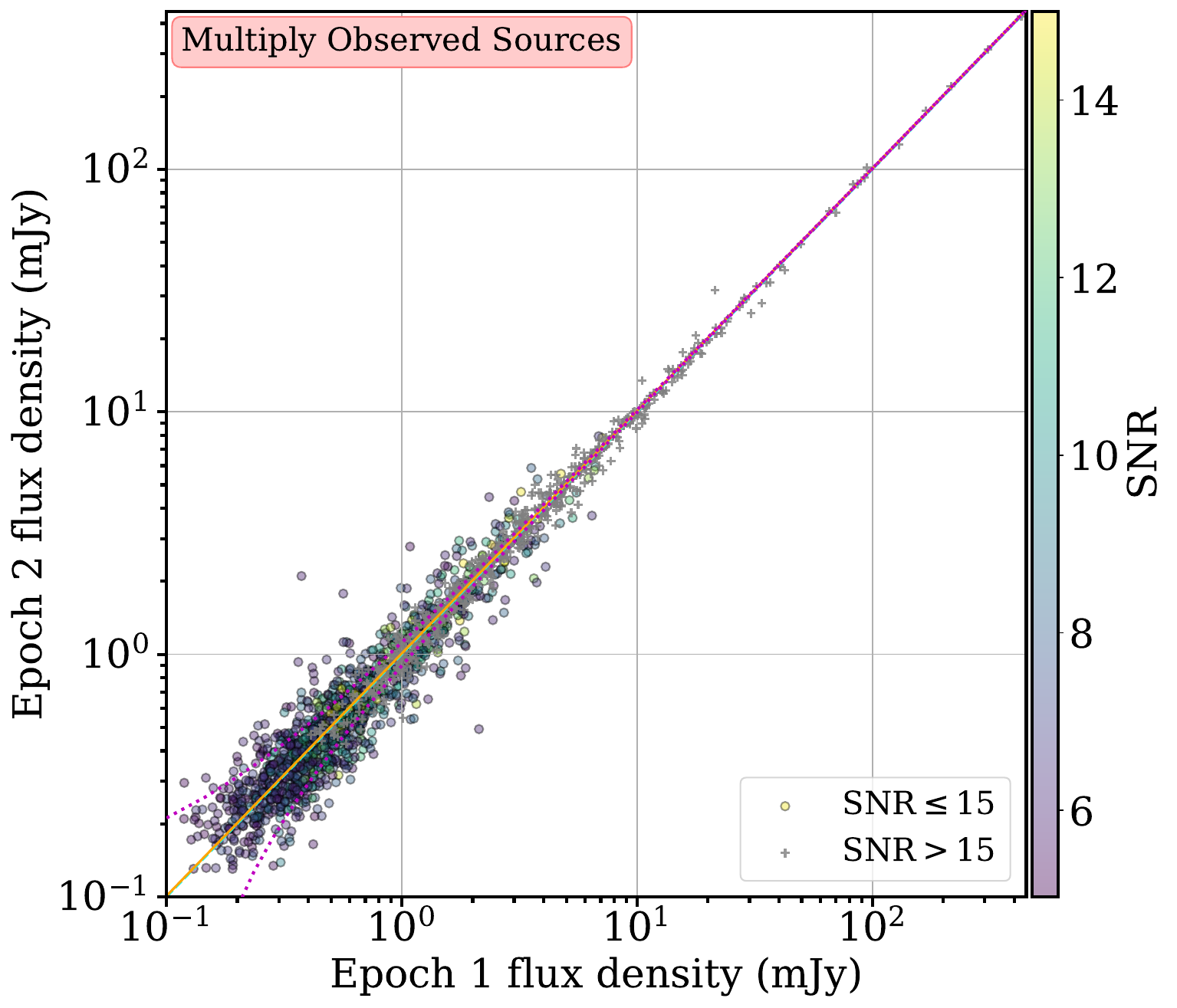}
       }
    \end{center}
    \caption{%
    Integrated flux density comparison at 1.38\,GHz of 345 central targets and 64 gain calibrators from MALS with NVSS \textit{(left)} and 1,150 compact sources (SNR$>$8) detected in four  twice-observed MALS fields \textit{(right)}.  The dashed lines mark median offsets.  In the \textit{right} panel, sources with $8\le$ SNR$\le$15 are color coded. 
    }%
    \label{fig:fratio_targets_MOSs}
\end{figure*}

In the left panel of Fig.~\ref{fig:fratio_targets_MOSs}, we compare the flux density measurements of MALS central targets and gain calibrators with NVSS. We reject three central targets with complex morphology in MALS (Section~\ref{sec:astrom}). Also, to account for the small frequency difference ($\Delta\nu \approx$ 20 MHz) between NVSS (at 1.40 GHz) and MALS-SPW9 catalogs (at 1.38 GHz), we scaled the NVSS flux densities to the frequency of our observations using a spectral index of $\alpha=-0.74$ (see Section~\ref{sec:uss}). 
Further, due to the coarser spatial resolution (FWHM = $45^{\prime\prime}$) a single NVSS source may split into multiple sources in MALS.  Among 345 central targets, 72 have additional radio sources in MALS within $60^{\prime\prime}$ radius and the remaining are isolated.  The median MALS-to-NVSS flux ratio considering all the central targets and gain calibrators is 1.00 (MAD = 0.04), implying that the flux densities of radio sources at the pointing center are in excellent agreement with NVSS.  
In all cases, the total flux density of additional sources is small (median$\sim$1\% of the NVSS flux) and, therefore, inconsequential to the sample statistics.  
Note that several gain calibrators are observed multiple times in MALS.  For comparison with NVSS, we have taken the average of their flux density measurements.  
In Fig.~\ref{fig:fratio_targets_MOSs}, the five outliers among gain calibrators are blazars, well known in the literature for their variability at radio wavelengths.
%

The right panel of Fig.~\ref{fig:fratio_targets_MOSs}  provides a comparison of  flux densities of 1,150 sources (SNR$>8$; $S\_code$ = `S') detected in four multiply observed MALS pointings.   
The median integrated flux density ratio is 1.01 (MAD = 0.08).
This increase in the scatter as compared to the central targets can be attributed to the low-SNR ($\leq$15) sources.  The latter when treated separately have a MAD of 12\%. In contrast, the high-SNR ($>$15) sources alone, exhibit a MAD of only 5\%, similar to the central sources ($\sim 4\%$). In conclusion, at low SNR, issues related to inaccurate modelling of source emission lead to a increased scatter in the distribution.

\begin{figure}
    \centering
    \includegraphics[width=1.0\linewidth]{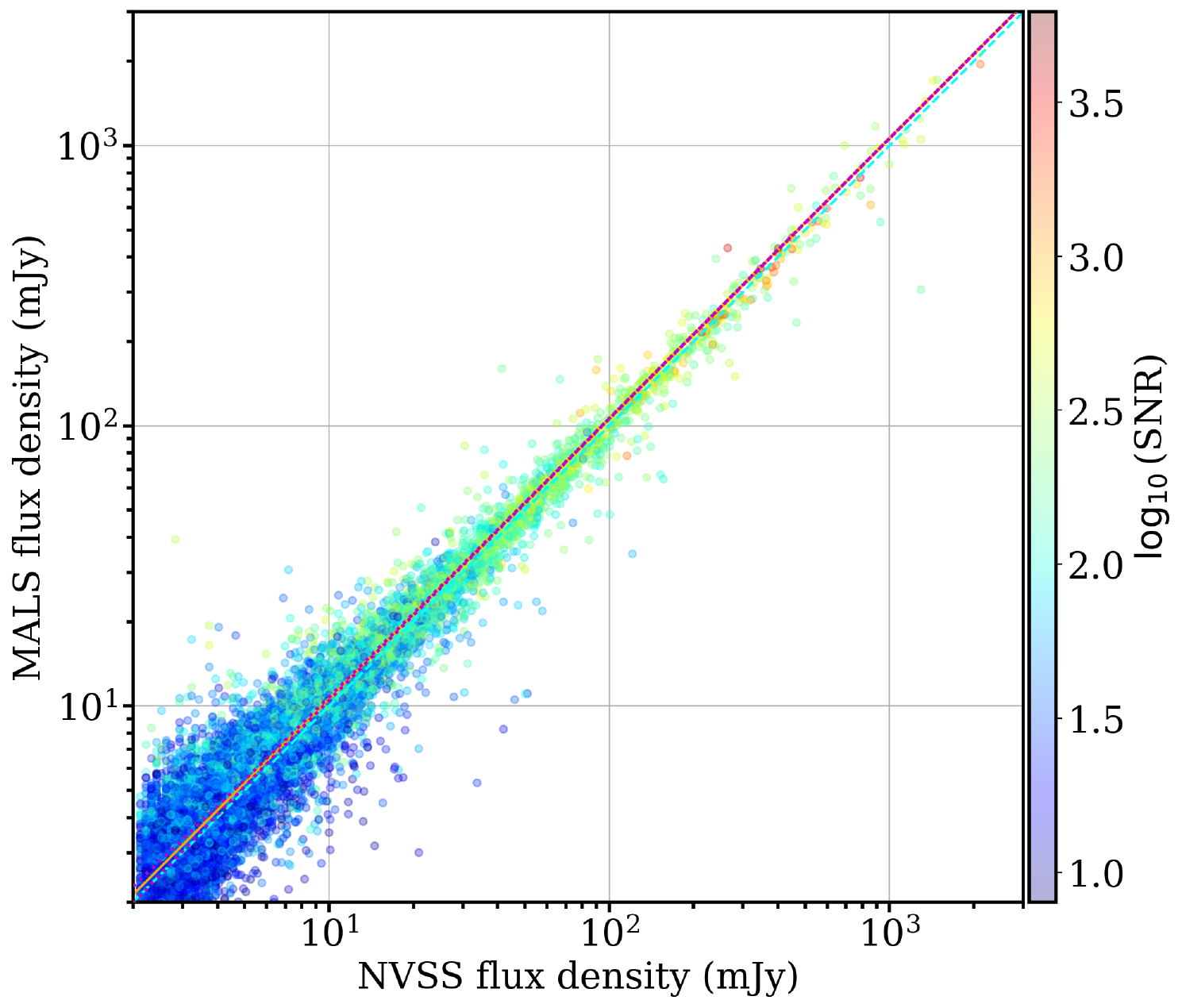}
    \caption{Flux density comparison at 1.38\,GHz of 15,834 compact and isolated sources from MALS with NVSS.}
    \label{fig:fratio_tgts_field}
\end{figure}

Next, we compared the MALS and NVSS flux densities of 15,834 compact and isolated sources detected over the entire MALS footprint (Fig.~\ref{fig:fratio_tgts_field}).  At 1.38\,GHz, the median MALS-to-NVSS ratio is 1.06 (MAD = 0.15). The MALS-to-FIRST ratio estimated using 5,990 compact sources is 1.12 (MAD = 0.15).
Restricting the comparison to brighter ($>$10\,mJy) 4,506 sources (30\% of 15,834) in NVSS, we find a median MALS-to-NVSS ratio of 1.03 (MAD = 0.09). 
This is very similar to the values obtained from the comparison of multiply observed sources presented in \textit{right} panel of Fig.~\ref{fig:fratio_targets_MOSs}. 
In conclusion, the overall observed flux density offset of 6-12\% between these surveys is well within the absolute flux density accuracy expected at these frequencies ($\sim$ 1 GHz). 
Thus, we do not apply any adjustment to the flux density scale of MALS sources.

The fitting errors ({\tt Total\_flux\_E\_fit}) on flux densities from {\tt PyBDSF} are likely underestimated.  The larger scatter in flux density comparisons at lower SNRs could be due to improper Gaussian modelling caused by confusion with adjacent noise pixels. 
The flux density comparison of MALS with NVSS and FIRST could also be affected by additional sources of error e.g., direction-dependent errors including the accuracy of primary beam correction and long-term variability of AGNs.  Therefore, we use the comparison between multiply observed sources in MALS to obtain an estimate of SNR dependent systematic uncertainty ({\tt Total\_flux\_E\_sys}) in the flux density measurement.
For this, we fit a simple power-law, $1.13\pm 0.01\times$ SNR$^{-0.743 \pm 0.002}$, to the scatter (1.483$\times$MAD) in the percentage variation observed in flux density measurements of these sources (Fig.~\ref{fig:fratio_targets_MOSs}; \textit{right} panel).   
The systematic error is then given by, {\tt Total\_flux\_E\_sys} = {\tt Total\_flux} $\times$ $1.13$ $\times$ SNR$^{-0.74}$.
The total error is then calculated as the quadratic sum of the fitting and systematic errors as, 
\begin{equation}
\begin{split}
   &{\tt Total\_flux\_E} = \\
   &\sqrt{ {\tt Total\_flux\_E\_fit}^2  + {\tt Total\_flux\_E\_sys}^2} .
\end{split}
\label{eq:fluxer}
\end{equation}
In MALS DR1, the same recipe has been used to derive total errors on flux densities for SPW2.

\begin{figure}
    \centering
    \includegraphics[width=1\linewidth]{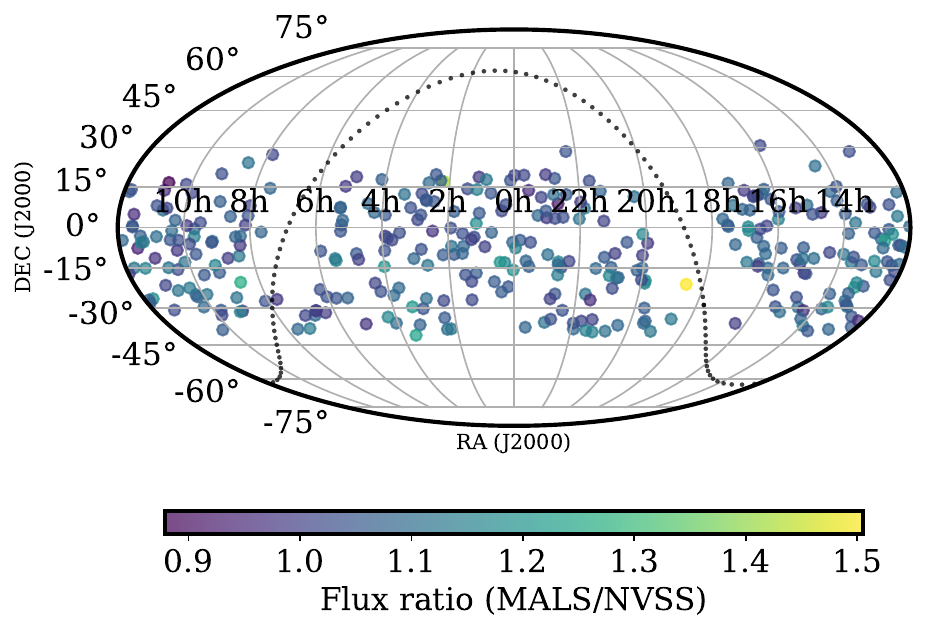}
    \caption{Sky distribution of MALS pointings for median flux density ratios between MALS and NVSS. The remaining details are same as in Fig.~\ref{fig:astrometry_mollweide}}.
    \label{fig:fratio_mollweide}
\end{figure}

Finally, in Fig.~\ref{fig:fratio_mollweide} we present the median flux density ratios of compact and isolated sources for each pointing. Clearly, there are no systematic trends across the survey footprint neither in RA nor in Dec for individual pointings or when grouped in bins of $10\degree$. 
However, two pointings i.e., J1833-2103 and J0211+1707 associated with central sources of $\sim$10\,Jy and $\sim$0.7\,Jy show extreme median offsets ($\sim$30\%).  In general, we do not find any relationship between the offsets and the flux density of the central source. 
%

\subsection{Accuracy of primary beam correction}    
\label{sec:pbaccur}  

\begin{figure}
    \centering
    \includegraphics[width=1.0\linewidth]{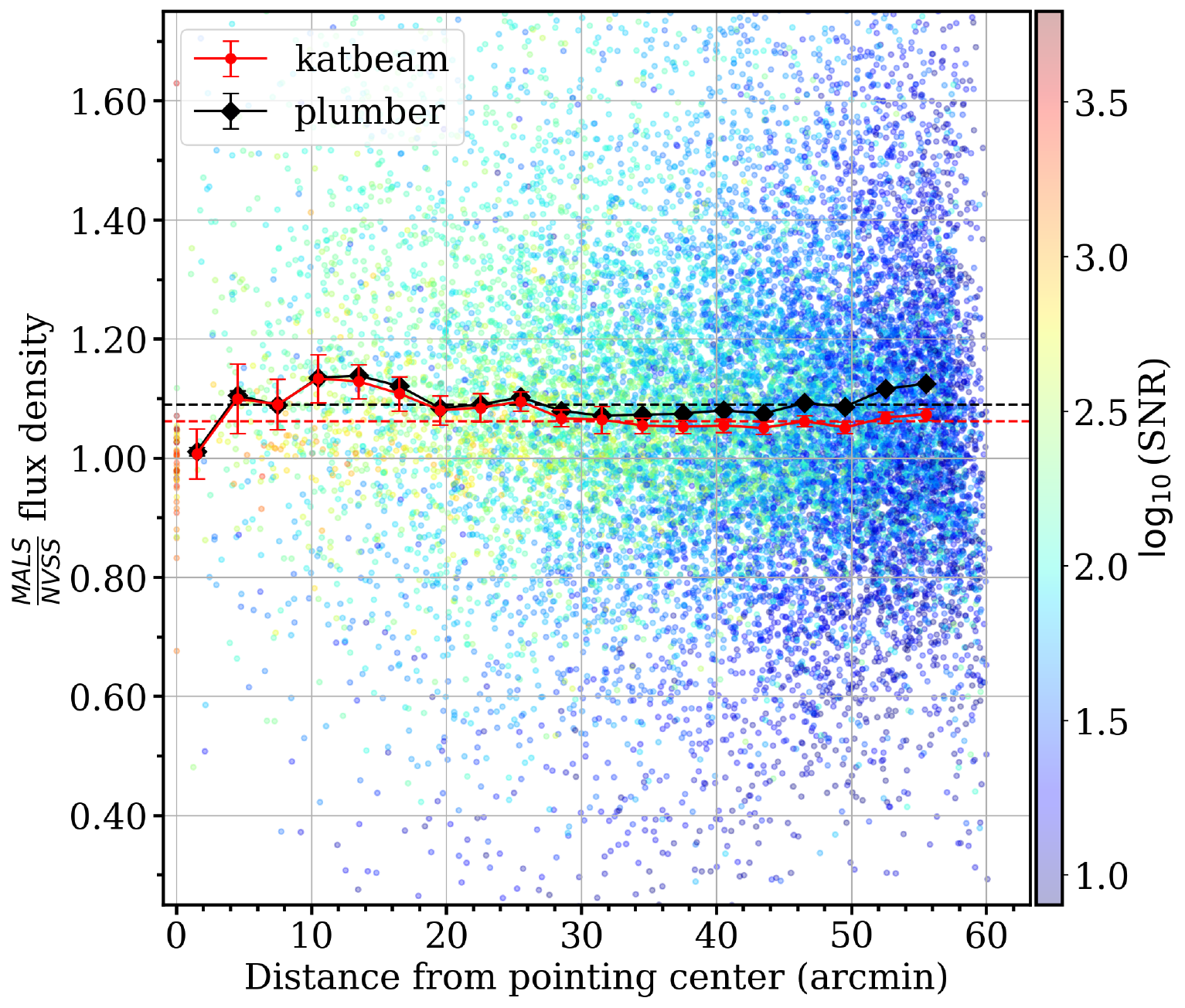}
    \caption{MALS SPW9 to NVSS flux density ratio for 15,834 sources. 
    The scatter plot is for flux densities corrected using the {\tt katbeam}.
    The circles and diamonds with error bars (1$\sigma$) correspond to ratios (3\,arcmin bins) based on beam  models from {\tt katbeam} and {\tt plumber}, respectively. The NVSS flux densities have been scaled to 1380 MHz using $\alpha$ = $-0.7$ for this analysis. 
    The horizontal dashed lines represent median values for the whole sample.
    }
    \label{fig:fratio_radial_KAT}
\end{figure}

In Fig.~\ref{fig:fratio_radial_KAT}, we plot the ratio of MALS and NVSS flux densities of compact and isolated sources as a function of distance from the pointing center.  
The median offset for this comparison involving the primary beam model from {\tt katbeam} is 1.06 (MAD = 0.15).
%
In general, the offsets are less than $\sim$10\%  implying that the {\tt katbeam} allows for a reasonable primary beam correction.  Note that MeerKAT's primary beam is elliptical \citep[][]{Mauch20}, and the {\tt katbeam} only provides a static beam for image domain correction at a position angle of $0\degree$.
We noticed that the peak of the {\tt katbeam} model for SPW9 is offset with respect to the center of the image by about $11^{\prime\prime}$.  As expected, adjusting for this offset does not lead to any significant change in the values.

\begin{figure*}[ht!]
     \begin{center}
        \subfigure{%
            \label{fig:first}
             \includegraphics[width=0.49\linewidth]{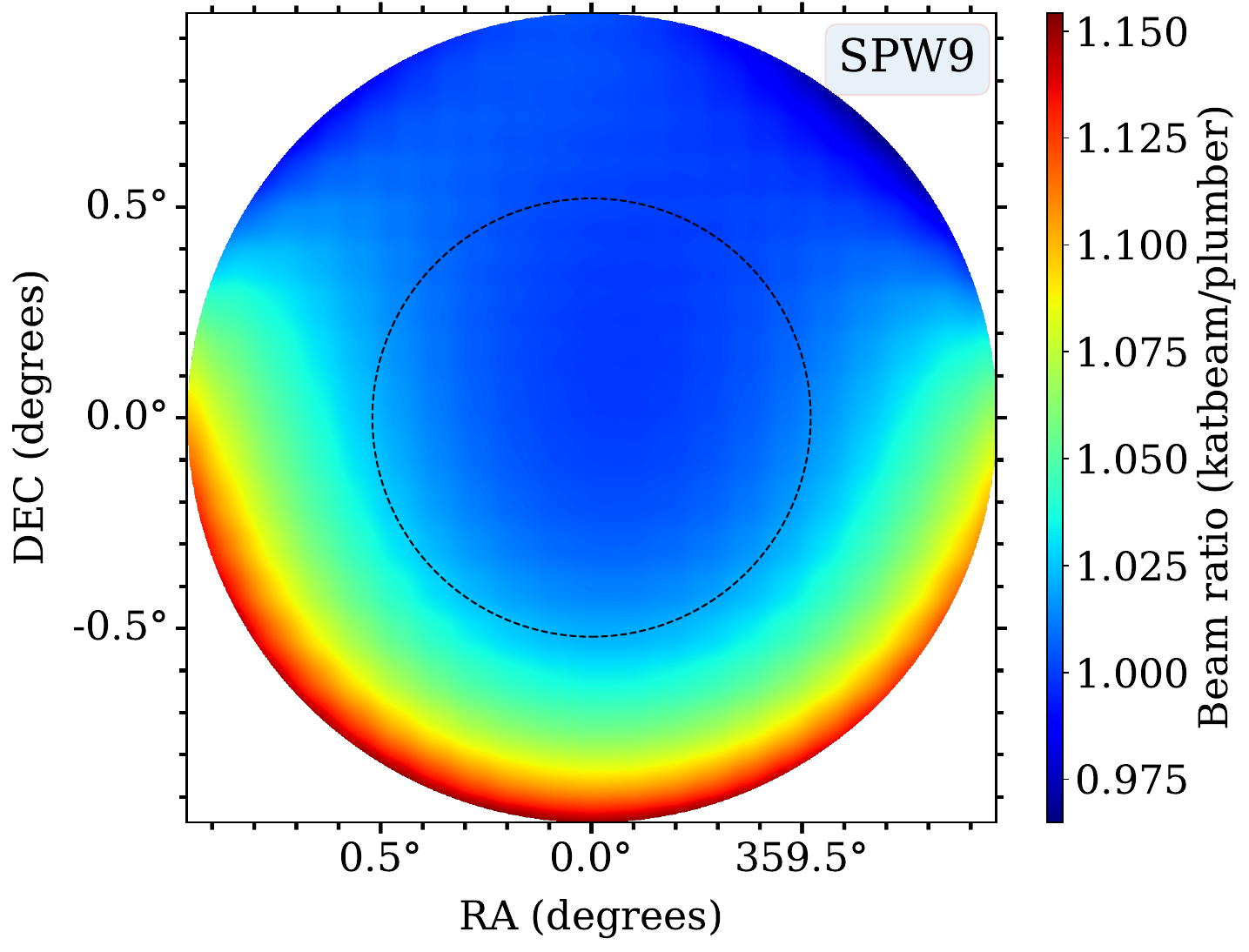}
         }%
        \subfigure{%
           \label{fig:second}
           \includegraphics[width=0.49\textwidth]{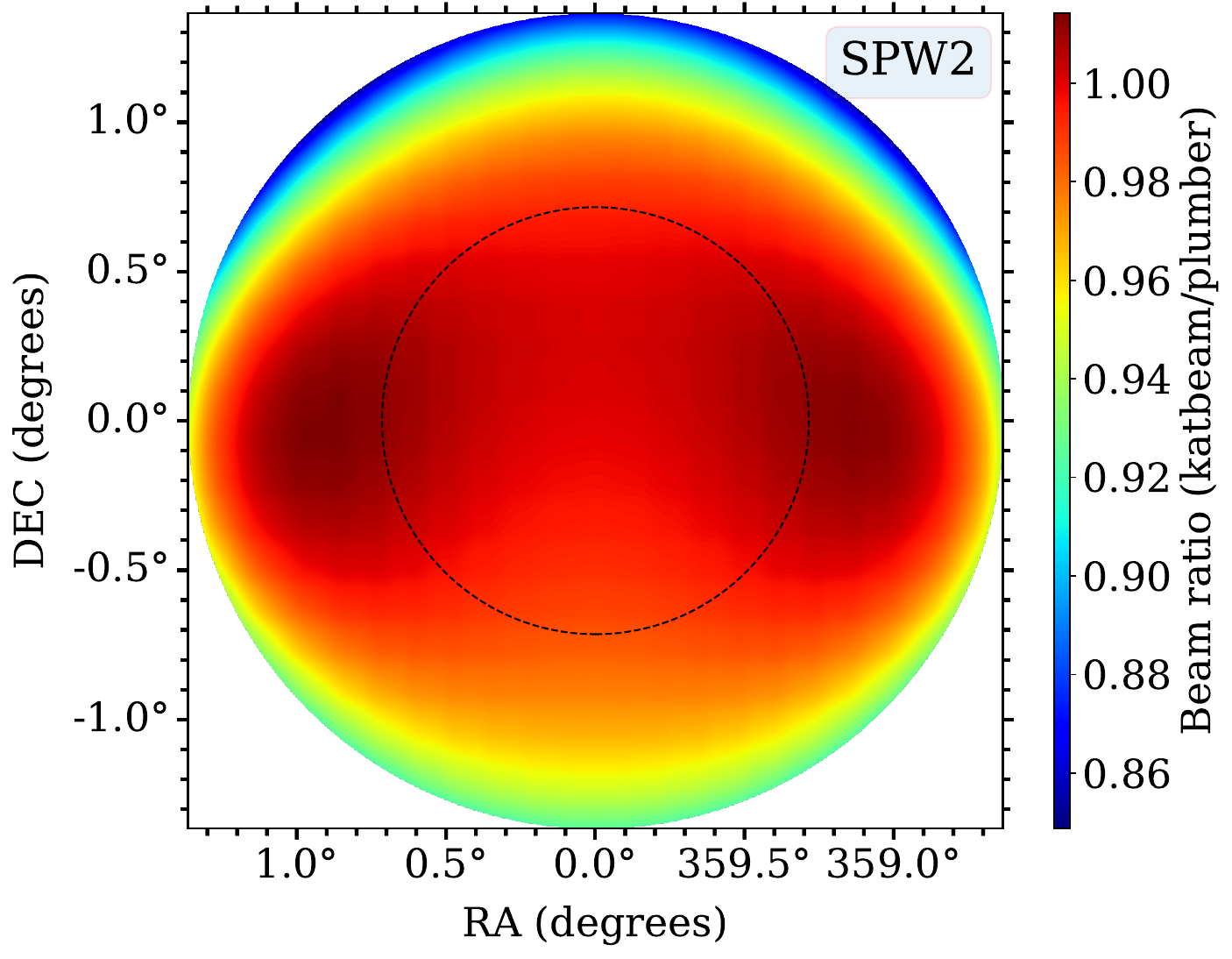}
       }
    \end{center}
    \caption{%
    Ratios ({\tt katbeam} / {\tt plumber}) of primary beams for SPW9 ({\it left}) and SPW2 ({\it right}). The extent of these images is same as the primary beam corrected MALS images. The dashed circles mark beam FWHMs of 62\farcm5 (SPW9) and 85\farcm7 (SPW2) from \citet[][]{Mauchc20}.}
    \label{fig:PB_ratio}
\end{figure*}

We also performed primary beam correction using {\tt plumber}\footnote{\url{https://github.com/ARDG-NRAO/plumber}} \citep[][]{Sekhar22}. {\tt plumber} generates primary beam models for radio interferometers using Zernike model coefficients of the antenna aperture illumination pattern \citep[also see][for holographic measurements]{deVilliers23}. 
The typical L-band observation of a MALS target are split into three scans at different hour angles over a duration of 3.5\,hrs. In such a situation, the primary beam correction ought to be applied over the range of parallactic angles using convolution kernels during the gridding of the visibilities \citep[][]{Bhatnagar13}.  In the image plane, one can at best use the illumination pattern at a specific orientation or averaged (smeared) over the range of parallactic angles traversed during the observation.  None of the image plane options are ideal, and at best are approximations of a visibility-plane primary beam correction.  Therefore, we adopted the simple approach of generating a beam model for each pointing at the parallactic angle at the center of observing run.

The ratios of {\tt katbeam} and {\tt plumber} beam models for SPW9 and SPW2 are shown in Fig.~\ref{fig:PB_ratio}. In Appendix~\ref{sec:PBratio} (Table~\ref{tab:PBratio}), we also provide annular averaged (2$^\prime$ bins) values for these. In general, in the inner region the two beam models follow each other and diverge in the outer regions.  The difference between the two models is much less dramatic at SPW2, and both are consistent within 3\% up to $\Delta\theta = $70$\arcmin$, where $\Delta\theta$ is the angular distance from the pointing centre.  Compared to SPW9, at SPW2 the {\tt katbeam} is narrower than the {\tt plumber} beam. Consequently, the spectral indices obtained using the former -- especially in the outer regions of the image -- are slightly steeper i.e., the median spectral index changes from $-0.70$ to $-0.74$ (see last column of Table~\ref{tab:PBratio} and Section~\ref{sec:uss}). 

For the data release presented here, we provide SPW9 and SPW2 flux densities i.e., {\tt Total\_flux}, {\tt Isl\_flux} and {\tt Peak\_flux} based on {\tt katbeam}.  The column {\tt Flux\_correction} provides the  multiplicative factor to be applied to these to obtain flux densities based on the {\tt plumber} model.  
For convenience,  total integrated flux densities for {\tt plumber} model are provided in {\tt Total\_flux\_measured}. 
For SPW9, over $0\arcmin < \Delta\theta < 40\arcmin$ the \texttt{plumber} corrected flux densities gradually increase from 0\% to 3\% relative to \texttt{katbeam} (Fig.~\ref{fig:fratio_radial_KAT}). In the outer regions, i.e., over $40\arcmin - 60\arcmin$, this increases steeply to 10\%.
Noticeably, the ratio based on the {\tt plumber} beam model remains flatter, and close to the overall median (1.09) as far as $50\arcmin$ from the pointing center.  
Therefore, the {\tt plumber} model may be a closer representation of MeerKAT's primary beam in the outer regions. 
However, overall the {\tt plumber} model yields a median flux density offset of 9\% with respect to NVSS with a MAD of 16\%, higher than the offsets obtained using {\tt katbeam}.
We anticipate further improvements by applying visibility plane primary beam corrections via AW projection \citep{Bhatnagar13}.

\section{Discussion}    
\label{sec:disc}  

\begin{table*}
\begin{center}
\caption{The MALS DR1 catalog summary for two SPWs.}
\begin{tabular}{lcc}
\hline

            & SPW9    & SPW2  \\
\hline
Number of sources                              &   240,321 & 495,325   \\
Median flux density (mJy)                      &    0.87   &    1.03   \\
Median angular size                            & 9\farcs 8 & 13\farcs 2\\
Median deconvolved angular size                & 2\farcs 8 &  3\farcs 9\\ 
Number of sources ({\tt S\_code} = `S')        &   215,328 & 441,988   \\     
Number of sources ({\tt S\_code} = `M')        &    24,993 &  53,337   \\
Number of Gaussian components                  &   285,209 &  586,290  \\ 
Number of sources with 2 components)           &    16,324 &  35,609   \\
Number of sources with $\ge$3 components)      &     8,669 &  17,728   \\
\hline 
\end{tabular}
\\
\tablecomments{
For a matching radius of $6^{\prime\prime}$, 205,435 sources are common between SPW2 and SPW9.
}
\label{tab:dr1src}
\end{center}
\end{table*}

\begin{figure*}[ht!]
     \begin{center}
        \subfigure{%
            \label{fig:FD_dist}
             \includegraphics[width=0.49\linewidth]{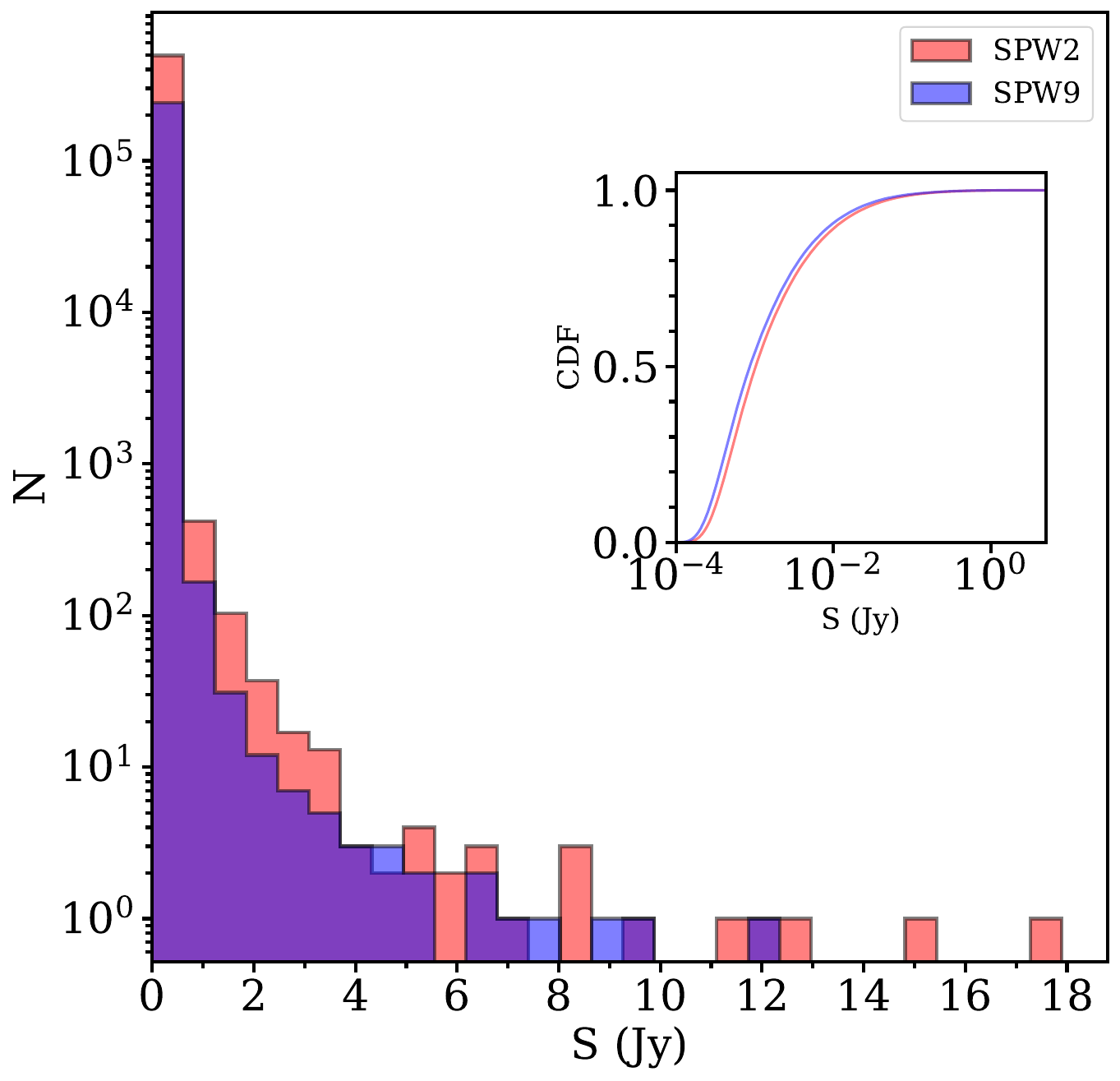}
                }%
        \subfigure{%
           \label{fig:size_dist}
           \includegraphics[width=0.49\textwidth]{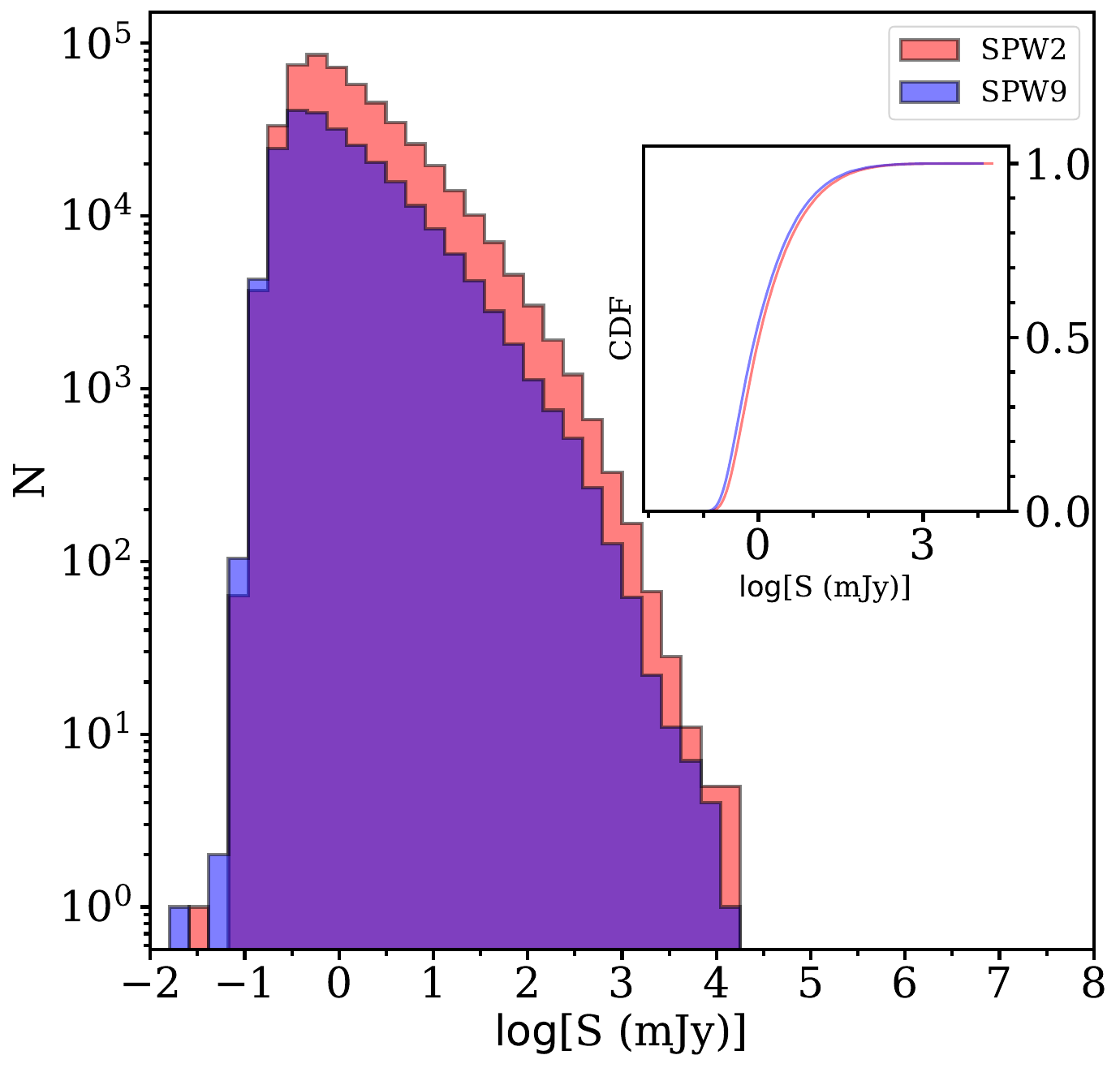}
                }
        \subfigure{%
           \label{fig:size_dist}
           \includegraphics[width=0.485\textwidth]{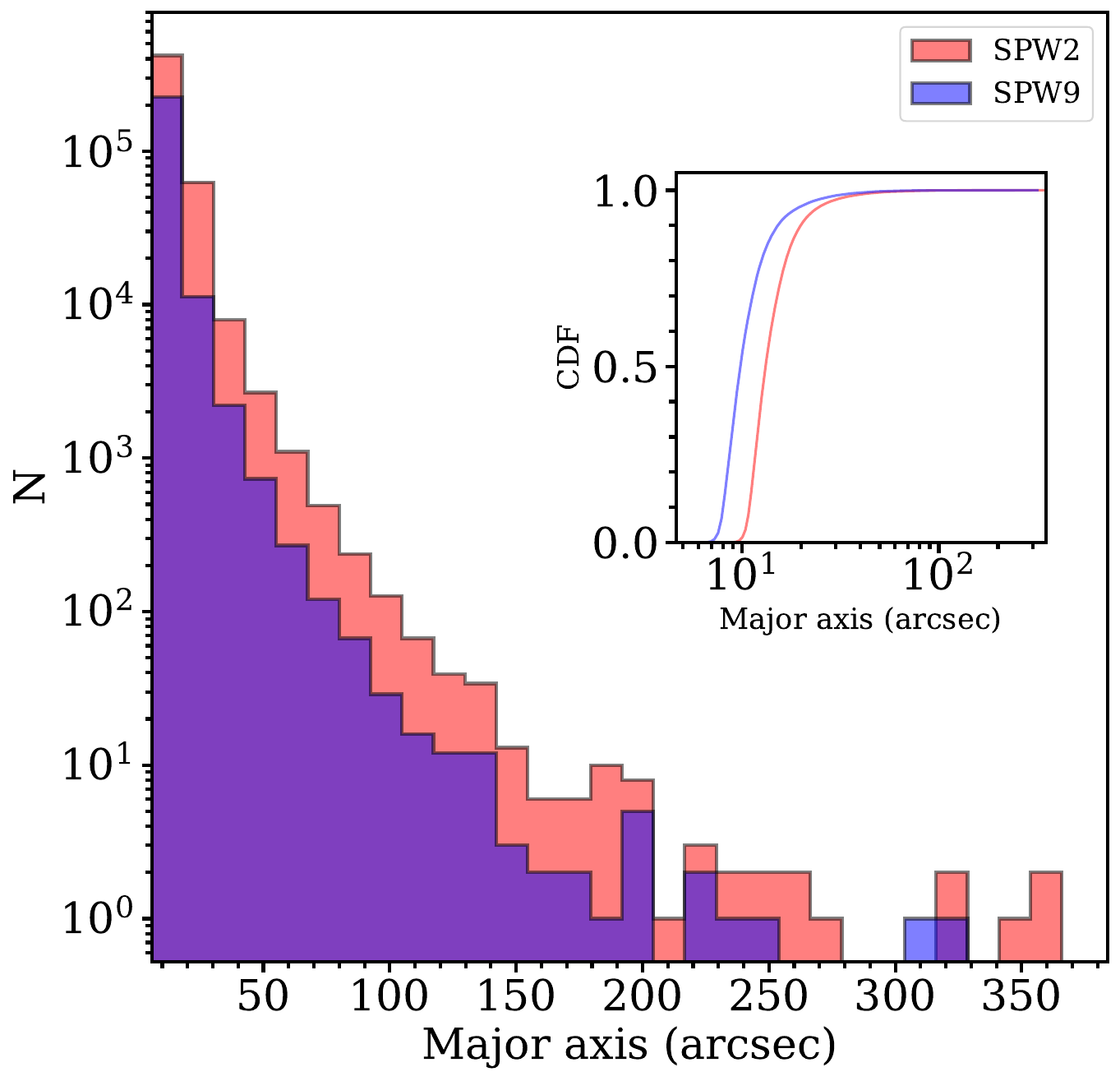}
                }
        \subfigure{%
           \label{fig:size_dist}
           \includegraphics[width=0.485\textwidth]{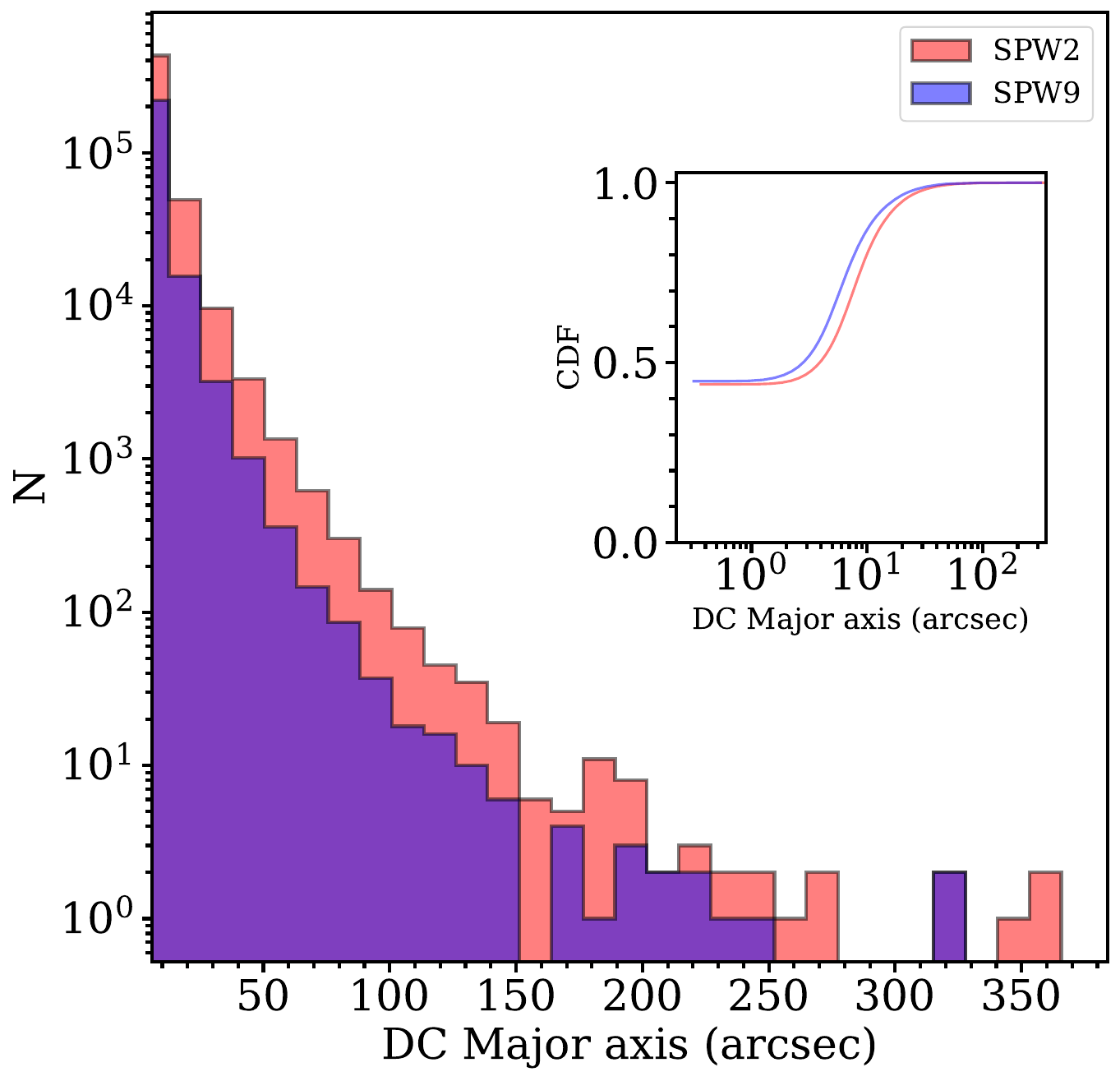}
                }
    \end{center}
    
    \caption{%
    Distributions of integrated flux densities (S) in linear and log scales (\textit{top panels}), and apparent and deconvolved (DC) major axis of sources (\textit{bottom panels}).   The insets show CDF.
    }%
    \label{fig:FD_src_size_dist}
\end{figure*}

In this section, we present the overall properties of SPW9 and SPW2 catalogs summarized in Table~\ref{tab:dr1src} and examine certain aspects to demonstrate their usage and utility.  The distribution of total flux densities ({\tt Total\_flux}) and source size ({\tt Maj}) are presented in Fig.~\ref{fig:FD_src_size_dist} (see also Table~\ref{tab:dr1src}). As expected for the spatial resolution of 8$^{\prime\prime}$-12$^{\prime\prime}$, the majority of sources ($\sim$90\%) are modelled with a single Gaussian component.  Only 4\% of the sources require 3 or more Gaussian components. Compared to SPW9, the radio source sizes are systematically larger at SPW2 (see bottom panels in Fig.~\ref{fig:FD_src_size_dist}).  The median value of the deconvolved major axis for SPW2 is $3\farcs9$, whereas for SPW9 it is $2\farcs8$.  The median angular separations between the Gaussian components of `M' type ($\tt {S\_code}$ = `M') sources are $15\farcs4$ (SPW2) and $12\farcs6$ (SPW9), respectively.  The median flux density is also larger at SPW2.  All these suggest that the larger SPW2 sizes are due to the excess of extended emission and not an artifact of coarser resolution.

At the extreme right end of the distributions in the \textit{bottom} panels of Fig.~\ref{fig:FD_src_size_dist} are the largest radio sources identified in the sample.
Contrary to the intuition, both the {\tt Total\_flux} and {\tt Isl\_Total\_flux} for these complex morphology sources modelled with more than 50 Gaussian components are in good agreement --  the two measurements for these differ by about $\sim$5\% \citep[see also][]{Wagenveld23}. 
The details of four unique largest radio sources ({\tt Maj}$\geq 300^{\prime\prime}$) detected in SPW2 are as following. 
Located at z = 0.056 \citep{Jones92}, J231757.32-421337.3 is a radio galaxy with a projected linear extent of about 350 kpc. J150726.21$+$082924.9, which has a linear extent of about 520 kpc, is at z = 0.079 \citep[][]{Abazajian09}. Another radio galaxy, J024105.35$+$084448.2, associated with NGC1044, is located at z = 0.021 \citep{Davoust95} and has a linear extent of about 150 kpc. The redshift of J034521.03$-$454816.7 (PKS 0343-459) is unknown. Overall, MeerKAT's high surface brightness sensitivity allows to detect large radio sources with faint diffuse lobes. Two of the four distinct large sources in SPW2 are outside the field of the SPW9 images, and another is divided into multiple sources due to the increased resolution. So, in SPW9, we essentially only observe one source with {\tt Maj}$\geq 300^{\prime\prime}$.  
%

In general, complex Fanaroff-Riley class I \citep[FRI; edge-darkened;][]{Fanaroff74} morphologies are well represented by the {\tt PyBDSF} Gaussian decomposition. But a single source with Fanaroff-Riley class II (FRII; edge-brightened) morphology in the lower frequency SPW2 image may be split into multiple sources in the SPW9 image due to {\it (i)} higher spatial resolution, and {\it (ii)} weaker jet emission linking the lobes.  
We crossmatch `M' type sources from SPW2 catalog with all the sources in SPW9 catalog.  We use a crossmatching radius equal to half the {\tt Maj} parameter of the SPW2 source.  Out of 34,103 matched sources, in 30,537 cases an SPW2 source is uniquely matched to a single source in SPW9.  In the remaining 3,566 cases, we find multiple matches in SPW9. About 90\% of these are fainter than 90\,mJy and form only a minuscule portion of the catalog. Nevertheless, the flux density and spectral index measurements for these could be miscalculated. 
The future MALS data releases will identify such missing linkages across the catalog through {\tt Source\_linked} parameter in Table~\ref{tab:cat_cols}.
%

In the following, we derive radio source counts and discuss the completeness of the MALS catalog (Section~\ref{sec:srccnt}).  We also derive spectral indices using SPW2 and SPW9 flux densities to understand the nature of detected radio source population (Section~\ref{sec:uss}).  Using TGSS ADR1 flux densities, we identify  ultra-steep spectrum (USS) sources as potential high-$z$ radio galaxies.  Finally, we investigate the variable and transient population of sources from the catalog (Section~\ref{sec:variability}).  Throughout these analyses, we take into account the above-mentioned complications caused by differing radio source morphology and spatial resolution at SPW2 and SPW9.  Therefore, besides sanity checks and value addition, these explorations also serve as demonstrations of the usage of MALS catalog.

\subsection{Differential source counts at 1.4\,GHz}
\label{sec:srccnt}

\begin{figure*}
    \centering
    \includegraphics[width=0.95\linewidth]{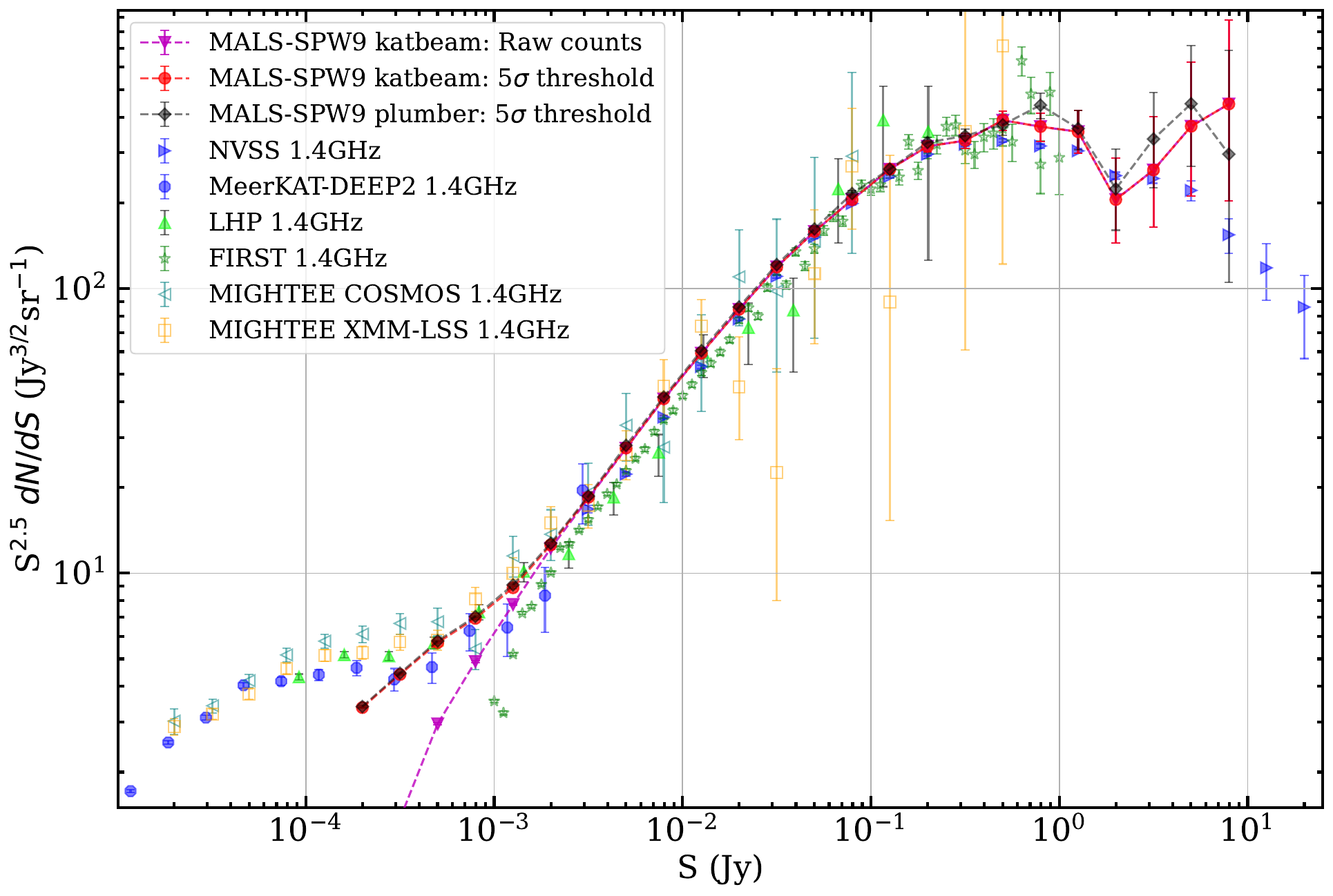}
    \caption{Differential source counts at 1.4\,GHz. The MALS source counts corrected for $5\sigma$ detection threshold, shown for both the {\tt katbeam} (red filled circles) and {\tt plumber} (black filled diamonds) beam models and the raw source counts (magenta filled triangles pointing downwards), have been scaled to 1.4 GHz using $\alpha = -0.74$. The counts from NVSS \citep[blue filled triangles pointing to the right;][]{Condon98} and MeerKAT-DEEP2 \citep[scaled from 1.266\,GHz using $\alpha = -0.7$ and displayed using blue filled circles;][]{Mauch20} presented in \citet[][]{Mathews21}, FIRST \citep[empty green asterisks;][]{White97first}, the Lockman hole project \citep[green filled triangles pointing up;][]{Prandoni18}, and the MIGHTEE COSMOS (empty teal triangles pointing left) and XMM-LSS (empty orange boxes) counts based on the modified SKADS model \citep[][]{Hale23} are also shown.
    }
    \label{fig:srccnt}
\end{figure*}

We estimate the differential source counts at 1.4\,GHz following standard recipes in the literature \citep[e.g.,][]{Condon98, White97first}.  For scaling the  integrated source flux densities ({\tt Total\_flux}) from the SPW9 catalog we adopt a spectral index of $-0.74$ (see Section~\ref{sec:uss}).
We binned these flux densities in logarithmic bins ($\Delta$S) of width 0.2 dex. The number of sources detected in each of these bins are normalized by the total survey area to obtain the differential source counts. These  are then multiplied by $S^{2.5}$, where $S$ is the mean of {\tt Total\_flux} corresponding to that bin.  The weighting by $S^{2.5}$ divides these by counts expected in a static Euclidean universe. These {\it raw} source counts are plotted in Fig.~\ref{fig:srccnt} (see also Table~\ref{tab:srccnt}).  The bright targets at the center of each pointing were selected as part of the survey design.  So, these have been excluded from the source count analysis.  We also exclude the regions with low reliability i.e., shaded regions shown in Fig.~\ref{fig:negative_sources}. 

The normalized source counts in each bin need to be corrected for the visibility function representing the area over which the source with a given flux density can be detected. We determine the visibility function by estimating the survey area over which the source with a given peak flux density can be detected at SNR$>$5 based on the rms maps (generated from {\tt PyBDSF} runs on primary beam corrected images). 
The corresponding {\it corrected} differential source counts at 1.4\,GHz for flux densities based on {\tt katbeam} ({\tt Total\_flux}) and {\tt plumber} ({\tt Total\_flux\_measured}) beam models are plotted in Fig.~\ref{fig:srccnt} and also provided in Table~\ref{tab:srccnt}.  The counts based on the two beam models agree within 2\%. 
Note that the relevant survey area i.e., column\,4 in Table~\ref{tab:srccnt} is nearly constant for sources brighter than 2\,mJy, and plummets to a few pointings below 0.1\,mJy. 
Also, above 1\,Jy only a handful of sources are  detected in MALS and the counts are highly uncertain.

Fig.~\ref{fig:srccnt} also presents 1.4\,GHz Euclidean-normalized differential source counts from various other surveys. It clearly demonstrates that the {\it corrected} source counts from MALS, within the scatter of various measurements, are in quite good agreement with the literature.  
The comparison between MALS and MeerKAT-DEEP2 counts shows that the SPW9 catalog is complete down to 2\,mJy. Below 0.5\,mJy level the completeness falls off steeply and at 0.1\,mJy it is only about 50\% complete.  
Between 0.5 and 200\,mJy the MALS sources counts are systematically higher ($\sim$10\%) than the source counts from NVSS and FIRST. The difference is, as expected, reduced to 3\% if the MALS flux densities are reduced by 6\% to account for the systematic offset with respect to NVSS noted in Section~\ref{sec:fluxscale}.
Over the same flux density range, the counts from MIGHTEE COSMOS field oscillate around MALS counts and may be affected by the cosmic variance.  The XMM-LSS counts over 10 - 50\,mJy are systematically lower which \cite{Hale23} suggest may likely be due to the incomplete grouping of emission components during source finding \citep[see][for details]{Hale23}.    

Overall, the slight offsets between source counts from various surveys in the 0.5 -- 200\,mJy range could originate from instrumental and analysis effects \citep[see also][]{Condon07, Prandoni18, vanderVlugt21}. 
In particular, the visibility function estimated here for MALS does not include corrections for Eddington and resolution biases. Eddington bias leads to redistribution of source counts in flux density bins in the presence of random noise and biases the detectability of unresolved sources near the detection threshold \citep[][]{Eddington13, Eddington40}.  The resolution bias leads to underestimation of extended sources in a flux density bin.  This is a consequence of the fact that the detection of a source depends on its peak flux density, therefore a larger source due to its lower peak flux density will drop below the detection threshold much sooner than a smaller source \citep[][]{Prandoni06, smolic17, vanderVlugt21, Mandal21}.

A detailed exploration of the above-mentioned issues will be presented in future papers involving MALS catalogs from more sensitive wideband images.
This will include simulations involving injection of radio sources of known flux densities and sizes in residual images and subjecting these to the same source finding procedures as used for cataloging to determine completeness as a function of rms and radio source morphology \citep[see e.g.,][]{Bonaldi21, Shimwell22, Hale23}. Indeed, wideband images of MALS exhibit large variations in completeness for compact and extended sources \citep[see Figs.~7 and 8 of ][]{Wagenveld23}.

\subsection{Spectral indices and ultra steep-spectrum sources}    
\label{sec:uss}  

Spectral indices provide useful information on the nature of radio sources and are helpful in disentangling various mechanisms responsible for the radio emission. In general, for a source the spectral index ($\alpha$) and the associated curvature ($\beta$) are related to its flux densities, $S_1$ and $S_2$ measured at $\nu_1$ and $\nu_2$, through the following relation:
\begin{equation}\label{eq:specidxfull}
    S_1 = S_2\Big(\frac{\nu_1}{\nu_2}\Big)^{\alpha+\beta\log(\nu_1/\nu_2)}.
\end{equation}
For the frequency coverage corresponding to SPW2 and SPW9, it is reasonable to ignore in-band curvature and use the simplified form of Eq.~\ref{eq:specidxfull} obtained by setting $\beta = 0$:
\begin{equation}\label{eq:specidx}
    S_1 = S_2\Big(\frac{\nu_1}{\nu_2}\Big)^\alpha.
 \end{equation}
The associated 1$\sigma$ uncertainties on $\alpha$ are calculated using:
\begin{equation}\label{eq:specidxerr}
    \sigma_\alpha = \Big|\frac{1}{\log(\nu_1/\nu_2)}\sqrt{\Big(\frac{\Delta S_1}{S_1}\Big)^2+\Big(\frac{\Delta S_2}{S_2}\Big)^2}\ \Big|,
\end{equation}
where $\Delta S_1$ and $\Delta S_2$ are uncertainties associated with $S_1$ and $S_2$.

We used {\tt Total\_flux} and {\tt Total\_flux\_E} in Equations~\ref{eq:specidx} and ~\ref{eq:specidxerr} to calculate spectral indices and errors of 125,621 sources (SNR $>8$) crossmatched using a radius of $6^{\prime\prime}$. The $6^{\prime\prime}$ radius minimizes the number of nearest neighbours for sources without a match in SPW9. This maximizes the number of sources for which spectral indices can be estimated, however, at the expense of spurious spectral indices in the sample.  
Therefore, we advise caution and imposition of additional cuts to reject spurious matches and obtain suitable samples  for various applications (for an example see Section~\ref{sec:uss2}). 

The spectral indices and errors are provided in columns {\tt Spectral\_index\_spwfit} and {\tt Spectral\_index\_spwfit\_E} of Table~\ref{tab:cat_cols}, respectively, of the SPW2 (reference SPW) catalog.
Note that {\tt Spectral\_index\_spwfit} is a two-element array, adopted to report both the spectral index and the curvature.  In the first data release, we report only spectral indices ($\alpha$), and leave the second element ($\beta$) blank. 
In the SPW2 catalog, we also provide upper and lower limits on spectral indices based on detection in either SPW2 or SPW9, respectively. The flux density for non-detection is taken as 5 times the local rms from the rms map. {\tt Spectral\_index\_spwfit\_E} is set to 999 or -999 to indicate whether the reported value in the SPW2 catalog is an upper or lower limit, respectively.

\begin{figure}
    \centerline{\hbox{
    \includegraphics[width=0.95\linewidth]{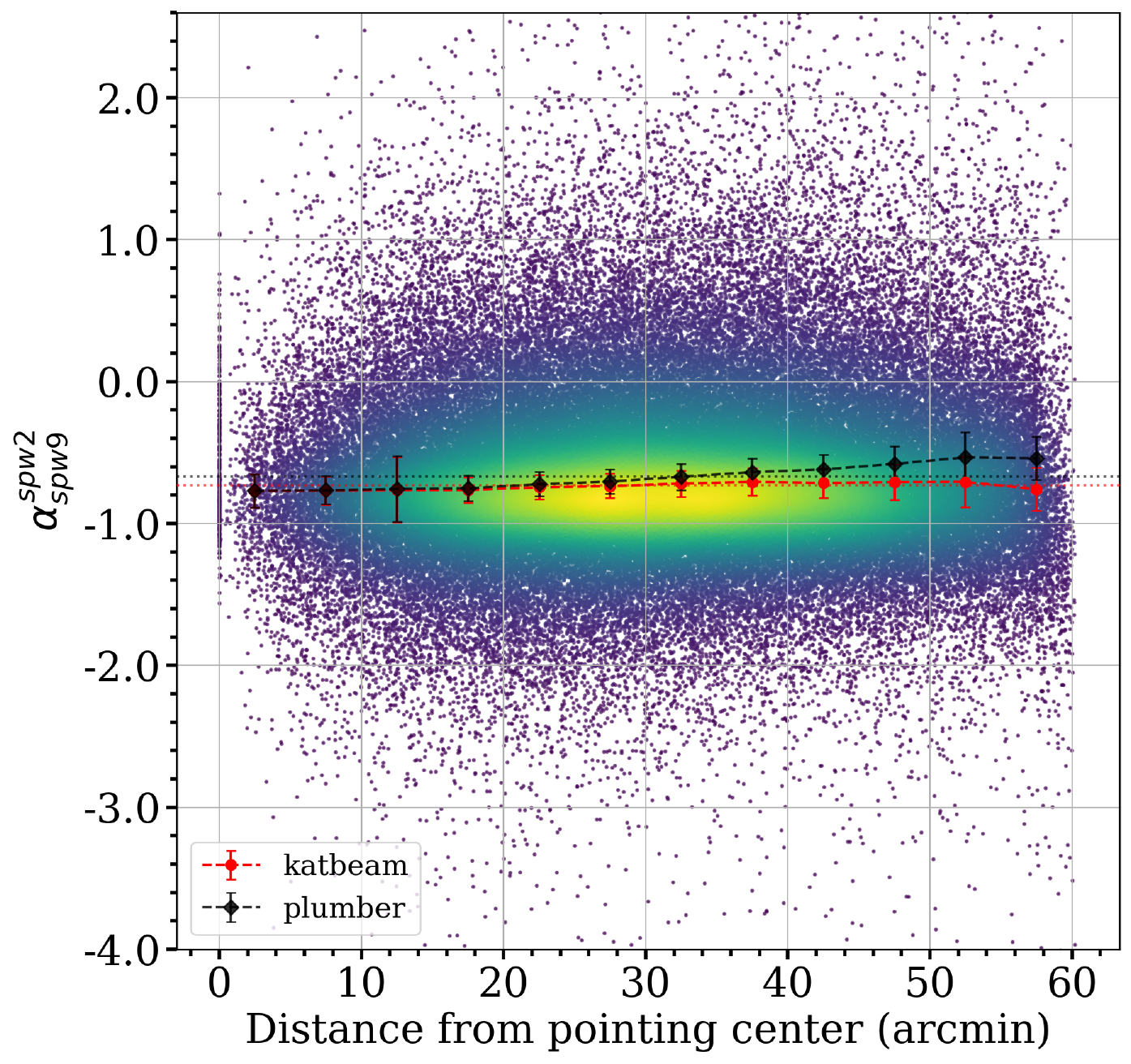}
    }}
    \caption{Spectral indices of 122,077 sources detected in both the SPWs versus the distance from the pointing center. The red (black) points mark median spectral indices derived from {\tt katbeam (plumber)} corrected flux densities in bins of 5\arcmin. The horizontal dotted lines indicate median spectral indices for the full sample corrected using {\tt katbeam} ($-0.732^{+0.003}_{-0.003}$) and {\tt plumber} ($-0.680^{+0.003}_{-0.003}$) beam models. The points are color coded according to their space density. The error bars defining 90\% confidence level estimated by bootstraping are enlarged by a factor of 10 for clarity.
    }
    \label{fig:spdrad}
\end{figure}

\subsubsection{Systematic uncertainty on $\alpha$}    
\label{sec:uss1}  

Fig.~\ref{fig:spdrad} shows $\alpha^{spw2}_{spw9}$ derived from {\tt katbeam} corrected flux densities  as a function of distance from the pointing center. We excluded from this analysis 12 pointings (marked as $\star$ in Table~\ref{tab:pointings}) based on unusually high rms in the SPW2 images. This led to a sample of 122,077 sources detected in both the SPWs with SNR$>$8. For comparison, the median spectral indices obtained using {\tt plumber} corrected flux densities are also shown. 
The {\tt katbeam} and {\tt plumber} based spectral indices diverge from the median in the outer regions. Therefore, there may be systematic uncertainties of the order of $\pm$0.05 in spectral indices beyond $45\arcmin$ from the pointing center (see last column of Table~\ref{tab:PBratio}). Further improvements in these will follow from better modeling of the frequency dependent behavior of the MeerKAT beam.  
%

Additionally, the $\alpha^{spw2}_{spw9}$ measurements could be affected by systematic uncertainties due to the splitting of a source in SPW2 image in to multiple sources in the higher spatial resolution SPW9 image.  The flux density measurement of a source may also be affected by blending with a nearby source in one of the spectral windows.  The extent of contamination due to blending depends on the complex interplay between intrinsic spectral index of a source and its position in the primary beam, making it less tractable. 
%

\subsubsection{$\alpha$ -- flux density  correlation}    
\label{sec:uss2}  

Several studies have reported flattening of spectral index with decreasing flux density \citep[e.g.,][]{Prandoni06, Gasperin18, Tiwari19} but counter examples have also been reported \citep[e.g.,][]{Ibar09}. The statistically large sample of spectral indices from MALS DR1 offers an opportunity to test this.
For deriving a suitable sample of spectral indices for this purpose, we consider the following cuts on the properties of radio sources:  
\begin{itemize}
\item For an `S' type detection in both SPWs we require that no other source is present within $6^{\prime\prime}$ radius.  

\item For an `M' type source detected in both the SPWs, this condition is modified to finding the same number of radio sources within a circle of  radius ($R_M$) defined by the distance of the farthest Gaussian component from the source position, plus the FWHM of the synthesized beam, taken to be $10^{\prime\prime}$ for all the cases. 
\item For an `M' type source detected only in one of the spectral windows, we require that no radio source is present within $R_M$ in the other spectral window.  

\item For an `S' type source detected only in one of the spectral windows, we follow a two-step validation to eliminate blending and confusion with nearby sources: {\it (a)} no Gaussian component is present within $6^{\prime\prime}$, and {\it (b)} the position of the source is outside the circle defined by the nearest `M' type source in the other spectral window.
\end{itemize}
These criteria exclude the majority of sources that may have been resolved into multiple sources in SPW9 due to higher spatial resolution or weaker diffuse emission connecting the two radio lobes.

\begin{figure}
    \centering

    \includegraphics[width=1\linewidth]{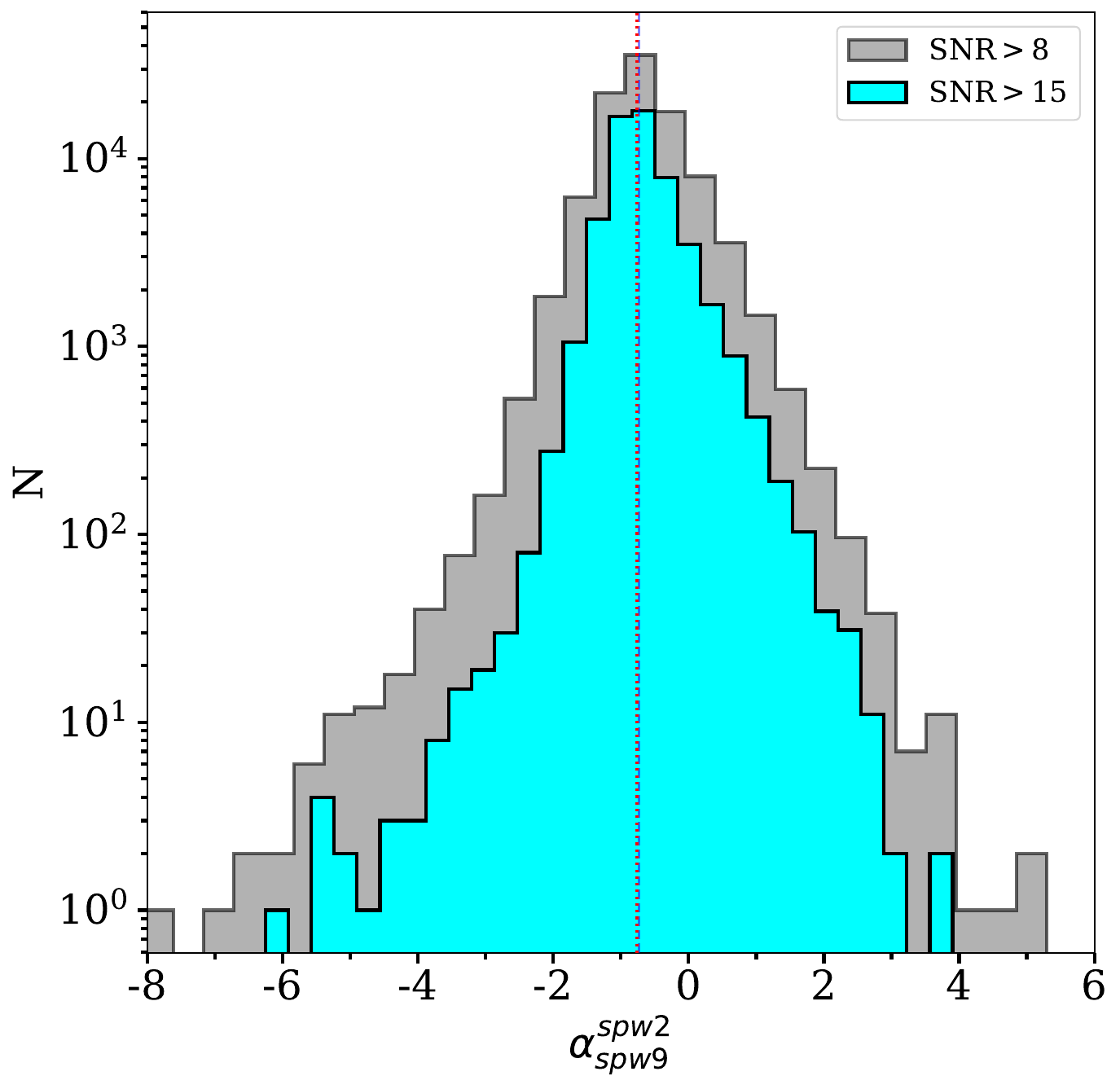}
    \caption{Spectral index ($\alpha^{spw2}_{spw9}$) distribution of sources detected at SNR $>$ $8$ (median $\alpha =  -0.736^{+0.003}_{-0.003}$) and at SNR $>$ $15$ (median $\alpha =  -0.757^{+0.003}_{-0.003}$).  
  The vertical lines mark median spectral indices.
}
    \label{fig:spdhist}
\end{figure}

\begin{figure*}
    \centerline{\hbox{
    \includegraphics[width=0.5\linewidth]{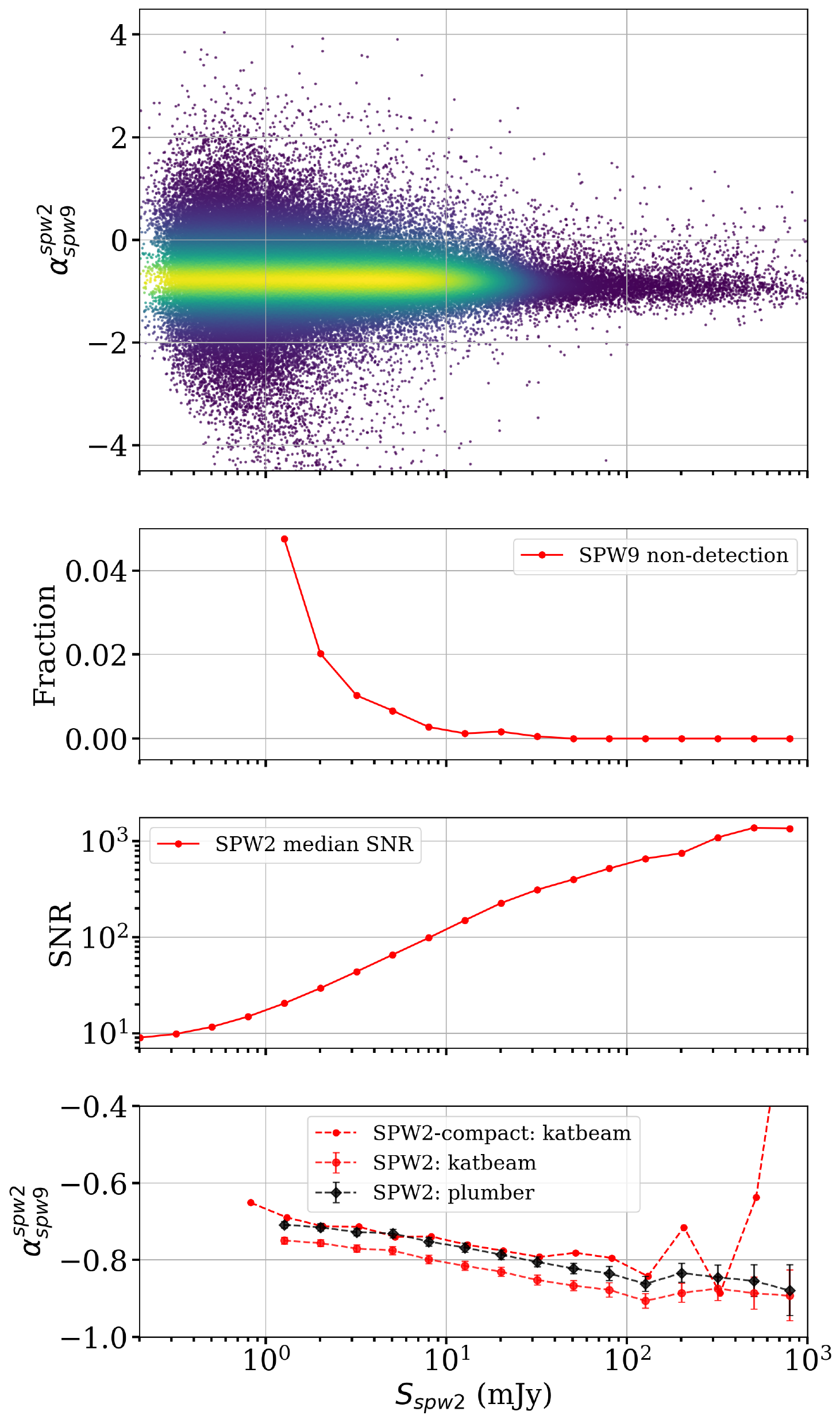}
        \includegraphics[width=0.5\linewidth]{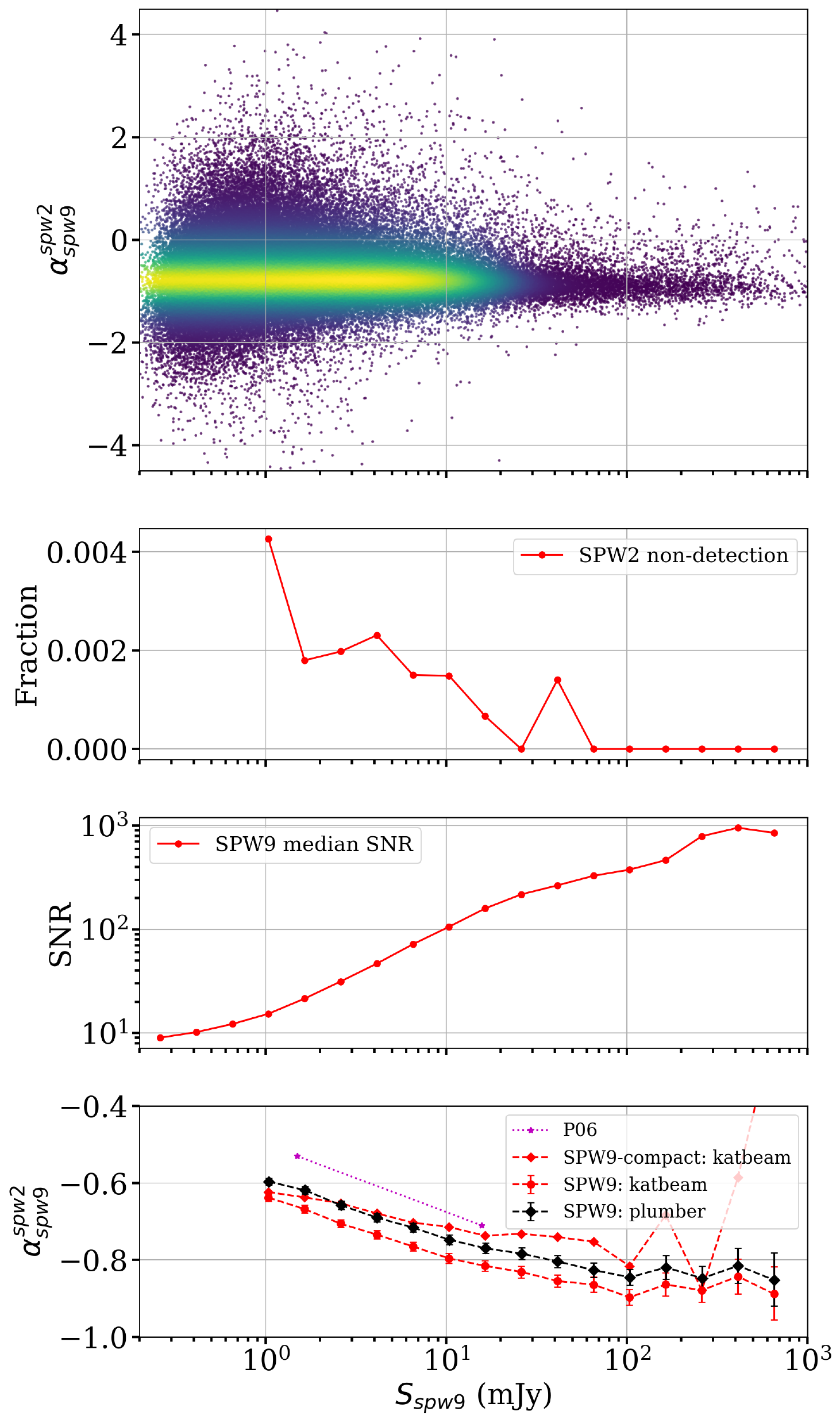}
    }}
    \caption{Spectral indices ($\alpha^{spw2}_{spw9}$) as a function of flux density in SPW2 ({\it left} panels) and SPW9 ({\it right} panels). In the {\it top} row, the points are color coded according to their space density. 
    The {\it second} row shows the fraction of upper ({\it left}) and lower ({\it right}) limits on spectral indices in each flux density bin. The {\it third} row shows the median SNR of sources in each bin. Note that the sources with SNR$<15$ were excluded from the analysis. 
    The median spectral indices calculated using survival analysis methods in bins of 0.2 dex in flux density are shown for both {\tt katbeam} and {\tt plumber} beam corrected flux densities ({\it bottom} panels).  The median spectral indices considering only compact sources for the case of {\tt katbeam} model are also shown.
    For reference, 1.4 GHz measurements from \cite[P06; ][]{Prandoni06} are plotted in the {\it bottom-right} panel. 
    }
    \label{fig:spdflux}
\end{figure*}

Through the above-mentioned selection cuts, we derive a sample of 98,832 sources that are detected in both the SPWs at SNR$>$8, 10,962 detected only in SPW2 and 822 only in SPW9, at a distance $\leq 45^\prime$ from the pointing center. We have again excluded the 12 pointings based on unusually high rms in the SPW2 images. Fig.~\ref{fig:spdhist} shows the distribution of $\alpha^{spw2}_{spw9}$ for sources detected in both the SPWs. The median spectral index is $-0.74$. 
The median spectral indices for sources detected at SNR$>$8 and SNR$>$15 are  $-0.736^{+0.003}_{-0.003}$ and  $-0.757^{+0.003}_{-0.003}$), respectively. The errors define 90\% confidence level estimated by bootstrapping. 
The median spectral index of `M' type sources in SPW2 is $-0.885^{+0.003}_{-0.003}$.  This is slightly steeper compared to the overall median value and is expected as the extended radio emission is primarily from older electrons associated with radio lobes or even relics. 
Visual inspection was done on the population of sources with extreme spectral indices in Fig.~\ref{fig:spdhist}. These are generally `M' type sources where the surrounding diffuse emission is brighter and hence well modelled in SPW2, however, in SPW9 only the brightest component's flux density is taken into consideration. Therefore, caution is advised while using spectral indices of `M' type sources.

Fig.~\ref{fig:spdflux} shows spectral index measurements ($\alpha^{spw2}_{spw9}$) versus SPW2 (\textit{left panel}) and SPW9 (\textit{right panel}) flux densities.   
The effects of relative sensitivity limits due to the two SPWs can be seen below $\sim$0.5\,mJy.
To examine the spectral index versus flux density relationship, we binned all the spectral index measurements into equally spaced logarithmic bins of flux densities, each consisting of $>$100 sources.  
Our initial investigation suggested that bins with SNR$<$15 are unsuitable for this analysis.
This is primarily due to a steep increase of upper or lower limits on spectral indices in these bins (see {\it second} and {\it third} row  of Fig.~\ref{fig:spdflux}).

Next,  using the {\tt ASURV}\footnote{ASURV Rev 1.2} package which implements the survival analysis methods discussed in \cite{Feigelson1985} and \cite{Isobe1985}, we estimate median spectral indices for bins with flux density larger than 1\,mJy. 
A clear flattening of spectral indices is seen with respect to decreasing flux densities at 1006\,MHz (SPW2) and 1381\,MHz (SPW9; {\it bottom} row in Fig.~\ref{fig:spdflux}).  
For the measurements based on {\tt katbeam}, the gradients can be modelled as $(-0.07 \pm 0.01) \times logS_{spw2}$  $-(0.74 \pm 0.01)$ and $(-0.12 \pm 0.01) \times logS_{spw9}$  $-(0.65 \pm 0.01)$, where $S$ is in mJy,   implying that the trend is less steep for sources selected at SPW2.
The {\tt plumber} based measurements are flatter by $\sim$0.05 and also exhibit the same trend but with an offset as expected from Fig.~\ref{fig:spdrad}.
Note that the systematic uncertainties due to the beam models are significantly larger than the errors ($\sim 0.003$) on mean spectral indices.

We also repeated the analysis for the subset of sources selected to be compact in both the SPWs using the envelope presented in Fig.~\ref{fig:comp_envlp}.  This selection primarily reduces the fraction of sources with extreme spectral indices ($\lvert\alpha\rvert>3$). 
The median spectral index ($-0.685^{+0.003}_{-0.003}$) corresponding to the compact sources are flatter ({\it dashed} lines in the bottom row) but the trends with respect to the SPW2 and SPW9 flux densities are still apparent.
In conclusion, a flattening of spectral indices at lower flux densities is indeed confirmed through our sample of 66,836 (SPW2) and 48,817 (SPW9) radio sources.  For reference we also show the median 1.4\,GHz spectral indices obtained in two bins, $\leq 4$ mJy and $>4$ mJy at 1.4 GHz, by \cite{Prandoni06} for a sample of 111 radio sources selected at 5\,GHz.  
Higher spatial resolution imaging is required to confirm that the observed trend is indeed due to higher abundances of FRI i.e., core dominated population of radio sources in lower flux density bins.

Overall, the distribution of $\alpha^{spw2}_{spw9}$ in Fig.~\ref{fig:spdhist} exhibits the presence of both flat ($\alpha^{spw2}_{spw9}$ $>$ -0.5) and steep ($\alpha^{spw2}_{spw9}$ $<$ -0.5) radio source populations.  The former corresponds to core-dominated AGN \citep{Antonucci85, Prandini22} and latter to the lobe-dominated AGN \citep{Saikia09, Sirothia13}.  A small fraction among these lobe-dominated AGN are young radio sources (age$<10^5$\,years) that are often embedded in gas rich environments, and may also exhibit a turnover at GHz frequencies, which is an indication of sub-kpc scale extent of the radio emission \citep{odea97, Orienti14, Liao20, Odea21}.  Also, present at the right end of the distribution in Fig.~\ref{fig:spdhist} are sources with inverted spectra, with radio SED peaking at higher frequencies. These high-frequency peakers may be even younger than steep-spectrum sources \citep{Orienti14, Stanghellini09}.  We will examine these aspects of the radio source population detected in MALS in the context of associated \hi\ 21-cm absorption in future papers.    

\subsubsection{Ultra steep-spectrum sources}    
\label{sec:uss3}  

Here, we focus on a special population of radio sources exhibiting ultra steep spectral indices (USS; $\alpha <$ -1.3) as prospective high-redshift radio galaxies \citep[HzRGs; $z>2$ ][]{Miley08, Bornancini07, Saxena18, Broderick22}. 
For this we crossmatch all the sources detected in SPW2 with the TGSS ADR1  \citep[][]{Intema17} at 147\,MHz. TGSS ADR1 has a spatial resolution of $25^{\prime\prime}$ (median rms noise $\sim$3.5\,mJy), therefore we use a crossmatching radius of $10^{\prime\prime}$ to maximize the coincidence of radio continuum peaks in the MALS and TGSS ADR1 images. We find counterparts for 34,735 sources of which 286  have $\alpha^{TGSSADR1}_{spw2} < -1.3$.  The median  SPW2 flux density for these sources is 5.5\,mJy.
The spectral indices and associated errors from this exercise are provided in columns {\tt Spectral\_MALS\_Lit} and {\tt Spectral\_MALS\_Lit\_E}, respectively.
%

It is widely accepted in the literature that HzRGs are young and have compact morphology \citep[][]{Miley68, Neeser95, Daly02, Morabito17}. To discard sources which are clearly resolved in our sample, we visually inspected their SPW2 and SPW9 cutouts. A total of 90 sources were found to have extended emission and therefore discarded. Further, following the reliability criteria discussed in section~\ref{sec:purity}, we rejected any candidate HzRG which was within 3\arcmin\, from the edge of the SPW2 primary beam. This led to a sample of 182 sources with spectral indices, $\alpha^{TGSSADR1}_{spw2}<-1.3$ and compact morphology on arc-seconds scales. 

In addition to having a steeper spectral index, non-detection in optical and/or infrared bands greatly enhances the probability of a source to be at higher redshifts. Therefore, we crossmatched the sample of remaining 182 candidate HzRGs with the $i$-band images from the Panoramic Survey Telescope and Rapid
Response System \citep[PS1;][]{Chambers16} and the AllWISE catalog from the Wide-field Infrared Survey Explorer \citep[WISE; ][]{Wright10, Cutri14}.  
The results of this crossmatching\footnote{30 sources at $\delta<-30\degree$ were excluded from the PS1 crossmatch but are treated as non-detections in optical bands.} by considering the nearest match within $2^{\prime\prime}$ radius, chosen conservatively based on the astrometric accuracy of the MALS catalog,  are reported in Table~\ref{tab:usslist}.  In the table, a detection is marked as `True' and a non-detection with `False'. For convenience, we have added a column `Flag' to the table, the value of which for each source is based on detection in PS1 and WISE. Flag = 1 denotes 27 sources not detected in PS1 but detected in WISE. Flag = 2 denotes 113 sources not detected in both PS1 and WISE.  Lower prospective candidates were given, Flag = 3: 14 sources detected in PS1 but not in WISE, and Flag = 4: 28 sources detected in both PS1 and WISE. 
The redshifts of 8 sources, all at $z<1$ and Flag = 4, are available from the NASA Extragalactic Database (NED; see last column of Table~\ref{tab:usslist}).

Overall, 140 sources with Flag 1 and 2 represents the prospective HzRG sample that needs further refinement through higher spatial resolution radio imaging, and then confirmation with infrared imaging and spectroscopy. 
In general, this candidate sample is expected to contain a mix of HzRGs and dust obscured AGN.  The subset that are at $z<1.5$ are expected to show \hi\ 21-cm absorption in MALS L- and UHF-band spectroscopy. In a future paper, we will present an expanded sample of HzRG candidates using both L- and UHF-band continuum images from MALS, and report on the results from \hi\ 21-cm absorption spectroscopy.

\subsection{Long term radio variability and transients}    
\label{sec:variability}  

The majority of NVSS observations were carried out between 1993 and 1996 i.e., about 26 years prior to MALS L-band observations presented here.  Here, we compare SPW9 catalog with NVSS to identify variable and transient radio sources. We define the former to be detected in both NVSS and MALS, whereas the latter are detected only in one of these. We expect the majority of variable and transient radio sources to be compact at the arcsecond scale resolutions of NVSS and MALS \citep[e.g.,][]{Thyagarajan11, Mooley2016}.  
For unresolved sources brighter than 3.4\,mJy the NVSS catalog is 99\% complete  and has an rms position uncertainty of $<3^{\prime\prime}$ \citep[see Figs.~30 and 32 of][]{Condon98}. Since MALS SPW9 catalog is also nearly complete at this threshold flux density (Fig.~\ref{fig:srccnt}), we adopt 4.0\,mJy as a stricter threshold to identify variable and transient sources.
%

For identifying variable sources, we consider 15,691 radio sources common between SPW9 and NVSS catalogs with the following properties: {\it (i)} brighter than 4.0\,mJy and compact in NVSS, and {\it (ii)} isolated sources detected at SNR$>8$ in SPW9 with {\tt Distance\_NN} $>60^{\prime\prime}$ and {\tt Distance\_pointing}$<45\arcmin$, and {\it (iii)} SPW9 - NVSS separation less than $3^{\prime\prime}$. These stringent criteria ensure that the positions of the selected sources are well-determined for the purpose of crossmatching with multi-wavelength catalogs and minimizes uncertainties due to a compact source being resolved in SPW9 or the presence of unrelated nearby source in either catalog, as already discussed in Section~\ref{sec:uss}.

\begin{figure*}
    \centering
    \hbox{
    \includegraphics[width=0.49\linewidth]{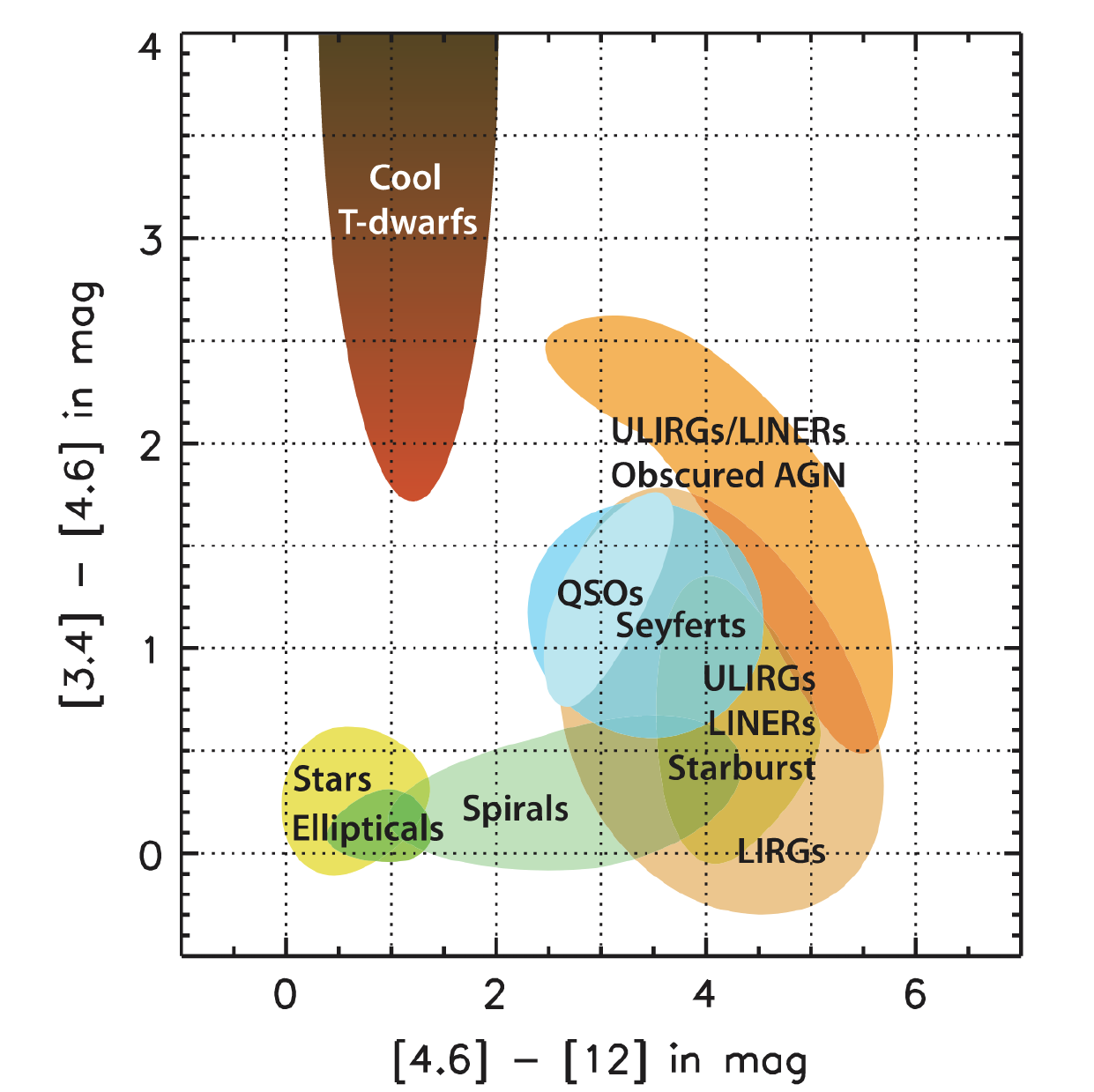}
    \includegraphics[width=0.49\linewidth]{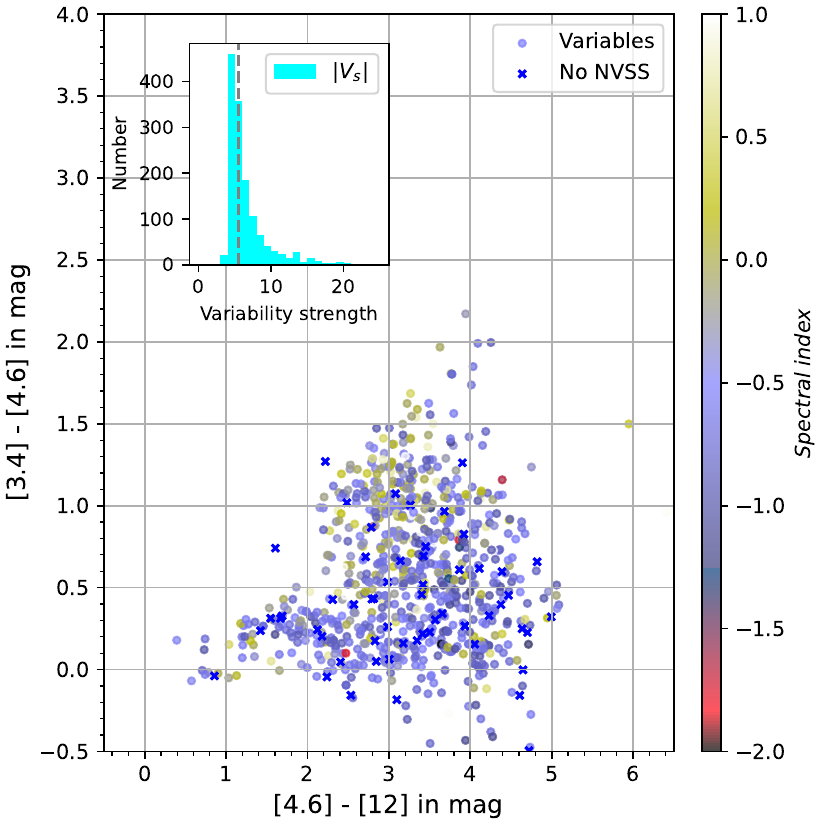}
    }
    \caption{
WISE color-color plot in Vega magnitudes of various classes of sources (\textit{left}), reproduced from \citet[][]{Wright10} with permission, and  variable radio sources from MALS (\textit{right}). The points have been color coded based on spectral index, $\alpha^{spw2}_{spw9}$. The inset in the \textit{right} panel shows the distribution of variability strength ($|V_s|$), with the vertical dashed line marking the median value. }
    \label{fig:vars}
\end{figure*}

We measure random variation in the flux density of radio sources using $\chi^2$ of the residuals around the mean flux density computed using the following equation:
\begin{equation}
    \chi^2_{obs} = \sum_{i=1}^{N} \frac{(S_i - \bar{S})^2}{\sigma_i^2},
    \label{eq:chisq}
\end{equation}
where
\begin{equation}
    \bar{S} = \frac{\sum_{i=1}^{N} S_i/\sigma_i^2}{\sum_{i=1}^{N} 1/\sigma_i^2}
    \label{eq:smean}
\end{equation}
and, $S_i$ are $N$ flux density measurements of a source with individual error $\sigma_i$. 
For MALS flux densities and errors, we adopted measurements based on the {\tt plumber} beam model. Therefore, the NVSS flux densities have been modified assuming a spectral index of $\alpha=-0.7$ (Fig.~\ref{fig:spdrad}). 
Note that the selection of variables at low fractional variability ($f_{var}$ = $S_{MALS}$ / $S_{NVSS}$) is affected by the MALS-to-NVSS offset of 1.06 (Section~\ref{sec:fluxscale}) and the choice of the beam model.  For {\tt katbeam} model, the sample would be smaller by $\sim$10\% at 1 $ < f_{var} <$ 1.15.
%

For comparison with previous work, in addition to $f_{var}$, we adopt the following two metrics to quantify the variability of the source.  The first is modulation index, $m$ = $\sigma/\bar{S}$, where $\sigma$ is the standard deviation of the flux density measurements.  For two epoch variability relevant here, we define \citep[][]{Mooley2016, Hajela2019, Ross2021}:
\begin{equation}
    m = |\Delta S| / \bar{S},
\end{equation}
where $\Delta S$ = $S_1 - S_2$, where $S_1$ and $S_2$ are two flux density measurements, and $\bar{S}$ is their mean as defined in Equation~\ref{eq:smean}.  
The second quantity we adopt is variability strength, $V_s$, defined as \citep[][]{Mooley2016, Hajela2019, Ross2021}:
\begin{equation}
    V_s = \frac{\Delta S}{\sqrt{\sigma_1^2+\sigma_2^2}}.
\end{equation}
$V_s$ is expected to be distributed according to student's t-distribution and may be preferred over $\chi^2$-statistics when the degrees of freedom are small \citep[][]{Mooley2016}. 

We computed the above mentioned quantities for all 15,691 sources to produce a list of 1960 variable sources applying a 99.9\% threshold based on $\chi^2$ statistics.  The arbitrary stringent threshold was adopted for practical reasons to generate a reasonable number of candidates which can then be subjected to visual examination. It also yields a sample that is less affected by measurement uncertainties at low SNR.
Indeed the crucial process of visual examination revealed 631 candidates to be false.  These are either imaging artifacts or the flux density comparison between MALS and NVSS is affected by the blending or proximity to a nearby source in NVSS or the splitting of a single source in NVSS into multiple components in MALS.  
We also omit 21 candidates corresponding to J0211+1707 and J1833-2103, the pointings with extreme flux density offsets (see Section~\ref{sec:fluxscale}).

The list of 1308 variable targets is provided in Table~\ref{tab:var}.  
The distribution of $|V_s|$ (median = 5.5) is shown in an inset in Fig.~\ref{fig:vars}.  
The distribution of $W_1$, $W_2$ and $W_3$ colors of 763 sources detected in the AllWISE catalog are also shown in the {\it right} panel of Fig.~\ref{fig:vars}.  The distribution of colors implies, as expected, that the majority of these variables are AGN, although a few could also be stars\footnote{Indeed seven of these with a counterpart within $1^{\prime\prime}$ in Gaia catalog show significant ($>3\sigma$) proper motion \citep[][]{Gaia22}.}.
The median radio spectral index ($\alpha^{spw2}_{spw9}$) of the sample based on MALS SPW2 and SPW9 is $-0.65^{+0.02}_{-0.02}$, which is slightly flatter as compared to whole MALS SPW9 catalog (see Section~\ref{sec:uss}).  The errors define 90\% confidence level estimated by bootstrapping.
The spectral indices are substantially flatter (median $\alpha^{spw2}_{spw9}$ = $-0.48^{+0.04}_{-0.04}$) for the sources with color overlapping with the locus of powerful AGN ($W_1 - W_2 >$ 0.5; also see \textit{left} panel of Fig.~\ref{fig:vars}), implying that a substantial fraction of these could be blazars.

The comparison between MALS and NVSS also revealed a subset of transients detected in NVSS but missing in the SPW9 catalog and vice versa.  
In summary, we identified 734 radio sources brighter than 4.0\,mJy (SNR$>8$ ; {\tt Distance\_pointing}$<45\arcmin$) in the SPW9 catalog but no counterparts within $60^{\prime\prime}$ in NVSS.  In NVSS, there are 123 such radio sources with no counterparts within $60^{\prime\prime}$ radius in the SPW9 catalog.  Through visual inspection, we found only 115/734 and 7/123 to be {\it true} transients.  The remaining are artifacts and false detections or missed-identifications in NVSS.  The median flux densities of radio sources explored among the transient candidates is 4.5\,mJy, implying a large fraction is close to the detection limit of NVSS and is severely affected by incompleteness and source finding inefficiencies.  This is also implied by much larger fraction of transients detected with no counterparts in NVSS. A large fraction of these could simply be variable sources that were fainter during the NVSS observations, and hence missed.  The distribution of 58 of these detected in the AllWISE catalog is shown in Fig.~\ref{fig:vars}, and is very similar to the colors of variables discussed above.

Further investigation of the variables and transients identified here requires inputs from multi-epoch optical images and spectra.  
Even though only 30\% of these have a counterpart within $2^{\prime\prime}$ in the Panoramic Survey Telescope and Rapid Response System \citep[][]{Chambers16}, besides being optically faint AGN, a small fraction of these could also be supernovae and GRBs.
This exploration is beyond the scope of this work and will be presented in a future paper.

\section{Summary and outlook}    
\label{sec:summ}  

Through the MeerKAT Absorption Line Survey \citep[MALS; ][]{Gupta17mals}, we have observed 391 telescope pointings at L-band (900 - 1670\,MHz) at declinations $\lesssim+20\degree$.  For spectral line processing, the L-band is split into 15 spectral windows (SPWs) labeled as SPW0 to SPW14. 
This paper presents radio continuum images and a catalog of 495,325 (240,321) radio sources detected over an area of 2289\,deg$^2$ (1132\,deg$^2$) at 1006\,MHz i.e., SPW2 (1381\,MHz i.e., SPW9).  This is the first of several data releases to come from MALS.  
%

The 1381\,MHz (SPW9) radio continuum images presented here have a spatial resolution of $8^{\prime\prime}$ and rms noise of $\sim$22\,$\mu$Jy\,beam$^{-1}$.  
The catalog released here is primarily constructed from the  cosine approximated analytic {\tt katbeam} model \citep[][]{Mauch20} but also provides measurements and corrections that can be used to obtain the values corresponding to the alternate beam model that implements holographic measurements through {\tt plumber} \citep[][]{Sekhar22}.  At 1381\,MHz, the outcomes from these two models are in excellent agreement, but tend to diverge by a few percent in the outer regions ($>45\arcmin$ from the pointing center). Thus, the measurements in the outer regions may have larger systematic errors and we advise caution in using these.  Further improvements will follow from the application of primary beam corrections in the visibility plane  via AW projection \citep[][]{Bhatnagar13}.

Through the analysis of 1,150 multiply-observed sources in MALS, we estimate the systematic uncertainties in astrometry and flux density scale ratio to be $<1^{\prime\prime}$ and $1$\% ($\sim 8\%$ scatter).
By comparing the positions and {\tt katbeam}-based flux densities with NVSS and FIRST at 1.4\,GHz, we establish the catalog's accuracy in astrometry and flux density scale to be better than 0\farcs8 and 6\% (15\% scatter), respectively. In comparison, with the {\tt plumber} model we find a median flux density offset of 9\% with respect to NVSS with a MAD of 16\%, higher than the flux density offsets obtained using {\tt katbeam}.

The majority ($\sim$90\%) of sources in the catalog are modelled with single Gaussian components, and only a few percent require 3 or more Gaussian components (median angular size $\sim9\farcs8$ in SPW9).
We derived radio source counts from the catalogs at 1381\,MHz and compared these with the existing measurements in the literature. Although not corrected for resolution and Eddington biases, the MALS counts show a good agreement -- within 10\% -- with literature counts and remain complete down to 2\, mJy, below which the counts rapidly decline. The slight offsets between source counts from various surveys  could originate from instrumental and analysis effects, and need further investigations \citep[][]{Condon07}.

For a matching radius of $6^{\prime\prime}$, 205,435 sources are common between SPW2 and SPW9.  We calculated spectral indices of 125,621 sources detected at SNR$>$8 in both the SPWs.  
Using a sample of 66,836 (48,817) sources at SPW2 (SPW9), we confirm the flattening of spectral indices with decreasing flux density.  This may be due to higher abundances of FRI i.e., core dominated population of radio sources in lower flux density bins.
Using MALS SPW2 and TGSS ADR1 flux densities, we identify 182  USS sources of which 140 due to optical-infrared properties are prime candidates for being HzRGs ($z>2$).
Through comparison with NVSS we have identified the long term variability (26 years) of radio sources.  We have determined 1308 variables (median variability strength, $|V_s|$ = 5.5) and 122 transients i.e., detected only in MALS or NVSS.  These are primarily AGN but may also comprise radio stars, supernovae and GRBs.  Further exploration of these will be presented in future papers.

The MALS SPW2 and SPW9 catalogs and primary beam corrected Stokes-$I$ images are available at \href{https://mals.iucaa.in}{https://mals.iucaa.in}.  We note that the calibration and imaging presented here, except for a static primary beam correction, does not correct for any direction-dependent errors.  This will be addressed using AW-Projection \citep[][]{Bhatnagar13} in future releases, which will also  provide continuum and spectral line data products from L- and UHF-bands.

\acknowledgments
We thank the anonymous referee for useful and detailed comments.
We thank the SARAO's team of engineers and commissioning scientists for years of intense and successful work towards delivering the wonderful MeerKAT telescope.  The MeerKAT telescope is operated by the South African Radio Astronomy Observatory, which is a facility of the National Research Foundation, an agency of the Department of Science and Innovation. 
The MeerKAT data were processed using the MALS computing facility at IUCAA (https://mals.iucaa.in/releases).
We thank Edward Wright and Chao-Wei Tsai for the permission to reproduce Fig.~12 from \citet[][]{Wright10}. 
KM and KK acknowledge support from the National Research Foundation of South Africa.
The Common Astronomy Software Applications (CASA) package is developed by an international consortium of scientists based at the National Radio Astronomical Observatory (NRAO), the European Southern Observatory (ESO), the National Astronomical Observatory of Japan (NAOJ), the Academia Sinica Institute of Astronomy and Astrophysics (ASIAA), the CSIRO division for Astronomy and Space Science (CASS), and the Netherlands Institute for Radio Astronomy (ASTRON) under the guidance of NRAO.
The National Radio Astronomy Observatory is a facility of the National Science Foundation operated under cooperative agreement by Associated Universities, Inc.  This research has made use of the NASA/IPAC Extragalactic Database (NED), which is funded by the National Aeronautics and Space Administration and operated by the California Institute of Technology. 
This work has made use of data from the European Space Agency (ESA) mission {\it Gaia} (\url{https://www.cosmos.esa.int/gaia}), processed by the {\it Gaia} Data Processing and Analysis Consortium (DPAC, \url{https://www.cosmos.esa.int/web/gaia/dpac/consortium}). Funding for the DPAC has been provided by national institutions, in particular the institutions participating in the {\it Gaia} Multilateral Agreement.

\facilities{MeerKAT}

\software{
ARTIP \citep[][]{Gupta21},
Astropy \citep{Astropy22},
ASURV \citep{Feigelson1985},
CASA \citep[][]{Casa22},
katbeam (\url{https://github.com/ska-sa/katbeam}),
Matplotlib \citep{Hunter07},
NumPy \citep{Harris20},
plumber \citep{Sekhar22},
and PyBDSF \citep{mohan2015}.
}

\clearpage


\def\aj{AJ}%
\def\actaa{Acta Astron.}%
\def\araa{ARA\&A}%
\def\apj{ApJ}%
\def\apjl{ApJ}%
\def\apjs{ApJS}%
\def\ao{Appl.~Opt.}%
\def\apss{Ap\&SS}%
\def\aap{A\&A}%
\def\aapr{A\&A~Rev.}%
\def\aaps{A\&AS}%
\def\azh{AZh}%
\def\baas{BAAS}%
\def\bac{Bull. astr. Inst. Czechosl.}%
\def\caa{Chinese Astron. Astrophys.}%
\def\cjaa{Chinese J. Astron. Astrophys.}%
\def\icarus{Icarus}%
\def\jcap{J. Cosmology Astropart. Phys.}%
\def\jrasc{JRASC}%
\def\mnras{MNRAS}%
\def\memras{MmRAS}%
\def\na{New A}%
\def\nar{New A Rev.}%
\def\pasa{PASA}%
\def\pra{Phys.~Rev.~A}%
\def\prb{Phys.~Rev.~B}%
\def\prc{Phys.~Rev.~C}%
\def\prd{Phys.~Rev.~D}%
\def\pre{Phys.~Rev.~E}%
\def\prl{Phys.~Rev.~Lett.}%
\def\pasp{PASP}%
\def\pasj{PASJ}%
\def\qjras{QJRAS}%
\def\rmxaa{Rev. Mexicana Astron. Astrofis.}%
\def\skytel{S\&T}%
\def\solphys{Sol.~Phys.}%
\def\sovast{Soviet~Ast.}%
\def\ssr{Space~Sci.~Rev.}%
\def\zap{ZAp}%
\def\nat{Nature}%
\def\iaucirc{IAU~Circ.}%
\def\aplett{Astrophys.~Lett.}%
\def\apspr{Astrophys.~Space~Phys.~Res.}%
\def\bain{Bull.~Astron.~Inst.~Netherlands}%
\def\fcp{Fund.~Cosmic~Phys.}%
\def\gca{Geochim.~Cosmochim.~Acta}%
\def\grl{Geophys.~Res.~Lett.}%
\def\jcp{J.~Chem.~Phys.}%
\def\jgr{J.~Geophys.~Res.}%
\def\jqsrt{J.~Quant.~Spec.~Radiat.~Transf.}%
\def\memsai{Mem.~Soc.~Astron.~Italiana}%
\def\nphysa{Nucl.~Phys.~A}%
\def\physrep{Phys.~Rep.}%
\def\physscr{Phys.~Scr}%
\def\planss{Planet.~Space~Sci.}%
\def\procspie{Proc.~SPIE}%
\let\astap=\aap
\let\apjlett=\apjl
\let\apjsupp=\apjs
\let\applopt=\ao
\bibliographystyle{aasjournal}
\bibliography{mybib}

\appendix

\section{List of pointings }
\label{sec:listpoint}

Here we present the list of 391 pointings observed during the first phase of MALS observations. 

\begin{longrotatetable}
\begin{deluxetable*}{lcccccccccc}
\tabcolsep=5pt
\tablecolumns{7}
\tablewidth{0pc}
\tablecaption{All L-band MALS pointings \label{tab:pointings}}
\tablehead{
        \colhead{Pointing ID}  & \colhead{Date and Start time}  & \colhead{$\sigma_1^{SPW9}$} & \colhead{$\sigma_2^{SPW9}$} & \colhead{S$^{SPW9}$} & \colhead{Class} &  \colhead{$\sigma_1^{SPW2}$} & \colhead{$\sigma_2^{SPW2}$} & \colhead{S$^{SPW2}$} & \colhead{$N_{src}^{SPW9}$} & \colhead{$N_{src}^{SPW2}$} \\
        \colhead{ } & \colhead{(UTC)} & \colhead{($\mu$Jy\,beam$^{-1}$)} & \colhead{($\mu$Jy\,beam$^{-1}$)} & \colhead{(mJy)} & \colhead{ } &  \colhead{($\mu$Jy\,beam$^{-1}$)} &  \colhead{($\mu$Jy\,beam$^{-1}$)} & \colhead{(mJy)} & \colhead{ } & \colhead{ } \\
        \colhead{(1)} & \colhead{(2)} & \colhead{(3)} & \colhead{(4)} & \colhead{(5)} & \colhead{(6)} & \colhead{(7)} & \colhead{(8)} & \colhead{(9)} & \colhead{(10)} & \colhead{(11)}\\
        }    
\startdata
J000141.57$-$154040.6 & 2020-07-16T01:40 & 29.4 & 22.8 & 449.1 & B & 36.1 & 23.9 & 609.7 & 563 & 1347 \\
J000234.52$-$052324.5 & 2020-08-19T20:40 & 27.2 & 23.7 & 248.8 & A.2 & 40.8 & 33.5 & 326.8 & 690 & 1154 \\
J000647.35$+$172815.4 & 2020-07-16T01:40 & 30.2 & 23.2 & 203.8 & A.1 & 45.2 & 28.1 & 225.8 & 527 & 968 \\
J001053.65$-$215703.9 & 2020-08-19T20:40 & 22.8 & 21.4 & 353.1 & A.3 & 26.8 & 22.5 & 350.7 & 819 & 1813 \\
J001508.49$+$082803.8 & 2020-07-16T01:40 & 29.1 & 24.0 & 428.9 & A.3 & 46.2 & 30.7 & 519.4 & 592 & 1048 \\
J001708.03$-$125624.9 & 2020-08-19T20:40 & 65.5 & 36.1 & 1065 & B & 73.8 & 43.3 & 1450 & 326 & 694 \\
J002232.46$+$060804.6 & 2020-08-19T00:10 & 28.0 & 22.8 & 615.5 & A.4 & 41.1 & 26.1 & 567.9 & 613 & 1224 \\
J002355.84$-$235717.8 & 2020-08-19T00:10 & 23.2 & 21.3 & 215.6 & A.1 & 29.0 & 23.6 & 285.2 & 744 & 1516 \\
J002546.66$-$124725.2$^\star$ & 2020-08-19T00:10 & 32.4 & 23.5 & 432.6 & B & 496.8 & 189.2 & 1428 & 568 & 128 \\
J003711.10$-$200300.1 & 2020-08-20T00:25 & 40.0 & 29.7 & 261.0 & B & 57.1 & 37.7 & 355.7 & 423 & 834 \\
J004057.62$-$014631.6 & 2020-08-20T00:25 & 55.3 & 47.4 & 599.2 & A.4 & 114.6 & 85.1 & 648.9 & 300 & 428 \\
J004243.06$+$124657.6 & 2020-08-20T00:25 & 30.7 & 24.3 & 630.1 & A.4 & 49.7 & 39.6 & 886.4 & 489 & 749 \\
J004959.40$-$573828.5 & 2021-01-17T15:37 & 35.8 & 24.4 & 1929 & A.4 & 33.1 & 22.7 & 2038 & 442 & 1405 \\
J005147.15$+$174710.3 & 2020-08-26T23:10 & 62.7 & 33.2 & 1369 & A.4 & 71.7 & 36.3 & 1918 & 306 & 710 \\
J005315.65$-$070233.4 & 2020-08-26T23:10 & 31.9 & 23.6 & 234.3 & B & 35.9 & 24.7 & 311.4 & 500 & 1238 \\
J010329.05$+$052131.2 & 2020-07-14T01:40 & 36.6 & 25.8 & 597.2 & A.4 & 48.3 & 32.3 & 798.7 & 445 & 902 \\
J010549.69$+$184507.1 & 2020-08-26T23:10 & 25.2 & 22.1 & 347.1 & A.3 & 36.7 & 26.1 & 450.6 & 571 & 1178 \\
J010838.75$+$013458.9 & 2020-07-14T01:40 & 57.3 & 44.8 & 3271 & A.4\_R & 62.9 & 42.3 & 3665 & 326 & 710 \\
J010838.77$+$140428.4$^\star$ & 2020-07-14T01:40 & 92.6 & 59.7 & 248.7 & B & 197.4 & 119.1 & 319.4 & 223 & 295 \\
J011035.13$-$164831.3 & 2020-09-16T22:40 & 35.1 & 25.9 & 115.6 & B & 50.4 & 35.3 & 117.0 & 530 & 785 \\
J011239.18$-$032843.3 & 2020-09-16T22:40 & 28.0 & 22.5 & 143.2 & A.1\_R & 32.2 & 24.7 & 150.6 & 689 & 1438 \\
J011727.84$-$042511.5 & 2020-08-22T23:45 & 26.4 & 22.3 & 420.1 & A.3 & 35.6 & 27.9 & 467.9 & 558 & 1150 \\
J011945.20$-$195827.9 & 2022-09-16T22:40 & 26.9 & 22.6 & 102.6 & B & 62.1 & 43.1 & 135.7 & 600 & 733 \\
J012156.62$-$265747.8 & 2020-08-22T23:45 & 26.4 & 22.1 & 246.9 & B & 28.4 & 22.2 & 332.3 & 551 & 1512 \\
J012613.24$+$142013.1 & 2020-06-14T04:35 & 34.4 & 24.6 & 594.4 & A.3 & 44.1 & 29.0 & 831.3 & 512 & 1055 \\
J013047.38$-$172505.6 & 2020-08-22T23:45 & 27.0 & 21.3 & 244.9 & B & 27.9 & 21.5 & 318.5 & 655 & 1578 \\
J013212.20$-$065236.8 & 2020-08-13T01:30 & 51.0 & 28.3 & 1299 & A.4 & 56.2 & 32.1 & 1754 & 361 & 848 \\
J013341.68$-$021013.5 & 2020-08-13T01:30 & 26.7 & 22.5 & 237.1 & A.1 & 32.4 & 25.6 & 309.9 & 682 & 1379 \\
J013435.66$-$093103.7 & 2020-08-07T03:12 & 37.7 & 27.6 & 836.0 & A.4 & 44.0 & 30.4 & 686.3 & 457 & 1091 \\
J014349.58$-$434148.4 & 2020-07-26T02:44 & 22.7 & 19.7 & 547.0 & A.1 & 26.5 & 20.2 & 754.6 & 710 & 1629 \\
J014922.36$+$055553.4 & 2020-08-13T01:30 & 52.9 & 30.1 & 954.2 & A.4 & 50.3 & 29.7 & 960.4 & 392 & 941 \\
J015002.78$-$095255.5 & 2020-09-12T22:45 & 30.8 & 23.8 & 211.8 & A.1 & 51.5 & 28.6 & 308.9 & 561 & 937 \\
J015408.54$-$065234.1 & 2020-08-07T03:12 & 33.8 & 27.6 & 100.8 & B\_R & 45.1 & 29.0 & 139.9 & 649 & 1316 \\
J015430.74$-$660417.1 & 2021-01-15T11:30 & 33.0 & 23.4 & 673.5 & A.4 & 31.9 & 22.8 & 847.9 & 490 & 1209 \\
J015454.37$-$000723.8 & 2020-07-26T02:44 & 35.6 & 25.2 & 245.0 & A.2 & 45.9 & 30.3 & 222.4 & 598 & 1080 \\
J015737.90$-$491517.7 & 2021-01-15T11:30 & 22.7 & 20.0 & 202.6 & A.1 & 31.9 & 23.3 & 220.1 & 720 & 1289 \\
J021000.18$-$100353.9 & 2020-09-12T22:45 & 31.3 & 23.6 & 262.6 & A.2 & 45.4 & 29.9 & 318.4 & 546 & 915 \\
J021148.77$+$170723.2$^\dag$ & 2020-09-12T22:45 & 47.1 & 34.5 & 762.0 & A.4\_R & 59.6 & 40.9 & 911.4 & 472 & 942 \\
J021231.86$-$382256.6 & 2020-08-29T23:17 & 21.3 & 19.1 & 247.8 & A.1\_R & 24.4 & 19.3 & 328.1 & 864 & 1924 \\
J021650.70$+$172404.9 & 2020-08-29T23:17 & 31.4 & 24.1 & 389.0 & A.3 & 45.6 & 30.6 & 499.7 & 519 & 1014 \\
J021715.31$-$271555.4 & 2020-08-29T23:17 & 20.5 & 19.1 & 165.6 & A.1\_R & 22.4 & 18.9 & 193.2 & 889 & 2238 \\
J022057.85$-$383302.4 & 2020-09-01T23:20 & 22.1 & 20.3 & 960.1 & A.3\_R & 23.2 & 19.9 & 1326 & 818 & 1997 \\
J022128.93$-$335827.0 & 2020-09-01T23:20 & 24.6 & 21.1 & 213.9 & A.1 & 29.9 & 22.1 & 291.9 & 660 & 1482 \\
J022613.72$+$093726.3 & 2020-09-01T23:20 & 31.0 & 24.3 & 406.1 & A.3 & 42.7 & 29.8 & 450.6 & 520 & 1023 \\
J022639.92$+$194110.1 & 2020-09-09T01:25 & 29.1 & 23.7 & 214.3 & A.1 & 35.6 & 25.9 & 269.5 & 577 & 1260 \\
J022807.66$-$011541.7 & 2020-09-09T01:25 & 25.9 & 23.0 & 478.9 & A.1\_R & 38.8 & 29.2 & 640.4 & 587 & 1087 \\
J023653.12$-$613616.1 & 2020-07-26T02:44 & 20.5 & 19.4 & 237.1 & A.1 & 26.5 & 21.2 & 271.0 & 917 & 1826 \\
J023838.89$+$163658.6 & 2021-01-17T15:37 & 27.1 & 21.6 & 835.2 & A.4 & 45.6 & 27.7 & 746.5 & 593 & 993 \\
J023939.11$-$135409.4 & 2020-09-09T01:25 & 27.8 & 22.3 & 141.0 & B & 37.0 & 25.1 & 122.2 & 554 & 1158 \\
J024027.19$+$095713.0 & 2020-09-14T23:04 & 45.4 & 29.4 & 509.3 & A.3 & 68.0 & 39.9 & 732.8 & 444 & 744 \\
J024122.37$+$101845.7 & 2021-01-11T17:00 & 47.5 & 33.4 & 269.8 & B & 85.3 & 46.0 & 381.2 & 426 & 735 \\
J024935.41$-$075921.0 & 2020-09-14T23:04 & 24.0 & 20.0 & 622.9 & A.4\_R & 29.9 & 21.5 & 876.0 & 695 & 1440 \\
J024939.93$+$044028.9 & 2020-09-14T23:04 & 26.3 & 21.6 & 427.1 & A.3 & 36.3 & 25.5 & 559.5 & 609 & 1201 \\
J024944.50$+$123706.3 & 2020-09-15T22:15 & 26.0 & 22.9 & 255.7 & A.2 & 31.4 & 24.0 & 310.4 & 689 & 1475 \\
J024948.24$-$544357.2 & 2021-01-11T17:00 & 32.2 & 25.2 & 302.7 & B & 50.2 & 29.9 & 399.7 & 585 & 1134 \\
J025035.54$-$262743.1 & 2020-09-15T22:15 & 27.5 & 24.5 & 282.6 & A.2 & 32.6 & 25.3 & 327.3 & 707 & 1558 \\
J025126.55$-$600005.3 & 2021-01-02T15:18 & 27.1 & 21.7 & 255.5 & A.2 & 35.1 & 23.5 & 269.4 & 620 & 1252 \\
J025509.79$+$025345.9 & 2020-06-14T04:35 & 33.3 & 24.0 & 803.7 & A.4 & 47.7 & 28.3 & 1087 & 528 & 981 \\
J030032.15$-$302243.8 & 2020-09-15T22:15 & 25.8 & 22.6 & 366.8 & A.3 & 29.1 & 23.4 & 511.3 & 631 & 1414 \\
J030413.80$-$112653.5 & 2020-08-22T03:31 & 25.6 & 21.1 & 347.5 & A.3 & 38.1 & 25.2 & 454.3 & 760 & 1303 \\
J030841.90$+$072043.4 & 2021-01-02T15:18 & 57.0 & 34.9 & 392.2 & A.3 & 59.4 & 38.1 & 461.0 & 296 & 702 \\
J031551.28$-$364450.8 & 2020-08-22T03:31 & 21.3 & 19.4 & 477.2 & A.3 & 36.3 & 32.5 & 631.1 & 812 & 1159 \\
J032128.78$-$294047.4 & 2020-08-22T03:31 & 26.9 & 21.1 & 2123 & A.4 & 31.1 & 21.9 & 2624 & 604 & 1333 \\
J032153.11$+$122114.0 & 2021-01-18T15:30 & 34.7 & 25.0 & 1910 & A.4 & 43.8 & 31.7 & 2067 & 429 & 874 \\
J032808.59$-$015220.2 & 2020-09-06T00:46 & 24.6 & 20.9 & 228.4 & A.1 & 29.6 & 23.1 & 300.9 & 699 & 1524 \\
J033145.66$-$515618.0 & 2020-07-12T07:27 & 20.8 & 19.3 & 177.8 & A.1 & 22.5 & 19.3 & 249.6 & 1023 & 2360 \\
J033157.66$-$405840.7 & 2020-08-07T03:12 & 23.9 & 20.6 & 131.9 & B\_R & 34.4 & 22.7 & 197.4 & 767 & 1458 \\
J033242.97$-$724904.9 & 2021-01-18T15:30 & 24.6 & 22.6 & 247.6 & A.2 & 23.7 & 20.2 & 263.7 & 793 & 2097 \\
J033940.22$-$222446.1 & 2020-09-06T00:46 & 24.2 & 20.2 & 743.7 & A.4 & 26.6 & 20.9 & 820.0 & 677 & 1591 \\
J034044.61$-$650711.8 & 2021-01-10T17:21 & 21.0 & 18.9 & 356.9 & A.2\_R & 28.5 & 21.2 & 549.7 & 870 & 1645 \\
J034211.45$-$043441.7 & 2020-09-06T00:46 & 24.1 & 20.3 & 340.4 & A.1\_R & 26.9 & 21.5 & 464.0 & 759 & 1677 \\
J034334.69$-$444530.7 & 2020-07-12T07:27 & 22.1 & 19.6 & 223.1 & A.1 & 25.3 & 19.5 & 310.4 & 811 & 2011 \\
J035016.86$-$035111.3 & 2020-09-17T23:00 & 29.5 & 21.6 & 251.1 & A.2 & 37.8 & 25.6 & 304.1 & 601 & 1129 \\
J035211.06$-$251449.8 & 2020-09-17T23:00 & 25.7 & 19.9 & 287.8 & A.2 & 28.4 & 20.0 & 306.4 & 713 & 1560 \\
J035414.70$-$030804.3 & 2020-09-17T23:00 & 27.7 & 23.9 & 735.5 & A.4 & 46.4 & 28.0 & 950.3 & 583 & 1054 \\
J035424.14$+$044107.4 & 2021-01-10T17:21 & 29.1 & 23.4 & 456.0 & A.3 & 44.9 & 31.0 & 463.1 & 515 & 861 \\
J035634.55$-$083121.3 & 2020-09-19T22:45 & 24.4 & 20.3 & 239.3 & A.2 & 31.3 & 25.0 & 297.4 & 670 & 1376 \\
J035721.82$-$481214.3 & 2020-08-02T04:00 & 22.4 & 19.4 & 337.4 & A.3 & 23.8 & 19.4 & 263.7 & 759 & 1984 \\
J040015.06$-$142328.0 & 2020-07-12T07:27 & 27.2 & 21.9 & 398.9 & A.3 & 36.6 & 23.9 & 533.7 & 676 & 1351 \\
J040353.79$-$360501.7 & 2020-09-19T22:45 & 22.6 & 19.3 & 1087 & A.4\_R & 24.3 & 19.7 & 976.6 & 648 & 1806 \\
J040729.20$+$175055.3 & 2021-01-10T17:21 & 28.2 & 22.1 & 542.1 & A.4 & 39.4 & 27.7 & 704.2 & 504 & 975 \\
J041437.75$+$053443.2 & 2020-08-02T04:00 & 44.8 & 29.1 & 1953 & A.4 & 54.2 & 31.9 & 2435 & 355 & 845 \\
J041530.42$-$234750.2 & 2020-09-19T22:45 & 21.5 & 18.8 & 260.6 & A.2 & 22.6 & 18.5 & 361.6 & 840 & 2069 \\
J041620.54$-$333931.3 & 2020-09-03T00:15 & 24.0 & 20.5 & 202.7 & B & 30.7 & 24.1 & 213.6 & 681 & 1399 \\
J041654.64$-$625101.7 & 2020-07-05T06:20 & 29.9 & 26.2 & 128.2 & B & 52.9 & 41.1 & 186.9 & 503 & 747 \\
J042047.23$-$571252.8 & 2020-08-02T04:00 & 21.4 & 19.2 & 345.3 & A.1 & 24.6 & 20.1 & 454.5 & 810 & 1927 \\
J042248.53$-$203456.6 & 2020-09-03T00:15 & 21.5 & 19.5 & 189.3 & A.1 & 23.8 & 20.2 & 219.4 & 871 & 2054 \\
J042725.05$+$085330.3 & 2020-07-05T06:20 & 52.8 & 31.4 & 981.4 & A.4 & 63.9 & 35.8 & 1316 & 361 & 777 \\
J044414.54$-$592453.3 & 2020-07-05T06:20 & 27.0 & 21.9 & 1266 & A.4 & 30.3 & 22.3 & 1754 & 660 & 1467 \\
J044604.72$-$114813.4 & 2020-09-03T00:15 & 22.3 & 20.1 & 264.1 & A.2 & 25.0 & 20.4 & 338.2 & 771 & 1879 \\
J044849.48$-$093531.3 & 2020-09-17T03:00 & 24.9 & 21.3 & 174.6 & B & 33.0 & 23.1 & 194.6 & 717 & 1444 \\
J045541.91$+$185010.9 & 2020-09-17T03:00 & 24.0 & 20.8 & 314.5 & A.2 & 30.8 & 23.8 & 422.9 & 654 & 1277 \\
J050725.04$-$362442.9 & 2020-09-17T03:00 & 21.8 & 20.0 & 224.6 & A.1\_R & 22.8 & 20.0 & 287.7 & 849 & 2014 \\
J051134.05$+$024416.0 & 2020-09-20T02:35 & 26.9 & 23.4 & 354.0 & A.3 & 44.7 & 35.3 & 466.9 & 696 & 1054 \\
J051240.99$+$151723.8 & 2020-09-20T02:35 & 21.9 & 19.3 & 491.3 & A.3\_R & 26.8 & 22.6 & 533.0 & 759 & 1601 \\
J051340.03$+$010023.6 & 2020-09-20T02:35 & 44.6 & 25.7 & 434.3 & B & 59.6 & 31.5 & 494.6 & 423 & 798 \\
J051511.18$-$012002.4 & 2020-09-21T00:15 & 27.5 & 21.2 & 293.4 & A.2 & 35.2 & 24.0 & 383.8 & 634 & 1212 \\
J051656.35$+$073252.7 & 2020-09-21T00:15 & 24.5 & 20.8 & 194.9 & A.1 & 32.9 & 24.4 & 181.6 & 755 & 1409 \\
J051700.70$+$071225.2 & 2020-09-21T00:15 & 30.8 & 24.9 & 201.7 & B & 40.4 & 29.6 & 269.1 & 595 & 1131 \\
J052318.55$-$261409.6 & 2020-09-27T01:51 & 25.3 & 20.1 & 1466 & A.4 & 26.7 & 19.8 & 1379 & 677 & 1726 \\
J052531.29$-$455755.4$^\star$ & 2020-08-01T04:10 & 43.0 & 36.8 & 830.6 & A.4 & 133.4 & 115.9 & 1141 & 383 & 351 \\
J052744.71$-$121153.7 & 2020-09-27T01:51 & 25.1 & 20.1 & 344.1 & A.3\_R & 36.7 & 25.1 & 445.3 & 685 & 1265 \\
J052905.55$-$112607.5 & 2020-08-01T04:10 & 51.6 & 46.0 & 452.2 & B & 87.1 & 63.2 & 594.8 & 378 & 605 \\
J054518.09$-$261857.3 & 2020-09-27T01:51 & 23.7 & 19.6 & 235.8 & B\_R & 29.1 & 20.2 & 336.9 & 780 & 1617 \\
J054606.57$-$515525.8 & 2020-08-01T04:10 & 22.4 & 20.7 & 244.2 & A.1 & 25.1 & 20.9 & 346.6 & 828 & 2021 \\
J055341.91$-$084003.3 & 2020-09-23T01:07 & 22.9 & 21.0 & 236.1 & A.2\_R & 30.4 & 23.9 & 265.1 & 777 & 1547 \\
J060248.09$-$254602.2 & 2020-09-23T01:07 & 23.2 & 20.5 & 541.2 & A.4\_R & 26.2 & 20.3 & 651.7 & 692 & 1711 \\
J061038.80$-$230145.6 & 2020-09-23T01:07 & 24.7 & 21.5 & 407.2 & A.3 & 29.8 & 23.3 & 410.0 & 642 & 1504 \\
J061856.02$-$315835.2 & 2020-09-27T23:30 & 24.7 & 21.4 & 333.4 & A.3 & 25.9 & 21.8 & 464.5 & 734 & 1966 \\
J063602.28$-$311312.5 & 2020-09-27T23:30 & 20.5 & 18.7 & 293.7 & A.2\_R & 23.1 & 18.7 & 316.1 & 877 & 1998 \\
J063613.53$-$310646.3 & 2020-09-27T23:30 & 20.3 & 18.5 & 205.5 & B\_R & 22.2 & 18.4 & 276.1 & 853 & 2183 \\
J064255.16$-$325917.6 & 2020-09-07T02:20 & 22.0 & 20.0 & 245.8 & A.2 & 25.1 & 20.9 & 342.8 & 800 & 1917 \\
J070249.30$-$330205.0 & 2020-06-27T07:30 & 21.6 & 19.7 & 312.8 & A.2 & 23.8 & 20.3 & 395.0 & 779 & 1986 \\
J071047.14$-$381346.1 & 2020-06-27T07:30 & 29.5 & 22.2 & 250.0 & B & 35.5 & 24.8 & 290.9 & 627 & 1372 \\
J073159.01$+$143336.3 & 2020-06-27T07:30 & 32.1 & 28.0 & 281.8 & A.2 & 66.7 & 52.3 & 282.0 & 494 & 717 \\
J073714.60$-$382841.9 & 2020-09-07T02:20 & 32.0 & 24.2 & 227.1 & B & 39.2 & 27.3 & 301.8 & 499 & 1018 \\
J073931.51$-$013729.7 & 2020-09-07T02:20 & 41.6 & 26.7 & 240.2 & B & 58.0 & 35.9 & 338.1 & 443 & 872 \\
J074155.69$-$264729.8 & 2020-09-01T04:30 & 28.3 & 25.5 & 497.9 & A.4 & 32.9 & 26.0 & 517.5 & 571 & 1280 \\
J075141.49$+$271631.8 & 2020-10-12T02:25 & 32.2 & 24.4 & 605.5 & A.4 & 42.2 & 27.4 & 802.9 & 457 & 904 \\
J080217.91$+$080525.4 & 2020-09-01T04:30 & 30.2 & 26.4 & 218.0 & A.1 & 38.8 & 29.1 & 288.8 & 542 & 1055 \\
J080356.46$+$042102.8 & 2020-06-28T08:15 & 30.9 & 23.4 & 346.7 & A.3 & 42.6 & 30.3 & 351.1 & 595 & 1101 \\
J080622.15$-$272611.5 & 2020-06-28T08:15 & 27.2 & 21.1 & 248.5 & B\_R & 30.1 & 22.6 & 339.5 & 660 & 1523 \\
J080804.34$+$005708.2 & 2020-06-28T08:15 & 24.6 & 21.0 & 336.2 & A.3 & 31.9 & 23.5 & 380.3 & 684 & 1359 \\
J081727.59$-$030737.7 & 2020-10-12T02:25 & 55.9 & 33.0 & 500.6 & B & 106.4 & 46.1 & 651.0 & 406 & 633 \\
J081936.62$-$063047.9 & 2020-10-12T02:25 & 25.0 & 22.1 & 296.2 & A.2 & 29.7 & 23.5 & 320.3 & 661 & 1546 \\
J082444.81$-$102943.6 & 2020-09-18T02:50 & 25.9 & 21.0 & 331.2 & A.3 & 32.6 & 24.1 & 405.1 & 684 & 1325 \\
J082638.44$-$032426.4 & 2020-07-18T08:10 & 25.0 & 20.7 & 220.6 & A.1 & 33.9 & 26.7 & 293.1 & 664 & 1259 \\
J083052.11$+$241058.7 & 2020-07-18T08:10 & 40.8 & 32.1 & 957.7 & A.4 & 63.4 & 44.0 & 966.8 & 363 & 632 \\
J083601.35$+$040636.0 & 2020-07-18T08:10 & 27.4 & 22.4 & 454.6 & A.3 & 38.9 & 27.6 & 573.6 & 654 & 1247 \\
J083639.27$-$201658.8 & 2020-09-18T02:50 & 83.2 & 49.6 & 2145 & B & 90.5 & 57.7 & 2362 & 189 & 481 \\
J084548.34$-$111800.1 & 2020-09-18T02:50 & 26.3 & 21.8 & 273.0 & A.2 & 36.3 & 24.5 & 372.6 & 617 & 1231 \\
J085826.92$-$260721.0 & 2020-09-01T04:30 & 52.6 & 37.5 & 393.5 & A.3 & 74.6 & 54.8 & 498.4 & 303 & 554 \\
J085838.19$+$093745.2 & 2020-09-06T04:50 & 32.0 & 25.0 & 413.5 & A.3 & 41.5 & 27.9 & 518.1 & 554 & 1043 \\
J090337.99$-$311739.0 & 2020-09-06T04:50 & 21.3 & 19.7 & 190.6 & A.1 & 23.5 & 20.1 & 201.7 & 925 & 2042 \\
J090910.66$-$163753.8 & 2020-09-06T04:50 & 24.7 & 20.5 & 323.3 & A.2 & 33.1 & 22.9 & 440.9 & 691 & 1412 \\
J090911.33$-$313334.1 & 2020-08-22T07:30 & 23.6 & 20.9 & 605.5 & A.1 & 28.1 & 22.3 & 816.8 & 710 & 1594 \\
J090916.99$-$050053.2 & 2020-08-22T07:30 & 27.1 & 21.6 & 448.7 & A.3 & 32.5 & 25.1 & 508.8 & 574 & 1318 \\
J091051.01$-$052626.8 & 2020-08-02T10:30 & 25.1 & 20.8 & 388.1 & A.3 & 27.7 & 21.9 & 409.1 & 677 & 1640 \\
J091133.51$+$195814.3 & 2020-08-02T10:30 & 42.2 & 35.0 & 262.1 & A.2 & 57.5 & 48.8 & 287.2 & 356 & 682 \\
J091253.34$-$135241.9 & 2020-08-02T10:30 & 28.9 & 22.7 & 270.1 & A.2 & 47.8 & 30.6 & 351.5 & 637 & 1108 \\
J092956.48$+$014843.8 & 2020-08-22T07:30 & 29.8 & 22.7 & 251.4 & B & 37.8 & 27.9 & 271.6 & 539 & 1191 \\
J093919.21$-$173135.4 & 2020-09-04T05:31 & 25.8 & 23.1 & 206.1 & A.1 & 31.1 & 23.4 & 227.6 & 698 & 1652 \\
J094040.47$-$002800.1 & 2020-09-04T05:31 & 30.9 & 25.4 & 329.8 & A.2 & 47.4 & 30.0 & 417.9 & 566 & 956 \\
J094146.01$-$032121.7 & 2020-09-04T05:31 & 25.3 & 21.5 & 356.5 & A.3 & 32.6 & 23.7 & 465.6 & 714 & 1405 \\
J094256.94$-$094800.5 & 2020-08-12T08:15 & 26.2 & 21.6 & 348.5 & A.3 & 34.1 & 24.8 & 477.0 & 644 & 1277 \\
J094532.32$-$305756.9 & 2020-08-12T08:15 & 23.3 & 21.4 & 210.1 & A.1 & 27.4 & 21.9 & 262.6 & 765 & 1741 \\
J095123.18$-$325554.8 & 2020-08-12T08:15 & 23.2 & 21.4 & 182.0 & A.1 & 26.3 & 22.6 & 244.8 & 821 & 1833 \\
J095138.34$-$060138.0 & 2020-09-13T05:50 & 29.9 & 24.2 & 391.1 & A.3 & 39.0 & 26.6 & 545.3 & 574 & 1185 \\
J095231.66$-$245349.1 & 2020-09-13T05:50 & 39.5 & 30.7 & 211.6 & A.1 & 58.2 & 34.9 & 223.9 & 469 & 947 \\
J095414.75$+$125222.6 & 2020-09-13T05:50 & 37.3 & 31.2 & 358.4 & A.3 & 49.0 & 34.1 & 460.0 & 449 & 940 \\
J095423.67$+$004350.5 & 2020-09-24T04:30 & 30.0 & 25.5 & 297.6 & A.1 & 79.9 & 52.3 & 439.2 & 544 & 597 \\
J100715.18$-$124746.7 & 2020-09-24T04:30 & 30.6 & 23.9 & 381.9 & A.3 & 43.4 & 28.2 & 410.0 & 570 & 1090 \\
J100954.53$+$160203.8 & 2020-09-24T04:30 & 24.3 & 21.5 & 199.5 & A.1 & 30.6 & 22.6 & 227.2 & 689 & 1462 \\
J101030.73$-$044006.7 & 2020-10-09T04:10 & 31.4 & 23.2 & 299.7 & A.2 & 37.6 & 24.5 & 349.0 & 528 & 1227 \\
J101313.10$-$254654.7 & 2020-10-09T04:10 & 27.2 & 23.2 & 258.1 & B & 32.6 & 24.5 & 263.1 & 612 & 1426 \\
J101357.73$-$245933.6 & 2020-06-25T14:32 & 26.5 & 23.0 & 266.5 & A.2 & 28.9 & 23.5 & 369.6 & 636 & 1579 \\
J101448.67$-$294433.7 & 2020-06-25T14:32 & 26.4 & 22.6 & 234.2 & A.2 & 30.4 & 24.1 & 318.7 & 777 & 1738 \\
J101549.11$+$024337.8 & 2020-06-25T14:32 & 32.3 & 23.1 & 305.2 & A.2 & 43.1 & 27.5 & 409.9 & 535 & 979 \\
J101555.33$-$082245.7 & 2020-10-09T04:10 & 25.8 & 22.3 & 258.6 & A.2 & 35.5 & 25.9 & 384.2 & 680 & 1286 \\
J101730.54$-$384643.7 & 2020-08-16T07:32 & 20.8 & 19.0 & 344.1 & A.2 & 22.5 & 19.2 & 510.2 & 802 & 1907 \\
J101838.68$-$222502.4 & 2020-08-16T07:32 & 30.8 & 22.9 & 291.1 & B & 34.8 & 22.8 & 357.5 & 546 & 1379 \\
J102159.78$-$011002.4 & 2020-08-16T07:32 & 36.2 & 25.6 & 513.9 & A.4 & 47.0 & 27.6 & 713.4 & 534 & 1067 \\
J102441.29$-$153025.2 & 2020-09-26T04:41 & 26.8 & 23.3 & 232.3 & A.2 & 40.8 & 29.0 & 293.4 & 681 & 1205 \\
J102548.78$-$042933.7 & 2020-09-26T04:41 & 29.2 & 22.9 & 471.6 & A.3 & 38.7 & 27.8 & 527.7 & 609 & 1188 \\
J103334.00$+$071126.3 & 2020-08-10T09:10 & 24.7 & 21.4 & 297.8 & A.2 & 27.6 & 21.8 & 294.2 & 737 & 1685 \\
J104015.57$+$062747.0 & 2020-09-26T04:41 & 54.5 & 36.9 & 641.0 & B & 64.8 & 45.4 & 881.5 & 369 & 748 \\
J104051.07$-$144919.4 & 2020-08-10T09:10 & 27.0 & 22.0 & 275.9 & A.2 & 31.5 & 23.2 & 367.6 & 687 & 1548 \\
J104314.53$-$232317.5 & 2020-09-20T06:25 & 28.1 & 23.5 & 206.1 & A.1 & 30.8 & 24.1 & 283.1 & 659 & 1485 \\
J104347.34$+$164106.9 & 2020-09-20T06:25 & 25.8 & 22.0 & 225.5 & A.1 & 43.7 & 31.2 & 283.3 & 574 & 905 \\
J104631.87$+$134911.1 & 2020-09-20T06:25 & 23.5 & 20.4 & 158.1 & A.1 & 32.3 & 23.6 & 225.3 & 701 & 1347 \\
J105024.43$-$031808.8 & 2020-08-01T12:15 & 30.7 & 22.3 & 195.0 & A.1 & 39.5 & 25.0 & 236.9 & 608 & 1209 \\
J105231.85$+$080609.1 & 2020-08-01T12:15 & 32.1 & 23.3 & 411.1 & A.3 & 42.5 & 26.6 & 549.8 & 571 & 1120 \\
J105955.36$-$111818.7 & 2020-08-01T12:15 & 26.8 & 21.0 & 323.0 & A.3 & 29.9 & 21.8 & 456.7 & 676 & 1610 \\
J110125.00$-$212429.4 & 2020-09-27T05:55 & 22.9 & 19.7 & 214.4 & A.1 & 35.9 & 24.6 & 282.0 & 674 & 1263 \\
J111457.57$-$051721.4 & 2020-09-27T05:55 & 28.4 & 21.7 & 328.1 & A.3 & 36.5 & 28.3 & 380.5 & 691 & 1306 \\
J111820.61$-$305459.0 & 2020-09-27T05:55 & 28.6 & 21.7 & 211.7 & A.1 & 30.6 & 21.8 & 153.0 & 676 & 1578 \\
J111917.36$-$052707.9 & 2020-08-17T11:27 & 41.7 & 28.9 & 1189 & A.4 & 52.2 & 31.2 & 1545 & 434 & 888 \\
J111925.22$-$030251.6 & 2020-08-17T11:27 & 57.3 & 34.7 & 1739 & A.4 & 73.7 & 42.3 & 2789 & 305 & 615 \\
J112402.56$-$150159.1 & 2020-09-20T10:05 & 25.5 & 21.6 & 431.2 & A.3 & 33.8 & 25.4 & 389.4 & 624 & 1238 \\
J112712.45$-$073512.7 & 2020-08-17T11:27 & 28.9 & 22.7 & 199.6 & B & 36.0 & 25.1 & 304.4 & 578 & 1189 \\
J113156.47$+$045549.3 & 2020-08-10T09:10 & 26.8 & 21.6 & 513.3 & A.4 & 45.3 & 26.8 & 675.3 & 666 & 1100 \\
J113303.12$+$001548.9 & 2020-06-26T15:35 & 36.5 & 29.6 & 359.3 & A.3 & 63.8 & 40.9 & 387.1 & 522 & 726 \\
J113320.08$+$004052.5 & 2020-10-03T09:00 & 21.5 & 17.7 & 306.8 & B & 41.9 & 31 & 330.1 & 778 & 1061 \\
                      & 2020-10-17T08:52 &      &      &       &   &      &    &       &     &      \\
J114226.58$-$263313.7 & 2020-10-03T09:00 & 20.5 & 15.6 & 313.1 & B & 23 & 16.4 & 361.9 & 924 & 2095 \\
                      & 2020-10-17T08:52 &      &      &       &   &      &    &       &     &      \\
J114450.43$-$145502.7 & 2020-10-03T09:00 & 20.0 & 15.6 & 427.5 & A.3 & 23.7 & 15.8 & 535.6 & 937 & 2114 \\
                      & 2020-10-17T08:52 &      &      &       &   &      &    &       &     &      \\
J114825.45$+$140449.7 & 2020-10-02T06:45 & 26.5 & 22.1 & 330.5 & A.3 & 37.9 & 28.6 & 442.2 & 604 & 1085 \\
J115043.70$+$131736.1 & 2020-10-02T06:45 & 43.2 & 29.1 & 260.0 & B & 68.0 & 43.4 & 340.0 & 407 & 623 \\
J115222.04$-$270126.3 & 2020-10-02T06:45 & 21.9 & 20.2 & 212.8 & A.1 & 24.8 & 21.0 & 266.6 & 838 & 1914 \\
J115306.72$-$044254.5 & 2020-10-17T05:16 & 26.9 & 22.3 & 680.5 & A.3 & 42.9 & 27.2 & 946.5 & 719 & 1123 \\
J115450.38$-$281249.7 & 2020-10-17T05:16 & 20.7 & 18.7 & 470.2 & A.3 & 22.3 & 18.3 & 578.8 & 842 & 2086 \\
J115855.48$-$170524.7 & 2020-09-05T11:00 & 22.2 & 19.5 & 314.2 & A.1\_R & 25.5 & 20.0 & 476.2 & 745 & 1633 \\
J120123.25$+$002829.4 & 2020-09-05T11:00 & 27.6 & 23.4 & 450.5 & A.3 & 32.8 & 23.9 & 522.7 & 709 & 1470 \\
J120632.23$-$071452.6 & 2020-09-05T11:00 & 32.4 & 25.0 & 722.4 & A.4 & 34.1 & 24.9 & 928.1 & 611 & 1428 \\
J120746.68$-$133201.0 & 2020-10-17T05:16 & 28.7 & 22.0 & 202.9 & A.1 & 45.2 & 25.5 & 271.3 & 682 & 1151 \\
J121059.72$-$162042.7 & 2020-10-24T05:46 & 22.6 & 19.9 & 337.1 & A.3 & 30.1 & 22.4 & 423.6 & 736 & 1474 \\
J121128.51$-$351003.4 & 2020-10-24T05:46 & 20.9 & 19.2 & 276.2 & A.2 & 25.2 & 20.8 & 330.4 & 789 & 1787 \\
J121157.77$-$192607.5 & 2020-10-24T05:46 & 28.8 & 21.9 & 468.8 & A.3 & 34.1 & 22.6 & 463.0 & 533 & 1253 \\
J121211.89$-$373826.9 & 2020-10-31T05:15 & 23.6 & 20.2 & 364.0 & A.3\_R & 24.2 & 20.6 & 439.4 & 746 & 1850 \\
J121332.13$+$130720.4 & 2020-10-31T05:15 & 30.9 & 23.3 & 1326 & A.4 & 48.7 & 28.8 & 1522 & 507 & 878 \\
J121514.42$-$062803.5 & 2020-10-31T05:15 & 26.3 & 22.6 & 315.5 & A.2 & 39.7 & 26.5 & 359.9 & 690 & 1211 \\
J121836.18$-$063115.8 & 2020-09-06T08:45 & 28.1 & 23.9 & 415.3 & A.3 & 41.2 & 28.0 & 524.0 & 662 & 1122 \\
J122131.10$-$135402.3 & 2020-09-06T08:45 & 24.4 & 21.2 & 180.6 & A.1 & 28.5 & 22.6 & 212.8 & 743 & 1705 \\
J122222.59$+$041317.3 & 2020-09-06T08:45 & 33.9 & 26.0 & 568.0 & A.4 & 45.1 & 39.4 & 692.5 & 524 & 951 \\
J123150.30$-$123637.5 & 2020-11-05T05:30 & 26.6 & 22.9 & 189.3 & A.1 & 32.9 & 24.2 & 216.7 & 680 & 1449 \\
J123200.13$-$022404.1 & 2020-06-26T15:35 & 61.9 & 38.5 & 1694 & A.4\_R & 84.7 & 47.8 & 2055 & 278 & 545 \\
J123410.08$-$332638.5 & 2020-11-05T05:30 & 28.1 & 22.9 & 284.5 & B & 35.6 & 25.0 & 364.8 & 634 & 1274 \\
J123428.07$+$082941.4 & 2020-11-05T05:30 & 35.8 & 28.5 & 172.0 & A.1 & 44.3 & 35.0 & 212.9 & 408 & 935 \\
J123906.45$-$003930.0 & 2020-09-01T10:25 & 32.7 & 24.6 & 409.8 & A.3 & 45.6 & 30.0 & 555.9 & 587 & 980 \\
J124448.99$-$044610.2$^\star$ & 2020-09-11T12:15 & 379.8 & 264.8 & 432.7 & B & 755.3 & 483.7 & 498.5 & 243 & 137 \\
J124553.81$-$161645.2 & 2020-09-01T10:25 & 25.9 & 22.6 & 508.5 & A.4 & 27.2 & 24.4 & 365.8 & 681 & 1823 \\
J124557.54$-$412846.1 & 2020-06-27T16:32 & 36.4 & 26.2 & 1583 & A.4 & 47.1 & 31.0 & 2074 & 394 & 812 \\
J124604.15$-$073047.4 & 2020-09-11T12:15 & 32.4 & 24.5 & 578.4 & A.4 & 40.7 & 28.4 & 543.7 & 589 & 1223 \\
J125203.09$-$040638.5 & 2020-09-11T12:15 & 39.8 & 31.7 & 418.3 & A.3 & 46.5 & 30.8 & 562.7 & 464 & 967 \\
J125442.98$-$383356.4 & 2020-09-01T10:25 & 23.7 & 21.9 & 303.8 & A.2 & 25.4 & 21.4 & 354.2 & 769 & 1912 \\
J125555.45$-$122254.7 & 2020-11-01T06:30 & 38.0 & 28.2 & 340.3 & A.3\_R & 56.4 & 40.5 & 562.5 & 494 & 830 \\
J125611.49$-$214411.7 & 2020-11-01T06:30 & 26.6 & 22.7 & 244.0 & A.2 & 34.7 & 24.6 & 308.2 & 608 & 1264 \\
J125950.95$-$293829.1 & 2020-08-10T12:56 & 20.7 & 19.3 & 249.4 & A.2\_R & 24.8 & 21.2 & 340.3 & 909 & 1876 \\
J130028.50$+$283010.5 & 2020-08-10T12:56 & 25.9 & 22.2 & 160.2 & A.1\_R & 31.1 & 22.8 & 128.6 & 564 & 1294 \\
J130457.29$-$015457.7 & 2020-08-10T12:56 & 29.1 & 22.9 & 343.8 & A.3 & 39.0 & 27.6 & 437.4 & 636 & 1195 \\
J130508.50$-$285042.0 & 2020-11-01T06:30 & 25.7 & 22.3 & 373.2 & A.3 & 26.9 & 22.2 & 353.4 & 708 & 1725 \\
J130958.32$+$073217.5 & 2020-06-27T16:32 & 24.5 & 21.2 & 196.7 & A.1 & 38.5 & 26.2 & 223.6 & 654 & 1138 \\
J131139.29$-$221641.2$^\star$ & 2020-10-18T06:46 & 81.5 & 38.6 & 5453 & A.4 & 110.0 & 54.1 & 8088 & 222 & 450 \\
J131207.86$-$202652.4 & 2020-06-26T15:35 & 26.8 & 21.5 & 769.1 & A.4 & 29.9 & 22.4 & 1004 & 542 & 1346 \\
J131301.41$-$272258.6 & 2020-10-18T06:46 & 25.4 & 21.3 & 124.8 & B & 33.2 & 22.8 & 150.5 & 702 & 1434 \\
J131403.21$+$084209.5 & 2020-10-18T06:46 & 26.6 & 21.8 & 274.7 & A.2 & 33.1 & 24.6 & 351.9 & 637 & 1362 \\
J131615.50$-$242824.1 & 2020-10-25T06:02 & 25.0 & 21.1 & 243.3 & A.2 & 28.5 & 22.6 & 344.1 & 803 & 1697 \\
J131736.55$-$134532.8 & 2020-10-25T06:02 & 28.9 & 24.5 & 306.5 & A.2 & 40.3 & 32.8 & 361.5 & 605 & 1125 \\
J132132.30$-$455401.8 & 2020-06-27T16:32 & 37.1 & 30.4 & 141.0 & B & 73.3 & 47.6 & 184.1 & 425 & 608 \\
J132846.53$-$485638.7 & 2020-06-30T15:30 & 22.9 & 21.1 & 470.0 & A.3 & 27.1 & 23.1 & 673.5 & 824 & 1699 \\
J133520.56$-$391144.6 & 2020-10-25T06:02 & 21.3 & 19.5 & 306.9 & A.2\_R & 23.7 & 20.7 & 337.8 & 803 & 2042 \\
J133621.59$-$105243.4 & 2020-10-26T07:05 & 50.1 & 44.9 & 385.0 & A.3\_R & 55.7 & 45.7 & 535.3 & 403 & 864 \\
J133739.83$-$125724.6$^\star$ & 2020-10-26T07:05 & 137.8 & 88.2 & 2790 & A.4 & 156.9 & 112.0 & 2728 & 216 & 396 \\
J133806.20$-$294439.1 & 2020-07-14T16:21 & 32.9 & 25.4 & 257.2 & B & 42.8 & 33.3 & 336.9 & 559 & 996 \\
J133932.48$+$014524.7 & 2020-07-14T16:21 & 30.5 & 26.5 & 825.6 & A.4 & 53.4 & 33.2 & 1144 & 518 & 960 \\
J134145.53$-$481815.9 & 2020-06-30T15:30 & 22.8 & 21.3 & 144.6 & B & 25.9 & 22.9 & 204.8 & 798 & 1806 \\
J134743.00$-$302754.6 & 2020-07-14T16:21 & 35.4 & 26.5 & 181.5 & B & 45.8 & 29.5 & 296.9 & 518 & 1090 \\
J135150.53$-$243921.3 & 2020-10-26T07:05 & 25.7 & 23.4 & 225.4 & A.2\_R & 37.0 & 26.6 & 297.4 & 781 & 1397 \\
J135157.84$-$020957.1 & 2020-06-30T15:30 & 39.6 & 25.9 & 277.3 & B\_R & 46.5 & 26.5 & 406.7 & 435 & 1002 \\
J135547.91$-$094756.5 & 2020-10-02T10:30 & 25.8 & 23.0 & 314.4 & A.3\_R & 37.3 & 29.2 & 328.9 & 681 & 1266 \\
J135547.99$-$370955.5 & 2020-10-02T10:30 & 22.5 & 21.1 & 298.9 & A.2 & 25.0 & 22.2 & 304.6 & 839 & 2037 \\
J135658.91$+$050505.2 & 2020-10-02T10:30 & 31.4 & 25.9 & 320.0 & A.1\_R & 44.7 & 33.2 & 439.5 & 531 & 991 \\
J140135.54$+$151325.2 & 2020-10-18T10:15 & 32.2 & 25.3 & 78.2 & A.1 & 41.3 & 29.4 & 87.6 & 581 & 1130 \\
J140807.52$+$095348.1 & 2020-08-14T14:00 & 28.2 & 21.8 & 254.1 & A.2 & 37.7 & 26.8 & 282.4 & 570 & 1242 \\
J141327.20$-$342235.1 & 2020-10-18T10:15 & 21.8 & 20.5 & 249.8 & A.2 & 24.9 & 21.9 & 231.3 & 885 & 2014 \\
J141558.82$+$132024.4 & 2020-10-16T10:29 & 38.2 & 26.4 & 1134 & A.4 & 48.3 & 32.9 & 1487 & 362 & 765 \\
J142155.55$-$310427.9 & 2020-08-14T14:00 & 24.4 & 20.7 & 727.2 & A.4\_R & 29.6 & 23.6 & 1105 & 673 & 1432 \\
J142213.87$-$091018.4 & 2020-08-14T14:00 & 26.2 & 21.4 & 406.9 & A.3 & 34.0 & 25.4 & 549.8 & 687 & 1375 \\
J142438.13$+$225600.7 & 2020-10-18T10:15 & 32.0 & 24.2 & 466.6 & A.3 & 38.9 & 29.3 & 306.2 & 488 & 980 \\
J143709.04$-$294718.5 & 2020-10-19T11:30 & 22.2 & 20.7 & 264.7 & A.1 & 25.1 & 22.2 & 316.6 & 813 & 1897 \\
J144032.09$+$053120.7 & 2020-10-19T11:30 & 24.3 & 21.6 & 213.9 & A.1 & 32.2 & 25.0 & 257.6 & 774 & 1482 \\
J144851.10$-$112215.6 & 2020-10-19T11:30 & 36.4 & 25.2 & 500.8 & A.4 & 41.3 & 30.0 & 642.0 & 498 & 1096 \\
J145212.06$-$255639.8 & 2020-09-24T12:25 & 22.4 & 20.7 & 233.5 & A.1 & 25.6 & 21.6 & 308.6 & 855 & 1884 \\
J145502.84$-$170014.2 & 2020-09-24T12:25 & 25.8 & 23.0 & 248.2 & A.2 & 30.4 & 25.0 & 280.4 & 659 & 1492 \\
J145625.83$+$045645.2 & 2020-09-24T12:25 & 31.6 & 24.7 & 301.0 & A.2 & 50.0 & 30.3 & 411.5 & 605 & 1042 \\
J145908.92$-$164542.3 & 2020-10-23T09:09 & 51.9 & 40.2 & 372.9 & B\_R & 32.6 & 25.6 & 507.1 & 338 & 1375 \\
J150425.30$+$081858.6 & 2020-10-27T09:46 & 27.6 & 24.3 & 170.5 & A.1\_R & 46.4 & 35.2 & 185.9 & 645 & 1010 \\
J150712.88$-$313729.6 & 2020-10-27T09:46 & 29.6 & 23.6 & 294.9 & B & 40.7 & 28.4 & 414.6 & 598 & 1112 \\
J150905.32$-$112646.3 & 2020-10-27T09:46 & 37.9 & 29.1 & 961.0 & A.4\_R & 44.1 & 30.5 & 1149 & 476 & 1077 \\
J151129.01$-$072255.3 & 2020-10-16T10:29 & 34.8 & 26.4 & 300.7 & A.2\_R & 46.9 & 31.6 & 374.6 & 554 & 1056 \\
J151204.81$-$132833.0 & 2020-10-16T10:29 & 24.7 & 22.3 & 300.8 & A.2 & 30.7 & 26.3 & 379.8 & 672 & 1459 \\
J151944.77$-$115144.6 & 2020-07-19T15:23 & 29.3 & 23.0 & 407.9 & A.3 & 42.8 & 27.6 & 391.8 & 544 & 1094 \\
J152225.50$-$293624.8 & 2021-01-14T02:59 & 28.0 & 22.4 & 620.7 & A.4 & 37.0 & 26.3 & 577.1 & 661 & 1289 \\
J152601.27$+$071525.5 & 2020-09-20T13:50 & 24.3 & 21.1 & 295.2 & A.2 & 31.1 & 24.7 & 421.3 & 662 & 1419 \\
J154044.18$-$145528.3 & 2020-07-19T15:23 & 25.9 & 22.0 & 318.4 & A.2 & 31.9 & 23.7 & 448.0 & 656 & 1406 \\
J154414.23$-$231200.8 & 2020-10-23T09:09 & 22.7 & 21.1 & 247.6 & A.2 & 25.6 & 22.6 & 270.1 & 754 & 1799 \\
J154912.59$+$304715.1 & 2020-09-28T12:51 & 31.8 & 23.6 & 1251 & A.4 & 45.0 & 27.0 & 1767 & 373 & 688 \\
J155212.77$-$073246.5 & 2020-10-23T09:09 & 27.9 & 23.2 & 220.3 & A.1 & 34.1 & 26.7 & 293.1 & 643 & 1388 \\
J155709.44$+$055216.1 & 2020-09-20T13:50 & 23.1 & 20.7 & 244.8 & A.2 & 34.0 & 26.2 & 313.7 & 697 & 1268 \\
J155930.95$+$030447.9 & 2021-01-14T02:59 & 27.3 & 23.3 & 532.8 & A.4 & 44.9 & 34.4 & 450.0 & 655 & 1079 \\
J160502.12$-$173415.9 & 2021-01-07T06:06 & 39.5 & 28.4 & 1530 & A.4 & 47.7 & 33.2 & 2111 & 363 & 844 \\
J160846.13$+$102908.2 & 2020-06-30T19:21 & 31.0 & 24.4 & 1040 & A.4 & 42.2 & 28.9 & 1018 & 524 & 1019 \\
J160903.37$-$145457.6 & 2020-06-30T19:21 & 25.9 & 23.3 & 231.7 & A.2 & 35.4 & 27.4 & 302.7 & 658 & 1272 \\
J161734.41$+$141137.9 & 2020-06-30T19:21 & 40.3 & 31.0 & 237.9 & B & 72.5 & 40.4 & 322.0 & 400 & 721 \\
J161749.32$-$771716.6 & 2020-07-19T15:23 & 28.7 & 22.1 & 3334 & A.4 & 33.0 & 23.3 & 3618 & 527 & 1194 \\
J161907.44$-$093953.1 & 2020-06-26T19:20 & 29.3 & 23.5 & 372.2 & A.3 & 38.9 & 25.7 & 468.5 & 630 & 1248 \\
J162047.94$+$003653.2 & 2020-06-26T19:20 & 28.5 & 22.4 & 234.4 & A.1 & 42.8 & 32.6 & 241.8 & 649 & 1086 \\
J162115.53$-$361136.3 & 2020-06-26T19:20 & 26.2 & 22.4 & 815.9 & A.4\_R & 31.0 & 25.7 & 1177 & 520 & 1211 \\
J162303.03$+$123958.4 & 2020-08-10T16:40 & 32.8 & 24.0 & 533.0 & A.4 & 44.6 & 30.1 & 760.2 & 465 & 951 \\
J162432.91$-$064949.9 & 2021-01-07T06:06 & 37.5 & 28.9 & 1017 & A.4 & 43.5 & 29.7 & 1188 & 386 & 955 \\
J162756.33$-$131958.7 & 2021-01-16T03:45 & 33.4 & 30.0 & 223.0 & A.1 & 51.8 & 40.7 & 286.4 & 530 & 805 \\
J162846.71$-$141541.8 & 2021-01-16T03:45 & 27.0 & 22.0 & 340.5 & A.3 & 39.0 & 24.7 & 363.8 & 578 & 1140 \\
J163145.29$+$115603.3 & 2020-08-10T16:40 & 36.3 & 25.4 & 1665 & A.4 & 43.8 & 29.1 & 2083 & 403 & 920 \\
J163257.77$-$003322.0 & 2020-09-28T12:51 & 28.0 & 21.4 & 227.4 & A.1 & 35.2 & 24.6 & 223.3 & 640 & 1308 \\
J163420.32$-$674420.9 & 2021-01-16T03:45 & 24.5 & 21.9 & 337.0 & B & 35.7 & 25.8 & 446.0 & 661 & 1288 \\
J163819.26$-$034005.4 & 2021-01-07T06:06 & 45.6 & 31.9 & 659.8 & A.4 & 74.5 & 48.7 & 665.3 & 320 & 550 \\
J163906.46$+$114409.2 & 2020-08-10T16:40 & 49.3 & 34.6 & 321.2 & B & 62.9 & 40.1 & 373.3 & 358 & 713 \\
J163956.36$+$112758.7 & 2020-04-01T00:30 & 37.2 & 27.3 & 156.6 & B & 64.7 & 41.3 & 193.5 & 392 & 665 \\
J164950.51$+$062653.3$^\star$ & 2020-10-26T10:54 & 34.1 & 27.0 & 373.3 & A.3 & 145.4 & 96.2 & 406.3 & 484 & 346 \\
J165038.03$-$124854.5 & 2020-10-26T10:54 & 31.1 & 26.2 & 327.3 & A.3 & 36.3 & 28.5 & 407.3 & 556 & 1324 \\
J165357.67$-$010214.2 & 2020-10-26T10:54 & 36.5 & 28.0 & 334.4 & A.3 & 53.8 & 36.4 & 448.9 & 505 & 921 \\
J165435.38$+$001719.5 & 2020-10-25T09:44 & 49.6 & 32.8 & 243.8 & B & 79.9 & 43.4 & 303.4 & 497 & 785 \\
J165441.78$+$134621.8 & 2020-09-28T12:51 & 25.4 & 21.3 & 425.9 & A.3 & 31.7 & 25.0 & 563.1 & 541 & 1148 \\
J170333.50$+$064539.7 & 2020-10-25T09:44 & 28.1 & 23.6 & 239.8 & A.2\_R & 41.5 & 29.5 & 313.0 & 619 & 1120 \\
J171627.09$-$061356.5 & 2020-10-25T09:44 & 33.3 & 24.6 & 202.0 & B & 38.7 & 30.3 & 256.6 & 564 & 1250 \\
J171844.67$-$625059.9 & 2020-08-17T15:01 & 23.1 & 21.2 & 145.1 & B\_R & 28.9 & 23.2 & 213.3 & 753 & 1578 \\
J171952.18$+$081703.3 & 2020-12-30T06:40 & 28.8 & 24.6 & 350.3 & A.3 & 42.6 & 33.0 & 303.4 & 600 & 1100 \\
J172105.79$+$162649.1 & 2020-08-17T15:01 & 27.9 & 23.0 & 503.9 & A.4 & 35.1 & 25.4 & 606.7 & 527 & 1146 \\
J173035.02$+$002438.4 & 2020-12-28T07:07 & 35.3 & 24.7 & 329.6 & A.3\_R & 52.3 & 30.6 & 380.6 & 585 & 980 \\
J173225.03$+$005941.4 & 2020-12-28T07:07 & 38.2 & 30.6 & 446.9 & A.2\_R & 65.2 & 38.9 & 620.7 & 496 & 803 \\
J174037.19$+$031149.6 & 2020-12-30T06:40 & 42.0 & 34.7 & 385.7 & A.3 & 66.3 & 42.0 & 348.5 & 390 & 710 \\
J174425.47$-$514443.1 & 2020-09-20T13:50 & 85.7 & 47.7 & 6830 & A.4 & 61.5 & 36.6 & 7269 & 187 & 635 \\
J183339.98$-$210339.9$^\dag$ & 2019-12-19T14:55 & 110.6 & 52.5 & 11902 & A.4 & 239.0 & 95.3 & 11287 & 194 & 261 \\
J183828.56$-$342741.6 & 2020-08-14T17:50 & 22.6 & 20.9 & 256.0 & A.2 & 31.3 & 24.8 & 256.0 & 835 & 1534 \\
J190923.40$-$392311.7 & 2020-09-07T14:00 & 24.6 & 21.7 & 210.3 & A.1 & 28.3 & 22.8 & 284.9 & 761 & 1722 \\
J191404.31$-$622317.1 & 2020-08-17T15:01 & 23.6 & 20.4 & 241.1 & A.1 & 28.3 & 21.6 & 327.3 & 807 & 1634 \\
J191623.16$-$621552.5 & 2020-06-20T00:01 & 27.2 & 22.2 & 130.7 & B & 33.1 & 23.4 & 175.5 & 654 & 1368 \\
J192956.85$-$351619.6 & 2020-09-07T14:00 & 28.8 & 23.8 & 262.4 & A.2 & 44.8 & 27.9 & 354.6 & 663 & 1142 \\
J193147.88$-$360659.1 & 2020-09-07T14:00 & 31.2 & 23.7 & 809.5 & A.4 & 38.8 & 26.4 & 1069 & 582 & 1233 \\
J194110.28$-$300720.9 & 2020-09-21T13:19 & 22.5 & 20.2 & 272.0 & A.2 & 25.8 & 21.9 & 277.4 & 722 & 1672 \\
J195243.90$-$193623.7 & 2020-09-21T13:19 & 30.2 & 23.9 & 722.6 & A.4 & 47.5 & 31.8 & 901.3 & 582 & 945 \\
J195302.23$-$203633.0$^\star$ & 2020-09-21T13:19 & 27.0 & 23.2 & 493.6 & A.4 & 404.2 & 238.2 & 662.7 & 584 & 289 \\
J195629.92$-$054246.9 & 2020-06-20T00:00 & 25.5 & 20.6 & 323.1 & A.2 & 30.5 & 23.2 & 440.3 & 651 & 1313 \\
J200008.58$-$192138.3 & 2020-09-14T14:00 & 27.7 & 22.7 & 856.1 & A.4 & 32.3 & 25.4 & 1064 & 588 & 1337 \\
J200209.37$-$145531.8 & 2020-09-14T14:00 & 29.2 & 22.6 & 612.2 & A.4 & 41.3 & 25.7 & 754.7 & 604 & 1054 \\
J200429.50$-$101433.1 & 2020-09-14T14:00 & 26.7 & 24.1 & 265.4 & A.1 & 44.4 & 34.3 & 324.4 & 673 & 1064 \\
J200608.40$-$022334.7 & 2020-12-20T09:38 & 70.6 & 40.7 & 2226 & A.4 & 79.5 & 40.5 & 2706 & 299 & 667 \\
J201115.72$-$154640.3 & 2020-12-20T09:38 & 28.0 & 21.8 & 655.5 & A.4 & 31.1 & 23.2 & 540.4 & 574 & 1328 \\
J201918.06$+$112712.9 & 2020-08-16T18:45 & 26.3 & 21.3 & 194.5 & A.1 & 39.6 & 24.9 & 195.3 & 611 & 1201 \\
J202346.21$-$365521.2 & 2020-04-01T00:30 & 24.1 & 21.5 & 404.6 & A.3 & 30.3 & 22.4 & 414.9 & 675 & 1565 \\
J203425.65$-$052332.2 & 2020-08-24T18:15 & 25.7 & 21.6 & 527.0 & A.4 & 36.8 & 26.6 & 561.7 & 676 & 1208 \\
J204439.43$-$182852.2 & 2020-08-24T18:15 & 27.0 & 21.8 & 217.5 & B & 30.3 & 22.7 & 293.8 & 620 & 1505 \\
J204608.17$-$391840.7 & 2020-08-24T18:15 & 24.2 & 22.1 & 264.3 & A.2 & 27.6 & 22.0 & 344.3 & 738 & 1879 \\
J204737.66$-$184141.6 & 2020-08-28T20:00 & 28.7 & 22.6 & 243.0 & B & 35.6 & 24.2 & 248.0 & 541 & 1238 \\
J205245.03$-$223410.6 & 2020-08-28T20:00 & 23.8 & 20.8 & 331.9 & A.2 & 25.5 & 20.8 & 427.1 & 767 & 1798 \\
J205456.08$-$093240.8 & 2020-08-28T20:00 & 24.8 & 20.9 & 321.2 & A.1 & 30.3 & 22.0 & 442.4 & 705 & 1492 \\
J205959.61$-$144043.1 & 2020-08-30T17:20 & 27.5 & 22.4 & 1020 & A.4 & 33.9 & 24.7 & 1368 & 521 & 1155 \\
J210143.29$-$174759.2 & 2020-08-30T17:20 & 31.9 & 23.0 & 959.2 & A.4 & 44.3 & 27.4 & 1320 & 519 & 1004 \\
J211311.86$-$393933.7 & 2020-08-30T17:20 & 26.0 & 22.2 & 566.1 & A.4 & 30.5 & 23.4 & 773.7 & 639 & 1484 \\
J211410.44$-$105558.3 & 2020-08-31T16:49 & 24.6 & 21.8 & 279.6 & A.2 & 32.3 & 28.4 & 349.4 & 698 & 1369 \\
J211629.36$-$643930.2 & 2020-06-20T00:01 & 22.0 & 20.0 & 135.2 & B & 33.9 & 25.2 & 191.5 & 757 & 1355 \\
J211650.79$+$022547.0 & 2020-08-16T18:45 & 25.0 & 22.0 & 116.2 & A.1 & 37.0 & 27.9 & 111.8 & 692 & 1258 \\
J211949.40$-$343938.6 & 2020-08-06T01:00 & 24.8 & 20.5 & 282.3 & A.2 & 30.7 & 24.7 & 382.7 & 783 & 1578 \\
J212042.48$+$132724.2 & 2020-08-06T01:00 & 24.7 & 20.6 & 375.5 & A.3 & 30.6 & 21.6 & 467.5 & 750 & 1491 \\
J212758.87$-$385856.3 & 2020-08-06T01:00 & 25.7 & 21.8 & 286.3 & A.2 & 39.0 & 26.1 & 336.6 & 697 & 1213 \\
J212821.83$-$150453.2 & 2020-08-31T16:49 & 27.3 & 22.6 & 250.0 & A.2 & 30.7 & 24.2 & 304.0 & 556 & 1396 \\
J212923.02$-$430828.5$^\star$ & 2020-06-29T02:35 & 24.2 & 21.7 & 181.4 & A.1 & 124.2 & 65.3 & 246.9 & 692 & 532 \\
J213135.55$-$270159.0 & 2020-08-31T16:49 & 25.4 & 21.9 & 152.9 & A.1 & 28.6 & 22.4 & 176.4 & 791 & 1781 \\
J213418.19$-$533514.7 & 2020-08-14T17:50 & 20.7 & 18.8 & 363.1 & B & 24.5 & 20.1 & 497.8 & 814 & 1808 \\
J213937.03$+$171826.9 & 2020-06-29T02:35 & 25.6 & 21.7 & 232.4 & A.1 & 32.7 & 23.9 & 226.1 & 605 & 1230 \\
J214120.76$-$351516.9 & 2020-08-01T00:15 & 20.6 & 18.7 & 378.7 & A.3\_R & 23.7 & 20.4 & 510.0 & 797 & 1878 \\
J214230.92$-$244440.1 & 2020-09-02T20:30 & 24.1 & 21.3 & 242.5 & A.1 & 29.5 & 22.2 & 247.4 & 777 & 1727 \\
J214439.28$-$750811.1 & 2020-06-29T02:35 & 26.7 & 22.9 & 307.6 & B & 37.6 & 24.5 & 406.4 & 625 & 1291 \\
J215131.44$+$070927.1 & 2020-08-01T00:15 & 35.2 & 31.0 & 885.5 & A.4 & 77.4 & 48.6 & 704.1 & 454 & 639 \\
J215445.08$-$382632.5 & 2020-08-01T00:15 & 21.1 & 18.9 & 794.9 & A.4\_R & 24.7 & 20.4 & 850.2 & 822 & 1865 \\
J215720.05$-$360411.1 & 2020-09-02T20:30 & 24.3 & 21.0 & 264.2 & A.1 & 26.9 & 21.4 & 379.6 & 625 & 1507 \\
J215800.88$+$092546.4$^\star$ & 2020-06-27T02:45 & 87.1 & 75.0 & 451.6 & A.3 & 102.8 & 83.9 & 536.7 & 186 & 467 \\
J215934.57$-$431237.8 & 2020-06-27T02:45 & 36.9 & 33.6 & 146.4 & B\_R & 43.6 & 34.6 & 185.7 & 445 & 1002 \\
J220127.50$+$031215.6 & 2020-08-08T00:10 & 23.9 & 21.1 & 186.2 & A.1 & 34.9 & 25.5 & 191.4 & 725 & 1241 \\
J221656.66$-$670337.9 & 2020-08-08T00:10 & 19.8 & 18.6 & 174.1 & A.1 & 22.0 & 19.7 & 272.6 & 861 & 1968 \\
J221811.04$+$282849.5 & 2020-09-02T20:30 & 26.4 & 23.2 & 32.7 & B & 35.7 & 25.7 & 115.2 & 569 & 1106 \\
J221843.50$-$122912.9 & 2020-09-13T21:30 & 26.8 & 22.7 & 254.9 & A.2 & 28.4 & 22.9 & 265.6 & 616 & 1588 \\
J222115.67$-$461537.2 & 2020-08-08T00:10 & 21.4 & 19.3 & 142.1 & A.1 & 31.3 & 21.9 & 195.6 & 908 & 1572 \\
J222332.81$-$310117.3 & 2020-09-13T21:30 & 22.4 & 20.0 & 258.3 & A.2 & 26.5 & 22.8 & 304.8 & 800 & 1775 \\
J222359.11$+$121338.9 & 2020-09-13T21:30 & 26.6 & 21.6 & 248.6 & A.2 & 36.0 & 27.9 & 220.0 & 588 & 1109 \\
J223050.19$+$034836.8 & 2020-09-08T21:40 & 31.9 & 24.8 & 246.7 & A.2\_R & 39.2 & 27.7 & 305.5 & 546 & 1032 \\
J223240.99$-$144226.7 & 2020-09-08T21:40 & 24.0 & 21.8 & 413.4 & A.3\_R & 30.6 & 24.1 & 512.7 & 732 & 1613 \\
J223816.27$-$124036.4 & 2020-09-08T21:40 & 32.3 & 23.9 & 224.2 & B & 42.8 & 28.1 & 288.8 & 486 & 975 \\
J223826.48$-$134422.6$^\star$ & 2020-09-04T21:52 & 26.3 & 22.5 & 243.2 & B & 1311.4 & 644.8 & 475.2 & 604 & 34 \\
J223831.50$-$173119.1 & 2020-07-22T01:40 & 36.3 & 25.0 & 347.6 & B & 36.9 & 28.7 & 462.9 & 400 & 956 \\
J223933.24$+$085043.7 & 2020-07-22T01:40 & 22.2 & 20.2 & 228.8 & A.1 & 27.5 & 21.2 & 321.3 & 699 & 1588 \\
J224111.48$-$244239.0 & 2020-07-22T01:40 & 21.4 & 19.8 & 188.8 & A.1 & 26.9 & 20.7 & 197.9 & 830 & 1803 \\
J224354.81$+$181445.9 & 2020-08-02T00:25 & 34.6 & 24.1 & 971.8 & A.4 & 47.4 & 29.4 & 1347 & 418 & 851 \\
J224500.21$-$343030.0 & 2020-09-04T21:52 & 22.9 & 20.7 & 429.0 & A.3 & 24.4 & 20.5 & 542.5 & 715 & 1770 \\
J224501.25$-$493132.3 & 2020-08-14T17:50 & 20.4 & 19.3 & 205.3 & A.1 & 23.0 & 19.3 & 280.6 & 856 & 1959 \\
J224528.27$+$032408.8 & 2020-08-02T00:25 & 32.1 & 25.4 & 462.2 & A.3 & 45.1 & 29.1 & 428.1 & 537 & 1041 \\
J224705.52$+$121151.4 & 2020-08-02T00:25 & 25.6 & 22.1 & 218.9 & A.1 & 42.9 & 35.3 & 284.9 & 622 & 944 \\
J224950.57$-$263459.6 & 2020-09-04T21:52 & 22.4 & 20.4 & 237.2 & A.2 & 24.6 & 20.3 & 305.1 & 802 & 1969 \\
J225630.38$-$425959.2 & 2020-06-28T02:50 & 20.9 & 19.6 & 436.8 & A.1 & 25.7 & 21.1 & 594.8 & 816 & 1790 \\
J225727.66$-$354528.1 & 2020-09-06T22:29 & 27.6 & 21.8 & 210.5 & B & 30.6 & 22.4 & 285.1 & 679 & 1552 \\
J225736.76$-$412012.3 & 2020-06-28T02:50 & 30.1 & 23.7 & 189.8 & B & 58.8 & 38.1 & 292.5 & 597 & 786 \\
J225813.47$-$095817.2 & 2020-09-06T22:29 & 28.3 & 21.8 & 231.3 & B & 37.6 & 23.3 & 313.6 & 551 & 1166 \\
J230036.41$+$194002.9 & 2020-06-28T02:50 & 29.2 & 22.5 & 203.1 & B & 36.7 & 24.6 & 268.5 & 471 & 1045 \\
J230040.87$+$033710.3 & 2020-08-05T00:25 & 30.7 & 22.2 & 451.4 & A.3 & 35.4 & 24.3 & 525.8 & 576 & 1269 \\
J230153.45$+$060912.5 & 2020-08-05T00:25 & 27.4 & 21.9 & 317.7 & A.2 & 34.2 & 25.4 & 309.5 & 622 & 1261 \\
J230733.06$-$531258.9 & 2020-08-16T18:45 & 20.2 & 19.1 & 214.0 & A.1 & 28.4 & 22.0 & 289.3 & 892 & 1678 \\
J231002.89$+$111403.6 & 2020-08-05T00:25 & 26.2 & 21.5 & 248.6 & A.2 & 31.8 & 24.2 & 334.5 & 552 & 1249 \\
J231621.00$-$433746.8 & 2020-07-05T02:30 & 31.5 & 22.8 & 1076 & A.4 & 38.6 & 25.9 & 1265 & 535 & 1196 \\
J231626.94$-$472925.5 & 2021-01-17T13:00 & 27.7 & 21.6 & 1054 & A.4 & 31.5 & 22.1 & 1375 & 578 & 1384 \\
J231634.61$+$042940.2$^\star$ & 2020-07-05T02:30 & 74.8 & 49.9 & 203.2 & B\_R & 142.9 & 91.1 & 199.8 & 268 & 351 \\
J231907.55$-$420640.5 & 2020-07-05T02:30 & 23.4 & 20.0 & 259.2 & A.2 & 29.5 & 23.1 & 293.8 & 650 & 1551 \\
J231915.61$-$423750.4 & 2020-06-22T02:55 & 27.7 & 21.5 & 493.1 & B & 31.3 & 22.1 & 640.2 & 640 & 1518 \\
J233004.06$-$052447.7 & 2020-09-06T22:29 & 27.7 & 22.2 & 303.1 & A.2 & 39.3 & 31.8 & 441.3 & 648 & 1166 \\
J233231.61$-$142318.8 & 2020-07-19T01:35 & 22.0 & 19.7 & 162.6 & B & 31.2 & 23.8 & 162.0 & 787 & 1407 \\
J233300.33$+$134635.2 & 2020-07-19T01:35 & 22.3 & 20.0 & 215.8 & A.1 & 28.5 & 22.0 & 282.8 & 676 & 1440 \\
J233635.99$-$374424.1 & 2020-07-19T01:35 & 21.1 & 18.8 & 232.6 & A.1 & 23.6 & 19.0 & 301.3 & 860 & 2114 \\
J233808.04$-$121851.4 & 2020-09-09T21:35 & 25.2 & 21.6 & 456.2 & A.3 & 28.1 & 21.6 & 617.5 & 703 & 1635 \\
J233913.22$-$552350.4 & 2020-06-14T04:35 & 15.3 & 14.1 & 184.3 & A.1 & 16.0 & 13.8 & 225.2 & 1292 & 3428 \\
                      & 2020-06-22T02:55 &      &      &       &     &    &      &       &      &      \\
J234118.78$+$192805.6 & 2021-01-17T13:00 & 24.8 & 20.9 & 381.6 & A.2 & 34.2 & 23.5 & 398.3 & 678 & 1205 \\
J234910.12$-$043803.2 & 2020-06-22T02:55 & 34.6 & 23.0 & 229.8 & A.1 & 36.1 & 25.1 & 216.7 & 561 & 1358 \\
J235722.47$-$073134.3 & 2020-09-09T21:35 & 27.8 & 22.4 & 247.2 & A.2 & 39.4 & 26.2 & 295.1 & 648 & 1190 \\
J235914.02$+$192420.6 & 2020-09-09T21:35 & 40.2 & 24.5 & 279.5 & B & 42.4 & 26.2 & 373.6 & 431 & 966 \\
\enddata
\tablecomments{
\small{
    Column\,1: Pointing ID based on R.A. and declination (J2000) of the bright radio source at the center of the pointing. Column\,2: Date and start time (UTC) of the observing run. Columns\,3-5: rms at primary beam FWHM, rms at twice of the primary beam FWHM and flux density of the central radio source, for SPW9. Column\,6: Class of the pointing as described in Section~\ref{sec:noise}. Column\,7-9: rms at primary beam FWHM, rms at twice of the primary beam FWHM and flux density of the central radio source, for SPW2. Columns\,10-11: Total number of sources detected in the primary beam corrected field for SPW9 and SPW2, respectively.} 
    }
\footnotesize{$^\star$ field removed from spectral index analysis (Section~\ref{sec:uss}) owing to problematic SPW2 images. For the 50 representative pointings, \_R is added to their class. $^\dag$ one of the two pointings with extreme median flux density offset (Section~\ref{sec:fluxscale}) and Fig.~\ref{fig:fratio_mollweide}}.\\
\end{deluxetable*}
\end{longrotatetable}
\clearpage

\section{Catalog columns}
\label{sec:cat_cols}
\begin{table*}
\movetabledown=5.2cm
\begin{rotatetable*}
\centering
\setlength{\tabcolsep}{1.7pt}
\caption{Initial 6 rows from MALS SPW based source catalog. The catalog column definitions are given in Table~\ref{tab:cat_cols}. The complete catalog and the images are available at \href{}{https://mals.iucaa.in}.} 
\label{tab:srlcat}
\tabletypesize{\footnotesize}
\begin{tabular}{lccccccc}
\\ \hline
Source\_name & Pointing & Obs & Obs & Obs & PBeamVersion & Fluxcal & Fluxscale\\
 & \_id & \_date\_U & \_date\_L & \_band &  &  & \\
 (1) & (2) & (3) & (4) & (5) & (6) & (7) & (8)\\
 \hline \hline
J000719.65-153823.5 & J000141.57-154040.6 & & ['2020-07-16T01:40'] & L & katbeam\_v0.1 & ['J1939-6342', 'J0408-6545'] & ['Stevens-Reynolds 2016', 'MANUAL'] \\
J000710.64-152927.6 & J000141.57-154040.6 & & ['2020-07-16T01:40'] & L & katbeam\_v0.1 & ['J1939-6342', 'J0408-6545'] & ['Stevens-Reynolds 2016', 'MANUAL'] \\
J000708.60-152055.5 & J000141.57-154040.6 & & ['2020-07-16T01:40'] & L & katbeam\_v0.1 & ['J1939-6342', 'J0408-6545'] & ['Stevens-Reynolds 2016', 'MANUAL'] \\
J000708.20-153644.3 & J000141.57-154040.6 & & ['2020-07-16T01:40'] & L & katbeam\_v0.1 & ['J1939-6342', 'J0408-6545'] & ['Stevens-Reynolds 2016', 'MANUAL'] \\
J000705.44-155332.3 & J000141.57-154040.6 & & ['2020-07-16T01:40'] & L & katbeam\_v0.1 & ['J1939-6342', 'J0408-6545'] & ['Stevens-Reynolds 2016', 'MANUAL'] \\
J000659.56-152433.2 & J000141.57-154040.6 & & ['2020-07-16T01:40'] & L & katbeam\_v0.1 & ['J1939-6342', 'J0408-6545'] & ['Stevens-Reynolds 2016', 'MANUAL'] \\
\hline
\end{tabular}

\setlength{\tabcolsep}{4.3pt}
\begin{tabular}{lcccccccc}
\\ \hline
SPW\_id & Ref\_freq & Maj\_restoring\_beam & Min\_restoring\_beam & PA\_restoring\_beam & Sigma\_1 & Sigma\_2 & Sigma\_20 & Distance\_poitning\\
 & (MHz) & (arcsec) & (arcsec) & (degree) & ($\mu$Jy\,beam$^{-1}$) & ($\mu$Jy\,beam$^{-1}$) & ($\mu$Jy\,beam$^{-1}$) & (arcmin) \\
(9) & (10) & (11) & (12) & (13) & (14) & (15) & (16) & (17)\\
\hline \hline
Lspw\_2 & 1005.94 & 10.8 & 8.9 & -5.9 & 36.9 & 23.9 & 68.2 & 81.4 \\
Lspw\_2 & 1005.94 & 10.8 & 8.9 & -5.9 & 36.9 & 23.9 & 68.2 & 80.0 \\
Lspw\_2 & 1005.94 & 10.8 & 8.9 & -5.9 & 36.9 & 23.9 & 68.2 & 81.2 \\
Lspw\_2 & 1005.94 & 10.8 & 8.9 & -5.9 & 36.9 & 23.9 & 68.2 & 78.7 \\
Lspw\_2 & 1005.94 & 10.8 & 8.9 & -5.9 & 36.9 & 23.9 & 68.2 & 79.0 \\
Lspw\_2 & 1005.94 & 10.8 & 8.9 & -5.9 & 36.9 & 23.9 & 68.2 & 78.3 \\
\hline
\end{tabular}

\setlength{\tabcolsep}{9.7pt}
\begin{tabular}{lccccccc}
\\ \hline
Distance\_NN & S\_Code & N\_Gauss & Maxsep\_Gauss & Maj & Maj\_E & Min & Min\_E \\
(arcmin) &  &    & (arcsec) & (arcsec) & (arcsec) & (arcsec) & (arcsec)\\
(18) & (19) & (20) & (21) & (22) & (23) & (24) & (25)\\
\hline\hline
3.2 & S & 1 & -1 & 0.003046 & 0.000065 & 0.002559 & 0.000046\\
3.9 & S & 1 & -1 & 0.004009 & 0.000526 & 0.003136 & 0.000343\\
4.2 & S & 1 & -1 & 0.004841 & 0.000856 & 0.002783 & 0.000331\\
3.2 & S & 1 & -1 & 0.003432 & 0.000721 & 0.002479 & 0.000384\\
3.8 & S & 1 & -1 & 0.002837 & 0.000275 & 0.002489 & 0.000211\\
2.1 & M & 2 & 10.4 & 0.004325 & 0.000047 & 0.002412 & 0.000016\\
\hline
\end{tabular}
\end{rotatetable*}
\end{table*}

\begin{table*}
\movetabledown=5.2cm
\begin{rotatetable*}
\centering
\setlength{\tabcolsep}{5pt}
\tabletypesize{\footnotesize}
\begin{tabular}{lccccccc}
\\ \hline
PA & PA\_E & DC\_Maj & DC\_Maj\_E & DC\_Min & DC\_Min\_E & DC\_PA & DC\_PA\_E\\
(degree) & (degree) & (arcsec) & (arcsec) & (arcsec) & (arcsec) & (degree) & (degree)\\
(26) & (27) & (28) & (29) & (30) & (31) & (32) & (33)\\
\hline\hline
160.6 & 4.6 & 0.000000 & 0.000065 & 0.000000 & 0.000047 & 0.1 & 4.6 \\
145.0 & 22.5 & 0.002831 & 0.000526 & 0.001658 & 0.000343 & 130.7 & 22.5 \\
59.3 & 14.0 & 0.004114 & 0.000855 & 0.000000 & 0.000331 & 63.0 & 14.0\\
167.2 & 25.9 & 0.001670 & 0.000721 & 0.000000 & 0.000384 & 160.3 & 25.9 \\
4.8 & 29.4 & 0.000000 & 0.000275 & 0.000000 & 0.000211 & 0.1 & 29.4\\
170.9 & 0.8 & 0.003105 & 0.000046 & 0.000000 & 0.000016 & 170.1 & 0.8\\
\hline
\end{tabular}

\setlength{\tabcolsep}{4.5pt}
\tabletypesize{\footnotesize}
\begin{tabular}{lccccccc}
\\ \hline
RA\_mean & RA\_mean\_E & DEC\_mean & DEC\_mean\_E & RA\_max & RA\_max\_E & DEC\_max & DEC\_max\_E \\
(degree) & (degree) & (degree) & (degree) & (degree) & (degree) & (degree) & (degree)\\
(34) & (35) & (36) & (37) & (38) & (39) & (40) & (41)\\
\hline\hline
1.83187868637453 & 0.00002066615631 & -15.6398776348322 & 0.00002698152798 & 1.83187868637453 & 0.00002066615631 & -15.6398776348322 & 0.00002698152798 \\
1.79433683090078 & 0.00017465334069 & -15.4910060485520 & 0.00020171832911 & 1.79433683090078 & 0.00017465334069 & -15.4910060485520 & 0.00020171832911 \\
1.78587019980654 & 0.00032185228765 & -15.3487510151291 & 0.00021979869525 & 1.78587019980654 & 0.00032185228765 & -15.3487510151291 & 0.00021979869525 \\
1.78419192604535 & 0.00017239690781 & -15.6123271992311 & 0.00030099512716 & 1.78419192604535 & 0.00017239690781 & -15.6123271992311 & 0.00030099512716 \\
1.77266691158313 & 0.00008978010356 & -15.8923209657271 & 0.00011654082226 & 1.77266691158313 & 0.00008978010356 & -15.8923209657271 & 0.00011654082226 \\
1.74818678200049 & 0.00001948799639 & -15.4092218248728 & 0.00000729532113 & 1.74844056895094 & 0.00001948799639 & -15.4092666854546 & 0.00000729532113 \\
\hline
\end{tabular}

\setlength{\tabcolsep}{5.3pt}
\tabletypesize{\footnotesize}
\begin{tabular}{lccccccc}
\\ \hline
Total\_flux & Total\_flux\_E & Total\_flux\_E\_fit & Total\_flux\_E\_sys & Total\_flux\_measured & Total\_flux\_measured\_E & Peak\_flux & Peak\_flux\_E \\
(mJy) & (mJy) & (mJy)  & (mJy) & (mJy) & (mJy) & (mJy\,beam$^{-1}$) & (mJy\,beam$^{-1}$) \\
(42) & (43) & (44) & (45) & (46) & (47) & (48) & (49)\\
\hline\hline
24.51 & 1.70 & 0.83 & 1.49 & 22.32 & 1.55 & 23.34 & 0.46 \\
6.88 & 1.96 & 1.22 & 1.53 & 6.27 & 1.78 & 4.06 & 0.49 \\
5.63 & 1.91 & 1.21 & 1.48 & 5.13 & 1.74 & 3.10 & 0.45 \\
2.59 & 1.13 & 0.79 & 0.81 & 2.38 & 1.04 & 2.26 & 0.41 \\
3.87 & 0.97 & 0.62 & 0.74 & 3.55 & 0.89 & 4.06 & 0.37 \\
91.95 & 3.02 & 1.13 & 2.80 & 84.72 & 2.78 & 57.40 & 0.44 \\
\hline
\end{tabular}
\end{rotatetable*}
\end{table*}

\begin{table*}
\movetabledown=5.5cm
\begin{rotatetable*}
\centering
\setlength{\tabcolsep}{5pt}
\tabletypesize{\footnotesize}
\begin{tabular}{lcccccccc}
\\ \hline
Isl\_Total\_flux & Isl\_Total\_flux\_E & Isl\_rms & Isl\_mean & Resid\_Isl\_rms & Resid\_Isl\_mean & Flux\_correction & Spectral\_index & Spectral\_index\_E \\
 (mJy)  & (mJy) & (mJy\,beam$^{-1}$) & (mJy\,beam$^{-1}$) & (mJy\,beam$^{-1}$) & (mJy\,beam$^{-1}$) &   &  & \\
(50) & (51) & (52) & (53) & (54) & (55) & (56) & (57) & (58)\\
\hline\hline
24.99 & 0.75 & 0.46 & 0.074 & 0.25 & 0.087 & 0.911 & \nodata & \nodata\\
5.93 & 0.69 & 0.45 & 0.003 & 0.14 & 0.008 & 0.911 & \nodata & \nodata \\
3.86 & 0.67 & 0.43 & 0.061 & 0.25 & 0.068 & 0.911 & \nodata & \nodata \\
2.39 & 0.46 & 0.40 & 0.174 & 0.02 & 0.174 & 0.918 & \nodata & \nodata\\
3.32 & 0.42 & 0.37 & 0.039 & 0.08 & 0.041 & 0.918 & \nodata & \nodata \\
90.50 & 0.85 & 0.44 & 0.005 & 0.30 & -0.032 & 0.921 & \nodata & \nodata\\
\hline
\end{tabular}

\setlength{\tabcolsep}{5pt}
\tabletypesize{\footnotesize}
\begin{tabular}{lcccccccc}
\\ \hline
Spectral\_index & Spectral\_index & Spectral\_index & Spectral\_index & Spectral\_index & Spectral\_index & Real & Resolved & Source \\
\_spwused & \_spwfit & \_spwfit\_E & \_MALS\_Lit & \_MALS\_Lit\_E & \_Lit & \_source & & \_linked \\
(59) & (60) & (61) & (62) & (63) & (64) & (65) & (66) & (67)\\ 
 \hline\hline
 & & & [-0.501] & [0.084] & [`TGSS-ADR1', `L:2'] & True & False & null \\
\nodata &\nodata  &\nodata  &  \nodata  & \nodata  & \nodata  & True & True & null \\
\nodata &\nodata  &\nodata  &\nodata  &\nodata  &\nodata  & True & True & null \\
\nodata &\nodata  &\nodata  &\nodata  &\nodata  & \nodata  & True & False  & null \\
\nodata &\nodata  &\nodata  & \nodata  & \nodata  & \nodata  & True & False  & null \\
\nodata &\nodata  &\nodata  & [-0.829] & [0.055] & [`TGSS-ADR1', `L:2'] & True & True  & null \\
\hline
\end{tabular}

\end{rotatetable*}
\end{table*}

\begin{table*}
\movetabledown=5cm
\begin{rotatetable*}
\centering
\setlength{\tabcolsep}{7pt}
\caption{Gaussian components ({\tt N\_Gauss}) for each source presented in Table~\ref{tab:srlcat}.   These are linked to Table~\ref{tab:srlcat} through the parameter {\tt Source\_name}. Note the multiple occurrences of the `M' type source in the table. The catalog column definitions are given in Table~\ref{tab:cat_cols}. The complete catalog and the images are available at \href{https://mals.iucaa.in}{https://mals.iucaa.in}.}
\label{tab:gaulcat}
\tabletypesize{\footnotesize}
\begin{tabular}{lccccccc}
\\ \hline
Source\_name & N\_Gauss & G\_RA & G\_RA\_E & G\_DEC & G\_DEC\_E & G\_Peak\_flux & G\_Peak\_flux\_E \\
& & (degree) & (degree) & (degree) & (degree) & (mJy\,beam$^{-1}$) & (mJy\,beam$^{-1}$)\\
(1) & (2) & (3) & (4) & (5) & (6) & (7) & (8)\\
\hline\hline
J000719.65-153823.5 & 1 & 1.83187868637453 & 0.00002066615631 & -15.6398776348321 & 0.00002698152798 & 23.34 & 0.46 \\
J000710.64-152927.6 & 1 & 1.79433683090078 & 0.00017465334069 & -15.4910060485520 & 0.00020171832911 & 4.06 & 0.49 \\
J000708.60-152055.5 & 1 & 1.78587019980654 & 0.00032185228765 & -15.3487510151291 & 0.00021979869525 & 3.10 & 0.45 \\
J000708.20-153644.3 & 1 & 1.78419192604535 & 0.00017239690781 & -15.6123271992311 & 0.00030099512716 & 2.26 & 0.41 \\
J000705.44-155332.3 & 1 & 1.77266691158313 & 0.00008978010356 & -15.8923209657271 & 0.00011654082226 & 4.06 & 0.37 \\
J000659.56-152433.2 & 2 & 1.74799553848509 & 0.00000781826425 & -15.4082879357922 & 0.00001157303107 & 56.79 & 0.45 \\
J000659.56-152433.2 & 2 & 1.74858297350623 & 0.00001354648886 & -15.4111268668004 & 0.00002044672516 & 30.58 & 0.44 \\
\hline
\end{tabular}

\setlength{\tabcolsep}{20pt}
\tabletypesize{\footnotesize}
\begin{tabular}{lccccc}
\\ \hline
G\_Maj & G\_Maj\_E & G\_Min & G\_Min\_E & G\_PA & G\_PA\_E \\
(arcsec) & (arcsec) & (arcsec) & (arcsec) & (degree) & (degree)\\
(9) & (10) & (11) & (12) & (13) & (14)\\
\hline\hline
10.964209 & 0.235017 & 9.214115 & 0.166663 & 160.6 & 4.6 \\
14.433258 & 1.894917 & 11.290056 & 1.235068 & 145.0 & 22.5 \\
17.427389 & 3.081403 & 10.016273 & 1.190850 & 59.3 & 14.0 \\
12.354763 & 2.594940 & 8.922901 & 1.383234 & 167.2 & 25.9 \\
10.212684 & 0.989746 & 8.958766 & 0.758822 & 4.8 & 29.3 \\
11.280227 & 0.098207 & 9.221270 & 0.066132 & 176.2 & 1.4 \\
10.861899 & 0.173346 & 8.841788 & 0.114819 & 0.5 & 2.8 \\
\hline
\end{tabular}

\setlength{\tabcolsep}{20pt}
\tabletypesize{\footnotesize}
\begin{tabular}{lcccccc}
\\ \hline
G\_DC\_Maj & G\_DC\_Maj\_E & G\_DC\_Min & G\_DC\_Min\_E & G\_DC\_PA & G\_DC\_PA\_E & G\_id \\
(arcsec) & (arcsec) & (arcsec) & (arcsec) & (degree) & (degree) & \\
(15) & (16) & (17) & (18) & (19) & (20) & (21) \\
\hline\hline
0.000000 & 0.235017 & 0.000000 & 0.166663 & 0.0 & 4.6 & 1 \\
10.193370 & 1.894917 & 5.968889 & 1.235068 & 130.7 & 22.5 & 2 \\
14.809919 & 3.081403 & 0.000000 & 1.190849 & 63.0 & 14.0 & 3 \\
6.011094 & 2.594940 & 0.000000 & 1.383233 & 160.3 & 25.9 & 4 \\
0.000000 & 0.989746 & 0.000000 & 0.758822 & 0.0 & 29.4 & 5 \\
3.232344 & 0.098207 & 2.361490 & 0.066132 & 17.9 & 1.4 & 6 \\
0.000000 & 0.173346 & 0.000000 & 0.114819 & 0.0 & 2.8 & 7 \\
\hline
\end{tabular}
\end{rotatetable*}
\end{table*}
\clearpage

\section{MALS SPW2 and SPW9 primary beam ratios}
\label{sec:PBratio}

Here we present the annular averaged MALS SPW2 and SPW9 primary beams generated using {\tt katbeam} and {\tt plumber}.

\begin{table*}
\begin{center}
\caption{MALS SPW2 and SPW9 annular averaged {\tt katbeam} and {\tt plumber}}
\label{tab:PBratio}
\begin{tabular}{lccccccc}
\hline
Distance & SPW2 & SPW2 & SPW2  & SPW9 & SPW9 & SPW9 & $\alpha_{\rm kat} / \alpha_{\rm plum}$\\
\_pointing & {\tt plumber} & {\tt katbeam} & {\tt $\frac{\text{katbeam}}{\text{plumber}}$} & {\tt plumber} & {\tt katbeam} & {\tt $\frac{\text{katbeam}}{\text{plumber}}$} &  \\
(1)  &  (2)   &  (3)  &  (4)    &  (5)   &  (6)   & (7) & (8) \\
\hline
1 & 0.999 & 0.999 & 1.000 & 0.999 & 0.999 & 1.000 & \nodata \\
3 & 0.997 & 0.997 & 1.000 & 0.993 & 0.993 & 1.000 & 1.000 \\
5 & 0.991 & 0.991 & 1.000 & 0.982 & 0.982 & 1.000 & 1.000 \\
7 & 0.983 & 0.983 & 1.000 & 0.965 & 0.966 & 1.001 & 0.944 \\
9 & 0.972 & 0.972 & 1.000 & 0.944 & 0.945 & 1.001 & 0.964 \\
11 & 0.959 & 0.959 & 1.000 & 0.917 & 0.919 & 1.002 & 0.951 \\
13 & 0.943 & 0.943 & 1.000 & 0.887 & 0.889 & 1.002 & 0.963 \\
15 & 0.925 & 0.925 & 1.000 & 0.852 & 0.854 & 1.003 & 0.971 \\
17 & 0.905 & 0.905 & 1.000 & 0.813 & 0.817 & 1.004 & 0.954 \\
19 & 0.883 & 0.883 & 1.000 & 0.772 & 0.776 & 1.005 & 0.962 \\
21 & 0.858 & 0.859 & 1.000 & 0.728 & 0.733 & 1.006 & 0.965 \\
23 & 0.832 & 0.833 & 1.001 & 0.682 & 0.688 & 1.008 & 0.962 \\
25 & 0.805 & 0.805 & 1.000 & 0.635 & 0.641 & 1.009 & 0.960 \\
27 & 0.776 & 0.776 & 1.000 & 0.587 & 0.594 & 1.011 & 0.958 \\
29 & 0.746 & 0.746 & 1.001 & 0.539 & 0.546 & 1.013 & 0.960 \\
31 & 0.714 & 0.715 & 1.001 & 0.492 & 0.499 & 1.015 & 0.966 \\
33 & 0.682 & 0.683 & 1.000 & 0.445 & 0.452 & 1.017 & 0.967 \\
35 & 0.650 & 0.650 & 1.000 & 0.399 & 0.407 & 1.019 & 0.959 \\
37 & 0.617 & 0.617 & 1.000 & 0.355 & 0.363 & 1.022 & 0.960 \\
39 & 0.583 & 0.583 & 1.000 & 0.313 & 0.321 & 1.025 & 0.959 \\
41 & 0.550 & 0.550 & 1.000 & 0.274 & 0.281 & 1.028 & 0.964 \\
43 & 0.517 & 0.517 & 0.999 & 0.237 & 0.244 & 1.032 & 0.963 \\
45 & 0.484 & 0.483 & 0.999 & 0.202 & 0.210 & 1.037 & 0.953 \\
47 & 0.452 & 0.451 & 0.998 & 0.171 & 0.178 & 1.041 & 0.956 \\
49 & 0.420 & 0.419 & 0.997 & 0.143 & 0.150 & 1.046 & 0.953 \\
51 & 0.389 & 0.387 & 0.996 & 0.118 & 0.124 & 1.051 & 0.954 \\
53 & 0.359 & 0.357 & 0.994 & 0.095 & 0.101 & 1.056 & 0.950 \\
55 & 0.330 & 0.327 & 0.992 & 0.076 & 0.081 & 1.062 & 0.950 \\
57 & 0.302 & 0.299 & 0.990 & 0.059 & 0.064 & 1.069 & 0.944 \\
59 & 0.275 & 0.272 & 0.988 & 0.045 & 0.049 & 1.076 & 0.947 \\
61 & 0.250 & 0.246 & 0.985 & \nodata & \nodata & \nodata & \nodata \\
63 & 0.225 & 0.221 & 0.981 & \nodata & \nodata & \nodata & \nodata \\
65 & 0.203 & 0.198 & 0.977 & \nodata & \nodata & \nodata & \nodata \\
67 & 0.181 & 0.176 & 0.972 & \nodata & \nodata & \nodata & \nodata \\
69 & 0.161 & 0.156 & 0.966 & \nodata & \nodata & \nodata & \nodata \\
71 & 0.143 & 0.137 & 0.959 & \nodata & \nodata & \nodata & \nodata \\
73 & 0.125 & 0.119 & 0.951 & \nodata & \nodata & \nodata & \nodata \\
75 & 0.110 & 0.103 & 0.942 & \nodata & \nodata & \nodata & \nodata \\
77 & 0.095 & 0.089 & 0.931 & \nodata & \nodata & \nodata & \nodata \\
79 & 0.082 & 0.075 & 0.919 & \nodata & \nodata & \nodata & \nodata \\
\hline 
\end{tabular}
\\
\tablecomments{
 Column 1: Distance from the pointing centre in arcminutes. Column 2, 3: annular averaged {\tt plumber} beam and {\tt katbeam}, respectively, for SPW2. Column 4: ratio between annular averaged {\tt katbeam} and {\tt plumber} beam for SPW2. Column 5, 6: annular averaged {\tt plumber} beam and {\tt katbeam}, respectively, for SPW9. Column 7: ratio of annular averaged {\tt katbeam} and {\tt plumber} beam for SPW9. Column 8: ratio of spectral indices corresponding to the two beam models.
}
\end{center}
\end{table*}
\clearpage


\section{MALS SPW9 source counts }
\label{sec:listsrccnt}

Here we present radio source counts derived from the SPW9 catalog.

\begin{table*}
\begin{center}
\caption{MALS 1.4\,GHz radio source counts.}
\label{tab:srccnt}
\begin{tabular}{lccccc}
\hline
$\Delta$ S            & $\log$[S (Jy)]      & N  & Area & $\frac{dN}{dS}$\,S$^{2.5}$ \\
(mJy)               &         &    & (deg$^2$) &          (Jy$^{1.5}$\,sr$^{-1}$)\\
\hline
0.16-0.25 & $-$3.7 & 18315 & 108.2 & 3.3708$^{+0.0249}_{-0.0249}$ \\
0.25-0.40 & $-$3.5 & 36764 & 331.8 & 4.4030$^{+0.0230}_{-0.0230}$ \\
0.40-0.63 & $-$3.3 & 39717 & 552.3 & 5.7011$^{+0.0286}_{-0.0286}$ \\
0.63-1.00 & $-$3.1 & 32942 & 751.5 & 6.9341$^{+0.0382}_{-0.0382}$ \\
1.00-1.58 & $-$2.9 & 26201 & 930.6 & 8.8867$^{+0.0549}_{-0.0549}$ \\
1.58-2.51 & $-$2.7 & 20713 & 1038.0 & 12.5661$^{+0.0873}_{-0.0873}$ \\
2.51-3.98 & $-$2.5 & 15548 & 1057.7 & 18.471$^{+0.148}_{-0.148}$ \\
3.98-6.31 & $-$2.3 & 11666 & 1063.0 & 27.516$^{+0.255}_{-0.255}$ \\
6.31-10.00 & $-$2.1 & 8726 & 1064.6 & 41.000$^{+0.439}_{-0.439}$ \\
10.00-15.85 & $-$1.9 & 6330 & 1065.3 & 59.306$^{+0.745}_{-0.745}$ \\
15.85-25.12 & $-$1.7 & 4524 & 1065.7 & 84.54$^{+1.26}_{-1.26}$ \\
25.12-39.81 & $-$1.5 & 3194 & 1065.9 & 119.07$^{+2.11}_{-2.11}$ \\
39.81-63.10 & $-$1.3 & 2133 & 1066.0 & 158.64$^{+3.44}_{-3.44}$ \\
63.10-100.00 & $-$1.1 & 1383 & 1066.0 & 205.24$^{+5.52}_{-5.52}$\\
100.00-158.49 & $-$0.9 & 887 & 1066.0 & 262.6$^{+8.8}_{-8.8}$\\
158.49-251.19 & $-$0.7 & 535 & 1066.0 & 320$^{+14}_{-14}$ \\
251.19-398.11 & $-$0.5 & 281 & 1066.0 & 330$^{+20}_{-20}$\\
398.11-630.96 & $-$0.3 & 166 & 1066.0 & 390$^{+30}_{-30}$\\
630.96-1000.00 & $-$0.1 & 79 & 1066.0 & 370$^{+42}_{-42}$\\
1000.00-1584.89 & 0.1 & 38 & 1066.0 & 360$^{+68}_{-58}$ \\
1584.89-2511.89 & 0.3 & 11 & 1066.0 & 210$^{+83}_{-61}$\\
2511.89-3981.07 & 0.5 & 7 & 1066.0 & 300$^{+100}_{-100}$ \\
3981.07-6309.57 & 0.7 & 5 & 1066.0 & 400$^{+300}_{-200}$\\
6309.57-10000.00 & 0.9 & 3 & 1066.0 & 400$^{+400}_{-200}$ \\
\hline 
\end{tabular}
\\
\tablecomments{
Column 1: Flux density range of each bin in mJy, Column 2: $\log$ of mean flux density of each bin in Jy.  Column 3: Number of sources detected in each bin. Column 4: Total survey area in deg$^2$ in which a source with the mean flux density of the bin can be detected at more than 5$\sigma$ threshold. Column 5: Eucledian normalized differential source counts.
}
\end{center}
\end{table*}
\clearpage

\section{List of USS as candidate HzRGs }
\label{sec:listuss}

Here we present the list of USS identified as promising HzRG candidates.  Note that the redshifts of 8 sources, all at $z<1$ and Flag = 4, are available from the literature.  For completeness, these are included in the table. The candidates with Flag 1 and 2 constitute the most promising candidates.

\begin{deluxetable*}{lcccccccc}
\tabcolsep=5pt
\tablecolumns{9}
\tablewidth{0pc}
\tablecaption{List USS as candidate HzRGs.  \label{tab:usslist}}
\tablehead{
        \colhead{ID (J2000)}  & \colhead{S\_Code}  & \colhead{Flux density} & \colhead{$\alpha_{SPW2}^{TGSSADR1}\pm \sigma_\alpha$} & \colhead{PS1} &  \colhead{WISE} & \colhead{Flag} & \colhead{$z$} & \colhead{Ref. $z$} \\
        \colhead{ } & \colhead{ } & \colhead{(mJy)} & \colhead{ } & \colhead{ }& \colhead{ } & \colhead{ } &  \colhead{ } &  \colhead{ } \\
        \colhead{(1)} & \colhead{(2)} & \colhead{(3)} & \colhead{(4)} & \colhead{(5)} & \colhead{(6)} & \colhead{(7)} & \colhead{(8)} & \colhead{(9)} \\
       }    
\startdata
J122204.80$-$134745.8 & S & 11.1 & -1.57$\pm$0.05 & False & False & 2 &  \nodata & \nodata \\
J231750.42$-$410627.7 & S & 4.29 & -1.53$\pm$0.07 & \nodata & True & 1 & \nodata & \nodata \\
J155624.36$+$031906.4 & M & 8.91 & -1.34$\pm$0.06 & False & False & 2 & \nodata & \nodata\\
J233125.38$+$141819.2 & S & 2.64 & -1.36$\pm$0.09 & False & False & 2 & \nodata & \nodata\\
J160615.40$-$153738.5 & S & 10.6 & -1.38$\pm$0.06 & False & False & 2 & \nodata & \nodata\\
J160530.70$-$142315.0 & S & 4.70 & -1.34$\pm$0.09 & False & True & 1 & \nodata & \nodata\\
J064045.47$-$320104.1 & S & 4.04 & -1.48$\pm$0.08 & \nodata & False & 2 & \nodata & \nodata\\
J122705.51$+$041654.3 & S & 31.9 & -1.38$\pm$0.07 & True & True & 4 & 0.305 & \citet{SDSS13}\\
J122107.70$+$031134.6 & S & 5.16 & -1.48$\pm$0.10 & False & False & 2 & \nodata & \nodata\\
J130812.61$-$010043.9 & S & 4.47 & -1.31$\pm$0.17 & True & False & 3 & \nodata & \nodata\\
J025329.44$+$033818.6 & S & 2.32 & -1.47$\pm$0.11 & True & True & 4 & 0.181 & \citet{SDSS12}\\
J035059.81$-$253629.4 & S & 9.09 & -1.32$\pm$0.06 & True & True & 4 & \nodata & \nodata\\
J012401.72$+$143719.8 & S & 1.56 & -1.67$\pm$0.12 & True & True & 4 & 0.578 & \citet{SDSS13}\\
J034643.37$-$252706.5 & S & 5.04 & -1.35$\pm$0.13 & False & False & 2 & \nodata & \nodata\\
J112228.54$-$142428.0 & S & 5.63 & -1.50$\pm$0.07 & False & False & 2 & \nodata & \nodata\\
J094313.12$-$023829.3 & S & 3.03 & -1.52$\pm$0.15 & True & True & 4 & \nodata & \nodata\\
J054229.19$-$264308.0 & S & 1.51 & -1.59$\pm$0.11 & False & False & 2 & \nodata & \nodata\\
J054015.04$-$264919.2 & S & 14.9 & -1.58$\pm$0.06 & True & True & 4 & \nodata & \nodata\\
J060859.26$-$223001.5 & S & 6.15 & -1.34$\pm$0.07 & False & False & 2 & \nodata & \nodata\\
J114145.63$-$143244.4 & S & 4.34 & -1.44$\pm$0.08 & False & False & 2 & \nodata & \nodata\\
J091348.61$-$161207.6 & S & 7.57 & -1.44$\pm$0.09 & False & False & 2 & \nodata & \nodata\\
J091257.09$-$165555.6 & S & 20.7 & -1.35$\pm$0.06 & True & True & 4 & 0.910 & \citet{Bornancini07}\\
J020224.56$-$484054.5 & S & 2.00 & -1.80$\pm$0.15 & \nodata & False & 2 & \nodata & \nodata\\
J020058.75$-$482928.3 & S & 1.78 & -1.65$\pm$0.15 & \nodata & True & 1 & \nodata & \nodata\\
J015732.10$-$485905.0 & S & 4.94 & -1.30$\pm$0.06 & \nodata & False & 2 & \nodata & \nodata\\
J155517.93$+$045342.9 & S & 4.98 & -1.44$\pm$0.08 & False & False & 2 & \nodata & \nodata\\
J090130.16$-$302809.6 & S & 3.85 & -1.47$\pm$0.08 & \nodata & False & 2 & \nodata & \nodata\\
J223620.38$-$183334.0 & S & 3.03 & -1.41$\pm$0.13 & True & True & 4 & \nodata & \nodata\\
J223440.87$-$175225.4 & S & 5.06 & -1.60$\pm$0.08 & False & True & 1 & \nodata & \nodata\\
J064045.36$-$320103.3 & S & 4.63 & -1.41$\pm$0.11 & \nodata & False & 2 & \nodata & \nodata\\
J210148.05$-$183455.4 & S & 3.96 & -1.46$\pm$0.12 & False & False & 2 & \nodata & \nodata\\
J053234.59$-$121819.7 & S & 3.37 & -1.47$\pm$0.14 & False & False & 2 & \nodata & \nodata\\
J052945.54$-$110554.1 & S & 14.9 & -1.31$\pm$0.07 & False & False & 2 & \nodata & \nodata\\
J055435.57$-$072925.2 & S & 4.24 & -1.33$\pm$0.12 & False & False & 2 & \nodata & \nodata\\
J162132.24$+$134823.5 & S & 2.90 & -1.42$\pm$0.17 & False & False & 2 & \nodata & \nodata\\
J100908.96$-$252513.1 & S & 4.15 & -1.32$\pm$0.11 & True & False & 3 & \nodata & \nodata\\
J075420.24$+$262834.1 & S & 16.3 & -1.46$\pm$0.06 & False & False & 2 & \nodata & \nodata\\
J100908.94$-$252513.2 & S & 3.82 & -1.36$\pm$0.09 & True & False & 3 & \nodata & \nodata\\
J235909.81$-$143353.1 & S & 6.84 & -1.30$\pm$0.09 & False & True & 1 & \nodata & \nodata\\
J080424.54$+$010309.5 & S & 2.84 & -1.63$\pm$0.08 & True & True & 4 & \nodata & \nodata\\
J102041.26$-$083911.4 & S & 3.23 & -1.80$\pm$0.14 & False & False & 2 & \nodata & \nodata\\
J101152.96$-$284020.1 & S & 24.0 & -1.52$\pm$0.06 & False & False & 2 & \nodata & \nodata\\
\enddata
\tablecomments{
\small{
    Column\,1: Source ID (J2000). Column\,2: S\_Code (defined in Table~\ref{tab:cat_cols}) from {\tt PyBDSF} catalog. Column\,3: Total integrated flux density from SPW2.  Column\,4: Spectral index and it's associated uncertainty based on TGSS ADR1 (147 MHz) and SPW2 measurements. Column\,5: Counterpart in PS1? Column\,6: Counterpart in WISE? Column\,7: Boolean flag based on PS1 and WISE matching; 1- detected in WISE but not in PS1, 2- not detected in both, 3- detected in PS1 but not in WISE, and 4- detected in both. Column\,8: Redshift measurements from NED. Column\,9: Reference for redshift measurement.
    }}
\end{deluxetable*}
\clearpage


\section{List of variable radio sources.}
\label{sec:listuss}

Here we present the list of variable radio sources.  

\startlongtable
\begin{deluxetable}{lccc}
\tabcolsep=5pt
\tablecolumns{7}
\tablewidth{0pc}
\tablecaption{List of variable radio sources providing variability strength ($V_s$), modulation index (m) and fractional variability ($f_{var}$). \label{tab:var}}
\tablehead{
        \colhead{ID (J2000)}  & \colhead{$V_s$}  & \colhead{m} & \colhead{$f_{var}$} \\
       }    
\startdata
J035422.58-252049.9  &  5.72 & 0.26 & 1.35 \\
J035130.75-243919.2  &  43.40 & 0.87 & 2.81 \\
J034906.59-252914.4  &  5.27 & 0.57 & 1.67 \\
J035318.51-024411.8  &  -6.88 & 0.29 & 0.77 \\
J035023.88-035626.3  &  6.96 & 0.34 & 1.50 \\
J034720.50-035418.4  &  6.69 & 0.20 & 1.24 \\
J162849.11-131055.4  &  5.21 & 0.30 & 1.42 \\
J162600.57-130147.1  &  4.24 & 0.30 & 1.41 \\
J163145.15-142111.2  &  4.90 & 0.19 & 1.23 \\
J162952.03-145721.4  &  5.69 & 0.17 & 1.19 \\
J162835.24-144855.1  &  4.64 & 0.37 & 1.56 \\
J162659.22-142652.8  &  13.80 & 0.34 & 1.48 \\
J122446.53+042001.1  &  6.23 & 0.21 & 1.27 \\
J122355.50+043104.0  &  6.72 & 0.39 & 1.60 \\
J122335.19+034204.4  &  7.24 & 0.30 & 1.38 \\
J122215.87+040029.0  &  4.83 & 0.27 & 1.35 \\
J122208.73+043503.4  &  5.49 & 0.19 & 1.24 \\
J122151.50+035731.0  &  6.09 & 0.39 & 1.63 \\
J122144.22+040148.8  &  6.85 & 0.24 & 1.31 \\
J122136.80+044403.5  &  5.39 & 0.20 & 1.25 \\
J122032.45+035245.8  &  6.03 & 0.18 & 1.21 \\
J122030.70+034440.1  &  6.11 & 0.30 & 1.39 \\
J121954.39+035230.5  &  4.93 & 0.14 & 1.16 \\
J121938.61+041820.4  &  6.31 & 0.30 & 1.40 \\
J122147.52-135157.6  &  5.69 & 0.25 & 1.33 \\
J121711.61-064901.9  &  5.14 & 0.23 & 1.29 \\
J025554.89+030205.6  &  9.38 & 0.36 & 1.54 \\
J112703.30-150617.9  &  7.06 & 0.45 & 1.74 \\
J112339.98-154201.2  &  4.34 & 0.13 & 1.14 \\
J112327.63-145234.0  &  6.31 & 0.39 & 1.60 \\
J112317.05-143931.5  &  16.10 & 0.37 & 1.56 \\
J112108.82-151331.4  &  7.68 & 0.44 & 1.72 \\
J161136.88-143653.2  &  4.09 & 0.19 & 1.22 \\
J160954.82-144656.5  &  -8.19 & 0.56 & 0.64 \\
J160828.50-141202.5  &  -8.63 & 0.62 & 0.61 \\
J162005.23+135218.9  &  4.92 & 0.34 & 1.43 \\
J161924.86+134000.0  &  4.32 & 0.58 & 1.68 \\
J161910.54+140759.9  &  7.70 & 0.58 & 1.92 \\
J161731.29+140821.5  &  6.51 & 0.21 & 1.26 \\
J161604.31+135053.2  &  6.86 & 0.25 & 1.33 \\
J161546.95+144615.1  &  4.29 & 0.18 & 1.20 \\
J161506.19+141645.3  &  6.38 & 0.40 & 1.59 \\
J160957.89+104415.2  &  5.45 & 0.22 & 1.28 \\
J160926.59+100855.6  &  4.61 & 0.18 & 1.22 \\
J160814.52+103059.4  &  12.50 & 0.68 & 2.14 \\
J160748.26+105447.5  &  7.50 & 0.40 & 1.61 \\
J160558.62+102446.4  &  11.20 & 0.62 & 2.10 \\
J125901.14+281140.4  &  4.26 & 0.17 & 1.20 \\
J125810.17+281945.6  &  4.57 & 0.27 & 1.36 \\
J125757.46+280339.9  &  5.10 & 0.53 & 1.67 \\
J130650.66-014642.7  &  4.16 & 0.23 & 1.30 \\
J130601.66-013529.9  &  4.49 & 0.19 & 1.23 \\
J130524.82-021259.9  &  4.24 & 0.21 & 1.26 \\
J130330.13-015411.4  &  -6.87 & 0.78 & 0.55 \\
J130007.99-293609.9  &  8.66 & 0.48 & 1.90 \\
J125752.17-295324.1  &  4.22 & 0.14 & 1.16 \\
J233555.44-380329.7  &  7.75 & 0.46 & 1.80 \\
J233211.62-150026.5  &  20.00 & 0.48 & 1.84 \\
J233052.30-135225.0  &  4.17 & 0.26 & 1.34 \\
J233017.67-144727.0  &  4.66 & 0.29 & 1.38 \\
J232950.87-141535.7  &  6.96 & 0.21 & 1.26 \\
J232947.41-140952.7  &  4.21 & 0.18 & 1.22 \\
J083626.87-205427.1  &  5.94 & 0.24 & 1.28 \\
J083606.30-202427.5  &  14.00 & 0.60 & 1.96 \\
J084729.82-112155.4  &  4.82 & 0.72 & 1.94 \\
J084357.98-113956.6  &  3.93 & 0.14 & 1.15 \\
J084338.43-114616.3  &  4.31 & 0.13 & 1.15 \\
J082732.17-104156.5  &  7.43 & 0.45 & 1.71 \\
J082725.26-102923.0  &  4.95 & 0.19 & 1.23 \\
J082436.97-104103.7  &  5.94 & 0.21 & 1.26 \\
J160151.45+024809.3  &  6.53 & 0.23 & 1.28 \\
J160015.32+031707.8  &  15.00 & 0.46 & 1.77 \\
J155943.47+030502.3  &  4.28 & 0.18 & 1.21 \\
J155829.41+032250.1  &  4.58 & 0.16 & 1.18 \\
J155637.90+030244.5  &  4.14 & 0.12 & 1.14 \\
J152145.66-294538.8  &  5.65 & 0.16 & 1.19 \\
J152132.16-291155.1  &  16.00 & 0.44 & 1.71 \\
J152027.82-293714.3  &  5.53 & 0.19 & 1.23 \\
J125356.39-042310.2  &  10.30 & 0.32 & 1.44 \\
J125258.56-034929.9  &  7.48 & 0.37 & 1.54 \\
J125258.48-033637.9  &  6.72 & 0.21 & 1.26 \\
J125229.92-040356.2  &  4.32 & 0.13 & 1.14 \\
J125225.24-034828.1  &  5.12 & 0.35 & 1.51 \\
J125220.61-042924.6  &  4.41 & 0.23 & 1.28 \\
J125209.91-040843.5  &  4.88 & 0.31 & 1.38 \\
\enddata
\end{deluxetable}
\clearpage


 \end{document}